\documentclass[
 reprint,
groupedaddress,
 amsmath,amssymb,
 aps,
 rmp,
floatfix,
nolongbibliography
]{revtex4-2}

\usepackage{comment}
\usepackage{graphicx,amsmath}
\usepackage{dcolumn}
\usepackage{bm}
\usepackage[skins,theorems]{tcolorbox}
\usepackage{tablefootnote}
\usepackage{slashed}
\usepackage{xcolor}
\usepackage{verbatim}
\usepackage{threeparttable,lipsum}
\usepackage{mathtools}

\usepackage{titlesec}
\titleformat*{\section}{\sffamily\bfseries\boldmath}

\usepackage[pagebackref]{hyperref}
\hypersetup{
    colorlinks,
    linktoc= {page},
    linkcolor={red!90!black},
    citecolor={blue!70!black},
    urlcolor={blue!90!black}
}
\usepackage[all]{hypcap}
\usepackage{aasmacros}
\usepackage{siunitx}
\usepackage{csvsimple}

\usepackage[T1]{fontenc}  

\usepackage{enumitem}

    \usepackage{pdfpages} 
    \usepackage{pgffor} 
    
    \makeatletter
    \AtBeginDocument{\let\LS@rot\@undefined}
    \makeatother
    
    \def\supplementfilename{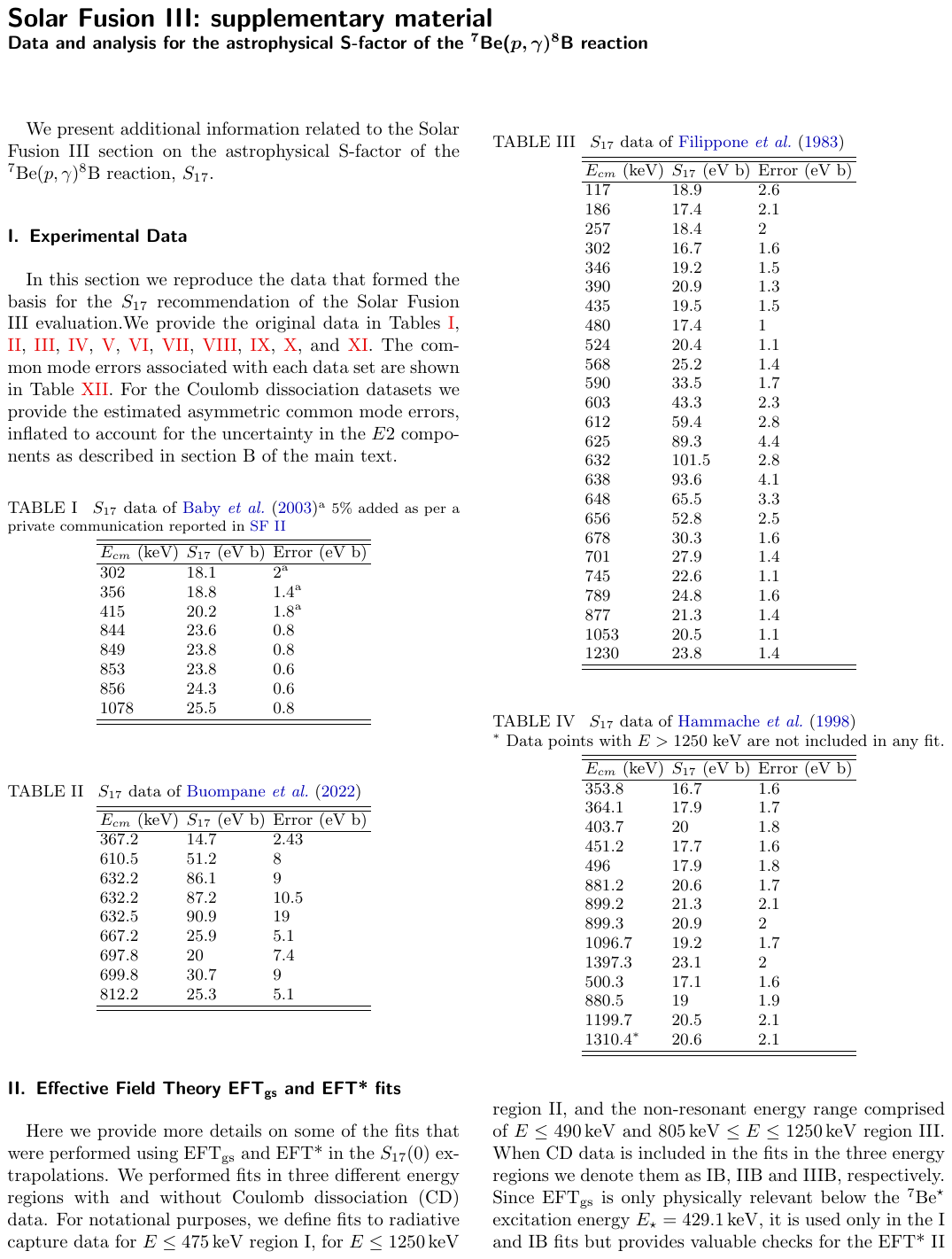}
    
    \pdfximage{\supplementfilename}
    \def\numbersupplementpages{\the\pdflastximagepages}
    
    \newif\ifarXiv
    \arXivtrue 

\DeclareSIUnit\barn{b}

\defcitealias{1998RvMP...70.1265A}{SF~I}
\defcitealias{2011RvMP...83..195A}{SF~II}

\tcbset{highlight math style={enhanced,
 colframe=cyan!10!gray,colback=gray!5!white,arc=4pt,boxrule=1pt,
  drop fuzzy shadow}}
  
\makeatletter
\g@addto@macro\bfseries{\boldmath}
\makeatother

  {\empheq[box=\tcbhighmath]{align}}
  {\endempheq}

\newcommand{\etal}{{\em et al.}}

\begin{document}
\title{Solar fusion III: New data and theory for hydrogen-burning stars.}


\author{ B. Acharya }
    \affiliation{Oak Ridge National Laboratory, Oak Ridge, TN 37831, USA}

\author{ M. Aliotta }
    \affiliation{ SUPA, School of Physics and Astronomy, University of Edinburgh, Edinburgh EH9 3FD, United Kingdom }

\author{ A. B. Balantekin }
     \affiliation{Department of Physics, University of Wisconsin-Madison, Madison WI 53706, USA }

\author{ D. Bemmerer }
    \affiliation{ Helmholtz-Zentrum Dresden-Rossendorf, 01328 Dresden, Germany }

\author{ C. A. Bertulani }
    \affiliation{Department of Physics and Astronomy,  Texas A\&M University-Commerce, Commerce, TX 75429-3011, USA }

\author{ A. Best }
    \affiliation{ University of Napoli Federico II, 80126 Napoli, Italy  }
    \affiliation{ Istituto Nazionale di Fisica Nucleare - Napoli, 80126 Napoli, Italy }

\author{ C. R. Brune }
    \affiliation{ Edwards Accelerator Laboratory, Department of Physics and Astronomy, Ohio University, Athens, Ohio 45701, USA }

\author{ R. Buompane }
    \affiliation{ Dipartimento di Matematica e Fisica, Universit\`a degli Studi della Campania ``Luigi Vanvitelli'', 81100 Caserta, Italy }
    \affiliation{Istituto Nazionale di Fisica Nucleare - Napoli, 80126 Napoli, Italy }

\author{ F. Cavanna }
    \affiliation{ Istituto Nazionale di Fisica Nucleare - Torino, 10125 Torino, Italy}

\author{ J. W. Chen }
    \affiliation{Department of Physics and Center for Theoretical Physics, National Taiwan University, Taipei 10617, Taiwan}
    \affiliation{Physics Division, National Center for Theoretical Sciences, Taipei 10617, Taiwan}

\author{ J. Colgan }
    \affiliation{ Los Alamos National Laboratory, Los Alamos, NM 87545, USA }

\author{ A. Czarnecki }
    \affiliation{Department of Physics, University of Alberta, Edmonton, Alberta T6G 2E1, Canada}

\author{ B. Davids }
	\affiliation{ TRIUMF, Vancouver, BC, V6T 2A3, Canada }
 	\affiliation{Department of Physics, Simon Fraser University, Burnaby, BC, V5A 1S6, Canada }

\author{ R. J. deBoer}
\affiliation{Department of Physics and Astronomy and the The Joint Institute for Nuclear Astrophysics, University of Notre Dame, Notre Dame, IN 46556, USA}

\author{ F. Delahaye }
    \affiliation{ Observatoire de Paris - PSL - Sorbonne Université, 5 place Jules Janssen, 92195 Meudon cedex, France }

\author{ R. Depalo }
    \affiliation{ Universit\`a degli Studi di Milano, 20133 Milano, Italy }
    \affiliation{ Istituto Nazionale di Fisica Nucleare - Milano, 20133 Milano, Italy }

\author{ A. Garc\'{i}a }
    \affiliation{ Department of Physics and Center for Experimental Nuclear Physics and Astrophysics, University of Washington, Seattle WA 98195, USA}

\author{ M. Gatu Johnson }
    \affiliation{Massachusetts Institute of Technology, Cambridge, MA 02139, USA}

\author{ D. Gazit }
    \affiliation{Racah Institute of Physics, The Hebrew University, The Edmond J. Safra Campus, Givat Ram, Jerusalem 9190401, Israel}    

\author{ L. Gialanella }
    \affiliation{ Dipartimento di Matematica e Fisica, Universit\`a degli Studi della Campania ``Luigi Vanvitelli'', 81100 Caserta, Italy }
\affiliation{Istituto Nazionale di Fisica Nucleare - Napoli, 80126 Napoli, Italy }

\author{ U. Greife }
    \affiliation{ Colorado School of Mines, Golden CO 80401,  USA  }

\author{ D. Guffanti}
    \affiliation{ Dipartimento di Fisica, Universit\`a degli Studi e INFN Milano-Bicocca, 20126 Milano, Italy}

\author{ A. Guglielmetti }
    \affiliation{ Universit\`a degli Studi di Milano, 20133 Milano, Italy }
    \affiliation{ Istituto Nazionale di Fisica Nucleare - Milano, 20133 Milano, Italy }

\author{ K. Hambleton }
    \affiliation{ Department of Physics, University of California, Berkeley, CA 94720, USA }

\author{ W. C. Haxton }
    \affiliation{ Department of Physics, University of California, Berkeley, CA 94720 }
    \affiliation{ Lawrence Berkeley National Laboratory, Berkeley, CA 94720, USA}

\author{ Y. Herrera }
    \affiliation{ Institute of Space Sciences, 08193 Cerdanyola del Vall\`es, Barcelona, Spain }
    \affiliation{ Institut d'Estudis Espacials de Catalunya, 08034 Barcelona, Spain }

\author{ M. Huang }
    \affiliation{ Department of Physics \& Astronomy, Iowa State University, Ames, IA 50011, USA }
    \affiliation{ Lawrence Livermore National Laboratory, P.O. Box 808, L-414, Livermore, CA 94551, USA }

\author{ C. Iliadis }
    \affiliation{Department of Physics \& Astronomy, University of North Carolina at Chapel Hill, NC 27599-3255, USA}
    \affiliation{Triangle Universities Nuclear Laboratory (TUNL), Duke University, Durham, North Carolina 27708, USA}

\author{ K. Kravvaris }
    \affiliation{Lawrence Livermore National Laboratory, P.O. Box 808, L-414, Livermore, CA 94551, USA}

\author{ M. La Cognata }
    \affiliation{ Istituto Nazionale di Fisica Nucleare - Laboratori Nazionali del Sud, 95123 Catania, Italy }

\author{ K. Langanke }
    \affiliation{GSI Helmholtzzentrum fuer Schwerionenforschung, Darmstadt, Germany }
    \affiliation{Institut fuer Kernphysik, Technical University Darmstadt, Darmstadt, Germany}

\author{ L. E. Marcucci }
    \affiliation{ Department of Physics ``E. Fermi'', University of Pisa, 56127 Pisa, Italy }
    \affiliation{ Istituto Nazionale di Fisica Nucleare - Pisa, 56127 Pisa, Italy }

\author{ T. Nagayama }
    \affiliation{ Sandia National Laboratories, Albuquerque, New Mexico 87123, USA }

\author{ K. M. Nollett }
    \affiliation{ Department of Physics, San Diego State University, San Diego, CA 92182, USA }

\author{ D. Odell }
    \affiliation{ Department of Physics \& Astronomy, Ohio University, Athens, Ohio 45701, USA }

\author{ G. D. Orebi Gann }
    \affiliation{University of California Berkeley, Berkeley CA 94720, USA }
    \affiliation{Lawrence Berkeley National Laboratory, Berkeley CA 94720, USA}

\author{ D. Piatti }
    \affiliation{ Università degli Studi di Padova, Via F. Marzolo 8, 35131 Padova, Italy}
    \affiliation{ INFN, Sezione di Padova, Via F. Marzolo 8, 35131 Padova, Italy}

\author{ M. Pinsonneault }
    \affiliation{ Department of Astronomy, The Ohio State University, 140 W. 18th Avenue, Columbus, OH 43210, USA} 
    \affiliation{Center for Cosmology and AstroParticle Physics, The Ohio State University, 191 W. Woodruff Avenue, Columbus, OH 43210, USA }

\author{ L. Platter}
    \affiliation{Department of Physics and Astronomy, University of Tennessee, Knoxville, TN 37996, USA}
    \affiliation{Physics Division, Oak Ridge National Laboratory, Oak Ridge, TN 37831, USA}

\author{ R. G. H. Robertson }
    \affiliation{ Department of Physics and Center for Experimental Nuclear Physics and Astrophysics, University of Washington, Seattle WA 98195, USA}

\author{ G. Rupak }
    \affiliation{ Department of Physics \& Astronomy and HPC2 Center for Computational Sciences, Mississippi State University, Mississippi State, MS 39762, USA }

\author{ A. Serenelli }
    \affiliation{ Institute of Space Sciences, 08193 Cerdanyola del Vall\`es, Barcelona, Spain }
    \affiliation{ Institut d'Estudis Espacials de Catalunya, 08034 Barcelona, Spain }

\author{ M. Sferrazza }
    \affiliation{ Universit\'e Libre de Bruxelles }

\author{ T. Sz\"ucs }
    \affiliation{HUN-REN Institute for Nuclear Research (HUN-REN ATOMKI), 4026 Debrecen, Hungary}
    
\author{ X. Tang }
    \affiliation{ IMP Chinese Academy of Science }

\author{ A. Tumino }
    \affiliation{ Dipartimento di Ingegneria e Architettura, Università degli Studi di Enna ``Kore'', Enna, Italy}
    \affiliation{ Istituto Nazionale di Fisica Nucleare - Laboratori Nazionali del Sud, 95123 Catania, Italy }

\author{ F. L. Villante }
    \affiliation{ Dipartimento di Scienze Fisiche e Chimiche, Universit\`a dell’Aquila, 67100 L’Aquila, Italy}
    \affiliation{Istituto Nazionale di Fisica Nucleare - Laboratori Nazionali del Gran Sasso, 67100 Assergi (AQ), Italy }

\author{ A. Walker-Loud }
    \affiliation{Nuclear Science Division, Lawrence Berkeley National Laboratory, Berkeley, CA 94720, USA }

\author{ X. Zhang }
    \affiliation{ Facility for Rare Isotope Beams, Michigan State University, East Lansing, MI 48824, USA }

\author{ K. Zuber }
    \affiliation{Institute for Nuclear and Particle Physics, Technical University of Dresden, 01062 Dresden, Germany }


\begin{abstract}
    ~In stars that lie on the main sequence in the Hertzsprung Russel diagram, like our Sun, hydrogen is fused to helium in a number of nuclear reaction chains and series, such as the proton-proton chain and the carbon-nitrogen-oxygen cycles. Precisely determined thermonuclear  rates of these reactions lie at the foundation of the standard solar model. 
    
    ~This review, the third decadal evaluation of the nuclear physics of hydrogen-burning stars, is motivated by the great advances made in recent years by solar neutrino observatories, putting experimental knowledge of the pp chain neutrino fluxes in the few-percent precision range. The basis of the review is a one-week community meeting held in  July 2022 in Berkeley, California, and many subsequent digital meetings and exchanges. 
    
    ~The relevant reactions of solar and stellar hydrogen burning are reviewed here, from both theoretical and experimental perspectives. Recommendations for the state of the art of the astrophysical S-factor and its uncertainty are formulated for each of them. 
    
    ~Several other topics of paramount importance for the solar model are reviewed, as well: recent and future neutrino experiments, electron screening, radiative opacities, and current and upcoming experimental facilities. In addition to reaction-specific recommendations, also general recommendations are formed.
\end{abstract}

\maketitle
\newpage
\clearpage

\tableofcontents

\section{Introduction}
\label{sec:Introduction}

The present review summarizes the state of our understanding, in the third decade of the 21$^{\rm st}$ century, of the nuclear reactions and decays taking place in solar and stellar hydrogen burning. It also addresses related issues, including solar neutrino detection, radiative opacities, electron screening of nuclear reactions, and the status of the current facilities for measuring cross sections and opacities. 
As was the case for two previous decadal reviews \cite{1998RvMP...70.1265A, 2011RvMP...83..195A}, the present review summarizes the progress made over the past decade in advancing the nuclear physics of main-sequence stars and makes recommendations for future work.

\subsection{Purpose}
\label{sec:Intro:Purpose}

Ray Davis's measurements of the flux of solar neutrinos \cite{1968PhRvL..20.1205D,2003RvMP...75..985D} showed a significant deficit with respect to the predictions of the standard solar model (SSM) \cite{1966PhRvL..17..398B}.  This discrepancy became known as the "solar neutrino problem." An intense debate ensued about possible explanations for this discrepancy, summarized in the entertaining 
review by \citet{1988Natur.334..487B}.
Suggestions included plausible flaws in the SSM that might produce a cooler solar core (e.g. \citealt{1981A&A....96....1S}; \citealt{1996PhRvL..77.4286C}), new particle physics such as neutrino oscillations (\citealt{1969PhLB...28..493G}; \citealt{1978PhRvD..17.2369W}), changes in the nuclear
physics governing He synthesis, such as an unidentified low-energy resonance in the $^3$He+$^3$He reaction affecting extrapolations of laboratory cross sections to solar energies \cite{1972Natur.238...24F,1972PhLB...40..602F}, or the presence of dark matter particles modifying energy transport in the solar core \citep{1985ApJ...294..663S}.  By the mid-1990s, with new results from Kamioka II and III (\citealt{1989PhRvL..63...16H}; \citealt{1996PhRvL..77.1683F}) and the SAGE/GALLEX (\citealt{1991PhRvL..67.3332A}; \citealt{1992PhLB..285..376A}) experiments revealed a pattern of neutrino fluxes inconsistent with the expected scaling of those fluxes with the solar
core temperature.  Attention increasingly turned to neutrino oscillations and other potential particle physics solutions.
Concerns about the nuclear physics of the SSM also evolved, focusing more on the uncertainties that might inhibit extraction of any such ``new physics.''

At that time, during a meeting on the Solar Neutrino Problem hosted by the Institute for Nuclear Theory (INT), University of Washington, Seattle, a suggestion was made to convene the community working on solar nuclear reactions, in order to reach consensus on the best current values of cross sections and their uncertainties. 
The INT hosted the proposed workshop in February, 1997, drawing representatives from almost every experimental group active in this area, as well as many of the theorists who were engaged in solar neutrino physics.  
The working groups that formed during this meeting worked over the following year to evaluate past work, determining the needed cross sections and their uncertainties.  
The results of this evaluation, which came to be known as Solar Fusion I or \citetalias{1998RvMP...70.1265A} \citep{1998RvMP...70.1265A}, became the standard for use in solar modeling over the following decade.

An update of \citetalias{1998RvMP...70.1265A} was launched with the January, 2009, workshop ``Solar Fusion Cross Sections II'', hosted again by the INT in Seattle.  Initial results from SuperKamiokande, the Sudbury Neutrino Observatory, and Borexino were
then in hand, and the conversion of approximately two thirds of solar electron neutrinos into other flavors had
been firmly established \cite{2002PhRvL..89a1301A}. Consequently, the motivation for the study was to ensure that
solar model predictions would be based on the most current nuclear physics, so that meaningful uncertainties
could be placed on neutrino parameters derived from solar neutrino measurements, such as the mixing angle $\theta_{12}$.
An additional source of uncertainty had arisen at the time, with the advent of three-dimensional (3D) radiative-hydrodynamic models of
the Sun's atmosphere for determining element abundances from photo-absorption lines \cite{2006NuPhA.777....1A,2009ARA&A..47..481A}.  This improved
analysis, though, lowered the inferred metallicities, leading to tension with the Sun's interior sound speed profile
determined from helioseismology \cite{2002RvMP...74.1073C}.  The discrepancy between higher metallicity models that accurately
reproduce the Sun's interior sound speed and lower metallicity models based on the most current treatment of the
photosphere was termed the ``solar composition problem''.  While the neutrino fluxes were not known to the precision
needed to distinguish between the competing models, it had been shown that the solar core metallicity
could be extracted directly from future carbon-nitrogen (CN) solar neutrino measurements, if the precision
of associated nuclear cross sections were improved  \cite{2008ApJ...687..678H}.  Reflecting the impact of 
new experimental work and improved theory, the \citetalias{2011RvMP...83..195A} study of \citet{2011RvMP...83..195A} significantly revised the \citetalias{1998RvMP...70.1265A} cross section for  
the driving reaction of the CNO-I cycle, and updated key pp-chain cross sections.

The present SF~III review began with a four-day workshop in July, 2022, ``Solar Fusion Cross Sections III.''
The meeting was hosted by Physics Frontier Center N3AS (Network for Neutrinos, Nuclear Astrophysics, and Symmetries)\footnote{\url{https://n3as.berkeley.edu/}}, a consortium of institutions involved in multi-messenger astrophysics, whose central hub is the University 
of California, Berkeley.  Sessions were
held at the David Brower Center, Berkeley, and the Physics Department, UC Berkeley, with the assistance of N3AS
staff.  The workshop was organized by N3AS, by Solar Fusion's long-term sponsor, the INT, and by the European ChETEC-INFRA Starting Community for Nuclear Astrophysics\footnote{\url{https://www.chetec-infra.eu/}}. 

The 47 workshop participants represented most of the leading experimental and theoretical research groups that are active in the field. Workshop participants were organized into nine working groups, who were charged with reviewing and
evaluating work completed since \citetalias{2011RvMP...83..195A}, to produce updated recommendations.  The structure of this review largely
follows that of the working groups. The timing of SF~III was driven in part by planned large-scale asteroseismic surveys of 
hydrogen-burning stars such as ESA's PLATO\footnote{\url{https://platomission.com/}} mission \citep{2014ExA....38..249R}, and the expectation that associated data analysis will require
large libraries of stellar models in which masses, ages, metallicities, and other parameters are varied. SF~III will
provide the best current nuclear astrophysics input for such modeling.  This review summarizes that input, based 
on the recommendations of the nine SF~III working groups.

\subsection{Terminology used}
\label{sec:Intro:Terminology}

The terminology used here follows recent practice, as summarized for example in the Nuclear Astrophysics Compilation of Reaction Rates (NACRE) compilation \cite{1999NuPhA.656....3A}. It is briefly described to aid the reader.

Nuclear reactions are designated as
\begin{equation}
    A(b,c)D,
\end{equation}
where $A$ is the target nucleus, $b$ is the bombarding particle, and $c$ and $D$ are the reaction products, 
with $D$ distinguished as the heavier product. Frequently in nuclear astrophysics, the reaction cross section $\sigma$
is re-expressed in terms of the S-factor $S_{ij}$, with S related to $\sigma$ as described below, and with $i,j$ the mass numbers of entrance channel nuclei $b$ and $A$.  
 
The relation between the  nuclear cross section $\sigma(E)$ and astrophysical S-factor is
\begin{equation} \label{eq:Sfactor}
    S(E) = \sigma(E) \, E \, \exp \left( 2\pi \frac{Z_A Z_b \alpha}{\beta} \right),
\end{equation}
with $Z_A$ and $Z_b$ the charge numbers of the target and projectile, $\alpha \sim 1/137$ the fine structure constant, $\beta \equiv v/c =\sqrt{2E/\mu c^2}$ the relative velocity of the interacting particles in units of $c$,
$\mu$ the reduced mass, and $E$ the center-of-mass energy. (In
this review we use energy units of MeV unless otherwise stated.)
This removes the sharp energy dependence from the cross section associated with s-wave Coulomb scattering off a point nucleus,
thereby accounting for the leading penetration effects associated with the Coulomb barrier. The quantity $ Z_A Z_b \alpha / \beta$ is known as the Sommerfeld parameter.  Consequently, unless there are resonances, $S(E)$ should
be a smooth function that at low energies can generally
be expressed as a low-order polynomial in $E$.

The S-factor in Eq. (\ref{eq:Sfactor}) is defined in terms of
the nuclear (or ``bare") cross section, while laboratory experiments are performed with targets consisting of neutral atoms. As electrons shield the nucleus, lowering the Coulomb barrier, their presence enhances the laboratory cross section.
The S-factors derived from laboratory data must be corrected for this effect.  As the energy is lowered, the screening correction becomes increasingly important.  Thus particular
care must be taken when extrapolating laboratory cross sections
to lower energies, as Fig. 4 of \citetalias{2011RvMP...83..195A} illustrates.
Unless otherwise noted, the S-factors discussed here are those for bare nuclei.

The rapid rise in the nuclear cross section with energy, reflecting the higher probability of penetrating the Coulomb barrier, competes with a rapid decrease in the probability of finding interacting particles with the requisite center-of-mass energy $E$ in the high-energy tail of the Maxwell-Boltzmann distribution. Consequently,
the thermonuclear reaction rate $\langle \sigma v \rangle$ for a given stellar temperature $T$ 
takes the form
\begin{equation}\label{eq:TNRR}
    \langle \sigma v \rangle = \sqrt{\frac{8}{\mu \pi}} (k_{\rm B}T)^{-\frac{3}{2}} \, \int_0^\infty S(E) \exp\left[-\frac{b}{\sqrt{E}}-\frac{E}{k_{\rm B}T}\right] dE,
\end{equation}
where $b=\pi Z_A Z_b \alpha \sqrt{2 \mu c ^2}$ is proportional to the Sommerfeld parameter of Eq. (\ref{eq:Sfactor}) and $k_{\rm B}$ is Boltzmann's constant. The maximum of the integrand
defines the Gamow peak -- the most probable energy for interactions -- while the $\sim 1 \sigma$ range around the peak is frequently termed the Gamow window.  

For the solar reactions of interest, the energy at the Gamow peak is well below the height of the Coulomb potential.  
The slowly varying $S(E)$ can be expanded as a power series around $E=0$, with the leading term being $S(0)$, and with corrections given by derivatives taken at $E=0$,
\begin{equation}
    \label{eq:Sfactor_expansion}
    S(E) \simeq S(0) + S'(0) ~ E + \frac{S''(0)}{2} ~ E^2 + \cdots \,.
\end{equation}
Most of the results presented in this review will be given in terms of $S(0)$ and its derivatives.

For most of the nuclear reactions studied here, the notation $S_{ij}$ is adopted, where $i$ and $j$ are the mass numbers of the projectile and target nuclei, respectively.  A summary of SF~III reactions and recommended astrophysical S-factors are presented in Table \ref{tab:Sfactors_summary}.

 Finally, we provide a short list of abbreviations that are used throughout the review in the List of Symbols and Abbreviations.

\begin{table*}[t]
  \begin{center}
    \caption{
        \label{tab:Sfactors_summary}
        List of nuclear reactions reviewed in SF~III. Denoting the astrophysical S-factor by $S_{ij}$, its value at zero-energy $S(0)$ is given along with, where applicable, derivatives parameterized in Eq. (\ref{eq:Sfactor_expansion}). See the corresponding section for uncertainties, higher precision values, and detailed discussion.
        }
    \begin{tabular}{l c c c c c}
        Reaction &~~~~ $S_{ij}$~~~~~ & ~~~~$S(0)$ (MeV b)~~~~~ &~~~~~ $S'(0)$ (b) ~~~~~&~~~~~ $S''(0)$ (MeV\textsuperscript{-1} b)~~~~ & ~~~Section~~~ \\
        \hline 
        \rule{-.3em}{2.3ex}
        $^{1}\text{H}(p,e^+\nu)^{2}\text{H}$ 
            & $ S_{11} $
            & $4.09 \times 10^{-25} $
            & $ 4.5 \times 10^{-24} $
            & $ 9.9  \times 10^{-23}$ 
            & \ref{sec:S11} \\
        $^{2}\text{H}(p,\gamma)^{3}\text{He}$ 
            & $ S_{12} $
            & $2.03 \times 10^{-7}$
            & see text
            & 
            & \ref{sec:S12} \\
        $^3\text{He}(^3\text{He},2p)^4\text{He}$
            & $ S_{33} $
            & $ 5.21  $
            & $-4.90 $
            & $ 22.42 $
            & \ref{sec:S33} \\
        $^3\text{He}(\alpha,\gamma)^7\text{Be}$
            & $ S_{34} $
            & $ 5.61 \times 10^{-4} $
            & $-3.03 \times 10^{-4}  $
            & $\cdots$
            & \ref{sec:S34} \\
        $^3\text{He}(p,e^+\nu)^4\text{He}$ 
            & $ S_\mathrm{hep} $
            & $ 8.6 \times 10^{-23} $
            & $\cdots$
            & $\cdots$
            & \ref{sec:hep} \\
        $^7\text{Be}(p,\gamma)^8\text{B}$
            & $ S_{17} $
            & $ 2.05 \times 10^{-5}$
            & $\cdots$
            & $\cdots$
            & \ref{sec:S17} \\
            
        $^{14}\text{N}(p,\gamma)^{15}\text{O}$
            & $ S_{1 \, 14} $
            & $ 1.68  \times 10^{-3}$
            & $\cdots$
            & $\cdots$
            & \ref{subsec:WG3:Total_Sfactors} \\  
            & & & & & \\

        \rule{-.3em}{2.3ex}
        $^{12}\text{C}(p,\gamma)^{13}\text{N}$ 
            &  $ S_{1 \, 12} $
            &  $ 1.44 \times 10^{-3}$
            &  $ 2.71 \times 10^{-3}$ 
            &  $ 3.74 \times 10^{-2}$  
            &  \ref{sec:OtherCNO:CNO1:c12pg} \\
        $^{13}\text{C}(p,\gamma)^{14}\text{N}$ 
            & $ S_{1 \, 13} $
            &  $ 6.1  \times 10^{-3}$
            &  $ 1.04 \times 10^{-2}$     
            &  $ 9.20 \times 10^{-2}$
            & \ref{sec:OtherCNO:CNO1:c13pg} \\        
        $^{15}\text{N}(p,\gamma)^{16}\text{O}$ 
            & $ S_{1 \, 15}^\gamma $
            &  $ 4.0   \times 10^{-2}$
            &  $ 1.07 \times 10^{-1}$     
            &  $ 1.84 $
            &  \ref{sec:OtherCNO:CNO2:n15pg} \\        
        $^{15}\text{N}(p,\alpha)^{12}\text{C}$ 
            & $ S_{1 \, 15}^\alpha $
            &  $ 73 $
            &  $ 3.37 \times 10^{2}$     
            &  $ 1.32 \times 10^{4} $
            &  \ref{sec:OtherCNO:CNO1:n15pa} \\        
        $^{16}\text{O}(p,\gamma)^{17}\text{F}$ 
            & $ S_{1 \, 16} $
            &  $ 1.09 \times 10^{-2}$
            &  $ -4.9 \times 10^{-2}$     
            &  $ 3.11 \times 10^{-1} $  
            &  \ref{sec:OtherCNO:CNO2:o16pg} \\   
         $^{17}\text{O}(p,\gamma)^{18}\text{F}$ 
            & $ S_{1 \, 17} $
            &  $ 4.7 \times 10^{-3}$   
            &  $\cdots$     
            &  $\cdots$
            &  \ref{sec:OtherCNO:CNO2:o17pg} \\           
          
        $^{18}\text{O}(p,\gamma)^{19}\text{F}$ 
            & $ S_{1 \, 18} $
            &  $ 2.30  \times 10^{-2}$
            &  $\cdots$    
            &  $\cdots$ 
            &  \ref{sec:OtherCNO:CNO3:o18pg} \\        
        $^{20}\text{Ne}(p,\gamma)^{21}\text{Na}$ 
            & $ S_{1 \, 20} $
            &  $ 6.78 $
            &  $\cdots$     
            &  $\cdots$ 
            &  \ref{sec:OtherCNO:NeNa:ne20pg} \\        
        $^{21}\text{Ne}(p,\gamma)^{22}\text{Na}$ 
            & $ S_{1 \, 21} $
            &  $ \approx 2.0 \times 10^{-2}$
            &  $\cdots$     
            &  $\cdots$  
            &  \ref{sec:21nepg} \\        
        $^{22}\text{Ne}(p,\gamma)^{23}\text{Na}$ 
            & $ S_{1 \, 22} $
            &  $ 0.415 $
            &  $\cdots$     
            &  $\cdots$  
            &  \ref{sec:OtherCNO:NeNa:ne22pg} \\        
        $^{23}\text{Na}(p,\gamma)^{24}\text{Mg}$ 
            & $ S_{1 \, 23} $
            &  $ 1.80 \times 10^{-2}$
            &  $ 0$     
            &  $ 0$  
            &  \ref{sec:OtherCNO:NeNa:na23pg} \\        
        \hline         \hline
    \end{tabular}
  \end{center}
\end{table*}

\subsection{Scope and structure of the review}
\label{sec:Intro:Scope}

\begin{figure}
    \centering
    \includegraphics[width=\columnwidth]{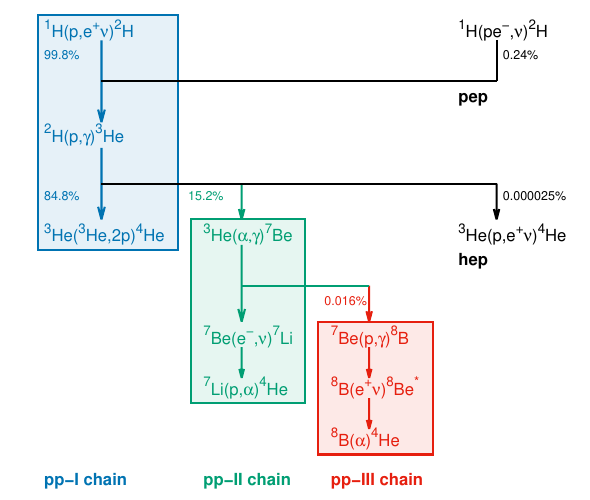}
    \caption{Nuclear reactions of the proton-proton (pp) chain. The percentage branchings are applicable for the \citet{2022A&A...661A.140M} solar composition.}
    \label{fig:pp-Chain}
\end{figure}

 In the Sun, the primary mechanism for the conversion of four protons into $^4$He 
 is the pp chain of Fig. \ref{fig:pp-Chain}, during which two charge-changing
 weak interactions take place, each converting a proton into a neutron with the emission of a neutrino.
 The pp chain consists of three main branches,
 ppI, ppII, and ppIII.  These branches
 are distinguished by their dependence on the solar core temperature and by
 the neutrinos they produce, with the pp, $^7$Be, and $^8$B neutrinos serving as ``tags''
 for the ppI, ppII, and ppIII chains, respectively.  The SSM
 predicts that the fluxes of pp, $^7$Be, and $^8$B neutrinos have an approximate {\it relative}
 temperature scaling of $\sim$ 1:T$_c^{10}$:T$_c^{22}$, where T$_c$
 is the solar core temperature \cite{annurev:/content/journals/10.1146/annurev.aa.33.090195.002331}.  By the mid-1990s, solar neutrino experiments established
 that the measured fluxes differed significantly from this expected pattern.
 This contributed to growing expectations that
 new neutrino physics might be the solution of the solar neutrino problem.  It also was
 an important motivation for \citetalias{1998RvMP...70.1265A}, as errors in the nuclear physics of the pp chain
 could distort the relationship between the various fluxes.

\begin{figure*}[htp]
    \centering
    \includegraphics[width=.8\textwidth,trim={30pt 10pt 30pt 10pt}]{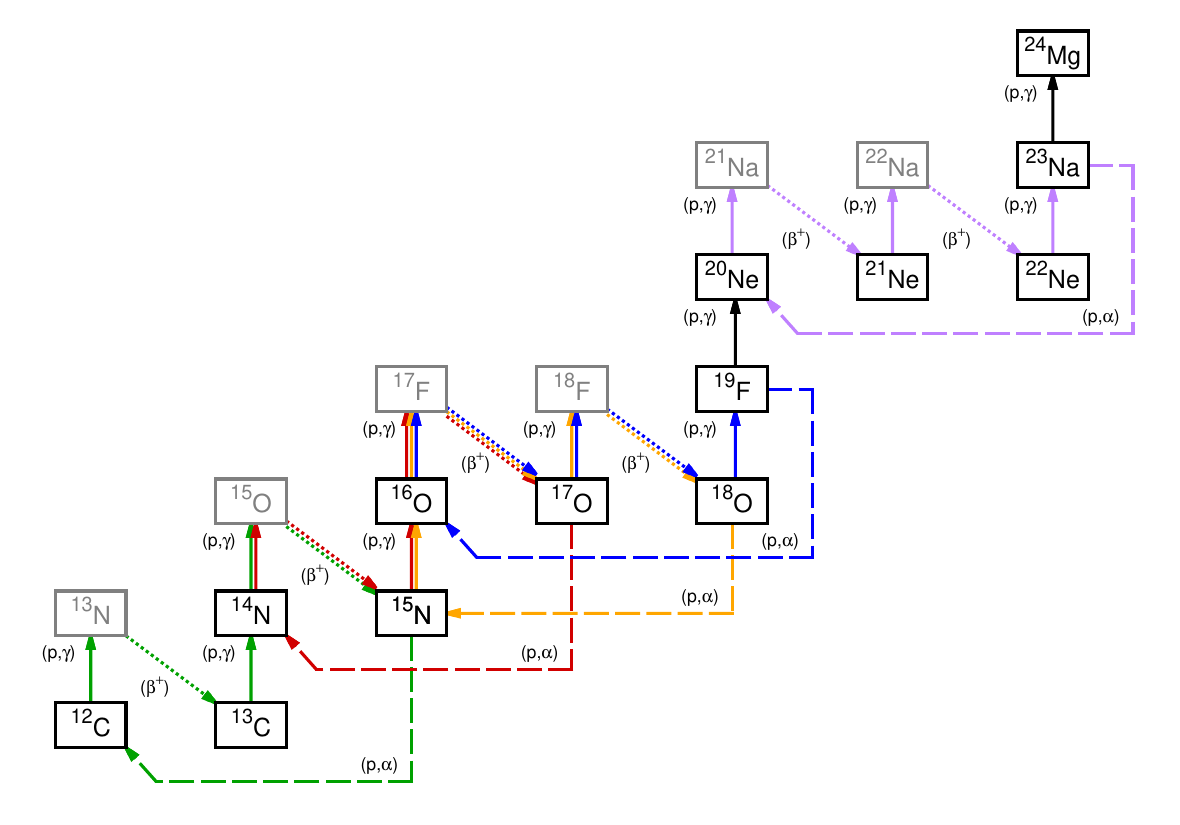}
    \caption{Nuclear reactions of the CNO-I (green line), CNO-II (red line), CNO-III (orange line), CNO-IV (blue line), and Ne-Na (purple line) cycles. Other reactions reported in this work are also present (in black).
    For visual aid, $(p,\gamma)$ reactions are styled using full arrows, while $(p,\alpha)$ reactions use dashed-line arrows.
    Greyed isotopes are short-lived and quickly $\beta^+$-decay (dotted arrows), resulting in $e^+ + \nu_e$ emissions. 
    }
    \label{fig:CNO_cycle}
\end{figure*}

 The pp chain accounts for 99\% of solar hydrogen burning and are the main focus of 
 this review. The remaining 1\% is generated through the CNO-I cycle, see Fig. \ref{fig:CNO_cycle}).  Motivated in part by Borexino's recent success in measuring the flux of solar CN neutrinos, this review also considers in more detail than past Solar Fusion reviews the reactions driving hydrogen burning in stars more massive than our Sun.

The review is structured as follows.  Section \ref{sec:Observations} begins with a summary
of past solar neutrino experiments, highlighting some of the open questions that remain.
It discusses the second-generation experiments Super-Kamiokande,
the Sudbury Neutrino Observatory, and Borexino that contributed in the resolution of the solar neutrino
problem. It also describes
Hyper-Kamiokande, SNO+, the Jiangmen Underground Neutrino Observatory (JUNO), and the Deep Underground Neutrino Experiment (DUNE), new detectors in various stages of construction.

The cross section of the driving reaction of the pp chain, $^{1}\text{H}(p,e^+\nu)^{2}\text{H}$, is too small to be measured
and thus must be taken from theory. This important reaction is reviewed in section \ref{sec:S11}. 
With the exception of the minor hep branch, all other reactions and decays have been studied in laboratories, 
with measurements then combined with theory to
predict rates under solar conditions. The $^2$H($p,\gamma)^3$He, $^3$He($^3$He,2$p$)$^4$He, and
$^3$He$(\alpha,\gamma)^7$B reactions are reviewed in Secs. \ref{sec:S12}, \ref{sec:S33}, and \ref{sec:S34}, respectively. The hep reaction is reviewed in Section \ref{sec:hep}, and the electron-capture reactions on pp and $^7$Be are discussed in Section \ref{sec:EC}. The $^7$Be($p$,$\gamma)^8$B reaction responsible for the high energy 
neutrinos measured in Super-Kamiokande is treated in Section \ref{sec:S17}. 

The CNO-I cycle is a modest contributor to solar energy generation, accounting for about 1\% of the sun's hydrogen burning.  It
is a potentially important probe of core metallicity as the burning is catalyzed
by the Sun's primordial C and N (Section \ref{sec:OtherCNO}).


Higher temperature pathways for hydrogen burning are of great importance in more massive
stars.  The associated nuclear reactions are an increasingly important focus of low-energy
nuclear astrophysics facilities.  Many of the contributing reactions
have been re-measured.  That progress is reviewed in Section \ref{sec:OtherCNO}. 

Charged-particle nuclear reactions occurring at energies below the Coulomb barrier are affected by electron screening, the shielding of the nuclear charge by electrons.  Because of their very different atomic
environments, laboratory reactions and those occurring in the solar plasma are affected
in distinct ways.  Section \ref{sec:Screening} describes the current status of efforts to
account for these differences, when solar cross sections are extracted from
laboratory measurements.

The energy transport inside the radiative zone is affected by the metals\footnote{In this review we follow the astrophysical terminology in which $metals$ refer to all chemical elements except for the two lightest ones, hydrogen and helium. } that were 
incorporated into the Sun when it first formed. While metals comprise less than 2\% of
the Sun by mass, they play an outsized role in determining the opacity.  Very few
of the needed radiative opacities have been measured, and the 
conditions under which measurements are made typically are not identical to those in the Sun. 
Section \ref{sec:Opacities} describes the current state of the art, and discusses 
connections between opacities and solar composition that have complicated efforts
to resolve the solar metallicity problem.

Much of the progress made in constraining the reactions of the pp chain and CNO-I cycles
has been possible thanks to new laboratories and the facilities they host. The underground LUNA laboratory is a prominent example \cite{2018PrPNP..98...55B}.  Section \ref{sec:Facilities} describes the capabilities
of selected laboratories involved in the measurements reviewed here. 



\section{Solar neutrino observations}

\label{sec:Observations}
Solar neutrinos offer a unique tool box for probing both the fundamental properties of these elusive particles, and their interactions with matter, as well as understanding their source: the fusion reactions that power our Sun.  The original motivation for observations of solar neutrinos was precisely the hope to probe solar fusion.  The first successful experiment beginning in 1967 at the Homestake mine \cite{1998ApJ...496..505C} offered the surprising result of a neutrino flux suppressed to approximately one third of expectation.  The intervening years have seen a tour de force of experimental efforts, leading to the resolution of the Solar Neutrino Problem and the confirmation of neutrino oscillations \cite{1998PhRvL..81.1562F,2001PhRvL..87g1301A,2002PhRvL..89a1301A,2003PhRvL..90b1802E,2004PhRvL..92r1301A}. These high-precision flux and spectral measurements mapped out the details of solar fusion, probed the structure of the Sun, and demonstrated
(together with atmospheric neutrino measurements) that neutrinos have masses and mix. Table \ref{tab:fluxes} lists the pp-chain and CNO cycle neutrino sources, their endpoint ($\beta$ decay) or line (electron capture) energies $E_\nu$, and predicted SSM fluxes.

\begin{table}[h!]
\caption{Solar neutrino sources, energies, and SSM flux predictions.  
All $\beta$ decay sources produce continuous spectra, while the pep and $^7$Be electron-capture sources produce line spectra. 
The fluxes are taken from the SSM calculations of \citet{B23Fluxes}, computed with nuclear reaction rates shown in Table \ref{tab:Sfactors_summary}, for the compositions of GS98 \cite{1998SSRv...85..161G} (high-Z), AAG21 \cite{2021A&A...653A.141A} (low-Z), and MB22p \cite{2022A&A...661A.140M} (high-Z) with
associated uncertainties indicated.}
\vspace{0.4cm}
\label{tab:fluxes}

\begin{tabular}{lccccc}
\footnotesize 
Source 
	& \parbox[c]{3.2em}{ E$_\nu$ \\[-0.5ex] {\scriptsize (MeV)} } 
	& GS98  
	& AAG21  
	& MB22p 
	&   \parbox[c]{3.8em}{ units \\[-0.5ex] {\scriptsize(cm$^{-2}$s$^{-1}$)} }  \\[1ex]
\hline
    \\[-.2cm]
pp  
	&  $\le$0.420 
	& 5.96\,(0.6\%) 
	& 6.00\,(0.6\%) 
	& 5.95\,(0.6\%) 
	& ~10$^{10}$ \\[0.1cm]
pep 
	&  1.442 
	& 1.43\,(1.1\%) 
	& 1.45\,(1.1\%) 
	& 1.42\,(1.3\%) 
	& 10$^8$ \\[0.1cm]
hep   
	& $\le$18.77 
	& 7.95\,(31\%) 
	& 8.16\,(31\%) 
	& 7.93\,(30\%) 
	& 10$^3$ \\[0.1cm]
$^7$Be 
	& \parbox[c]{3em}{0.862\textsuperscript{(a)} \\[-0.5ex] 0.384\textsuperscript{(b)}}
	& 4.85\,(7.4\%) 
	& 4.52\,(7.3\%) 
	& 4.88\,(8.1\%) 
	& 10$^9$ \\ [0.2cm] 
$^8$B 
	&  $\lesssim$ 15  
	& 5.03\,(13\%) 
	& 4.31\,(13\%) 
	& 5.07\,(15\%) 
	& 10$^6$ \\[0.1cm]
$^{13}$N 
	&  $\le$1.198 
	& 2.80\,(16\%) 
	& 2.22\,(13\%) 
	& 3.10\,(15\%) 
	& 10$^8$ \\[0.1cm]
$^{15}$O 
	&  $\le$1.732 
	& 2.07\,(18\%) 
	& 1.58\,(16\%) 
	& 2.30\,(18\%) 
	& 10$^8$ \\[0.1cm]
$^{17}$F 
	&   $\le$1.738 
	& 5.35\,(20\%) 
	& 3.40\,(16\%) 
	& 4.70\,(17\%) 
	& 10$^6$ \\[1ex]
\hline 
\end{tabular}
 \scriptsize{
 \textbf{Notes:}
 (a) 90\% and (b) 10\%  of the $^7$Be neutrino flux, respectively.
 }
\end{table}

\subsection{Open questions}
 The solar neutrino spectrum has been measured for energies above 233 keV, with some fluxes determined to a precision equivalent to or better than theoretical predictions. 
 However, improved measurements can offer further insight on several aspects.  
 For instance, more precise determinations of the fluxes of neutrinos from the CNO cycle (Fig. \ref{fig:CNO_cycle}) could determine the 
 contemporary abundances of N and O in the solar core \citep{2007A&A...463..755C,2011ApJ...743...24S,2022A&A...667L...2K}. 
 While these abundances differ significantly from the
 primordial values that are input into the SSM --the primordial abundances are altered by out-of-equilibrium CNO-cycle burning
 that occurs in the early Sun \cite{Roxburgh85}-- this burning redistributes the C, N, and O, but does not alter the net number of such nuclei.
 Consequently, CNO neutrino flux measurements might allow one to experimentally address the solar composition problem discussed below.
 
 Solar neutrino flux measurements can also constrain physical processes in stars such as chemical mixing.  The fluxes are sensitive to solar/stellar model inputs, including the radiative opacities (see Sect. \ref{sec:Opacities}).  Measurements of  the neutrino fluxes from $^8$B and $^7$Be can offer a handle on the temperature of the solar core, and other environmental factors~\cite{1996PhRvD..53.4202B}.  
 
 A measurement of the pp flux with percent-level precision would provide a test of the luminosity constraint, further probing solar power generation mechanisms~\cite{2002PhRvC..65b5801B,2021JPhG...48a5201V}.  The one branch of the pp chain that yet remains undetected, the hep flux, has both the highest energy but also the lowest flux of neutrinos from the Sun.  The spectrum extends beyond that of the $^8$B neutrinos, offering a small window for unequivocal observation.  Limits have been placed on this flux \cite{2020PhRvD.102f2006A}, but a definitive observation will likely require a large, next-generation detector such as Hyper-Kamiokande (HyperK) \cite{2018arXiv180504163H} or the Deep Underground Neutrino Experiment (DUNE) \cite{2019PhRvL.123m1803C}.


Finally, the observation of monochromatic neutrinos produced by electron capture reactions on $^{13}$N, $^{15}$O and $^{17}$F (which we refer to as ecCNO neutrinos) \cite{1990PhRvD..41.2964B, 2004PhRvC..69a5801S, 2015PhLB..742..279V} could be rewarding in terms of physical implications.
Indeed, they can be used as probes of the metallicity of the solar core. 
Moreover, they provide a measure of the electron neutrino survival probability at specific neutrino energies $E_\nu \sim 2.5$ MeV 
within the broad
transition region between vacuum and matter-enhanced oscillations, where currently we have no constraints.
This would constitute a new test of the large-mixing-angle (LMA) Mikheyev-Smirnov-Wolfenstein (MSW)
flavor oscillation solution. The detection of this subdominant component of the CNO-cycle is extremely difficult but it could be within reach of future very large ultra-pure liquid scintillator detectors~\cite{2015PhLB..742..279V}.

Solar neutrinos provide an opportunity to understand the interaction of neutrinos with matter.  The effect of MSW oscillations has a significant impact on the observed solar neutrino spectrum: with vacuum oscillation dominating at low energy, below approximately 1 MeV, where the survival probability is roughly one half, while matter effects further suppress the flux to a survival probability of roughly one third above approximately 5 MeV.  The transition region between these two regimes offers an extremely sensitive probe of the details of the interactions of neutrinos with matter, including the potential to search for new physics such as sterile neutrinos, or non-standard interactions, by looking for distortions to the expected spectral shape~\cite{2019arXiv190700991B,2020JHEP...02..038B,2023JHEP...08..032C}.  

The same theory of matter effects predicts a small regeneration of electron neutrinos during the night time as they propagate through the bulk of the earth.  This so-called ``day/night effect'' has been sought after by both Super-Kamiokande (SuperK) and the Sudbury Neutrino Observatory (SNO) \cite{2014PhRvL.112i1805R,2002PhRvL..89a1302A}, but a significant observation is still limited by statistics.  Future data from HyperKamiokande or DUNE may be needed to confirm our understanding of this prediction of the MSW effect.

The slight tension between measurements of the mass splitting parameters, $\Delta m^2_{12}$ in solar neutrino experiments and terrestrial data sensitive to the same parameter, from the KamLAND reactor experiment has now disappeared with the addition of new data on day-night effects and spectral shapes \cite{Esteban2020TheFO,2018PrPNP.102...48C}.  Further tightening of the oscillation parameters can be expected from JUNO, which will obtain highly precise data on both the solar mass splitting and the mixing angle~\cite{2016JPhG...43c0401A}.

\subsubsection{The solar composition problem}

Solar photospheric abundances, determined with spectroscopic techniques, are a fundamental input for the construction of SSMs, but also they are used as input to nearly all models in astrophysics, including stellar evolution, protoplanetary disks, and galactic chemical evolution. Over the last two decades, the development of three dimensional radiation hydrodynamic models of the solar atmosphere \citep{2009MmSAI..80..711L,2000A&A...359..729A} and of techniques to study line formation under non-local thermodynamic conditions \citep{2005ARA&A..43..481A}, together with improved descriptions of atomic properties (e.g. transition strengths \citealt{2022A&A...661A.140M,2021MNRAS.508.2236B}), have led to a significant revision of solar abundances.
Initial results based on these more sophisticated methods favored a markedly lower solar metallicity \cite{2009ARA&A..47..481A}, particularly in the CNO elements, than obtained in the 90s from older techniques \cite{1993oee..conf...15G, 1998SSRv...85..161G}. 

Recently, two groups have revisited the solar photospheric abundances using modern methods \citep{2021A&A...653A.141A,2022A&A...661A.140M}. While \citet{2021A&A...653A.141A} obtain results consistent with their previous findings, \citet{2022A&A...661A.140M} find an O abundance intermediate between that of Asplund's group and those from the 90s, in agreement with another 3D-based determination by \citet{2011SoPh..268..255C}. Interestingly, \citet{2022A&A...661A.140M} find higher C and N abundances and, indirectly, a higher Ne abundance due to the larger Ne to O ratio measured in the solar corona \cite{2018ApJ...855...15Y}. This leads to a combined metal-to-hydrogen ratio that is by chance comparable to those from \citet {1993oee..conf...15G} and \citet{1998SSRv...85..161G}, albeit with a different mixture of elements.

 Considering that uncertainties in element abundances are difficult to quantify, it has become customary to consider two canonical sets of abundances, which we refer to as high metallicity (HZ) and low metallicity (LZ) solar admixtures.  See, for example, \citet{2017ApJ...835..202V} for SSM reference values. In this context, the new solar abundance determinations by \citet{2021A&A...653A.141A} fall into the LZ category, while those from \citet{2022A&A...661A.140M} are HZ.
 Solar models employing the LZ abundances fail to reproduce most helioseismic probes of solar properties. This disagreement constitutes the so-called solar composition problem \cite{2004ApJ...606L..85B,2005ApJ...618.1049B,2006ApJ...649..529D} that has defied a complete solution. All proposed modifications to physical processes in SSMs offer, at best, only partial improvements in some helioseismic probes, see e.g. \cite{2005ApJ...627.1049G,2007A&A...463..755C,2008PhR...457..217B,2010ApJ...713.1108G, 2011ApJ...743...24S}. The same conclusions are obtained with SSMs computed with the newest LZ \cite{2021A&A...653A.141A} and HZ \cite{2022A&A...661A.140M} abundances as discussed also there. 
 An alternative possibility is to consider modifications to the physical inputs of SSMs at the level of the constitutive physics, radiative opacities in particular. This is possible because most helioseismic probes depend actually not directly on the solar composition, but on the radiative opacity profile in the solar interior, i.e. on the combination of solar composition and atomic opacities (see \S~\ref{sec:Opacities}). The same can be said for solar neutrinos from the pp chain. 
Early work \cite{2005ApJ...621L..85B,2004ESASP.559..574M} already suggested that a localized increase in opacities could solve or, at least, alleviate the disagreement of low-Z solar models with helioseismology, and \citet{2009A&A...494..205C} and \citet{2010ApJ...724...98V} showed that a tilted increase in radiative opacities, with a few percent increase in the solar core and a larger (15-20\%) increase at the base of the convective envelope could lead to LZ SSMs that would satisfy helioseismic probes equally well as HZ SSMs.

 The degeneracy between solar composition and opacities can be broken using CNO solar neutrinos, e.g., following the methodology developed in \citet{2008ApJ...687..678H}. Such a study was recently carried out by the Borexino collaboration \citep{2020Natur.587..577B,2022PhRvL.129y2701A,2023PhRvD.108j2005B}.  The Borexino measurement of the CNO-I cycle neutrino flux (the $^{13}$N and $^{15}$O fluxes) were used to determine the C+N core abundance. Results show a $\sim 2\sigma$ tension with LZ metallicity determinations, while being in better agreement with HZ mixtures. While the error budget is presently dominated by the uncertainty of the Borexino CNO neutrino measurement, a significant contributor to the error ($\sim 10\% $) is nuclear, due to uncertainties in $S_{114}$, $S_{34}$, and $S_{17}$. 
 The interpretation of future improved CNO neutrino measurements will be impacted, unless these nuclear physics uncertainties are reduced. 

Very recently, \citet{2024JHEP...02..064G} presented a new global determination of all solar neutrino fluxes using all available experimental data, including the latest phases of Borexino. Results from this global analysis are in line with those from Borexino, although the added $^{13}$N and $^{15}$O fluxes are about 10\% lower than the Borexino result alone. A comparison of the solar fluxes with SSM calculations \cite{B23Fluxes} shows that HZ SSMs are in better agreement with solar neutrino fluxes than the LZ SSMs, pointing toward a C+N solar core abundance consistent with HZ abundances. The discrimination that solar neutrino fluxes can offer between solar compositions is at most, however, of the order of 2$\sigma$. (See Table~2 in \citet{2024JHEP...02..064G}). 

Helioseismology also offers the potential to determine the total solar metallicity, i.e. without disentangling individual element abundances, rather independently from opacities. This relies on using the so-called adiabatic index $\Gamma_1$, which deviates from 5/3 in regions of partial ionization. The underlying technique has been used widely to determine the helium abundance in the solar envelope and it has also been extended to determine solar metallicity. Attempting the latter is difficult because the imprint of partial ionization of metals is quite subtle. Previous work along this line \citep{2006ApJ...644.1292A} found an overall metallicity consistent with HZ abundances, but new work \citep{2024A&A...681A..57B} claims to favor LZ values. While this method depends very weakly on radiative opacities, the abundance determination is degenerate with the equation of state. Independent confirmation of these results would be desirable. 
In short, the controversy related to the solar composition is far from being resolved, currently with different indicators showing contradictory results.

\subsubsection{The Gallium anomaly}
\label{sec:galliumanomaly}

\begin{figure}[tb]
    \centering
    \includegraphics[width=3.3in]{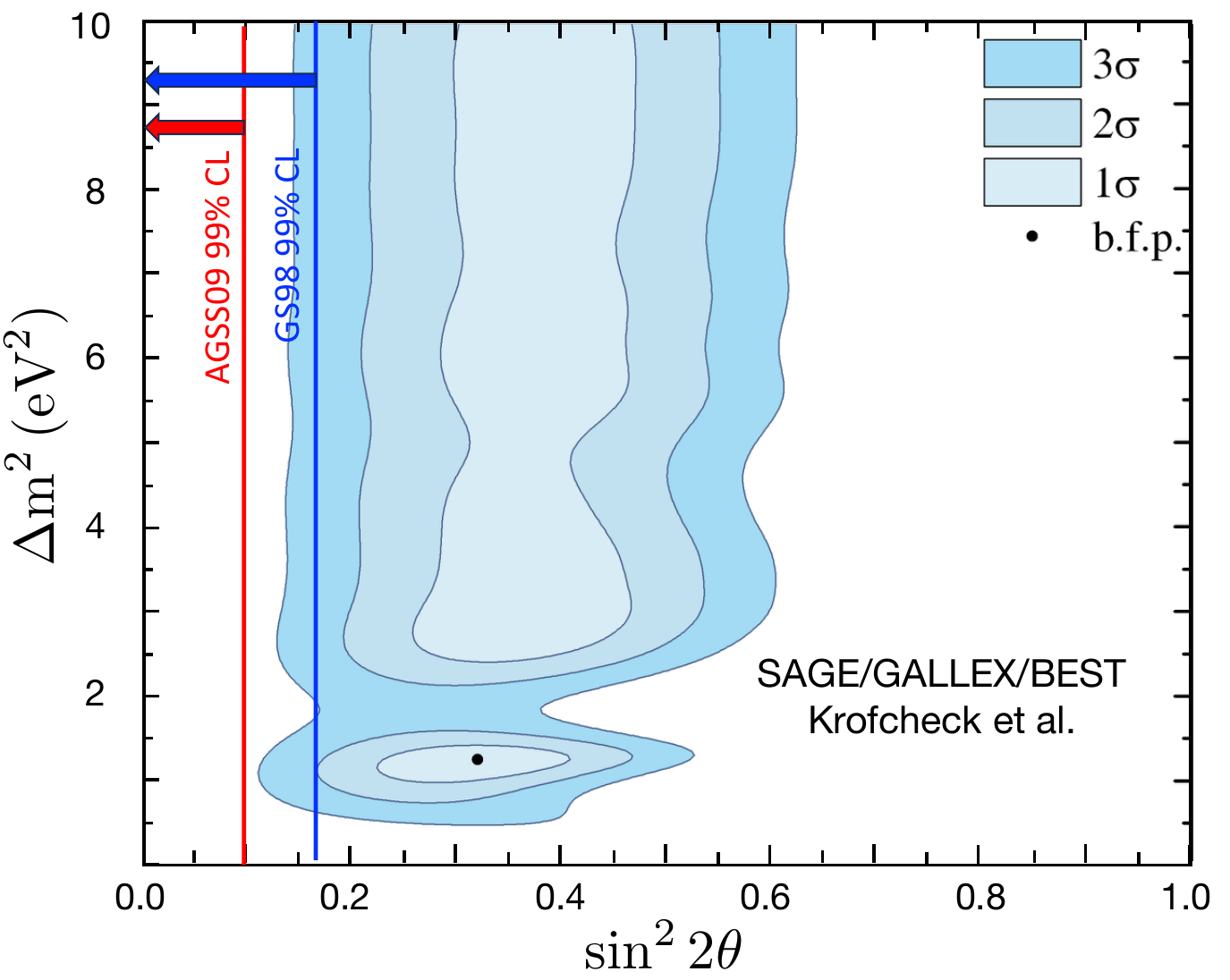}
    \caption{The contours are the sterile neutrino interpretation of the Ga anomaly, from~\cite{PhysRevC.108.035502}. Limits at the 99\% confidence level (CL) on the mixing angle from solar neutrinos, from \cite{2022EPJC...82..116G}, are indicated at left for two solar models, GS98 and AGSS09, and essentially exclude the indicated sterile-neutrino space.}.  
    \label{fig:solarsterile}
\end{figure}

With the aid of very intense radioactive sources of $^{51}$Cr and $^{37}$Ar, tests have been made of the rate of production of $^{71}$Ge by neutrino interactions on $^{71}$Ga, the basis of radiochemical measurements of the low-energy solar neutrino flux. Initial experiments showed lower rates than expected, with interesting but inconclusive statistical precision.  Very recently the BEST experiment \cite{2022PhRvL.128w2501B,2022PhRvC.105f5502B,2018PhRvD..97g3001B} has confirmed this anomaly at more than 4$\sigma$.   Many possible explanations have been explored, but for each there are contradictions~\cite{2023EPJC...83..578K,2022JHEP...11..082A,2023PhLB..84237983G,2023JHEP...05..143B}.  A sterile neutrino explanation is disfavored~\cite{2022JHEP...10..164G,2022EPJC...82..116G}, particularly because of conflict with solar neutrino limits (see Fig.~\ref{fig:solarsterile}), as well as cosmological bounds~\cite{2021PhRvD.104l3524H}.  

The sterile neutrino contribution to the solar flux is fundamentally limited by the luminosity constraint, that the Sun's total energy output is the result of nuclear reactions that produce neutrinos, whether active or sterile.  If precise neutral-current data were available across the solar spectrum, a completely model-independent limit on a possible sterile component could be determined.  The present neutral-current data at low energies (from Borexino) are imprecise but can be supplemented by much more precise charged-current data together with 3-flavor oscillation physics.  Still better limits, at the cost of some model dependence, can be obtained with fits of experimental data to solar models, as shown in Fig.~\ref{fig:solarsterile} above, as well as Fig.~1 of \citet{2022EPJC...82..116G}.

The precisely known rate of electron capture on $^{71}$Ge to $^{71}$Ga, for which the half-life is 11.43(3) d~\cite{1985PhRvC..31..666H}, places an important constraint
on the neutrino absorption rate \cite{2023PhLB..84237983G,PhysRevC.108.035502}.  The half-life has now been remeasured~by two groups, \citet{2024PhRvC.109e5501N} and \citet{PhysRevC.108.L021602}, yielding values of 11.468(8) d
and 11.46(4) d, consistent with the previous determination
while increasing the precision. The allowed matrix element that governs neutrino capture to the ground state of $^{71}$Ge can be extracted
to a precision of $\sim$ 1\% from the electron capture rate.  The ground-state transition by itself generates a
significant anomaly: inclusion of the excited-state contributions doubles the effect.  In Appendix A we describe some of the details of both the 1997 extraction of the allowed (Gamow Teller)
strength by \citet{1997PhRvC..56.3391B} and the recent update of  \citet{PhysRevC.108.035502,ELLIOTT2024104082}. 

 \citet{PhysRevC.108.035502,ELLIOTT2024104082} find the ground-state cross section for absorbing $^{51}$Cr neutrinos is 2.5\% lower than Bahcall's value. This work includes the contributions of weak magnetism and radiative corrections, which are 
 shown to be sub-1\% effects.  An improved extraction of the $\sim 6\%$ excited-state contribution to neutrino absorption was performed using data from forward-angle (p,n) scattering. This led to a slight increase in that contribution relative to \citet{1997PhRvC..56.3391B}, thereby reducing the 2.5\% difference above by about half,
 for the total $^{51}$Cr and $^{37}$Ar capture cross sections.
 
 The procedures followed by \citet{PhysRevC.108.035502,ELLIOTT2024104082} could be extended
 to the solar neutrino $^{71}$Ga capture cross section, including the contribution from the high-energy $^8$B neutrinos. That has not been done, but as noted in Appendix A,
 could help resolve small discrepancies noted there.
 
 Results from the SAGE and GALLEX/GNO experiments remain part of the solar neutrino database used in various global fits to neutrino parameters.  A resolution of the anomaly that
 emerged from BEST and the four gallium detector calibration experiments is important as it would increase confidence in these data.

\subsubsection{The Boron-8 neutrino spectrum}
The neutrino spectrum from $^{8}{\rm B}$, extending to approximately 17 MeV, has been the most accessible part of the solar neutrino spectrum~\cite{2016PhRvD..94e2010A,2020PhRvD.102f2006A},  playing  an important  role in disclosing the physics of neutrino oscillation and testing the SSM. Knowing it with precision, particularly at the high energy end, has renewed interest in the context of observing the hep neutrinos~\cite{2006ApJ...653.1545A,2020PhRvD.102f2006A}, which could be accessible with future observatories~\cite{2022arXiv220212839A}. 

In \citetalias{2011RvMP...83..195A} it was decided to recommend the spectrum calculated by Winter et al.~\cite{2003PhRvL..91y2501W, 2006PhRvC..73b5503W, Win:2007} based on their measurement of the alpha spectrum from $^{8}{\rm B}$. This spectrum showed excellent agreement with an independent experiment \cite{2006PhRvC..73e5802B}.
Both radiative and recoil-order corrections were included by Winter et al. The radiative corrections are relatively small, due to a cancellation between the real and virtual contributions~\cite{1995PhRvD..52.5362B}. The recoil-order corrections are dominated by the weak magnetism part, which has been deduced from measurements of the analog electromagnetic decays~\cite{1995PhRvC..51.2778D}.

\begin{figure}[tb]
    \centering
    \includegraphics[width=3.3in]{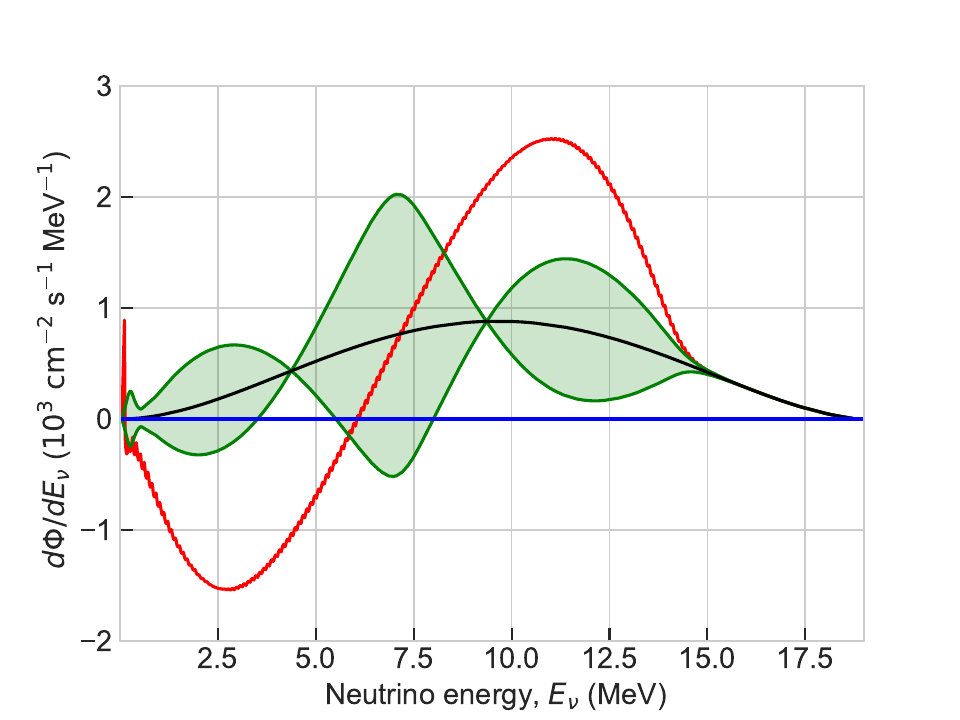}
    \caption{The hep spectrum is in black, and the uncertainties in $^8$B from Longfellow \etal\,\cite{2023PhRvC.107c2801L} span the region between the green lines.  The red line indicates the difference between Longfellow \etal~and Winter \etal\,\cite{2006PhRvC..73b5503W} deduced $^8$B spectra.  The hep spectrum is used here as a convenient metric for the size of the uncertainties in the $^8$B spectrum.}  
    \label{fig:8bspectrum}
\end{figure}

Since then three additional measurements~\cite{2011PhRvC..83f5802K,2012PhRvL.108p2502R,2023PhRvC.107c2801L} of the $^{8}{\rm B}$ $\beta$-decay alpha spectrum have been performed. All three find that the peak of the alpha spectrum appears about 20 keV lower than determined by Winter et al. 
Overall, these new measurements yield differences in the $^{8}{\rm B}$ neutrino spectrum of $\lesssim$ 5\% below $E_\nu=15$~MeV, the energy above which the $^{8}{\rm B}$ contribution becomes small compared to the hep component.  Figure~\ref{fig:8bspectrum} shows the $^8$B spectral uncertainties on the scale of the hep spectrum.  With resolution effects  considered, it will be difficult to extract the hep flux from solar neutrino measurements without further reduction in $^8$B uncertainties.

In producing a recommendation for the $^8{\rm B}$ neutrino spectrum we considered the following:
\begin{enumerate}
    \item Apparently the uncertainty estimations for the neutrino spectrum in \citet{2012PhRvL.108p2502R} are not quite correct. \citet{2023PhRvC.107c2801L} show that the uncertainty due only to the weak magnetism part, for which all authors follow similar prescriptions, is larger in the 0-12 MeV range than the overall uncertainty estimated by Roger et al. We estimated the uncertainties due to the weak magnetism term only and found agreement with those of Longfellow et al.
    \item All three of the recent efforts~\cite{2011PhRvC..83f5802K,2012PhRvL.108p2502R,2023PhRvC.107c2801L} used detectors (Si strip) more complicated than those used by Winter et al.~(a Si surface barrier detector). The strip detectors allow for better position resolution and reduced beta summing. However, for detection from sources external to the detectors (as opposed to detection from sources implanted deep into a detector), the complicated pattern of dead layers needs to be measured and taken into consideration for accurate calibrations. Effects of partial charge collection in-between strips are also significant. The effects are different for a calibration source than from the emission from $^8{\rm B}$ trapped as an ion. 
\end{enumerate}
Based on these considerations, we recommend the spectrum of \citet{2023PhRvC.107c2801L}, but suggest that conclusions sensitive to the choice of $^{8}{\rm B}$ neutrino spectrum, such as evidence for hep neutrinos, should be evaluated using both the Winter et al. and the Longfellow et al. spectra.

Given the importance of the weak magnetism contributions, it would be good to have a new experiment with reduced uncertainties that could be compared directly to the results of \citet{1995PhRvC..51.2778D}.
Recoil-order effects in the $A=8$ system were most recently measured by \citet{2011PhRvC..83f5501S}, making use of the alignment-$\beta$ correlation. Using combinations of their measurements and $\beta-\alpha$ correlation data from \citet{1980PhRvC..22..738M}, they were able to extract some of the recoil-order matrix elements. Calculations of these matrix elements in the symmetry-adapted no-core shell model were recently performed by \citet{2022PhRvL.128t2503S}. 
The comparison to the experimentally derived matrix elements of \citet{1980PhRvC..22..738M}, \citet{1995PhRvC..51.2778D}, and \citet{2011PhRvC..83f5501S} is not straightforward because the experiments quote averages over the whole beta spectrum as opposed to fits for each of the four levels used in the calculation. A more detailed comparison of theory to experiment that takes into account the averaging would be helpful.
An additional point of comparison that has been used previously is the beta spectrum from $^8{\rm B}$~\cite{1996PhRvC..54..411B}. An experiment with improved statistics and systematics that extends over the  full range of beta energies would be very useful. The existing data from \citet{1987PhRvC..36..298N} is in good agreement with the alpha spectrum from Winter et al. But given the apparent 20 keV difference that the newer alpha-spectrum experiments described above report, it would be good to have a modern beta spectrum measurement specifically designed to address this difference.

\subsection{Experimental program on solar neutrinos}

A broad range of technology can be used to interrogate solar neutrinos.  Radiochemical experiments utilizing chlorine and gallium played a crucial role in the early days of the solar neutrino problem.  Large monolithic detectors have achieved great success in real-time measurements -- from the water Cherenkov detectors such as Kamiokande, SNO, and SuperK, which could use the unique topology of Cherenkov light to point back to the neutrinos' origin, to liquid scintillator detectors such as KamLAND and Borexino, whose high light yield and low threshold allow for precision spectroscopy.

\subsubsection{Super-Kamiokande and Hyper-Kamiokande}
The first real-time detection of solar neutrinos was achieved by the Kamiokande experiment \cite{1989PhRvL..63...16H}, which detected the neutrinos via elastic scattering of electrons, a process in which the outgoing electron's direction is highly correlated to that of the incoming neutrino.  This allowed Kamiokande to directly point back to the neutrinos' origin: our Sun.  The successor experiment, Super-Kamiokande, is a 50-kton volume of pure water, surrounded by over 10,000 photon detectors, which has achieved an unparalleled program of neutrino observations and other physics over its several decades of operation.  This program has included the highest-precision measurement of the elastic scattering signal from solar neutrino interactions, as well as sensitive searches for the day/night distortions of the spectral shape \cite{2016PhRvD..94e2010A}.  Now filled with a gadolinium additive to enhance neutron capture, SuperK's primary current focus is the search for the Diffuse Supernova Neutrino Background (DSNB) via inverse beta decay \cite{2023ApJ...951L..27H}.  It also continues its atmospheric neutrino program: 
SuperK measurements made 25 years ago demonstrated the neutrino oscillations were responsible for the puzzling zenith-angle-dependence 
of this flux \cite{PhysRevLett.81.1562}.

In parallel to ongoing operation of SuperK, an even larger sister project is under construction.  At over 250 kton in total mass, and with improved light collection relative to SuperK, Hyper-Kamiokande (HyperK) will impact statistics-limited searches the involve the higher energy $^8$B solar neutrinos, including day/night effects, the search for hep neutrinos, and the shape of the $^8$B neutrino spectrum.
After 10 years of operations, HyperK will reach a sensitivity to day/night effects $\gtrsim$ 4 (8) $\sigma$, given the oscillation parameters deduced from reactor (solar) experiments.  HyperK will serve as the far detector for JPARC's
long-baseline neutrino program.

\subsubsection{Sudbury Neutrino Observatory and SNO+}
SNO's unique use of heavy water as a target medium offered the ability to detect solar neutrinos via two additional interactions besides elastic scattering. In the charged current interaction an electron neutrino interacts with the neutron in the deuteron, producing an electron and a proton.  This interaction path is sensitive only to electron-flavor neutrinos at the few-MeV energy scale of solar neutrinos, providing a measurement of the pure $\nu_e$ flux.  SNO also had access to the neutral current interaction, in which a neutrino of any flavor interacts with the deuteron, breaking it apart into its constituent nucleons.  Being equally sensitive to all active flavors, the neutral current interaction yields a measurement of the total, flavor-blind neutrino flux.  It was this capability that allowed SNO to resolve the solar neutrino problem, demonstrating that most of the $\nu_e$s produced by the Sun were arriving at earth in a different flavor state \cite{2001PhRvL..87g1301A,2002PhRvL..89a1301A,2004PhRvL..92r1301A}.

SNO ceased data taking in 2006.  Analysis of the data set has continued, and produced a number of new results, including constraints on non-standard effects such as Lorentz violation and neutrino decay \cite{2018PhRvD..98k2013A,2019PhRvD..99c2013A}, limits on the fluxes of hep neutrinos and the diffuse supernova neutrino background (DSNB) \cite{2020PhRvD.102f2006A}, and neutron production from cosmogenic muons and from atmospheric neutrino interactions \cite{2019PhRvD..99k2007A,2019PhRvD.100k2005A}.

After decommissioning, the detector was repurposed for the SNO+ experiment, in which the target material was replaced with a pure organic liquid scintillator: linear alkyl benzene (LAB), loaded with 2.2 g/L of the secondary fluor, PPO.  The high light yield of this scintillator, coupled with the location in SNOLAB --at 6-km water equivalent, one of the deepest underground laboratories in the world-- and an extremely well-understood detector, allow for a range of high-precision measurement programs.  In a preliminary water phase, SNO+ demonstrated detection of neutron captures on hydrogen \citep{PhysRevC.102.014002}, an impressive technical achievement in an unloaded water detector; a low background measurement of $^8$B solar neutrinos \citep{PhysRevD.99.012012}; several searches for invisible modes of nucleon decay \citep{PhysRevD.99.032008,PhysRevD.105.112012}; and detection of antineutrinos from reactors over 200~km away \citep{PhysRevLett.130.091801}.  The low backgrounds achieved and the detector's demonstrated technical capabilities set the stage for future measurements.  In a ``partial-fill'' stage, in which the upper half of the detector was filled with liquid scintillator while the lower half still contained water, the first demonstration of event-level direction reconstruction was achieved for $^8$B neutrinos in a scintillator detector \citep{PhysRevD.109.072002}.  This was facilitated by a lower loading of PPO at that time, which results in lower scintillation yield and a slower time profile, effectively enhancing the Cherenkov component, which can be leveraged for directional information.

Now fully filled with liquid scintillator, SNO+ is the deepest, largest operating liquid scintillator detector in the world.  The future program will include measurements of several solar neutrino fluxes, as well as the $^8$B neutrino energy spectrum.

\subsubsection{Borexino \label{subsec:WG2:Borexino}}
Water Cherenkov experiments such as Kamiokande, SuperK and SNO provided the first real-time measurement of solar neutrinos but their relatively high energy threshold made them sensitive only to a small fraction of the solar neutrino flux. To study in real time the bulk of solar neutrino emission a different detector technology was required.
The Borexino experiment, located deep underground at the INFN Laboratori Nazionali del Gran Sasso, used an organic liquid scintillator target made of pseudocumene with \SI{1.5}{g/L} PPO to detect the elastic scattering of solar neutrinos off electrons. The high light yield of the liquid scintillator made it possible to significantly lower the energy threshold, but given the small signal rate and the lack of a clear signature to separate it from the background (such as the direction indicated by the emission of Cherenkov radiation), the experiment required a long preparatory phase to develop the most advanced techniques to suppress the background, especially the one due to radioactive contamination of the liquid scintillator itself.

The detector's extreme radiopurity was key to its success, and over its 14 years of data taking, ending in October 2021,  Borexino proved itself capable of covering the entire solar neutrino spectrum. 
At the beginning of its data taking in 2007 the level of $^{238}$U and $^{232}$Th contamination in Borexino were lower than \SI{2e-17}{} and \SI{7e-18}{g/g} respectively \cite{2008PhLB..658..101B}, paving the way for the first measurement of the $^{7}$Be sub-MeV solar neutrinos, followed by a low-threshold measurement of the $^{8}$B flux and a first indication of the pep neutrinos. 
After a purification campaign that further reduced the liquid scintillator contamination, Borexino aimed at improving the accuracy of its results in the measurement of the pp-chain solar neutrinos. 
The flux of neutrinos produced in the pp fusion process was first reported in \citet{2014Natur.512..383B} and later improved, along with all the other fluxes from the pp chain, apart from the hep neutrinos \cite{2018Natur.562..505B,2019PhRvD.100h2004A, 2020PhRvD.101f2001A}.
The pp flux was determined to an uncertainty of 11\%, improving the neutrino-based estimate of solar luminosity, while $^{7}$Be neutrinos were measured with an uncertainty of $2.7\%$, half that of
the SSM prediction.
Furthermore, using the SSM to constrain the flux of CNO neutrinos, 
the pep neutrino signal was established with 
a significance exceeding $5\sigma$ for the first time, and the $^{8}$B flux 
was  measured with a threshold as low as 3\,MeV. 

After completing the investigation of the pp-chain neutrinos, Borexino reported the first detection of solar neutrinos produced in the CNO-cycle \cite{2020Natur.587..577B}, demonstrating directly that this mode of hydrogen burning operates in stars.
Profiting from a larger exposure and a better understanding of the radioactive backgrounds enabled by the unprecedented thermal stability of the detector, Borexino further improved its measurement of the CNO neutrino flux \cite{2022PhRvL.129y2701A,2023PhRvD.108j2005B}, where the new CNO result was used in combination with existing solar neutrino results to probe solar composition.

Borexino's physics program was not limited to solar neutrinos: its outstanding radiopurity made it a excellent detector for geoneutrinos \cite{2020PhRvD.101a2009A} and for searches for various rare processes.
Borexino data were used to constrain exotic properties of neutrinos, such as their magnetic moments \cite{2017PhRvD..96i1103A} and non standard interactions with matter \cite{2020JHEP...02..038B}.

\subsubsection{JUNO}
JUNO \cite{2015arXiv150807166A} is a large liquid scintillator detector currently under construction in an underground laboratory with a vertical overburden of about 650~m (roughly 1800~m water equivalent) in Jiangmen city in Southern China.
JUNO is located 52.5~km from two nuclear power plants, a
baseline optimized for JUNO's primary goal, the determination of the neutrino mass
ordering [28, 29].
To achieve this, JUNO requires a large target mass (20 kton) and excellent energy resolution, attributes also import in
solar neutrino detection.  The relatively shallow overburden limits the solar neutrino program, due to cosmogenic activity, but the low threshold and large detector mass may allow measurements of day/night effects, non-standard neutrino interactions affecting the $^8$B spectral shape, and $^7$Be neutrinos, as well as providing a new solar determination of $\Delta m^2_{12}$.  JUNO's reactor data will constrain $\Delta m^2_{12}$ to extremely high precision, allowing comparisons between the solar and reactor determinations of this parameter.  To the extent that oscillation effects can be treated with greater confidence, the connections between solar neutrino flux measurements and the solar fusion reactions generating those fluxes will be sharpened \cite{2016JPhG...43c0401A}.

\subsubsection{DUNE}
As recommended in the 2023 P5 report \cite{P5Report}, 
the DUNE Collaboration seeks to construct two 10-kton fiducial liquid argon time projection chambers (LArTPC), deep underground in the Homestake mine in South Dakota \cite{2016arXiv160102984A}.  A third LArTPC will follow in Phase II, along with other upgrades. The newly developed Sanford Underground Research Facility (SURF) offers 4800-m water equivalent overburden, and will form the far site for a long baseline neutrino program utilizing a high-energy neutrino beam directed from Fermilab to SURF.  The DUNE LArTPC detectors may be sensitive to measurements of the high-energy solar neutrinos via charged current interactions on argon, which offers good precision on the spectral shape and, thus, the potential to measure day/night effects \cite{2019PhRvL.123m1803C}.  The solar neutrino program at DUNE is limited primarily by backgrounds (the detector is optimized for GeV-scale physics) and energy threshold. Multiple technologies are under consideration for a fourth detector module, with the motivation of expanding the
physics program, including improved sensitivity to solar neutrinos.  Use of underground argon, or alternative technologies such as organic or
water-based scintillators, could preserve the long baseline neutrino sensitivity while also opening up a rich program of low-energy physics.

\subsubsection{Future prospects}
A ``hybrid'' detector that could utilize both Cherenkov and scintillation light simultaneously could achieve unprecedented levels of particle and event identification and, hence, background rejection \cite{2014arXiv1409.5864A}.  The Cherenkov signature offers directional information, while the high light yield scintillation offers precision energy and vertex reconstruction.  A full waveform analysis of detected light offers yet more information, based on the pulse shape of the scintillation, which is subject to species-dependent quenching effects, and the impact of the Cherenkov threshold.  As a result, both the shape of the waveform and the ratio of the two signals will differ for different particle types.  A full-scale detector utilizing novel scintillators along with fast and spectrally sensitive photon detectors, such as the proposed \textsc{Theia} experiment \cite{2020EPJC...80..416A}, or the Jinping detector \cite{2017ChPhC..41b3002B} could achieve percent-level precision on the CNO neutrino flux, as well as improving precision across the suite of solar neutrino measurements, such as the $^8$B spectral shape and hep neutrino flux.

As experiments grow larger, and capabilities increase, there are also opportunities to apply detectors designed for other purposes to the detection of solar neutrinos.  Noble liquid detectors built primarily for dark matter searches and long-baseline neutrino experiments have now reached the size and background levels to be sensitive to solar neutrinos~\cite{2024arXiv240802877A}.  The detection of neutrinos from the high-energy tail of the $^8$B spectrum via coherent elastic neutrino-nucleus scattering (``CE$\nu$NS'') is particularly interesting because it is a purely neutral-current process sensitive to the total flux of active neutrinos (above the experimental threshold, about 8 MeV) from the Sun.  Only the SNO experiment has had such a capability. Initial results reported in~\cite{2024arXiv240802877A}  are in agreement, with limited statistical precision.   Solid state detectors built purposely for observing CE$\nu$NS have observed neutrinos from stopped pion decay at the theoretically expected rate \cite{doi:10.1126/science.aao0990}.



\section{The \texorpdfstring{$^1$H$(p,e^+\nu)^{2}$H reaction ($S_{11}$)}{1H(p,e+ν)2H reaction (S11)}}
\label{sec:S11}

The cross section for the initial reaction in the pp chain (Figure \ref{fig:pp-Chain}),  $p+p\rightarrow \mathrm{d}+e^{+}+\nu _{e}$, is too small to be measured in the laboratory. It must be calculated from the standard theory of weak interactions. 

\subsection{Introduction and terminology}

Near the Gamow peak energy $E \sim$ 6 keV for temperatures characteristic of the solar core, the first and second derivatives of the astrophysical $\mathrm{S}$ factor at zero energy, ${S}^{\prime }_{11}(0)$ and ${S}^{\prime \prime}_{11}(0)$, generate $\sim$ 7\% and $\sim$ 0.5\% corrections, respectively, in Taylor's series expansion of $S_{11}(E)$ around $E=0$.  Higher derivative terms are neglected in this review since they only contribute at the $10^{-4}$ level.
The recommended values for ${S}^{\prime }_{11}(0)$ and ${S}^{\prime \prime}_{11}(0)$ are discussed in Section~\ref{subsec:WG1:Sprime}. Here we focus on $S_{11}(0)$.

At zero relative energy, $S_{11}(0)$ can be
written as \cite{1968ApJ...152L..17B,1969ApJ...155..501B}, 
\begin{equation}
\label{S11_def}
S_{11}(0)=6\pi ^{2}m_{p}\alpha \ln 2\,{\frac{\overline{\Lambda }^{2}}{\gamma ^{3}
}}\left( {\frac{G_{A}}{G_{V}}}\right) ^{2}{\frac{f_{pp}^{R}}{
(ft)_{0^{+}\rightarrow 0^{+}}}},
\end{equation}
where $\alpha=1/137.04$ is the fine-structure constant, $m_{p}=938.272$ MeV is the proton mass, 
$\gamma =(2\mu B_{d})^{1/2}=0.23161$~fm$^{-1}$ is the deuteron
binding wave number, $\mu $ being the proton-neutron reduced mass and $B_{d}$
the deuteron binding energy,
$G_{V}$ and $G_{A}$ are the Fermi vector and axial-vector weak coupling
constants. Finally,  $f_{pp}^{R}$ is the phase-space factor for
the pp reaction with radiative corrections, $(ft)_{0^{+}\rightarrow 0^{+}}$
is the $ft$ value for superallowed $0^{+}\rightarrow 0^{+}$ transitions, and  $
\overline{\Lambda }$ is proportional to the transition matrix element
connecting the pp and deuteron states.

\subsection{Adopted parameters for this review}

For the phase-space factor $f_{pp}^{R}$, we use the same value $f_{pp}^R=0.144(1\pm 0.001)$ 
as in \citetalias{2011RvMP...83..195A}. It comes from the value without
radiative corrections, $f_{pp}=0.142$ \cite{1969ApJ...155..501B}, increased
by 1.62\% to take into account radiative corrections to the cross section
\cite{2003PhRvC..67c5502K}. The main source of uncertainty in $f_{pp}^R$ arises from neglected
diagrams 
in which the lepton exchanges a weak boson and a photon with different
nucleons. These diagrams are estimated to modify $f_{pp}^R$ by $\sim 0.1\%$, based
on scaling the similar nucleus-dependent correction in superallowed $\beta $
decay \cite{2003PhRvC..67c5502K}. Direct computations of these diagrams were recommended in Solar 
Fusion II. Here we urge again that this computation be carried out.  

For $G_{A}/G_{V}$, we use the value from the Particle Data Group (PDG) compilation \cite{2022PTEP.2022h3C01W}, 
$1.2754\pm 0.0013$, whose central value is larger than the 2008 PDG value, $1.2695\pm 0.0029$, used in \citetalias{2011RvMP...83..195A} by $0.45\%$ (or 1.9 $\sigma$). 
Naively, this would lead to a $\sim 1\%$ increase in the central value of $S_{11}(0)$ according to Eq.~(\ref{S11_def}). Its effect will be discussed below.
For $(ft)_{0^+\rightarrow 0^+}$, we take $3072.24\pm 1.85$ s from the most updated comprehensive analysis of experimental rates with the radiative and Coulomb
effects corrected \cite{2020PhRvC.102d5501H}. This value is consistent with $3071.4\pm 0.8$ s \cite{2009PhRvC..79e5502H}
used in \citetalias{2011RvMP...83..195A} with a larger error. 

The dominant uncertainty in $S_{11}(0)$ comes from the normalized Gamow-Teller
(GT) matrix element $\overline{\Lambda }$. Reducing this uncertainty has been the main focus of theoretical work since \citetalias{1998RvMP...70.1265A}. 
In \citetalias{1998RvMP...70.1265A}, $\overline{\Lambda }$ was written as $\overline{\Lambda }=\Lambda+\delta \Lambda$, 
i.e. the sum of the one- and two-body current matrix elements, $\Lambda $ and $\delta \Lambda$, respectively, with their uncertainties estimated independently. 
In \citetalias{2011RvMP...83..195A}, two major steps had contributed to reducing the uncertainty on $\overline{\Lambda }$. The first was a much deeper understanding
of the correlation between the uncertainties in $\Lambda $ and $
\delta \Lambda $:  the overall uncertainty in $\overline{\Lambda }$
could be described by a universal parameter that could be fixed by a single measurement.  
The study of \citet{1998PhRvC..58.1263S} demonstrated this
phenomenologically in the context of potential-model approaches,
while later analysis via effective field theory (EFT) provided a more formal
justification \cite{2001PhRvC..63c5501B,2003PhRvC..67e5206P}. 
The second step was the
use of the precisely known tritium $\beta $\ decay rate $\Gamma _{\beta
}^{T}$  to fix this universal
parameter, as first proposed by \citet{1991PhRvC..44..619C}. This has been done in both potential models \cite%
{1998PhRvC..58.1263S} and in the hybrid EFT approach \cite{2003PhRvC..67e5206P} as explained below. 

In \citetalias{2011RvMP...83..195A}, $\overline{\Lambda }$ was determined with three approaches. 
The first one was the potential model approach. In the most elaborate 
calculation for the pp fusion process, a comparison of the results for five representative modern potentials designed to accurately reproduce nucleon-nucleon scattering data was carried out \cite{1998PhRvC..58.1263S}. After adjusting the unknown strength of
the two-body exchange currents to reproduce $\Gamma _{\beta }^{T}$, 
the variation in $S_{11}(0)$ that otherwise would come from
the choice of the phenomenological potential was largely removed. 
Predictions for five representative high-precision phenomenological
potentials fell in a narrow interval $7.03 \lesssim \overline{\Lambda }^{2} \lesssim 7.04$. There were additional uncertainties in the three-body potentials and three-body currents in $\Gamma _{\beta }^{T}$, of the order of $\sim 0.8\%$, and a 0.5\% uncertainty due to effective range parameters for nucleon-nucleon scattering. Hence, the recommended $S_{11}(0)$ value from the potential model approach was $S_{11}(0)=4.01(1\pm0.009)\times 10^{-25}\,{\rm MeV~b}$.

The second and third approaches were both based on EFT. The second one was a hybrid EFT (EFT*), which used the current operators derived from EFT in conjunction with the initial and final state wave functions generated by a potential model \cite{2003PhRvC..67e5206P}. For pp fusion, the relevant two-body current contained only one unknown low-energy constant (LEC) $\hat{d}^{R}$ which parameterized the 
contact axial coupling to two nucleons
\cite{2003PhRvC..67e5206P}.
A weakness of this approach was the mismatch between the operators and wave functions. However, it was argued that the mismatch only happened for short distance physics which could be absorbed by the LECs. Hence, when the ultraviolet cutoff was changed over a physically reasonable range, the residual cutoff dependence of physical observables provided a measure of the model dependence of the EFT* calculation. By combining the $0.8\%$ error from changing the cutoff $\Lambda_{NN}$ in the range of $500-800$ MeV and the $\sim$ 0.4\% higher order correction, obtained by multiplying the $1.8\%$ contribution of the highest calculated order with the small expansion parameter $m_{\pi}/\Lambda_{NN} \sim 1/4$, \citet{2003PhRvC..67e5206P} provided 
the value $S_{11}(0)=4.01(1\pm0.009)\times 10^{-25}\,{\rm MeV~b}$,
in perfect agreement with the one obtained within the phenomenological approach.

The third study was performed with the pionless EFT ($\slashed{\pi}$EFT) approach. It is a framework applicable to processes with the 
characteristic momentum $p$ much smaller than the pion mass $m_{\pi }$, such that the pion field can be ``integrated out'' and becomes non-dynamical 
(\citealp{1996APS..DNP..BA03K}; \citealp{1999PhRvL..82..463B}; \citealp{1999NuPhA.653..386C}).
In this approach, all nucleon-nucleon
interactions and two-body currents are described by point-like contact
interactions with a systematic expansion in powers of $p/m_{\pi }$. 
For all the deuteron weak breakup processes (e.g. $\nu d$ and $\bar{\nu} d$ scattering) and their inverse processes, including the pp fusion, only one two-body current (with coupling $L_{1,A}$) is needed up to next-to-next-to-leading order (N$^2$LO) \cite{2001PhRvC..63c5501B}. 
Therefore a single measurement will fix $L_{1,A}$ and the rates of all such processes. 
This feature is shared by the other approaches discussed above.
The computation of $\overline{\Lambda }$ in $\slashed{\pi}$EFT was carried out to
the second order in the $p/m_{\pi }$ expansion by \citet{2001PhRvC..64d4002K} 
and then to the fifth order by \citet{2001PhLB..520...87B}. Constraints on $L_{1,A}$ from two-nucleon systems \cite{2002PhLB..549...26B,2003PhRvC..67b5801C} yielded  
$S_{11}(0)=3.99(1\pm0.030)\times 10^{-25}\,{\rm MeV~b}$.

Based on the consistent results of the above three approaches, \citetalias{2011RvMP...83..195A} recommended $S_{11}(0)=4.01(1\pm0.009)\times 10^{-25}\,{\rm MeV~b}$. The \citetalias{1998RvMP...70.1265A} value
${S}^\prime_{11}(0)=S_{11}(0) (11.2\pm0.1)\,{\rm MeV}^{-1}$ from \citet{1969ApJ...155..501B} was not re-computed in \citetalias{2011RvMP...83..195A}. However, with $S_{11}(0)$ reaching a $1\%$ accuracy, new calculations of $S^{\prime}_{11}(0)$ and $S^{\prime \prime}_{11}(0)$,
together with the full pionful chiral EFT ($\chi$EFT) computations to remove the unknown systematics of the hybrid EFT,    
were called for in \citetalias{2011RvMP...83..195A}. These two challenges have been met in SF~III and will be described below.

\subsection{Experimental progress on muon capture of the deuteron}

In \citetalias{2011RvMP...83..195A} it was also recommended to carry out the experimental determination of the 
muon capture rate on deuteron, as proposed in the MuSun experiment \cite{2003nucl.ex...4019K,2017APS..DNP.CG007S}. This quantity could be used to
constrain $S_{11}(0)$
without the need to rely on the three-body calculation of $\Gamma_\beta^T$. The theoretical calculations for the muon capture rate on deuteron have been carried out in $\slashed{\pi}$EFT \cite{2005PhRvC..72f1001C} (it is possible to impose neutron energy cut to isolate the low-energy neutron events so $\slashed{\pi}$EFT is applicable \cite{2003nucl.ex...4019K}), chiral hybrid EFT* \cite{2002PhLB..533...25A}, phenomenological potential model \cite{2011PhRvC..83a4002M}, and recently in $\chi$EFT 
(\citealp{2012PhRvL.108e2502M,2018PhRvL.121d9901M}; \citealp{2018PhRvC..98f5506A}; \citealp{2023FrP....1049919C}; \citealp{2023PhRvC.107f5502B};  \citealp{PhysRevC.109.035502}).
However, the MuSun result is yet to be released.
On the other hand, we should notice that the muon capture processes happen at a rather large momentum transfer compared to pp fusion. The momentum transfer dependence of the single nucleon axial coupling constant $g_A(q^2)$, with $g_A(q^2=0)\equiv g_A\equiv G_A/G_V$, has been recently studied by \citet{2018RPPh...81i6301H}, who have provided an experimental determination for the axial charge radius, given by $r_A^2=0.46(16)$ fm$^2$, where $r_A$ is defined by the relation 
$g_A(q^2)=g_A(1-r_A^2 q^2/6)$ for small $q^2$. The $\sim 30$\% uncertainty on $r_A^2$ has an impact on the ability of the MuSun experiment alone to directly constrain $S_{11}(0)$~\cite{2018PhRvC..98f5506A,2023FrP....1049919C,2023PhRvC.107f5502B,PhysRevC.109.035502}.

In the next few years, lattice Quantum Chromodynamics (QCD) calculations of $g_A(q^2)$ are expected to reduce the $r_A$ uncertainty by a factor of two or more (see \citet{2023slft.confE.240M} for the most recent review of the lattice results). Currently, lattice QCD results are consistent with \citet{2018RPPh...81i6301H} on $r_A$, often with comparable uncertainties. At larger $q^2$ ($|q^2|\gtrsim 0.25$ GeV$^2$), there is a growing tension between lattice QCD predictions of $g_A(q^2)$ and the phenomenological determination from older neutrino-deuterium bubble chamber data \cite{2016PhRvD..93k3015M} with the lattice QCD results yielding a 30\% larger neutrino-nucleon cross section over a large range of $q^2$ \cite{2022ARNPS..72..205M}. 

In any case, it is evident that accurate experimental determinations of muon capture rates on the deuteron and other light nuclei, which can be addressed theoretically \textit{ab initio} approaches, represent fundamental tests for the theoretical approaches themselves, either within $\chi$EFT or $\slashed{\pi}$EFT, and might be able to provide the necessary experimental information to fix the unknown parameters of the theory.

\subsection{Progress in \texorpdfstring{$S_{11}(0)$}{S11(0)} calculations since SF~II}
\label{subsection:WG1:S11}
Below we summarize the $S_{11}(0)$ calculations performed after \citetalias{2011RvMP...83..195A} using different approaches. 

\subsubsection{\texorpdfstring{$\chi$EFT}{χEFT}}
\label{subsubsec:WG1:chiEFT}

The pioneering work of \citet{2013PhRvL.110s2503M} used the next-to-next-to-next-to-leading-order (N$^3$LO) chiral two-nucleon potential 
\cite{2003PhRvC..68d1001E,2011PhR...503....1M} augmented with higher order ($\mathcal{O}(\alpha^2)$) two-photon and vacuum-polarization electromagnetic interactions. These $\mathcal{O}(\alpha^2)$ corrections reduced $S_{11}(0)$ by $\sim 0.8$\% mainly due to the vacuum-polarization-induced pp wave function distortion. This was consistent with the 0.84\% first found in the potential model calculation of \citet{1998PhRvC..58.1263S}. This correction was also included in the EFT* calculation of \cite{2003PhRvC..67e5206P}.
The relevant LECs 
were fitted to reproduce the $A=3$ binding energies, magnetic moments, and GT matrix element in $\Gamma _{\beta}^{T}$ to obtain $S_{11}(0)=4.030(1\pm0.006)\times 10^{-25}\,{\rm MeV~b}$, with the $p$-wave initial state contributing at $\sim 1\%$, which was about the accuracy level of the calculation
\cite{2013PhRvL.110s2503M}. 
However, using
$\slashed{\pi}$EFT at next-to-leading-order (NLO), 
\citet{2019PhRvC.100b1001A} later found that $p$-wave only contributed at 
the order of $10^{-30}\,{\rm MeV~b}$. 
Re-examining the computer programs, \citet{2013PhRvL.110s2503M} found an error in the determination of one of the $p$-wave reduced matrix elements (associated with the longitudinal multipole operator). Consequently, \citet{2019PhRvL.123a9901M} reported in the Erratum that 
$S_{11}=4.008(1\pm0.005)\times 10^{-25}\,{\rm MeV~b}$.

In more recent work, \citet{2016PhLB..760..584A} used chiral interactions and consistent currents up to N$^3$LO (called next-to-next-to-leading-order (N$^2$LO) in the original literature since the second order vanished) and developed
a robust procedure for the error quantification. In particular, they analyzed a family of 42 interactions \cite{2016PhRvX...6a1019C} with 7 different cutoff values from 450 to 600 MeV. The 26 LECs were fitted to 6 different pools of input data including nucleopn-nucleon ($NN$) and $\pi N$ scatterings, as well as the binding energies and charge radii of
$^{3}$H and $^3$He,
the quadrupole moment of $^2$H, and $\Gamma _{\beta}^{T}$.  This thorough study yielded $S_{11}(0)=4.047(1^{+0.006}_{-0.008})\times 10^{-25}\,{\rm MeV~b}$.

In both of the $\chi$EFT calculations mentioned above, a widely used relation first proposed by \citet{2009PhRvL.103j2502G} that linked the two-body axial current LEC $\hat d_R$ with the LEC $c_D$ from the $\pi NN$ vertex was employed.  
However, later \citet{2018PhRvL.121d9901M} found that there was a factor $-1/4$ missing in this $\hat d_R$-$c_D$ relation which was then acknowledged in the Erratum of \citet{2019PhRvL.122b9901G}. Fortunately, this error was unimportant in muon capture on deuteron~\cite{2018PhRvL.121d9901M} and
it only affected $S_{11}(0)$ at the $0.1\%$ level \cite{2023arXiv230403327A}.

In the most recent and comprehensive $\chi$EFT study, \citet{2023arXiv230403327A} compared the above calculations of \citet{2019PhRvL.123a9901M} and \citet{2016PhLB..760..584A}  in detail. In addition to the $0.1\%$ increase of $S_{11}(0)$ from using the correct $\hat d_R$-$c_D$ relation, updating the input parameters to their most recent values increased $S_{11}(0)$ by $\sim 1\%$ in both calculations, mainly due to the $0.45\%$ increase of $G_{A}/G_{V}$ from its \citetalias{2011RvMP...83..195A} value mentioned above. Furthermore, \citet{2019PhRvL.123a9901M} received a $\sim 1\%$ increase from removing the truncation error of the basis functions which effectively cut off the long distance part of the wave functions, as \citet{2017PhRvC..95c1301A} advocated. 
After these corrections, the $\sim 1\%$ difference between \citet{2019PhRvL.123a9901M} and \citet{2016PhLB..760..584A} was reconciled and the combined result was found to be \cite{2023arXiv230403327A}
\begin{equation}
S_{11}(0)=4.100(1\pm0.007)\times 10^{-25}\,\,{\rm MeV~b} \,\, .
\label{ChEFTS11}
\end{equation}

\citet{2023arXiv230403327A} obtained consistent values of $S_{11}(0)$ using four different $\chi$EFT models for the nuclear interaction. This value is also in agreement with the result obtained in $\slashed{\pi}$EFT by \citet{2022arXiv220710176D} (see below). In estimating the order-by-order convergence, however, \citet{2023arXiv230403327A} used one of the $\chi$EFT models, which is able to nicely reproduce the deuteron properties already at leading order. Therefore, the $\chi$EFT error of Eq.~(\ref{ChEFTS11}) is likely to be an underestimate, warranting the enlarged error advocated in Section \ref{subsubsec:WG1:final}. 

Finally, we would like to mention the work of \citet{2022PhRvC.106e5501L}, where the power counting of the $\chi$EFT weak current operator involved in the pp reaction is revisited using renormalization group (RG) invariance as the guideline. In particular, it is argued that the contact two-body axial current proportional to the $\hat d_R$ LEC must appear one order lower than assessed by naive dimensional analysis. Then it can be shown that RG invariance is fulfilled at N$^2$LO. However, the estimate for ${\overline{\Lambda}}$ obtained by \citet{2022PhRvC.106e5501L} does not use a value for $\hat d_R$ obtained by fitting $\Gamma_\beta^T$, but extracted in order to match the value of ${\overline{\Lambda}}$ obtained in \citetalias{2011RvMP...83..195A}. Therefore, we will not consider the work of \citet{2022PhRvC.106e5501L} in the present $S_{11}(0)$ evaluation.

\subsubsection{\texorpdfstring{$\slashed{\pi}$EFT}{Pionless EFT}}
\label{subsubsec:WG1:nopiEFT}

The universal two-body current coupling $L_{1,A}$ was determined using $\Gamma _{\beta}^{T}$ in $\slashed{\pi}$EFT for the first time
by \citet{2019PhRvC.100e5502D}. 
This calculation was carried out up to NLO using the dibaryon formulation of \citet{2001NuPhA.694..511B}, which partially resumed higher order effective range contributions to improve the convergence. Then, this result was used by \citet{2022arXiv220710176D}, with updated input parameters, to obtain
\begin{equation}
\label{nopiEFTS11}
S_{11}(0)=4.12(1\pm0.015)\times 10^{-25}\,{\rm MeV~b} , 
\end{equation}
where the $\mathcal{O}(\alpha^2)$ electromagnetic correction was not calculated but assumed to be the same as the potential model value, $0.84\%$ \cite{1998PhRvC..58.1263S}. While this number was not model independent, the model dependence was believed to be well below the assigned $1.5\%$ error.

The small error assigned to this NLO result has been justified by drawing an analogy from the corresponding electromagnetic processes. Using the same approach, the $np\to d\gamma$ matrix element at threshold was predicted at NLO within 0.5\% to the experimental value after the electromagnetic two-body current $L_1$  was fit to the magnetic moments of $^3$He and $^3$H. This indicates that the contribution of the three-body current at N$^2$LO is small in this electromagnetic case. The weak sector is shown to follow the same operator structure and hence provides support for the calculation procedure and the uncertainty estimate.

It is important to note that the reason that the three-body current is an N$^2$LO effect in $\slashed{\pi}$EFT is related to the non-trivial renormalization of the non-derivative three-body contact interaction, which shows up at LO to absorb the cutoff dependence of Feynman diagrams. The subleading two-derivative three-body contact interaction is expected to show up at N$^2$LO. This interaction, combined with the one-body current, renormalizes the three-body current. Hence, the three-body current should also appear at the same order, N$^2$LO.
However, if the non-derivative three-body contact interaction were counted as higher order, such as N$^3$LO, as in $\chi$EFT (because cutoff independence is not strictly enforced order by order in $\chi$EFT), then the three-body current would contribute at much higher order. Although the $\chi$EFT power counting indeed yields good convergence in the expansions, it is unsatisfactory that one can not remove the cutoff dependence at each order of the expansion. In addition, the uncertainty estimate of the $\chi$EFT still lacks a broad inspection of the specific nuclear $\chi$EFT potential implementation. As a consequence, we consider the $\slashed{\pi}$EFT 1.5\% error a better estimate for the $S_{11}(0)$ theoretical uncertainty.

\subsubsection{Lattice QCD and Lattice EFT}
\label{subsubsec:WG1:LQCD}

Ideally lattice QCD would provide a first principles prediction of the pp fusion rate and the GT matrix element of $\Gamma _{\beta }^{T}$, for both pure QCD and with QED effects incorporated.
However, such calculations are quite challenging and not yet available at the required precision.
A proof-of-principle calculation was carried out by \citet{2017PhRvL.119f2002S} using a background field method to determine both the pp fusion GT matrix element and $\Gamma _{\beta }^{T}$.  This exploratory calculation utilized a single pion mass at the SU(3)-flavor--symmetric point with $m_\pi\approx806$~MeV, a single and relatively coarse lattice spacing of $a\approx0.145$~fm, and a single volume.
The calculation was performed under the assumption that the two- and three-nucleon systems were deeply bound.  Without this assumption, matrix elements computed in the finite volume can be significantly different from those in infinite volume, due to Lellouch-L\"uscher factors~\cite{2001CMaPh.219...31L} that lead to power-law finite volume corrections.  These can range from the few-percent level to $\mathrm{O}(1)$~\cite{2013PhRvD..88i4507B,2015PhRvL.115x2001B}.

More recent lattice QCD calculations have found that two-nucleon systems at heavy pion masses are in fact not bound
(\citealp{2016JHEP...10..101I}; \citealp{2019PhRvD..99g4505F}; \citealp{2021PhRvC.103a4003H}; \citealp{2021arXiv210810835A}).
These efforts, which employ interpolating operators more sophisticated than those of \citet{2017PhRvL.119f2002S}, suggest there could be large systematic uncertainties affecting the conclusions of \citet{2017PhRvL.119f2002S} stemming from misidentification of the spectrum and inaccurate Lellouch-L\"uscher factors.
In addition, \citet{2021PhRvL.127x2003G} found that the two-baryon spectrum may be particularly sensitive to discretization effects, which would also have important implications for the continuum extrapolation of the matrix elements.  These issues are discussed in some detail in \citet{2022FBS....63...67T}.  Finally, results with pion masses $m_\pi \lesssim 300$~MeV are needed for accurate extrapolation to the physical pion mass.

Given these unresolved systematic issues, the result of \citet{2017PhRvL.119f2002S} is not included in the present $S_{11}(0)$ evaluation -- even though the
extracted value $S_{11}(0)=4.07(1\pm0.008)\times 10^{-25}\,{\rm MeV~b}$ is consistent with our recommended range quoted in Section~\ref{subsubsec:WG1:final}.

The lattice EFT computation of \citet{2015PhLB..741..301R} performed the pp fusion calculation by implementing $\slashed{\pi}$EFT on a spacetime lattice. The purpose of this leading-order study was to demonstrate that lattice EFT could reproduce the infinite volume and continuum result of $\slashed{\pi}$EFT such that it could be applied to various reactions of astrophysical interest in the future. Therefore, for 
$S_{11}(0)$, this result is considered as a subset of the $\slashed{\pi}$EFT calculation. 

\subsubsection{Final Recommendation of \texorpdfstring{$S_{11}(0)$}{S11(0)}}
\label{subsubsec:WG1:final}

The above discussions show that determinations of $S_{11}(0)$ from $\chi$EFT in Eq.(\ref{ChEFTS11}), $\slashed{\pi}$EFT in Eq.(\ref{nopiEFTS11}), and lattice QCD (although with unquantified systematics) are all consistent with each other. 
Furthermore, these values are also consistent with the recommended value of \citetalias{2011RvMP...83..195A}, provided the central value is increased\footnote{To know the precise shift requires an explicit calculation. However, if the shift is within the range of 0.8-1.0 $\%$, 
the averaged value and $\chi^2$ remain the same within the significant digits.} by $0.9 \%$
to $S_{11}(0)=4.05(1\pm0.009)\times 10^{-25}\,{\rm MeV~b}$ to account for  the $G_A/G_V$ update. Averaging this value with the $\chi$EFT value in Eq.(\ref{ChEFTS11}) and the $\slashed{\pi}$EFT value in Eq.(\ref{nopiEFTS11}) yields $S_{11}(0)=4.09(1\pm0.005)\times 10^{-25}\,{\rm MeV~b}$ with $\chi^2$ per degree of freedom to be $0.9$. This shows that the $\chi$EFT, $\slashed{\pi}$EFT and \citetalias{2011RvMP...83..195A} estimates of $S_{11}(0)$ are all mutually consistent.

In addition, we would like to advocate adding an additional correlated error to account for any input that would tend to move all results in a coordinated way. For example, from \citetalias{2011RvMP...83..195A} to SF~III, we experienced the $\sim 1\%$ shift due to  
the update of the PDG value of $G_A/G_V$. 
It is not inconceivable that $G_A/G_V$ or other input parameters or physics could change again by similar amounts: the large PDG inflation factor of 2.7 reflects the tension that continues to exist among $G_A/G_V$ measurements \cite{2022PTEP.2022h3C01W}.  Therefore, assigning an additional $1\%$ correlated error seems reasonable. 
This is also in line with the subtleties discussed in Secs.\ref{subsubsec:WG1:chiEFT} and \ref{subsubsec:WG1:nopiEFT} which call for an enlarged error. 
Therefore, our final recommended value for $S_{11}(0)$ is
\begin{equation}
\label{FinalS11}
S_{11}(0)=4.09(1\pm0.015)\times 10^{-25}\,{\rm MeV~b} , 
\end{equation} 
where we have added the correlated and uncorrelated errors linearly to be conservative.

\subsection{Progress in \texorpdfstring{$S_{11}^\prime(0)$}{S'11(0)} and \texorpdfstring{$S_{11}^{\prime\prime}(0)$}{S''11(0)}}
\label{subsec:WG1:Sprime}
Using $\slashed{\pi}$EFT, \citet{2013PhLB..720..385C} computed $S_{11}(E)$ analytically with all partial waves included up to N$^2$LO. The Fermi matrix element only contributed at the $10^{-4}$ level and was neglected compared with the GT matrix element. 
The energy dependence of the phase factor $f_{pp}^{R}$ of Eq.~(\ref{S11_def}) was the dominant effect in $S_{11}^\prime(0)$  and $S_{11}^{\prime\prime}(0)$, the energy dependence of pp scattering was subdominant, while the $L_{1,A}$ contribution was much less important in these derivatives than in $S_{11}(0)$. Therefore, these derivatives could be predicted more reliably than $S_{11}(0)$. Furthermore, 
the derivatives were computed analytically and were free from errors of fitting $S_{11}(E)$ to a polynomial. The result was $S_{11}^\prime(0)/S_{11}(0)=(11.3 \pm 0.1)$ MeV$^{-1}$ and $S_{11}^{\prime\prime}(0)/S_{11}(0)=(170 \pm 2)$  MeV$^{-2}$.

In $\chi$EFT, $S_{11}(E)$ for $E<100$ keV was fit to polynomials of $E$ \cite{2019PhRvL.123a9901M}. Depending on using a quadratic or quartic fit,  
$S_{11}^\prime(0)/S_{11}(0)$ changed from 12.23 to 10.82 MeV$^{-1}$ and $S_{11}^{\prime\prime}(0)/S_{11}(0)$ changed from 178.4 to 317.4 MeV$^{-2}$. \citet{2016PhLB..760..584A} used cubit fit and $E<30$ keV to obtain $S_{11}^\prime(0)/S_{11}(0)$=10.84(2) MeV$^{-1}$ and $S_{11}^{\prime\prime}(0)/S_{11}(0)$=317.8(13) MeV$^{-2}$. 
Recently
\citet{2023arXiv230403327A} repeated the calculation of \citet{2019PhRvL.123a9901M} with the same energy range and cubic fit as \citet{2016PhLB..760..584A}, obtaining a consistent result with \citet{2016PhLB..760..584A}: $S_{11}^\prime(0)/S_{11}(0)=10.83$ MeV$^{-1}$ and $S_{11}^{\prime\prime}(0)/S_{11}(0)=313.72$  MeV$^{-2}$. We will take these as the recommended values from $\chi$EFT.

Although the face values of $S_{11}^\prime(0)/S_{11}(0)$ and $S_{11}^{\prime\prime}(0)/S_{11}(0)$ from $\slashed{\pi}$EFT and $\chi$EFT look quite different, we would like to remark that they actually agree on $S_{11}(E)/S_{11}(0)$ better than 0.1\% below the $\sim 6$ keV Gamow peak. For massive stars with central temperatures $\sim 15$ keV, the agreement is better than 0.8\% to second order in the derivatives and 0.5\% if $S_{11}^{\prime\prime\prime}(0)/S_{11}(0)=-5382$ MeV$^{-3}$ is included in the $\chi$EFT result. Hence, we take the average of the $\slashed{\pi}$EFT and $\chi$EFT results as the recommended value:    
\begin{eqnarray}
S_{11}^\prime(0)&=&S_{11}(0)(11.0\pm0.2){\rm MeV}^{-1} , \nonumber \\
S_{11}^{\prime\prime}(0)&=&S_{11}(0)(242\pm 72){\rm MeV}^{-2} .
\label{FinalS'S"}
\end{eqnarray}

Finally, we comment on the work of \citet{2019PhRvC.100c5805G} which performed a study of $S_{11}(E)$ in a wide energy range. The main focus of this work was to perform a proper energy-dependence analysis of the pp process,
in order to reliably extract $S_{11}^\prime(0)/S_{11}(0)$
and $S_{11}^{\prime\prime}(0)/S_{11}(0)$. However, the calculation
was performed within a phenomenological approach,
using a quite simplified model for the nuclear currents (i.e.\
not including two-body currents, which are well known to be significant)
and structure (i.e.\ neglecting the $d$-wave components in the
deuteron wave function). They found results for $S_{11}^\prime(0)/S_{11}(0)$ and 
$S_{11}^{\prime\prime}(0)/S_{11}(0)$ compatible with those of Eq.~(\ref{FinalS'S"}). 
However, since these values were not obtained with state of the art calculations, they have not been considered in the determination of $S_{11}^\prime(0)/S_{11}(0)$ and 
$S_{11}^{\prime\prime}(0)/S_{11}(0)$.



\newcommand*{\dpg}{$^2$H($p$,$\gamma$)$^3$He }

\section{The \texorpdfstring{$^2\text{H}({p},{\gamma})^3\text{He}$ reaction ($S_{12}$)}{2H(p,γ)3He reaction (S12)}} \label{sec:S12}

\subsection{Introduction}
\label{subsec:WG4:Introduction}

The \dpg reaction is the second step in the pp chain (Figure \ref{fig:pp-Chain}). Compared with the reactions mediated by the weak interaction, this reaction occurs much more rapidly. 
Consequently, on the time scales relevant to solar energy generation, deuterium is effectively converted to $^3$He instantaneously and thus it is only sensitive to the $Q$ value of the reaction and not the uncertainties of the rate.

However, the \dpg reaction plays an important role in the development of protostars, because the onset of deuterium burning slows down the protostars contraction and heating, increasing their lifespan. Accurate knowledge of the \dpg reaction rate, particularly within the few keV range corresponding to the Gamow peak in protostars, is vital for modeling protostellar evolution effectively \cite{1988ApJ...332..804S}.

Another astrophysical scenario where the \dpg reaction plays a key role is big bang nucleosynthesis (BBN), responsible for the production of light elements during the first few minutes of the Universe. Among these elements, deuterium is an excellent indicator of cosmological parameters because its primordial abundance is the most sensitive to the baryon density and critically depends on the radiation density of the early Universe, see for example a review of BBN~\cite{2016RvMP...88a5004C}. The reactions involved in the synthesis of deuterium are: production via the well known $p(n,\gamma$)$^2$H process and destruction via the $^2$H($^2$H,$n$)$^3$He, $^2$H($^2$H,$p$)$^3$H and \dpg reactions \cite{2020JCAP...03..010F}.

Since the comprehensive review performed in \citetalias{2011RvMP...83..195A}, there have been both new measurements and advances in the theoretical and phenomenological analysis of the \dpg reaction.  The new experimental results have been determined with accelerator-based measurements of the cross section \cite{2019EPJA...55..137T,2020Natur.587..210M,2021PhRvC.103d5805T}, as well as plasma-based, intertial confinement fusion measurements \cite{2020PhRvC.101d2802Z,2022FrP....10.4339M} and are reviewed in Section \ref{subsec:WG4:Data_sets} below.
On the theoretical side, there have been advances in \textit{ab initio} calculations, where nucleons are the fundamental degrees of freedom interacting among themselves and with the external electromagnetic probe, see for example~\cite{2023FrP....1129094E}. Finally, Bayesian analysis methods have been used to model the energy dependence of the S-factor, starting from the \textit{ab initio} predictions and applying a polynomial approximation to it.     
We review these updates and provide recommended values and uncertainties of the S-factor over the energy range of interest for Solar fusion, based upon a Bayesian averaging of  various models.

\subsection{Data sets used for the present review} 
\label{subsec:WG4:Data_sets}

The \dpg reaction has a Q value of 5.5 MeV and proceeds through the %
radiative capture of a proton on deuterium.
Different experimental approaches were followed to measure its cross section.  \citet{2019EPJA...55..137T} irradiated deuterated titanium targets with a proton beam and detected the $\gamma$-rays with two high-purity germanium (HPGe) detectors placed at different angles. The final S-factor is provided at four energies in the 47 - 210~keV range, with approximately 15\% statistical uncertainty. More recently, the LUNA Collaboration performed a measurement in the energy range of $32-263$~keV at the underground in the Gran Sasso Laboratories, exploiting the six orders of magnitude suppression of the cosmic radiation background \cite{2021FrASS...7..119Z,2018IJMPA..3343010C}. A windowless deuterium gas target was used and the $\gamma$-rays emitted by the \dpg reaction were detected by a large HPGe detector at 90$^\circ$ with respect to the beam axis. Great care was taken to minimize all sources of systematic uncertainties in the S-factor
overall uncertainty
at the 3\% level \cite{2020EPJA...56..144M, 2020Natur.587..210M}. These new results provided stringent constraints on cosmological parameters obtained by comparing the precise primordial deuterium abundance predictions of the standard BBN model with astronomical observations \cite{2018ApJ...855..102C}. 
A deeper discussion of the LUNA results and their implications\footnote{After the closing of the present review, the \citet{2020Natur.587..210M} data were re-analyzed to extract the experimental $^2$H($p,\gamma$)$^3$He $\gamma$-ray angular distribution \cite{2024PhRvC.110c2801S}. The data confirm the ab initio angular distributions used in the original analysis by \citet{2020Natur.587..210M}.} can be found in \cite{2021MNRAS.502.2474P, 2021JCAP...04..020P, 2021JCAP...03..046Y,2021ApJ...923...49M}
%

Finally, a new measurement was performed at the Helmholtz-Zentrum Dresden-Rossendorf in the 300-1000\,keV energy range using implanted deuterium targets on tantalum backings and two HPGe detectors~\cite{2021PhRvC.103d5805T}.  The resulting S-factors show $\approx1-2\sigma$ tension with the analysis of the LUNA results extrapolated to $E\gtrsim300$~keV~\cite{2020Natur.587..210M}.  However, they are affected by large systematic uncertainties.  A new measurement of the $^2{\rm H}(p,\gamma)^3{\rm He}$ reaction is planned at the Felsenkeller laboratory in Germany~\cite{2019sone.conf..249B}.

In addition to the new accelerator-based results, two recent sets of measurements \cite{2020PhRvC.101d2802Z,2022FrP....10.4339M} have also been performed using the inertial confinement fusion plasma-based platform \cite{2017PhPl...24d1407G,2022FrP....10.4339M,2023FrP....1180821G}, which has recently begun to be exploited for this type of work 
\citep[e.g.,][]{2016PhRvL.117c5002Z,2017NatPh..13.1227C}. Both measurements, which were performed at the OMEGA laser facility (see section \ref{subsec:WG8:Plasma}), used laser-driven implosions of spherical plastic-shell capsules filled with H$_2$D$_2$ gas, and measured the emitted $\gamma$-rays using a gas-Cherenkov-detector \cite{2014RScI...85kE124H} that was calibrated applying the technique described in \citet{2019RScI...90l3504Z}. The initial experiment \cite{2020PhRvC.101d2802Z} obtained good statistics at an energy of 16 keV by making several repeated measurements, with a final statistical uncertainty of 6\% and a systematic one of 17\% (dominated by uncertainty in the absolute calibration of the detector). The second experiment \cite{2022FrP....10.4339M} obtained data at three different energies in the 17-37 keV region, with comparable systematic uncertainty, but with larger statistical uncertainty due to fewer repeated measurements. The results obtained on this unique platform agree within error bars with the accelerator-based measurements.

\subsection{Theoretical studies} 
\label{subsec:WG4:Theoretical}

Nuclear reactions of astrophysical interest in general, and the \dpg in particular, are of great importance in nuclear theory because the available experimental data can be used to test the adopted theoretical framework.
The \dpg reaction has the great advantage of involving only $A\leq 3$ nuclear systems, and can be addressed with a microscopic \textit{ab initio} study. This means that the nuclear systems involved in the process are viewed as made up of $A$ nucleons, interacting among themselves and with the external electromagnetic probes. Within such an approach, the following ingredients are essential for the calculation: (i) realistic models for the nuclear interactions and currents, possibly rooted in Quantum Chromodynamics (QCD); (ii) a numerical technique able to solve the $A$-body bound and scattering state problem, including the Coulomb interaction without approximation. Such a technique is usually referred to as an \textit{ab initio} method. The agreement (or disagreement) between \textit{ab initio} theoretical predictions and experimental data represents a validation (or indicates the necessity of improvement) mostly for ingredient (i), 
i.e. are the models of the nuclear interactions and currents accurate within the precision specified by the ab initio method?
This is why few-nucleon reactions can be used as an ``ideal'' laboratory, where the \textit{ab initio} framework can be stringently tested in systems under more theoretical control.

The most recent \textit{ab initio} calculation of the \dpg reaction is that of \citet{2016PhRvL.116j2501M}. Here the pair-correlated Hyperspherical Harmonics \textit{ab initio} method was used to calculate the $A=3$ initial scattering and final bound state wave functions (see ~\citet{2008JPhG...35f3101K} and \citet{2020FrP.....8...69M} for details). 
The nuclear interaction model adopted to describe the $A=3$ nuclear state in \citet{2016PhRvL.116j2501M} consists of a two-nucleon term, the Argonne $v_{18}$ (AV18) potential ~\cite{1995PhRvC..51...38W}, augmented by a three-nucleon contribution, the Urbana IX (UIX) potential~\cite{1995PhRvL..74.4396P}. The AV18 potential can reproduce the large two-nucleon database with a $\chi^2$/datum $\sim 1$ \cite{1995PhRvC..51...38W}, while the combination AV18/UIX can describe quite accurately the properties of $^3$He, the spectra of light p-shell nuclei \cite{2001ARNPS..51...53P}, and $p-d$ scattering observables (see for instance \citet{2002PhRvC..65c4002W}). 
The electromagnetic current operator used in \citet{2016PhRvL.116j2501M} includes, in addition to the non-relativistic one-body operator, two- and three-body terms required by gauge invariance in a system of interacting particles.
These terms were constructed in \citet{2005PhRvC..72a4001M}.  The model was then tested against various electromagnetic observables, to access the quality of its predictions.
As a potential model, however, there is no systematic procedure for assigning uncertainties for observables whose values are unknown.

The results of \citet{2016PhRvL.116j2501M} have been found to be about 10\% higher than the experimental data of \citet{2020Natur.587..210M}. Given the lack of a procedure for quantifying errors in calculations based on phenomenological interactions and currents, it is difficult to access the significance of this discrepancy. This leads us to make several recommendations to the theory community. First, the \dpg reaction should be studied within the framework of chiral effective field theory, which has reached a degree of accuracy and predictive power comparable to potential-based phenomenology.  This approach is formally \textit{ab initio} and rooted in QCD, and as an operator expansion will provide an estimate of the theoretical uncertainty. 
Work along this line is currently underway. Second, the angular distribution of the \dpg capture reaction, and possibly also polarization observables, should be both calculated and measured.  This would be valuable even if measurements were limited to higher energies, where they are less difficult. Such measurements would provide a further test of the predictive power of the theory.

\subsection{Phenomenological and Bayesian Analyses}
\label{subsec:WG4:Reviews}

In the energy range of astrophysical interest, the \dpg reaction does not have any resonance or coupled channels that can give rise to non-trivial energy dependence, and the S-factor can be modelled by a low-order polynomial in energy \cite{1967ARA&A...5..525F}.
Moreover, the recent results from LUNA~\cite{2020Natur.587..210M}, combined with previous measurements, 
place stringent constraints on the cross section, reducing the uncertainty over the range of interest to Solar fusion and BBN~\cite{2002NuPhA.706..203C,2020Natur.587..210M, 2021MNRAS.502.2474P, 2021JCAP...04..020P, 2021JCAP...03..046Y}.

A significant change since \citetalias{2011RvMP...83..195A} has been the wide adoption of Bayesian analysis methods to evaluate thermonuclear reaction rates and provide more rigorous uncertainty estimates in a statistical sense  \cite{2016ApJ...831..107I}.  This approach has been followed in the analysis
presented here, which closely follow work on the \dpg reaction reported in \citet{2021ApJ...923...49M}, and references therein.  The mathematical details of the
Bayesian method will be found in Appendix~\ref{AppendixWG4}:  Here we focus on the application to \dpg by
the authors cited above.

Both the \textit{ab initio} prediction from \citet{2005PhRvC..72a4001M} as well as a third-order polynomial in the energy were used to constrain the data, where we define an $n^{th}$-order polynomial as
\begin{equation}\label{eq:polynomial}
S_n(E;\lambda) = \sum_{i=0}^n \lambda_i E^i\, .
\end{equation}
As noted in the literature, the energy dependence of \citet{2005PhRvC..72a4001M} is more reliable than the absolute normalization.
Furthermore, the updated prediction in \citet{2016PhRvL.116j2501M}, as noted above, is $\approx10\%$ larger than that of \citet{2005PhRvC..72a4001M}.
Therefore, \citet{2021ApJ...923...49M} modeled the prediction of \citet{2005PhRvC..72a4001M}, denoted $S_{\rm nuc}(E)$, as
\begin{equation}\label{eq:S_pheno}
S(E;\lambda) = 
    a S_{\rm nuc}(E) + b\, ,
\end{equation}
where 
$a$ and $b$ are an unknown scale factor and offset to be determined in the analysis.
While $S_{\rm nuc}(E)$ was not determined with theoretical uncertainty, model uncertainty is introduced through the parameters $a$ and $b$.
The resulting mean values and uncertainties of $S(E;\lambda)$ determined from Eq.~\eqref{eq:S_pheno} were found to be comparable to those with the third-order polynomial.


\begin{figure*}
    \centering
    \includegraphics[width=\textwidth]{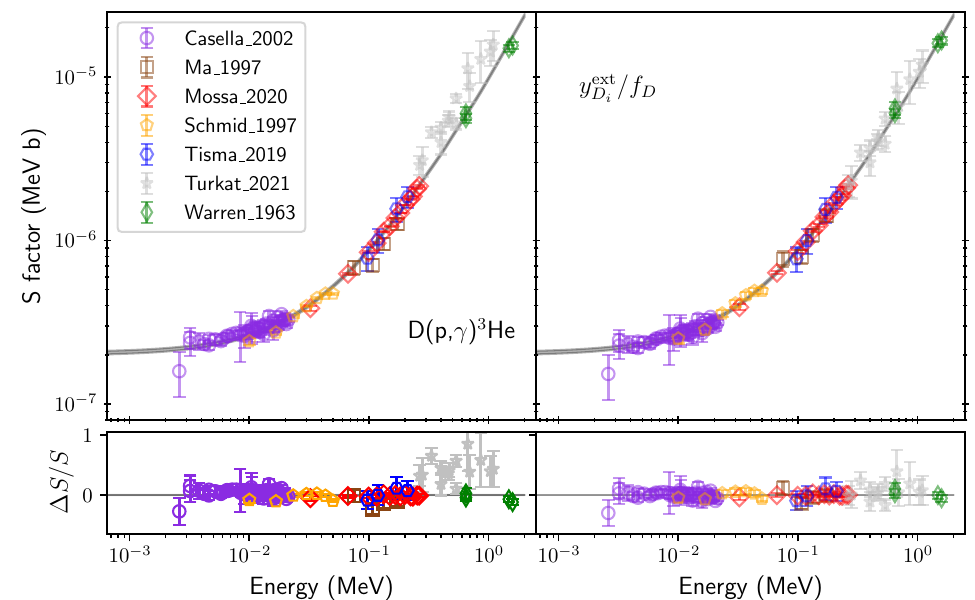}
    \caption{
    S-factor determined via Bayes model averaging (BMA) of the \dpg data sets \cite{2002NuPhA.706..203C,1997PhRvC..55..588M,2020Natur.587..210M,PhysRevC.56.2565,2019EPJA...55..137T,2021PhRvC.103d5805T,PhysRev.132.1691} analysed as described in the text.
    The left panel shows the original data with statistical uncertainties only.
    The right panel shows the data after normalizing by BMA systematic uncertainty posteriors of $f^{-1}_D$ for each data set, and the uncertainties correspond to the statistical, systematic and extrinsic uncertainties all added in quadrature.
    In most cases, the quoted statistical uncertainty is the dominant uncertainty.
    The residual is defined as $\Delta S / S = (S_{\rm data} - S_{\rm fit}^{\rm BMA}) / S_{\rm fit}^{\rm BMA}$ and is plotted for the original data with statistical uncertainty only on the left, and the adjusted data as described above, on the right.
    \label{fig:S_12_BMA}
}
\end{figure*}

\subsection{Summary and Recommendations} 
\label{subsec:WG4:Summary}

\begin{table}[]
\caption{Logarithm of the Gaussian approximation to the Bayes factor, and corresponding weight, Eq.~\eqref{eq:P(M|D)}, for each model under investigation.}\label{tab:S_12_Bayes}
\begin{ruledtabular}
\begin{tabular}{lcc}
Model& logGBF & weight  \\
  \cline{1-3} \\
  
\textit{ab initio}, $a S_{\rm nuc}(E)$ &    1478.7 & 0.585\\
\textit{ab initio}, $a S_{\rm nuc}(E) +b$ & 1478.3 & 0.414\\
3$^{rd}$ order polynomial & 1470.2 & 1.2 $\times 10^{-4}$\\
4$^{\rm th}$ order polynomial & 1469.2 & 4.6 $\times 10^{-5}$\\
5$^{\rm th}$ order polynomial & 1468.2 & 1.6 $\times 10^{-5}$\\
6$^{\rm th}$ order polynomial & 1464.1 & 2.9 $\times 10^{-7}$
\end{tabular}
\end{ruledtabular}
\end{table} 

In this review, we take the further step of applying Bayes Model Averaging to a number of reasonable models that describe the data, which systematically captures additional theoretical ``model selection uncertainty''.  
Detailes of the Bayesian analysis and model averaging procdure are presented in Appendix~\ref{AppendixWG4}.
The models we explore include Eq.~\eqref{eq:S_pheno} with both $a$ and $b$, as done in \citet{2021ApJ...923...49M}, as well as setting $b=0$, and then we consider polynomials in energy of order\footnote{\label{footnote:leaner}The implementation of the Bayesian analysis presented here is different than that of \citet{2021ApJ...923...49M} but it was verified to produce the same results when the same implementation of the systematic uncertainties was made.} $n=3,4,5,6$.

With such a Bayes model averaging, we can quantitatively compare and contrast the polynomial parameterizations of \dpg along with \textit{ab initio} results predicted by \citet{2005PhRvC..72a4001M, 2016PhRvL.116j2501M},
including the extra variance that arises from this set of reasonable models, e.g., using polynomials of different order, as well as the phenomenological model of Eq.~\eqref{eq:S_pheno}. Results are listed
in Table~\ref{tab:S_12_Bayes}. The first column lists the model (i.e., polynomial or \textit{ab initio}), the second column the natural logarithm of the Gaussian approximation to the Bayes factor (BF), and the third column the corresponding weight in the model averaging.
From this model averaging, the resulting prediction for the \dpg S-factor at a few representative energies is provided in Table~\ref{tab:S_12_sample_E}, where the first uncertainty arises from the first term in Eq.~\eqref{eq:BMA_Var} and the second uncertainty is from the second term, which we denote as \textit{model selection uncertainty}.
The resulting Bayes model average prediction of the S-factor over the entire kinematic range considered is depicted in Fig.~\ref{fig:S_12_BMA}, with the gray band representing the 68\% coverage probability.
For this particular reaction, it is interesting to note that the Bayesian analysis strongly favors the phenomenological models of Eq.~\eqref{eq:S_pheno} over the polynomial approximations.
One reason for this might be that the scaled \textit{ab initio} models have only one or two free parameters, and the energy dependence of $S_{\rm nuc}(E)$ given in \citet{2005PhRvC..72a4001M,2016PhRvL.116j2501M} is sufficient to accurately describe the various data sets. On the other hand, polynomial approximations are  disfavored as they require more parameters to capture the energy dependence. Also of note, in the model of Eq.~\eqref{eq:S_pheno} with $b=0$ and using \citet{2016PhRvL.116j2501M} for $S_{\rm nuc}(E)$, the scale factor is given by
\begin{equation}
a = 0.921(19)\, ,
\end{equation}
indicating that the prediction in \citet{2016PhRvL.116j2501M}  overestimates the \dpg data by 7.9\%, consistent with expectations noted above.
In comparison, the third order polynomial fit predicts values of $S(E)$ that are 
$1\sigma$ higher at $E=0$ and $\frac{2}{3}\sigma$ lower at $E=91$~keV as compared to those in Table~\ref{tab:S_12_sample_E}.

The analysis in Table~\ref{tab:S_12_Bayes} and Fig.~\ref{fig:S_12_BMA} can be reproduced with the code linked in Table~\ref{tab:S_12_sample_E}.  The code also can provide a prediction of the mean value and uncertainty of $S_{12}(E)$ at any energy over the same kinematic range.

\begin{table*}[]
\centering
\caption{\label{tab:S_12_sample_E}
The S-factor of the \dpg at a few selected energies determined from data using Bayesian model averaging as described in the text.  The first uncertainty is the statistical and the second is the model selection uncertainty.
The value at any energy in the fitted range can be obtained by running the analysis provided at \url{https://github.com/nrp-g/leaner}.
}
\begin{ruledtabular}
\begin{tabular}{cccccccc}
\multicolumn{8}{c}{$S_{12}(E) [10^{-7}{\rm MeV b}]$}\\
& &&&&&& \\[-.2cm]
\hline
$E=0$ keV & $E=10$ keV & $E=20$ keV & $E=40$ keV & $E=80$ keV & $E=91$ keV & $E=100$ keV & $E=120$ keV\\[.1cm]
\hline
 &&&&&&&\\[-.25cm]
2.028(51)(9)& 2.644(60)(8)&
3.276(70)(7)& 4.579(94)(5)& 
7.31(15)(0)& 8.11(16)(0)& 
8.77(18)(0)& 10.24(21)(0)
\end{tabular}
\end{ruledtabular}

\end{table*}    



\section{The \texorpdfstring{$^3\text{He}(^3\text{He},{2{p}})^4\text{He}$ reaction $(S_{33}$)}{3He(3He,2p)4He reaction (S33)}}
\label{sec:S33}

The $^3$He($^3$He,$2p$)$^4$He reaction terminates the pp-I chain. The ratio of its rate to that of the $^3$He($\alpha,\gamma$)$^7$Be reaction controls the branching to the pp-II and pp-III chains, so that in \citetalias{1998RvMP...70.1265A} increasing $S_{33}$ was discussed as a potential solution to the Solar Neutrino Problem. Subsequent experiments, notably a very low energy measurement at the LUNA \mbox{50 kV} accelerator deep underground in Gran Sasso \cite{1999PhRvL..82.5205B} and a complementary experiment at somewhat higher energies \cite{2004PhRvC..69a5802K} ruled out such an increase in  $S_{33}$, as summarized in \citetalias{2011RvMP...83..195A}.

\subsection{Shape of the particle spectrum}

Since no new absolute measurements of $S_{33}$ have been reported since \citetalias{2011RvMP...83..195A}, we consider the same four experiments~\cite{1999PhRvL..82.5205B,1998PhRvC..57.2700J,2004PhRvC..69a5802K,1987NuPhA.467..273K}. However, there is new information on the energy spectrum of the emitted protons. This spectrum has recently been measured using inertial confinement fusion plasmas for a Gamow peak energy of 165 keV~\cite{2017PhRvL.119v2701Z}. The results show significant structure, indicating the presence of a sequential reaction mechanism passing through the unbound ground state of ${}^5{\rm Li}$. This spectrum is important for all of the $S_{33}$ measurements since they determined cross sections by detecting only the protons above an energy threshold. 

The efficiency correction by which \textcite{1987NuPhA.467..273K} account for the threshold is not well documented, but the more recent measurements~\cite{1999PhRvL..82.5205B,1998PhRvC..57.2700J,2004PhRvC..69a5802K} utilized the {\sc genbod} event generator~\cite{James68-GENBOD}. It employs a simple reaction model without final state interactions, the Pauli principle, or Coulomb effects that are important near the spectrum endpoint. The only angular correlations are those required by energy and momentum conservation, and together these simplifications give simple ellipses for the singles energy distributions of the emitted nuclei. Published proton spectra obtained using accelerator beams do exist (Fig.~2 of \textcite{1974PhRvC...9..805D} and Fig.~3 of \textcite{1987NuPhA.467..273K}), and although they are not corrected for instrumental effects, they do not appear to be well described by ellipses.

The $^3$He($^3$He,$2p$)$^4$He reaction may proceed via several sequential mechanisms, including $p+{}^5{\rm Li}$ (with ${}^5{\rm Li}$ in its ground or first excited state) and $\mbox{di-proton}+\alpha$ (where the di-proton is two correlated protons in a singlet state)~\cite{2015PhRvC..92a4003B}. More complicated three-body decay channels, sometimes called direct decays, are also possible. Only a relatively narrow intermediate state (here, the 1 MeV wide $^5$Li ground state) could produce a peak in the energy spectrum, and precise classification of the reaction mechanism is in general both experimentally and theoretically ambiguous. When coincident detection of reaction products is used [as in \textcite{1999PhRvL..82.5205B}], possible angular correlations between the reaction products also matter. A reaction through the $3/2^-$ ground state of ${}^5{\rm Li}$ would emit the second proton preferentially either along or opposite the direction of the first proton, while di-proton emission would tend to send both protons in the same direction~\cite{2015PhRvC..92a4003B}. 

\begin{figure}[tb]
\includegraphics[width=\columnwidth]{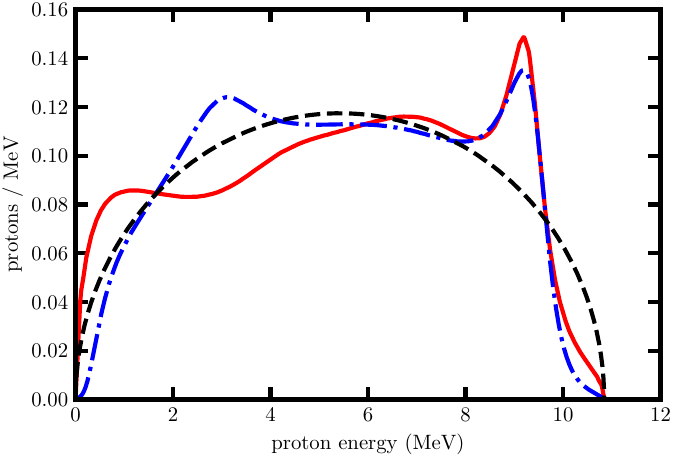}
\caption{\label{fig:WG7_S33_Protons} Calculated proton energy spectra in the c.m.\ frame for the $^3$He($^3$He,$2p$)$^4$He reaction at $E_{\rm c.m.}=165$~keV. The solid (red on line) and dot-dashed (blue online) curves are models~1 and~2, respectively, from Fig.~3(a) of \textcite{2017PhRvL.119v2701Z}. The dashed black curve is the elliptical spectrum. The spectra are normalized to unit area.}
\end{figure}

The solid curves in Fig.~\ref{fig:WG7_S33_Protons} show two different R-matrix models of the proton energy spectrum at $E_{\rm c.m.}=165$~keV, fitted to the spectrum measured by \textcite{2017PhRvL.119v2701Z} for $E_p\gtrsim 6$~MeV. The measured spectrum differs significantly from the elliptical spectrum of \textsc{genbod}. Based on their Fig.~6, the experiment of \textcite{1998PhRvC..57.2700J} had a detection threshold of about 5~MeV proton energy. The measurements by \textcite{1999PhRvL..82.5205B} required a coincidence between two detectors, with a detection threshold of 2~MeV proton energy in each detector. Finally, Fig.~19 of \textcite{2004PhRvC..69a5802K} indicates a detection threshold of about 4~MeV proton energy. We have estimated their sensitivities to the assumed proton spectrum by integrating the curves shown in Fig.~\ref{fig:WG7_S33_Protons} above energy thresholds of 2, 4, and~6~MeV. These integrals vary from 3\% below to 6\% above the result from an elliptical spectrum, depending on the specific threshold and the assumed spectrum.

\subsection{Recommendation}
\label{sect:S33_recommendation}
Based on these investigations, our recommended $S_{33}$ fit is the one from \citetalias{2011RvMP...83..195A}, but with an additional 4\% systematic uncertainty typical of the 3\% to 6\% corrections to the extrapolated proton spectra estimated above to take the uncertain spectral shape into account. This gives 
\begin{eqnarray} 
S_{33}^\mathrm{best}(E) = 5.21 - 4.90 \left(\frac{E}{\mathrm{MeV}}\right)
+ 11.21 \left( \frac{E}{\mathrm{MeV}} \right)^2\mathrm{MeV~b} \nonumber \\
\delta S_{33}(E) = \left[0.118 - 1.516 \left( \frac{E}{\mathrm{MeV}} \right)
+ 14.037 \left( \frac{E}{\mathrm{MeV}} \right)^2 \right.~~~ \nonumber \\
\left. -15.504 \left(\frac{E}{\mathrm{MeV}} \right)^3 +
 71.640 \left(\frac{E}{\mathrm{MeV}} \right)^4 \right]^{1/2}\mathrm{MeV~b}. \nonumber
\end{eqnarray}
Only the larger constant term in the uncertainty differs from the previous evaluation, and it is due to the spectral shape discussed above.

New measurements of this reaction could provide more accurate absolute cross sections by using lower energy thresholds and improve our understanding of the proton energy spectra and angular correlations. The analysis of new measurements should use Monte Carlo simulations considering a variety of plausible energy spectra and angular correlations, in order to estimate the sensitivity to these effects.  We note finally that the proton energy spectrum may depend on $E_{\rm c.m.}$, as this has been found to be the case for neutrons from the mirror reaction ${}^3{\rm H}(t,2n)\alpha$~\cite{2018PhRvL.121d2501G}.

\section{The \texorpdfstring{$^3\text{He}(\alpha,\gamma)^7\text{Be}$ reaction ($S_{34}$)}{3He(α,γ)7Be reaction (S34)}}
\label{sec:S34}

The $^3$He($\alpha,\gamma$)$^7$Be reaction proceeds via non-resonant capture to the ground and 429~keV first excited states of $^7$Be. It has been studied experimentally using three main methods, with references given in subsections \ref{subsec:WG7:S34_SF2} and \ref{subsec:WG7:S34Expt}: First, by detecting at least two of the three $\gamma$-rays from the reaction and taking the angular correlation with the alpha beam direction into account (prompt-$\gamma$ method). Second, by detecting the induced $^7$Be \cite{2002NuPhA.708....3T} radioactivity (activation method). Third, by counting the $^7$Be recoils (recoil method).

\subsection{Previous \texorpdfstring{$S_{34}$}{S34} recommendation in SF~II}
\label{subsec:WG7:S34_SF2}

The $^3$He($\alpha,\gamma$)$^7$Be S-factor recommended in \citetalias{2011RvMP...83..195A} was developed in multiple steps.  First, a model was selected for the shape of $S(E)$, based on existing nucleon-level calculations \cite{1986NuPhA.460..559K,2001PhRvC..63e4002N}.  A rescaling of the curve was fitted to data, but only at $E\leq$1.002 MeV center-of-mass to minimize the role of short-distance physics in the models.  

In \citetalias{1998RvMP...70.1265A}, a possible systematic discrepancy between data from the two methods previously used was discussed. By the time of \citetalias{2011RvMP...83..195A}, in-depth studies from two groups using both activation and prompt methods were available, from LUNA \cite{2006PhRvL..97l2502B,2007PhRvC..75f5803C,2007PhRvC..75c5805G} and Seattle \cite{2007PhRvC..76e5801B} groups. These studies did not find any discrepancy between activation and prompt-$\gamma$ data in direct comparisons. \citetalias{2011RvMP...83..195A} opted to limit the fitting to the data by the activation and recoil methods.  The prompt-$\gamma$ data were left out because of their somewhat larger common mode errors, concerns about how well the $\gamma$-ray angular distribution was known, and avoidance of the correlated errors between activation and prompt data from the same experiment.

In \citetalias{2011RvMP...83..195A}, only data published after 1998 were included. 
%
The \citetalias{2011RvMP...83..195A} recommended value was $S_{34}(0) = 0.56 \pm 0.02_{\rm expt} \pm 0.02_{\rm theor}$, based on the data by the Weizmann \cite{2004PhRvL..93z2503S}, LUNA \cite{2006PhRvL..97l2502B,2007PhRvC..75f5803C,2007PhRvC..75c5805G}, Seattle \cite{2007PhRvC..76e5801B}, and ERNA \cite{2009PhRvL.102w2502D} groups. 

\subsection{Theory progress on \texorpdfstring{$S_{34}$}{S34}}
\label{subsec:WG7:S34_Theory}

Significant theoretical work on this reaction  has occurred since \citetalias{2011RvMP...83..195A}, but the basic understanding of its mechanism remains unchanged from the 1960s \cite{1963PhRv..131.2582T}.  It is dominated by external non-resonant capture into the two bound states, and most of the dipole strength at low energy arises beyond the range of nuclear interaction.  The external capture part of the cross section is determined by the ANCs of the bound states and by scattering phase shifts; near threshold, most of the strength lies at $\sim 5$--20 fm \cite{2011PhRvL.106d2502N}.  In models with explicit wave functions the shorter-range strength largely cancels out due to effects of nucleon-exchange antisymmetry.  All models feature a shallow minimum of $S(E)$ near \mbox{1.25 MeV}, where capture from $d$-waves becomes comparable to that from $s$-waves.

The first fully \textit{ab initio} calculation \cite{2011PhRvL.106d2502N} appeared just after the \citetalias{2011RvMP...83..195A} analysis concluded; it used the fermionic molecular dynamics (FMD) method  and a softened representation of the Argonne $v_{18}$ interaction \cite{1995PhRvC..51...38W}.  It agrees well with both the scale and the energy dependence of the modern S-factor data, and it is very close to the energy dependence assumed in \citetalias{2011RvMP...83..195A}.  Neff's elastic-scattering phase shifts also agree well with experiment.  Notably, Neff found that the dipole strength distribution departs significantly from pure external capture at $^3$He-$^4$He separations as large as 9 fm, compared with 3-4 fm in potential models.

Another \textit{ab initio} model \cite{2016PhLB..757..430D,2019PhRvC.100b4304V} has also appeared, based on the no-core shell model with continuum (NCSMC) method and the chiral interaction of \citet{2003PhRvC..68d1001E}, ``softened'' by the similarity renormalization group (SRG) procedure.  Freedom to choose the SRG stopping point allowed exact reproduction of the $^7$Be breakup energy, which corrects for the main effect on the external capture of omitting the three-nucleon potential.  The results agree well with the overall scale of the modern data, but their energy dependence and phase shifts depart from experiment, possibly due to the omission of explicit three-body forces.

An important advance since \citetalias{2011RvMP...83..195A} has been the application of halo effective field theory (halo  EFT) methods to astrophysical capture reactions \cite{2002NuPhA.712...37B,2018EPJA...54...89H,2020EPJA...56..166P,2020JPhG...47e4002Z}.  EFTs are valid for systems with a natural separation between the momentum scales $Q$ probed in low-energy experiment (e.g., corresponding to a binding energy or a projectile energy) and the much larger momentum scale $\Lambda$ where the low-energy degrees of freedom are no longer valid: the word ``halo'' here refers to nuclei with only shallow bound states.  In halo EFT the only degrees of freedom are the initial- and final-state nuclei treated as point particles, plus photons.  For $S_{34}$, $\Lambda \sim 200$ MeV corresponds to the momentum needed to separate a proton from $^3$He or to excite $^4$He;  the thresholds to separate the two $^7$Be bound states into $^3$He and $^4$He correspond to momenta $Q = 71.4$ MeV and 60.9 MeV.

Given a sufficient separation of scales, one constructs a Lagrangian that respects the system's symmetries and known qualitative features systematically, organized by powers of $Q/\Lambda$.  This series can be truncated (at leading order or LO, next-to-leading order or NLO, next-to-next-to-leading-order or NNLO, etc.), and the precision of the resulting theory depends on the sizes of the omitted terms. The precision of a calculation and the energy where it breaks down can be estimated by assuming coefficients of the first omitted term to have a ``natural'' size.  Coupling constants of the Lagrangian must be fitted to data, and consistency of the power counting scheme (identification of powers of $Q/\Lambda$ for the main operator terms of a system) is tested by whether the fitted constants have natural sizes.   Low-order coupling constants in a halo EFT can often be identified with familiar quantities like ANCs, scattering lengths, and effective ranges; the description of elastic scattering in halo EFT reproduces the (Coulomb-modified) effective-range expansion \cite{2022JPhG...49d5102P,2000NuPhA.665..137K,2008NuPhA.809..171H}.
When a separation of scales exists, halo EFT is well-suited to data extrapolation because it avoids the tacit, hard-to-test, and unavoidable prior assumptions present in the short-range parts of models based on explicit wave functions.

Two groups have studied $S_{34}$ in halo EFT \cite{2018EPJA...54...89H,2020EPJA...56..166P,2020JPhG...47e4002Z}.   The sizes of the Coulomb interaction and the large $s$-wave scattering length make the correct power counting tricky to establish for this system.  The work of \citet{2018EPJA...54...89H} and \citet{2020EPJA...56..166P} includes a careful examination of possible power-counting schemes and strict adherence to a power counting once established, up to terms of NLO or NNLO, respectively, in two different power-countings.  These authors fitted EFTs to S-factor data, and they examined effects of including or excluding elastic scattering constraints in their fits.  When they included scattering phase shifts from \citet{1972NuPhA.195..241B} in their fits, they concluded that the large-scattering-length power counting of the NLO theory was favored.  This result promotes two-body currents (i.e., contributions not equivalent to external non-resonant capture) to leading order in the theory.  The $d$-wave contribution that becomes important above \mbox{1 MeV} first appears at NLO.  In addition to data-fitting errors, these theories have errors estimated to be 10\% from EFT truncation.

The work of \citet{2020JPhG...47e4002Z} followed a different approach to power counting up through NLO.
At $E< 2$ MeV, their derived expression for $S_{34}(E)$ is essentially the same as that from the \citet{2020EPJA...56..166P} NLO theory.  
\citet{2020JPhG...47e4002Z} also developed some \textit{ad hoc} (i.e., not systematically developed) higher-order EFT terms, referred to as partial-N4LO, to test for their impact on the fitting.  The additional terms proved not to be required by the data and did not improve the fit; this result was taken to indicate that corrections from omitted terms are not large compared with experimental errors below 2 MeV.  Scattering data were not considered in this work apart from very broad priors on scattering length and effective range; the correlated errors in the S-factor data were taken into account.  

Like halo EFT, the phenomenological R-matrix approach avoids a model of nuclear interactions and uses a systematic parameterization to fit data \cite{2010RPPh...73c6301D}.  At the time of \citetalias{2011RvMP...83..195A}, $S_{34}$ had been the subject of very little R-matrix fitting, apparently consisting only of the simple treatment in \citet{2004ADNDT..88..203D} that focused on BBN energies.  After \citetalias{2011RvMP...83..195A} an R-matrix analysis using the AZURE2 code \cite{2010PhRvC..81d5805A,azure2-manual} was carried out on both elastic scattering and S-factor data in conjunction with the Notre Dame experiment \cite{2013PhRvC..87f5804K}, and a more elaborate analysis using Monte Carlo sampling to estimate errors was later reported in \citet{2014PhRvC..90c5804D}.  In the latter analysis both the fitted value and error bars were heavily influenced by the numerous scattering data of \citet{1964NucPh..50..629B}.  Very recently, the BRICK software package has been constructed to carry out Bayesian parameter estimation for AZURE2 and applied to both capture and elastic scattering at all energies in the $^7$Be system by \citet{2022FrP....10.8476O}. In this work it was shown that markedly different $S_{34}(0)$ values result from inclusion or not of the older scattering data \cite{1964NucPh..50..629B} alongside the very recent SONIK scattering data of \citet{2024PhRvC.109a5802P}.

A small amount of additional theoretical work in more traditional frameworks has appeared since \citetalias{2011RvMP...83..195A}.  This includes fits of potential models to S-factor data \cite{2018PhRvC..97c5802T,2021NuPhA100622108T} and use of resonating group methods \cite{2017PhRvC..96f4605S,2019PhRvC..99e4618S}.

\subsection{Experimental progress on \texorpdfstring{$S_{34}$}{S34}}
\label{subsec:WG7:S34Expt}

Since \citetalias{2011RvMP...83..195A}, five new experiments have been reported: Four from the Madrid and ATOMKI groups by the activation technique, at relatively high center of mass energies \cite{2012PhRvC..86c2801C,2013NuPhA.908....1B,2019PhRvC..99e5804S,2023PhRvC.108b5802T}, and one from the Notre Dame group using the prompt-$\gamma$ method \cite{2013PhRvC..87f5804K}. 

Following the approach adopted in \citetalias{2011RvMP...83..195A}, we again only use the recoil and activation data for the $S_{34}$ fits below. The $\gamma$-ray angular distribution is not known experimentally \cite{2007PhRvC..76e5801B}, and the resultant uncertainty increases the common-mode error for the prompt-$\gamma$ studies somewhat. (An experiment to address previously-raised concerns about angular distribution \cite{2007PhRvC..76e5801B} has recently concluded at Felsenkeller Dresden. The data suggest a higher than expected anisotropy but are so far only available in the form of a PhD thesis \cite{Turkat23-PhD}. If confirmed, they may lead to corrections of a few percent for some of the in-beam experiments where only one angle was instrumented.) In addition, all of the ``modern'' works reporting prompt-$\gamma$ data except for Notre Dame \cite{2013PhRvC..87f5804K} also include data obtained with other methods, leading to partial correlations between data sets that would complicate fitting.

Following this restriction, our data selection proceeds as follows. First, the four data sets previously used in \citetalias{2011RvMP...83..195A} are carried over here: Weizmann \cite{2004PhRvL..93z2503S}, LUNA \cite{2006PhRvL..97l2502B,2007PhRvC..75f5803C,2007PhRvC..75c5805G} (only the activation data), Seattle \cite{2007PhRvC..76e5801B}  (only the activation data), and ERNA \cite{2009PhRvL.102w2502D}. 

Two of the archival data sets excluded by the 1998 cutoff date in \citetalias{2011RvMP...83..195A} merit some further discussion here: The data of \citet{1982PhRvL..48.1664O,1984NuPhA.419..115O} consist of two points measured by activation in a $^3$He gas cell. However, their uncertainties are not separated into statistical and systematic components. Including them would have required an uncertain guess on how to divide the error bars, and the fit result would barely change, so we left them out.  The work by \citet{1983PhRvC..27...11R} consists of one data point that was obtained by averaging two separate activation measurements with a $^3$He and a $^4$He gas cell, respectively. We do not use this result because we lack details to verify the background subtraction, an issue that was raised during our meeting by one of its authors.   We also choose not to use the activation study by \citet{1983ZPhyA.310...91V}. In that experiment there were thick entrance foils and the $^4$He beam was completely stopped inside a high-pressure $^3$He gas cell.  This gave an integrated measurement over a wide energy range, so that the analysis depends strongly on the assumed shape of the $S_{34}$ curve, and the result was reported only as an extrapolated $S_{34}(0)$.

We now consider the new data since \citetalias{2011RvMP...83..195A}. An activation experiment in Madrid \cite{2012PhRvC..86c2801C} used a $^3$He beam incident on a $^4$He gas cell, reporting three data points. Another activation experiment was reported by the ATOMKI group, using a $^4$He beam on $^3$He gas cells \cite{2013NuPhA.908....1B}. Two higher-energy campaigns at ATOMKI were again performed using the activation method. These latter data are at $E$ = 2.5-4.4 MeV \cite{2019PhRvC..99e5804S} and $E$ = 4.3-8.3 MeV \cite{2023PhRvC.108b5802T}, respectively, above the energy range suitable for halo EFT, and therefore not included.  

A detailed study by the Notre Dame group reported 17 data points \cite{2013PhRvC..87f5804K}, using the primary $\gamma$-ray from ground state capture and the secondary $\gamma$-ray from the deexcitation of the 429 keV first excited state of $^7$Be. No activation data are reported in that work, and the $\gamma$-ray detector was placed at just one angle, 90$^\circ$. Since  only prompt-$\gamma$ data are reported, we did not include this data set. In order to test the effects of this decision, we repeated some of our fits (see below, Section \ref{subsec:WG7:S34Fits}) with the modern prompt-$\gamma$ experiments included: the 17 points from \citet{2013PhRvC..87f5804K}, 3 prompt-$\gamma$ points from LUNA \cite{2007PhRvC..75f5803C} and 8 from Seattle \cite{2007PhRvC..76e5801B}. The extrapolated S-factor changed by less than 1\%, well within the error bars for the recommended $S_{34}$.

Finally, an indirect experiment using the $^6$Li($^3$He,d)$^7$Be reaction and the ANC technique has recently been reported \cite{2020PhLB..80735606K}. It is left out of our fits due to the additional normalization and theory uncertainties involved in determining an ANC from a transfer experiment, which are larger than the errors in the S-factor measurements used for the present fits.
All the data used in the present fits are summarized in Table \ref{tab:WG7_S34_Data}.

As in \citetalias{2011RvMP...83..195A}, we model data uncertainties as consisting of a component that is independent for each point and a common-mode component that applies to all data from a given experiment as a multiplicative factor.  This separation is well-documented for all of the modern data and is also shown in Table \ref{tab:WG7_S34_Data}.  Except at the lowest energies, the common-mode error typically dominates.

\begin{table*}
\caption{\label{tab:WG7_S34_Data} Experimental data used for the $S_{34}$ fit. See text for details. For each experimental data set, the rescaling factors $s_\alpha$ are determined for the \citet{2018EPJA...54...89H}, \citet{2020JPhG...47e4002Z}, and BRICK (\citep{2022FrP....10.8476O}) without or with the inclusion of elastic scattering. }
\begin{tabular}{p{60mm} D{-}{ - }{5} r r D{.}{.}{4} r r r r} \hline \hline
Group and references & \multicolumn{1}{l}{Energy} & \multicolumn{2}{c}{Data points} &     \multicolumn{1}{l}{Common}   &   \multicolumn{4}{c}{Rescaling factor $s_\alpha$}   \\ \cline{3-4}\cline{6-9}
    &   \multicolumn{1}{l}{range [keV]}  &   used &     total    & \multicolumn{1}{l}{mode uncert.}   & Higa EFT  & Zhang EFT & BRICK  & BRICK+S \\ \hline
Weizmann \cite{2004PhRvL..93z2503S} & 420-950 & 4   & 4 & 2.2 \% &   1.03(2)     &   1.02(2) &   1.02(2) &   1.03(2) \\
LUNA \cite{2006PhRvL..97l2502B,2007PhRvC..75f5803C,2007PhRvC..75c5805G} & 93-170 & 7 & 7 & 3.0 \%    &   1.02(2) &   1.04(2) &   1.05(2) &   1.01(2) \\
Seattle \cite{2007PhRvC..76e5801B} & 327-1235 & 8   &   8 & 3.0 \%   &   0.96(1)   & 0.95(2) &   0.95(2) &   0.98(2) \\
ERNA \cite{2009PhRvL.102w2502D} & 650-2504 & 22     &   51  & 5.0 \%  &   0.96(3) &   0.94(2) &   0.96(2) &   0.99(2) \\
Madrid \cite{2012PhRvC..86c2801C} & 1054-2804 & {1}   &   3   & 5.2 \%     &   0.99(3) &   0.99(3) &   0.97(2)     &   0.99(2) \\
ATOMKI \cite{2013NuPhA.908....1B} & 1473-2527 &     {2}   &   5 & 5.9 \%   &   1.01(3) &   1.00(3) &   1.01(3) &   1.04(3)\\
\hline\hline
\end{tabular}
\end{table*}

\subsection{Data fitting}
\label{subsec:WG7:S34Fits}

We base our recommended $S_{34}$ on fits to the  halo EFT and R-matrix parameterizations discussed above.  These avoid tacit assumptions present in potential-models and conceptual difficulties involved in combining \textit{ab initio} constraints with data.  The fits presented here differ from the previously published fits of \citet{2020EPJA...56..166P}, \citet{2020JPhG...47e4002Z}, and \citet{2022FrP....10.8476O} mainly in the uniform use of the agreed-upon capture data and uncertainties from Section~\ref{subsec:WG7:S34Expt} across all fits.  We restricted fitting to the $E < 2$ MeV range of validity for the NLO halo EFT expressions.  Despite uniform handling of capture data, the fits in each framework handle scattering inputs differently for reasons discussed below.

In addition to the total capture cross section and scattering data, the fitted data also include branching ratios for capture into the two $^7$Be bound states, taken from \citet{2007PhRvC..76e5801B}, \citet{2009PhRvL.102w2502D}, \citet{2007PhRvC..75f5803C}, and \citet{2013PhRvC..87f5804K}.  These are necessarily from prompt-$\gamma$ experiments and suffer from the concerns about angular distribution discussed in Section~\ref{subsec:WG7:S34Expt}.  However, their inclusion simplifies the fitting considerably by breaking parameter degeneracies between ground- and excited-state transitions (especially in the fitted ANCs), probably without strong impact on $S_{34}(0)$.

In constructing fits we split common-mode and point-to-point errors and ``float the norms'' of data sets using the cost function
\begin{equation}
\chi^2=\sum_{\alpha=1}^{N_{\text{set}}}\sum_{i=1}^N\frac{(y_i-\mu_i/s_\alpha)^2}{\sigma_{\alpha,i}^2}+\sum_{\alpha=1}^{N_{\text{set}}}\frac{(1-s_\alpha)^2}{\omega_{\alpha}^2}
\end{equation}
to describe goodness of fit \cite{1994NIMPA.346..306D}.

A rescaling factor $s_\alpha$ is fitted for each data set $\alpha$, with its deviation from unity penalized by the common-mode errors $\omega_\alpha$ given in Table \ref{tab:WG7_S34_Data}.  The index $i$ sums over all points within a given data set; $y_i$ is a measured cross section, $\mu_i$ is a predicted cross section, and $\sigma_{\alpha,i}$ is the point-to-point error of the $i$th point in data set $\alpha$.   

We carried out both frequentist fits that minimize $\chi^2$ and Bayesian fits based on the posterior probability distribution of a likelihood function computed from $\exp(-\chi^2/2)$.     In the Bayesian fits parameters, extrapolated $S_{34}(0)$, and their errors were determined by Monte Carlo sampling of the parameter space.  The relatively large number of model parameters and the significant fitting degeneracies between some of them make Bayesian analysis a natural choice for finding best fit and confidence intervals for the multi-parameter models applied here. Two of the groups ran into serious difficulty with the frequentist fits due to parameter degeneracies and shallow local minima; during this study they were unable to produce frequentist fits in which they had confidence.  

We performed multiple fits in the NLO halo EFT of \citet{2018EPJA...54...89H} and \citet{2020EPJA...56..166P}.  These fits were carried out both with and without scattering constraints, which mainly impact $S_{34}(0)$ by removing parameter degeneracies that would otherwise leave the $s$-wave scattering length poorly constrained.  The code base for this version of halo EFT incorporates scattering data through phase shifts.  These were taken from the partial-wave analysis of \citet{1972NuPhA.195..241B}, which at low energy are based mainly on the data of \citet{1964NucPh..50..629B}.

One set of fits for the \citet{2018EPJA...54...89H} halo EFT was produced by $\chi^2$ minimization, proceeding in two steps: first LO parameters were fitted to $E\leq 1000$ keV data, and then the results were taken as initial values in the search to minimize $\chi^2$ over all parameters in the NLO theory for the full set of data.  These fits gave $S_{34}(0)=0.566\pm 0.025$ \mbox{keV b} with the scattering constraint and $S_{34}(0)=0.588\pm 0.015$ \mbox{keV b} without (uncertainties being propagated in linear approximation using covariances and partial derivatives).  Formally the EFT truncation error from stopping at NLO corresponds to an additional theoretical error that can shift S-factors by 10\%.  However, any fitted curve is constrained by low-energy data, which in some sense become effectively renormalization conditions of the field theory; the error on extrapolated $S_{34}(0)$ should probably be smaller than 10\% by an amount that is hard to estimate. 

A second set of fits to the \citet{2018EPJA...54...89H} halo EFT was carried out using Bayesian methods.  Priors for the EFT parameters were developed based on previous experience, and data rescaling factors were incorporated as additional priors.  The Bayesian results (including only experimental error) are $S_{34}(0)=0.561^{+0.017}_{-0.018}$ \mbox{keV b} including the phase shifts by \citet{1972NuPhA.195..241B} and $S_{34}(0)=0.559^{+0.018}_{-0.019}$ \mbox{keV b} excluding them, consistent with the $\chi^2$-minimization. 

Searches of the parameter space to minimize $\chi^2$ for the NLO halo EFT of \citet{2020JPhG...47e4002Z} ran into difficulties with parameter degeneracy and local minima in the $\chi^2$ surface;  for this formalism we report only Bayesian results.  No experimental information about scattering was used, but flat priors on scattering length and effective range were chosen over a 5$\sigma$ range around recent experiment.   This fit differs from \citet{2020JPhG...47e4002Z} mainly by excluding the Notre Dame data and including the ERNA activation data, and it gives $S_{34}(0)= 0.581\pm 0.016$ \mbox{keV b}.  The error associated with EFT truncation at NLO is estimated in this approach by separately fitting the partial-N4LO theory discussed above.  The result suggests that EFT truncation at NLO affects extrapolation from the data to threshold by $\sim 2$--3\%. 

Our R-matrix fits are based mainly on sampling Bayesian posterior probabilities with the BRICK code.   We also produced frequentist fits, but we were unable to estimate their errors convincingly.  The R-matrix fits used capture data both alone and in combination with the SONIK scattering data  \cite{2024PhRvC.109a5802P} -- the latter being chosen because of concerns with the \citet{1964NucPh..50..629B} data that are discussed in \citet{2022FrP....10.8476O}.  Since BRICK fits elastic differential cross sections directly, it was not feasible to use the same phase-shift-based scattering constraints as our halo EFT fits.  The R-matrix fit that includes scattering data is essentially the ``CS'' fit of \citet{2022FrP....10.8476O}, but with a restriction to only the capture data described in Table \ref{tab:WG7_S34_Data}, only scattering data below 2 MeV center-of-mass, and only the R-matrix parameters relevant below 2 MeV (no $7/2^-$ level or radiative widths for $d$-wave background poles).

\begin{table}
  \caption{Main results of the Bayesian fits performed on the $S_{34}$ data. The ANC refers to the sum of the ANCs for the ground and excited state transitions. See text for details.}
  \label{tab:fitted-S34}
  \begin{tabular}{lcccl}
    \hline\hline
    Fit & $S_{34}(23)$  & $S'_{34}/S_{34}$  & $a_0$     &   $C^2$ \\ 
        & [keV b] & [MeV$^{-1}$]    &  [fm]     &   [fm$^{-1}$]   \\ \hline 
%
\citet{2018EPJA...54...89H} EFT-NLO & 0.554 & -0.56  &  30$^{+5}_{-2}$ & $26.9^{+4.5}_{-5.1}$ \\
\citet{2020JPhG...47e4002Z} EFT-NLO & 0.573 &   -0.61   &    46(6)  &   29(3)\\
BRICK R-matrix          &    0.562 & -0.57  &   &  24.4(9) \\
    BRICK R-matrix+scattering &   0.531   &   -0.47   &   & 22.6(7)  \\ \hline 
    ANC \cite{2020PhLB..80735606K}  &   &   & & $34\pm1.6$ \\
    \hline \hline
  \end{tabular}
  
\end{table}

\begin{figure}
    \includegraphics[width=\columnwidth]{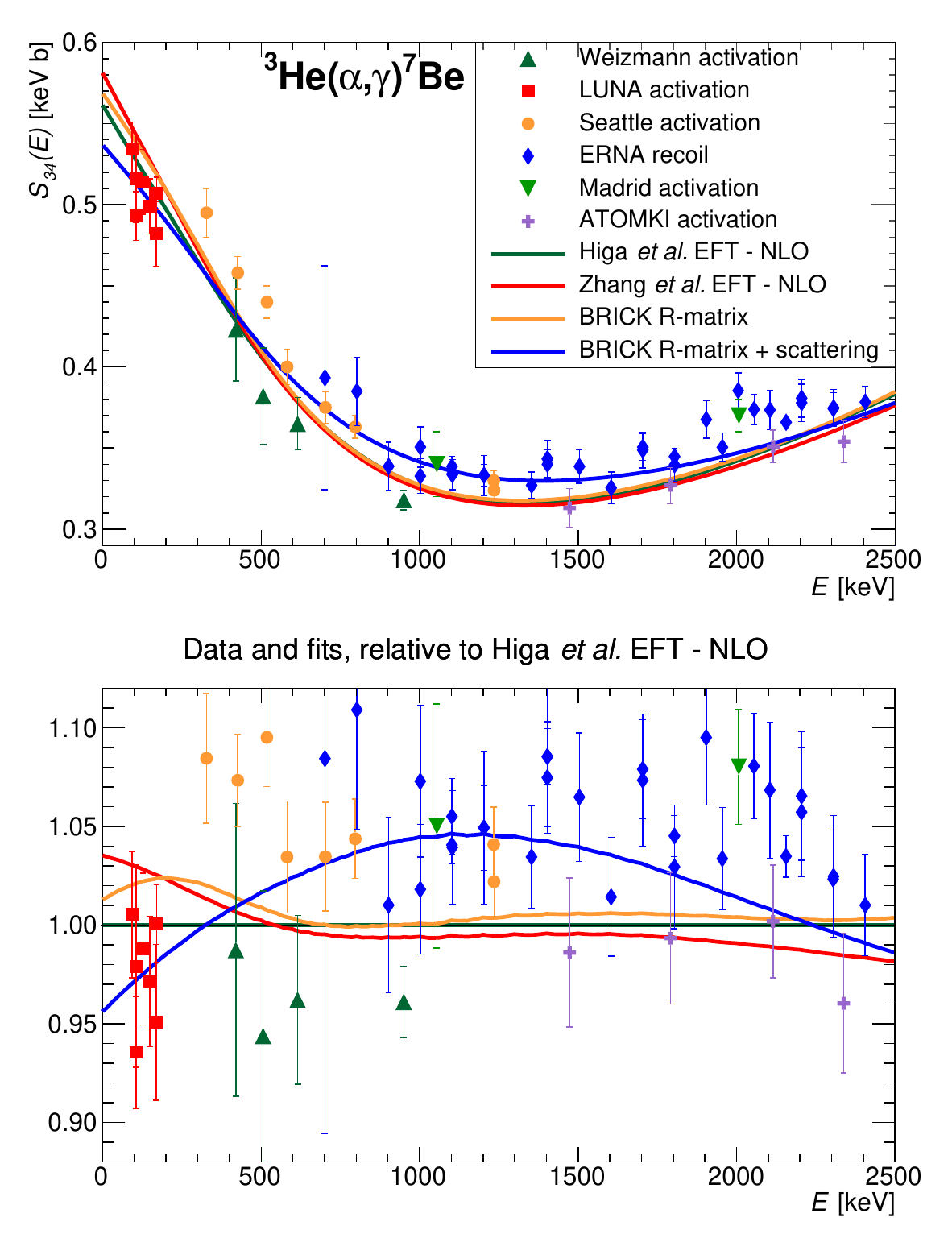}
    \caption{
    Top panel: $^3$He($\alpha,\gamma$)$^7$Be astrophysical S-factors from the experiments used for the fitting of the excitation function: Weizmann \cite{2004PhRvL..93z2503S}, LUNA \cite{2006PhRvL..97l2502B,2007PhRvC..75f5803C,2007PhRvC..75c5805G}, Seattle \cite{2007PhRvC..76e5801B}, Madrid \cite{2012PhRvC..86c2801C}, and ATOMKI \cite{2013NuPhA.908....1B}. The fit curves from two different EFT based fits (approaches based on \citet{2018EPJA...54...89H} and \citet{2020JPhG...47e4002Z}) and from two R-matrix fits (BRICK \citet{2022FrP....10.8476O}, without or with elastic scattering data) are also shown. 
    Bottom panel: Experimental data and fits, normalized to the \citet{2018EPJA...54...89H} based EFT fit. See Table \ref{tab:WG7_S34_Data} and text for details.
    }
    \label{fig:WG7:S34} %
\end{figure}

Several of the fits are compared in Fig. \ref{fig:WG7:S34}.  There it is apparent that they all agree within $\leq$5\% in the 0--2000 keV energy range, and within $\leq$4\% in the astrophysically relevant range 0--500 keV.  The fit to the \citet{2018EPJA...54...89H} NLO EFT lies in the middle of the range, so we adopt it as a reference in the lower panel of the figure.  Relative to that fit, the fits to the NLO EFT of \citet{2020JPhG...47e4002Z} and the R-matrix fit without scattering rise up as much as 4\% higher at 0--200 keV.

The most striking difference among the curves is between the BRICK R-matrix fit including the SONIK scattering data \cite{2024PhRvC.109a5802P} and all the other fits.  While the R-matrix fit with scattering finds rescaling factors somewhat closer to unity than the other fits do, its energy dependence in the 0--1000 keV range is qualitatively different, as is especially visible in the lower panel of Fig. \ref{fig:WG7:S34} where residuals relative to the fitted \citet{2018EPJA...54...89H} EFT are shown.  The BRICK fit without the scattering data is much closer to all the halo EFT fits, so the recent scattering data apparently have a large impact.

\subsection{Recommended \texorpdfstring{${{S}}_{34}$}{S34} value}
\label{subsec:WG7:Recommended}

Because the fit to the NLO EFT of \citet{2018EPJA...54...89H} gives the central result among those attempted here, we adopt it for our recommended $S(E)$.  For ease of adoption, it is noted that this recommended $S_{34}$ curve can be empirically parameterized using the same shape as in \citetalias{2011RvMP...83..195A}, by the numerical equation (following a customary form in the past literature \cite{1987ApJ...319..531K,
2011RvMP...83..195A})
\begin{eqnarray} \label{eq:WG7_S34}
    S_{34}(E) & = & \left(0.5610\ \mathrm{keV\ b}\right) \, \exp \left( -0.5374 \, E \right) \\ 
    & & \times    \left[ 1 - 0.4829 \, E^2 + 0.6310 \, E^3 - 0.1527 \, E^4    \right]. \nonumber
\end{eqnarray}
Here $E$ is the center of mass energy in MeV. Equation (\ref{eq:WG7_S34}) reproduces our recommended curve within 0.3\% and is only applicable for $E$ = 0--1600 keV.  

Propagation of uncertainties from measurement to extrapolated $S_{34}(0)$ is relatively straightforward and unambiguous (at least in the Bayesian fitting).  However, the uncertainty due to theoretical approximation is more complicated, and error estimation on extrapolated quantities in EFTs remains an open area of research; estimates from the groups working on $S_{34}$ are discussed above.   We estimate the overall theoretical uncertainty on our recommended S-factor to be the measured by the 4\% dispersion among fitted curves; this is comparable to the more formal estimates.

Our recommendations for the zero energy astrophysical S-factor and the low-energy slope are then
\begin{eqnarray}
\label{eq:S34_Recommendation}
    S_{34}(0) & =   &   0.561 \pm 0.018_{\rm exp} \pm 0.022_{\rm theor} \, \mathrm{keV\ b}  \\
    \frac{S_{34}'(0)}{S_{34}(0)}  &   = & -0.54\pm0.07 \, \mathrm{MeV}^{-1}. 
\end{eqnarray}
The given value of $S_{34}'/S_{34}$ applies over $E$ = 0-500 keV.  Its considerable uncertainty is given by the full span of slopes among the four fits discussed above, which is dominated by the inclusion of elastic scattering in one of the R-matrix fits. The $S_{34}(E)$ curve is quite straight at low energy in all models; its second derivative is consistent with zero and even difficult to compute for a given fit because of numerical cancellations.

The first error bar for $S_{34}(0)$ is given by the fit uncertainty, conservatively using the largest value among the various fits, and it encodes the experimental uncertainty. The second error bar reflects the 4\% dispersion among theoretical formalisms. The two errors can be added in quadrature for applications.  Our recommendation is essentially the same as in \citetalias{2011RvMP...83..195A}, but derived from completely different theoretical approaches that are much more directly connected to the data and much less dependent on model assumptions. It is also based on data over a wider energy range.  The dispersion of $S_{34}$ and $S_{34}^\prime$ among models at the solar Gamow peak is illustrated in Table \ref{tab:fitted-S34}, along with some indication of how fitted scattering lengths $a_0$ and squared ANCs $C^2$ vary among frameworks.  (The latter are highly degenerate with short-range parameters in the fitting; in the \citet{2018EPJA...54...89H} EFT they are sensitive to small variations in the $p$-wave effective ranges that are fitted instead of ANCs.)

\subsection{Recommendations for future work on \texorpdfstring{${{S}}_{34}$}{S34}}
\label{subsec:WG7:FutureWork}

Here we present recommendations for further work that would improve understanding of the $^3$He($\alpha,\gamma$)$^7$Be reaction.  Our first recommendation for experiment concerns elastic $^3$He+$^4$He scattering. New experiments over a wide energy range including at least part of the astrophysical range are needed to address the marked impact of the recent SONIK scattering data \cite{2024PhRvC.109a5802P} on fitting, for independent confirmation.  Data below \mbox{2 MeV} would be most useful for consistent fitting of halo EFTs.

Second, new measurements (besides the preliminary data of \citet{Turkat23-PhD}) are needed for the angular distribution of the emitted $\gamma$-rays.  This would relieve concerns about including prompt-$\gamma$ data, most notably the Notre Dame \cite{2013PhRvC..87f5804K} that do not include activation measurements, in future fits. These experiments should also be designed to provide new angle-corrected branching ratios for capture into the two bound states. 

Our third recommendation is a study of the astrophysical S-factor over a wide energy range, 50-1500 keV and extending to even lower energies when technically feasible, with small point-to-point errors.  Even with a large common-mode error, this would significantly constrain theoretical curve shapes near threshold, and providing an independent check on the influence of scattering data on $S(E)$.

On the theoretical side, further development of \textit{ab initio} reaction calculations is desirable, improving on points of current difficulty like explicit inclusion of three-body potentials.  Calculations with the same potential but by different methods would help to separate computational issues from the impact of potential choice.  Further exploration of how \textit{ab initio} inputs like ANCs and scattering lengths could serve as fitting priors would also be a useful avenue of research.

Authors of future theoretical calculations should consider including $\gamma$-ray angular distributions in their results; 
until very recently angular distributions were only available from two potential models \cite{1981PhRvC..23...33K,1963PhRv..131.2582T}.  New calculations would inform analysis of experiments, and the angular distribution measurements we recommend would provide tests of the calculations (or fitting constraints).

A source of complication here has been the need for a good meeting point between scattering data and theory.  Given the difficulties of constructing a unique phase shift analysis without an underlying theory, and the recent arrival of modern scattering data, theorists should consider including elastic differential cross sections directly as outputs and fitting constraints. In addition, efforts should be made to extend EFT fits to higher energies, 
in order to take into account the 3 MeV resonance.



\section{The \texorpdfstring{\lowercase{hep} reaction ($S_\text{{hep}}$)}{hep reaction (Shep)}} 
\label{sec:hep}

The reaction $p +\,^3\text{He}\rightarrow\,^4\text{He}+e^++\nu_e$,
also known as the hep reaction, is one of the possible processes
taking place after the $^2{\rm H}(p,\gamma)^3{\rm He}$,
and it produces the most energetic neutrinos,
with an endpoint energy of 18.8 MeV.
It is however the least probable, with its S-factor,
and consequently its rate,
even beyond the reach of current
experiments. This is due to the fact that
the hep reaction is induced by the weak interaction and further suppressed
by Coulomb barrier and by the fact that the
leading one-body axial (Gamow-Teller) current operator cannot connect the main
$s$-state components of the
$p+^3{\rm He}$ and $^4{\rm He}$ initial- and final-state wave functions.
As a consequence, the one-body axial current non-zero contribution
is due to the small components of the $^3{\rm He}$ and $^4{\rm He}$ wave
functions. Furthermore, other contributions, such as weak magnetism and other one-body corrections to the vector current, and
two-body vector and axial currents,
and the $p$-waves in the $p+^3{\rm He}$ initial state, normally
suppressed at solar temperatures, are significantly enhanced. 
This is further complicated by cancellations between the one-body and two-body axial current contributions.
As a consequence, similarly to the pp fusion, the hep reaction cross section is too
small to be measured in laboratory and only theoretical predictions are available.

The most recent studies of this reaction, already 
evaluated in \citetalias{2011RvMP...83..195A}, 
are those of \citet{2000PhRvC..63a5801M} and \citet{2003PhRvC..67e5206P}.
The first one
was performed within a phenomenological approach, similar to the calculation of 
\citet{1998PhRvC..58.1263S} for the pp fusion. In this particular case,
the $p+^3{\rm He}$ and $^4{\rm He}$ initial and
final nuclear wave functions were obtained with the 
correlated Hyperspherical Harmonics (CHH) variational
method~\cite{1995FBS....18...25V,1998PhRvL..81.1580V}, using 
the Argonne $v_{18}$ (AV18) two-nucleon potential~\cite{1995PhRvC..51...38W}
augmented by the Urbana IX (UIX) three-nucleon
interaction~\cite{1995PhRvL..74.4396P}.
In this way, the binding energies of $^3$He and $^4$He, and the singlet and
triplet $p+^3{\rm He}$ scattering lengths were in good agreement with
experiment. The weak vector and axial transition operators were obtained
within a phenomonological approach. In particular, the largest two-body
contribution to the axial current arising from the excitation of intermediate
$\Delta$-isobar degrees of freedom was included using the
transition-correlation operator scheme
of \citet{1991PhRvC..44..619C,1992PhRvC..45.2628S}
and fixing the
nucleon-to-$\Delta$ axial coupling constant to reproduce $\Gamma^T_\beta$ as in 
the pp case~\cite{1998PhRvC..58.1263S}.

The second study of \citet{2003PhRvC..67e5206P} was performed within
the same hybrid EFT approach used for the pp funsion (EFT$^*$), with the same 
nuclear weak current operators obtained
within EFT as discussed in Section~\ref{sec:S11}, 
the LEC $\hat{d}^{R}$ being fixed to reproduce $\Gamma^T_\beta$. 
The initial and final wave functions were derived using
the CHH method with the AV18/UIX potential,
as in \citet{2000PhRvC..63a5801M}. 

Following the results of \citet{2000PhRvC..63a5801M} and \citet{2003PhRvC..67e5206P}, \citetalias{2011RvMP...83..195A} recommended for
the zero-energy S-factor, $S_{\rm hep}(0)$, the 
value $(8.6 \pm 2.6) \times 10^{-20}$ keV b.
Three recommendations were also made:
(i) to perform new studies with a broad
spectrum of Hamiltonian models, in order to properly
access the theoretical uncertainty; (ii) to study other weak reactions for which experiments can test the predictions made by employing the same theoretical ingredients as those used for the hep reaction calculation; (iii) to further understand the relation
between the hep reaction and the so-called hen process
($n+\,^3{\rm He}\rightarrow\,^4{\rm He}+\gamma$). Of these
recommended works, only the second one has been performed so far,
applying the theoretical framework used for the hep study to the muon capture
reactions on light nuclei~\cite{2011PhRvC..83a4002M}.

Given the lack of further studies on the hep reaction after \citetalias{2011RvMP...83..195A}, we decide to maintain  
\begin{equation}
    S_{\rm hep}(0)=(8.6\pm 2.6)\times 10^{-20}\,\,{\rm keV~b}
    \label{eq:S_hep}
\end{equation}
as recommended value. Notice that we could have increased $S_{\rm hep}(0)$ by $\sim 1$\%, in analogy with $S_{11}(0)$, due to the updated values of input parameters (mostly $G_A/G_V$). However, we have decided not to apply such an increase, because it does not derive from a direct calculation, and, most of all, because such an increase lies well within the quoted $\sim 30$\% uncertainty. 
However, given the advances in \textit{ab initio} studies of the
last decade and in the development of the EFT framework,
we update the recommendations of \citetalias{2011RvMP...83..195A}
as follows: (i) the hep process should be revisited within a fully
consistent $\chi$EFT approach, similarly to what has been done in these years
for the pp fusion and other processes, as, for instance, muon captures
on light nuclei \cite{2012PhRvL.108e2502M,2018PhRvL.121d9901M,2018PhRvC..98f5506A,2023FrP....1049919C,2023PhRvC.107f5502B,PhysRevC.109.035502};
(ii) the initial and final wave functions should be calculated with the
most accurate \textit{ab initio} methods, as the uncorrelated hyperspherical harmonic (HH)
method~\cite{2008JPhG...35f3101K,2020FrP.....8...69M}, which has been
proven to provide more
accurate $A=4$ bound and scattering wave functions than the CHH
method, using both local and
non-local interactions; (iii) the relation between the hep and the hen
process should be understood. The hen process has been studied within $\chi$EFT and the HH method
by \citet{2022PhRvC.105a4001V}, but a consistent
parallel study of both $A=4$ reactions is still missing.

A relatively recent SNO Collaboration analysis of data from all three of the detector's running
phases yielded a one-sided confidence-interval bound on the hep neutrino flux of 
\citet{2020PhRvD.102f2006A}

\begin{equation}
\phi_\mathrm{hep} < 30 \times 10^3 ~ \mathrm{cm}^{-2} \mathrm{s}^{-1} ~(90\% ~\mathrm{CL}),
\end{equation}
a result in agreement with the SSM prediction.  The lack of a definite measurement,
however, has meant that there is no substantive experimental test of the predicted $S_\mathrm{hep}(0).$  Various global analyses of
solar neutrino data have provided weak evidence for a nonzero flux, e.g.,
from \citet{Bergstrom:2016cbh}
\begin{equation}
\phi_\mathrm{hep}\Big|_\mathrm{global~analysis} = 19^{+12}_{-9}   \times 10^3 ~ \mathrm{cm}^{-2} \mathrm{s}^{-1},   
\end{equation}
which is consistent with the experiment bound given above.

An extraction of the hep flux from experiment will likely require a detailed shape analysis to separate
hep neutrinos from the high-energy tail of the $^8$B spectrum and the low-energy tail of atmospheric neutrinos.
Typically the hep spectrum is assumed to have an allowed shape: indeed, in the work reported here spectra are not provided, where $p$-waves, weak magnetism, and similar corrections have been included. Because these corrections
are unusually large for hep neutrinos, future studies should provide spectra and evaluate the impact of such
corrections on the shape.
\vspace{3ex}

\section{Electron capture by {\textit{\MakeLowercase{p}+\MakeLowercase{p}}} and \texorpdfstring{$^7\text{Be}$}{7Be}} 

\label{sec:EC}

Electron capture reactions are the sources of lines found in the solar neutrino spectrum. In particular,
electron capture on $p+p$, i.e., the process $p+p+e^-\rightarrow $ d$+\nu_e$,
also known as the pep reaction, competes with
$p+p$ fusion and depends on the same nuclear matrix element.
Therefore, the ratio between the pp and pep rates is independent
of nuclear physics. Based on this consideration,
in \citetalias{2011RvMP...83..195A} the result of \citetalias{1998RvMP...70.1265A}, which is based on the work of
\citet{1969ApJ...155..501B}, was multiplied by the radiative corrections
calculated by \citet{2003PhRvC..67c5502K}, leading to the final result
given by
\begin{eqnarray}
  R({pep})&=&1.130(1\pm0.01)\times 10^{-4} (\rho/\mu_e)
  \nonumber \\
  &\times& T_6^{-1/2}[1+0.02(T_6-16)]\,R(pp)\label{eq:rate_pep}\; ,
\end{eqnarray}
where $\rho$ is the density in units of g/cm$^3$, $\mu_e$ is the mean
molecular weight per free electron, and $T_6$ is the temperature in units of
megakelvins. The range of validity is $10<T_6<16$. Given that no new evaluation
of the pep rate has been performed since \citetalias{2011RvMP...83..195A}, 
the result of Eq.~(\ref{eq:rate_pep}) represents also the present
recommended value.

Competition between electron and proton capture on $^7$Be fixes the branching ratio of
the ppII and ppIII chains, and thereby the $^7$Be and $^8$B neutrino
fluxes. Superallowed electron capture on the $J^\pi=3/2^-$
ground state of $^7$Be leads to the $J^\pi=3/2^-$ ground state or the $J^\pi=1/2^-$, 478~keV first excited state of $^7$Li with a measured (terrestrial) branching ratio of 10.44(4)\% \cite{2002NuPhA.708....3T}. It is customary to calculate the solar decay rate 
in terms of the measured terrestrial decay rate, which has the advantage of
removing the nuclear physics dependence, but it requires a proper
calculation of the densities of electrons at the
nucleus. This has usually been done by considering continuum and bound electrons separately, using the Debye-H\"uckel approximation pioneered by \citet{1954AuJPh...7..373S} 
for the screening of bound electrons in the solar plasma; the capture rate of continuum electrons is not significantly influenced by electron screening.
Furthermore, it is necessary to calculate 
the atomic probability densities governing the $K$ and $L$ terrestrial
electron capture rates. \citet{2020PhRvL.125c2701F} recently measured the ratio of $L$ to $K$ capture using superconducting tunnel junctions.
The recommendation of both \citetalias{1998RvMP...70.1265A} and \citetalias{2011RvMP...83..195A} was
\begin{widetext}
  \begin{equation}
    R(^7\mathrm{Be}+e^-) = 5.60 (1 \pm 0.02) \times 10^{-9}~~ (\rho/\mu_e)
    ~~T_6^{-1/2}[1+0.004(T_6-16)] ~\mathrm{s}^{-1}, \label{eq:EC-7Be-rate}\ 
  \end{equation}
\end{widetext}
which is valid for $10 < T_6 < 16$. This expression is based on the continuum capture rate calculated by \citet{1969ApJ...155..511B}, the average ratio of the total capture rate to the continuum rate calculated using three solar models in \citet{1994PhRvD..49.3923B}, and the current value of the half-life, 53.22(6)~d \cite{2002NuPhA.708....3T}. It agrees within 1\% with a density matrix calculation of the rate that makes no assumptions regarding the nature of electronic quantum states in the solar plasma and allows for aspherical fluctuations in the spatial distributions of plasma ions \cite{1997ApJ...490..437G}. The estimated uncertainty of 2\% accounts for possible corrections to the Debye-H\"uckel approximation due to thermal fluctuations in the small number of ions in the Debye sphere, and breakdowns in the adiabatic approximation \cite{1992ApJ...392..320J}. 

Although calculations have appeared in the literature claiming new plasma effects based on 
model calculations (see \citet{2013ApJ...764..118S,2019A&A...623A.126V}), we continue to regard the arguments in
\citet{2002A&A...383..291B} as definitive. That reference presented five distinct derivations demonstrating that the
Salpeter formula for screening corrections is valid for pp-chain reactions and solar conditions, up to corrections
on the order of a few percent.



\section{The \texorpdfstring{$^7\text{Be}(p,\gamma)^{8}\text{B}$ reaction ($S_{17}$)}{7Be(p,γ)8B reaction (S17)}} \label{sec:S17}

\def\pBe{$^7\mathrm{Be}(p,\gamma)^8\mathrm{B}$}
\newcommand*\xbar[1]{%
	\hbox{%
		\vbox{%
			\hrule height 0.5pt 
			\kern0.3ex
			\hbox{%
				\kern-0.1em
				\ensuremath{#1}%
				\kern-0.1em
			}%
		}%
	}%
}

\subsection{Introduction} \label{subsec:WG5:Intro}
	Radiative proton capture on $\rm ^7Be$ ($\rm J^\pi$=3/2$^-$) at solar energies proceeds through non-resonant capture to the ground state of $\rm ^8B$ ($\rm J^\pi$=2$^+$). This capture occurs predominantly at separations well beyond the range of the strong interaction via $E1$ transitions from $s$ and $d$ partial waves, as the $M1$ amplitude and contributions from higher partial waves are negligible in the energy range of interest \cite{1961NucPh..24...89C}. The importance of this reaction for the determination of the high energy solar neutrino spectrum and the experimental data reviewed by \citetalias{1998RvMP...70.1265A} inspired several experiments that, taken together, provided a consistent picture of the energy dependence of the reaction cross section, despite discrepancies in absolute scale. Prior to the review of \citetalias{2011RvMP...83..195A}, most direct measurements were performed using a radioactive target and an intense proton beam. In spite of the great care taken in evaluating the systematic energy-independent uncertainties, in the following referred to as common-mode errors (CMEs), a remarkable discrepancy persisted even in the most recent experiments, that was finally handled in \citetalias{2011RvMP...83..195A} by inflating the stated experimental uncertainties. 
	Indirect measurements of the cross section of $^7\mathrm{Be}(p,\gamma)^8$B based on Coulomb dissociation were considered, but not included in the final recommendation of \citetalias{2011RvMP...83..195A} for the zero-energy astrophysical S-factor $\rm S_{17}(0)$. However, despite the experimental issues, the uncertainty on the recommended value of $\rm S_{17}(0)$ in \citetalias{2011RvMP...83..195A} is dominated by the theoretical contribution. Therefore, our analysis focused on these aspects (discrepancy of experimental data, indirect measurements, theoretical uncertainties), as discussed in the next sections.

	\subsection{Experimental Data}\label{subsec:WG5:Data}
	
	\subsubsection{Direct Measurements}
	The thorough review of published work done in \citetalias{2011RvMP...83..195A} investigated the influence of beam-target overlap, target stoichiometry, beam energy loss, and the backscattering of $\rm ^8$B recoils on CMEs and led to the selection of a homogeneous group of well-documented data sets. Since then, no further information became available on these aspects of direct measurements. 
	The data that formed the basis of the \citetalias{2011RvMP...83..195A} recommendation were those of \citet{2003PhRvC..67f5805B}, \citet{2003PhRvL..90b2501B}, \citet{1998PhRvL..80..928H}, \citet{2001PhRvL..86.3985H}, \citet{1999PhLB..462..237H}, \citet{2001NuPhA.696..219S}, and the BE3 data set of \citet{2003PhRvC..68f5803J}. 
	We considered the complete set of radiative capture measurements that were the basis of the \citetalias{2011RvMP...83..195A} recommendation for $\rm S_{17}(0)$, where we adopted the revised BE3 data presented by \citet{2010PhRvC..81a2801J}, which were reanalyzed using a more  sophisticated model of the target and more accurate stopping powers. In addition, we included in our analysis
	 the only new experiment reported thereafter of \citet{2022PhLB..82436819B}, which used a radioactive $\rm ^7Be$ ion beam and the recoil mass separator ERNA to detect the $\rm ^8$B recoils. While these new results contribute to the determination of $S_{17}$ with different, well controlled systematics \cite{2018EPJA...54...92B}, their large statistical uncertainty relative to the other experiments, due to the low counting statistics, limits their impact.
	
	New scattering data have been published. Elastic and inelastic scattering cross sections at relative kinetic energies between 474~keV and 2.74~MeV were measured and $s$ wave scattering lengths were inferred by \citet{2019PhRvC..99d5807P}. At higher energies between 1.6 and 3.4~MeV, thick target elastic and inelastic scattering excitation function measurements were used to infer the existence of new resonances above 1.8~MeV \cite{2013PhRvC..87e4617M}. These high lying resonances have no direct influence on $S_{17}(0)$ because of their small partial widths in the $^7\mathrm{Be}_\text{gs}+p$ channel and/or because they cannot be populated by s- or d-wave capture at low energy.

		\subsubsection{Indirect Measurements}
		
		In \citetalias{2011RvMP...83..195A}, Coulomb dissociation measurements based on the formalism developed by \citet{1986NuPhA.458..188B} were considered \cite{PhysRevLett.83.2910,PhysRevLett.73.2680,KIKUCHI1997261,1998EPJA....3..213K,PhysRevLett.86.2750,PhysRevC.63.065806, PhysRevLett.90.232501,2006PhRvC..73a5806S}, where in some cases the most recent publications supersede the previously published ones. Finally, the results of these experiments were not included in the determination of $S_{17}(0)$ due to disagreements over whether the analyses of the various measurements properly accounted for the contributions of $E2$ transitions.
		We disagree with these reservations and find the agreement of these measurements with radiative capture measurements significant. However, in order to include these consistent data in formulating our recommendation for $S_{17}(0)$, we inflated the CMEs of the Coulomb breakup measurements to adequately account for the uncertainty in the E2 components assumed in each analysis.
		
		The common mode errors are obtained from a linear sum of those given in the manuscripts and an additional common mode error due to the estimated size of the $E2$ component in each measurement. This additional error is applicable in one direction only, downward for the \citet{1998EPJA....3..213K} and \citet{2006PhRvC..73a5806S} analyses of the RIKEN and GSI measurements, respectively, and upward for the \citet{2003PhRvC..68d5802D} analysis of the Michigan State University (MSU) measurement, according to whether the $E2$ component was included or excluded. Though the \citet{2003PhRvC..68d5802D} analysis did include a common mode error contribution of 2.5\% due to the $E2$ component (which represented 5\% of the total cross section), we have included an additional 2.5\% contribution to the MSU experiment. The size of the additional common mode errors for the RIKEN and GSI measurements was based on scaling this 2.5\% according to the predicted size of the $E2$ contribution to the cross section in these measurements relative to that in the MSU measurement, evaluated using first-order perturbation theory  and the $E1$ and $E2$ S-factors calculated in a potential model of $^8$B \cite{1996ZPhyA.356..293B}. Thus the additional single-sided common mode errors added to the GSI, MSU, and RIKEN measurements are -2.1\%, 2.5\%, and -4.3\%, respectively.

		\subsection{Theory}\label{subsec:WG5:Theory}

		The size of the theoretical contribution to the uncertainty in the \citetalias{2011RvMP...83..195A} estimate of $S_{17}(0)$ was twice that of the experimental contribution. Since then, significant progress has been made.  A new potential model calculation was performed by \citet{2019NuPhA.983..175D}. The large-scale computational demands of calculations using high-quality nucleon-nucleon (NN) and three-nucleon (3N) interactions beyond $A$=5 had hindered the \textit{ab initio} approach in the past. Nevertheless progress has been rapid and it has been possible to extend \textit{ab initio} calculations to $^7\mathrm{Be}(p,\gamma)^8$B, albeit without 3N forces. \citet{2011PhLB..704..379N} performed such a calculation using the no-core shell model in the continuum with a nucleon-nucleon interaction derived from chiral EFT at fourth order. The calculation converged with respect to the model space size and reproduces experimental data for the S-factor despite having virtually no free parameters (though the similarity renormalization group evolution parameter was tuned to reproduce the binding energy of $^8$B). The final value of $S_{17}(0)$=19.4(7)~eV~b agrees with the \citetalias{2011RvMP...83..195A} recommendation and, as will be shown later, with our analysis. More recently, a new approach was introduced to combine experimental measurements with \textit{ab initio} predictions, resulting in an ``\textit{ab initio}-informed evaluation”~\citet{Kravvaris23-PLB} that arrived at a value of $S_{17}(0)=19.8(3)$~eV~b.  
		
		A significant theoretical advance was the application of the formalism of halo EFT developed by \citet{2002NuPhA.712...37B} to $^7$Be($p,\gamma)^8$B calculations by \citet{2015PhLB..751..535Z}, \citet{2018PhRvC..98c4616Z}, and \citet{2022PhRvC.106a4601H}. It treats the incoming $p$ and $^7$Be as point-like particles and the $^8$B final state as a bound state of the two. The calculation and the associated errors depend upon the accuracy with which one is able to describe the incoming scattering states, the final bound state and the relevant electromagnetic currents. The relative kinetic or c.m. energy range $E\lesssim 1$ MeV, corresponding to a relative momentum $p\lesssim 40$ MeV/c, is considered within the domain of applicability of halo EFT. The two cluster physical description of \pBe~ is expected to hold at $^3$He-$\alpha$ relative momenta below a physical cutoff $\Lambda\sim 70 $ MeV/c set by the threshold (binding momentum) for breaking the $^7$Be core into these constituents. Further, at energies above the excitation energy of the first excited state, $E_\star =0.4291$ MeV, including the contribution of the excited core $^7\text{Be}^\star$ is imperative in the EFT.  
		
		\citet{2015PhLB..751..535Z} and \citet{2018PhRvC..98c4616Z} include such an excited $^7\text{Be}$ contribution.
		\citet{2022PhRvC.106a4601H} provides a calculation without the  $^7\text{Be}^\star$ contribution below $E_\star$ and one with it above $E_\star$ that also includes an $M1$ contribution to the $1^+$ resonance. The two groups' halo EFT calculations with the $^7\text{Be}^\star$ contributions differ in several respects that are elaborated in \citet{2022PhRvC.106a4601H}. 
		Differences in the final bound state calculation affect the treatment of the short-range interactions and divergences in the EFTs. The interpretation of the asymptotic normalization constants (ANCs) in terms of elastic scattering parameters is affected. However, the momentum dependence of the capture cross section is not impacted by the ANCs, which are fitted to capture data. Thus there is no effect on the cross section. In the incoming channel, \citet{2022PhRvC.106a4601H} use a more general form of the $s$-wave short-ranged  interaction involving the $^7\text{Be}^\star$ core based on the low-energy symmetry that is included only at NNLO. In contrast, \citet{2015PhLB..751..535Z} and \citet{2018PhRvC..98c4616Z} include a short-ranged interaction at LO that could make a difference in the momentum dependence of the cross section. However, as shown in an order-by-order calculation~\cite{2022PhRvC.106a4601H}, the short-ranged interaction is a sub-leading effect so the two halo EFTs should have similar low-momentum dependence. The third important difference is in the treatment of two-body currents, which affects EFT error estimates. \citet{2015PhLB..751..535Z} and \citet{2018PhRvC..98c4616Z} include this at NLO whereas \citet{2022PhRvC.106a4601H} estimate it to be an N$^3$LO effect. Bayesian estimates of the two-body current \cite{2015PhLB..751..535Z,2018PhRvC..98c4616Z} are not strongly constrained by capture data, consistent with this being a higher-order contribution.

		Another widely used approach for describing the cross sections of low-energy nuclear reactions is R-matrix theory. In R-matrix theory the $^7$Be($p,\gamma)^8$B cross section can be parametrized in terms of a few parameters representing either ``real'' or ``background'' poles. The real poles correspond to the observed states and resonances, while the background terms correspond to the ``mean-field'' effects that are of the non-resonant type. In the present case for example, a background pole in the 1$^-$ channel is needed so as to reproduce the non-resonant $E1$ part of the capture. The parameters of the real pole corresponding to the bound state are determined by the binding energy and the two ANCs, while the 1$^+$ resonance parameters are determined by its energy and partial decay widths (both proton and radiative). The R-matrix analysis of this reaction carried out by \citet{1995NuPhA.588..693B} arrived at a value of $S_{17}(0)=17(3)$~eV~b; however, it was based on data that were excluded in \citetalias{2011RvMP...83..195A} and here, with the exception of those of \citet{1983PhRvC..28.2222F}.

		We fit data using R-matrix theory and halo EFTs, with both frequentist and Bayesian approaches, investigating the impact of different choices for the energy range in which data were fitted, and obtained consistent results. Our recommended $S_{17}(0)$ value and its uncertainty are based on a Bayesian fit of both direct and Coulomb dissociation data taken at energies $E \leq \SI{1250}{\kilo\eV}$ using the \citet{2022PhRvC.106a4601H} halo EFT$_\star$. 
		This theory provides a good description of capture data over a wide energy range including the $M1$ contribution from the $1^+$ resonance. The theoretical uncertainties are well understood. Further, the contributions of the $S=1$ and $S=2$ spin channels to the capture cross section are self-consistently parametrized with the elastic scattering information contained in the known ANCs for this system. 
		
		The Bayesian formalism naturally incorporates prior information concerning the ANCs in the fits.  
		The upper limit on the energy in the fit was determined to be high enough to include some data from all the modern experiments, and yet low enough that the contribution of the wide $3^+$ resonance at $E\approx\SI{2456}{\kilo\eV}$ does not exceed $1\%$ of the total cross section at any energy, as estimated by our R-matrix analysis. The fitting procedure is described in the next section, whereas the details, as well as the results of the fits done in different energy ranges and with different theories are reported in the supplemental material.

		\subsection{Fitting Procedures}\label{subsec:WG5:Fit}
		
		In \citetalias{2011RvMP...83..195A}, it was noted that the different $S_{17}(E)$ data sets had similar energy dependences. The discrepancies among the data sets were primarily due to different absolute normalizations. We address the normalization issue by introducing scaling factors associated with the CMEs.   This contrasts with the \citetalias{2011RvMP...83..195A} approach, in which the errors were inflated by the factor $\sqrt{\chi^2/\chi^2(P = 0.5)}$. The procedure for handling discrepant data sets adopted here is mathematically rigorous and aligns naturally with the method of estimating theoretical uncertainty discussed below. We prefer this approach to the  inflation factor method used in \citetalias{2011RvMP...83..195A}, since the absolute scales of the different data sets can be used as constraints for each other, while there is no information available for a reevaluation of the CMEs. The procedure we used, described below, can be applied to both frequentist and Bayesian analyses.
  
Although our final recommendation is based on Bayesian analysis alone, both $\chi^2$ minimization (a frequentist approach) and Bayesian posterior probability distribution function evaluation were performed to fit EFTs to the data; in the R-matrix analysis, only the former was used. The likelihood function, defined as the conditional probability $P(D|\bm{\theta})$ for the data $D$ given theoretical parameters $\bm{\theta}$, enters both fitting procedures. It can be derived from the $\chi^2$ expression given in \citet{1994NIMPA.346..306D}
		\begin{multline}\label{eq:cost}
			P(D|\bm{\theta},\bm s)= \prod_{\alpha=1}^{N_{\text{set}}} \frac{1}{\sqrt{2\pi \omega_\alpha^2}}
			\exp\left[-\frac{(1-s_\alpha)^2}{2\omega_\alpha^2}\right]\\
			\times\prod_{i=1}^{N_\alpha}\frac{1}{\sqrt{2\pi \sigma_{\alpha,i}^2}}
			\exp\left\{-\frac{[y_{\alpha,i}-\mu_{\alpha,i}(\bm\theta)/s_\alpha]^2}{2\sigma_{\alpha,i}^2}
			\right\}\, , 
		\end{multline}
		with $\alpha=1,\dots, N_{\text{set}}$ labeling the different experimental data sets and $i=1,\dots,N_\alpha$ labeling the individual data points of each experiment. In the second term, we divide theoretical predictions  $\mu_{\alpha,i}(\bm\theta)$ for the data points of the experiment $\alpha$ by a scaling factor $s_\alpha$ and consider the combination as our full model for $y_{\alpha,i}$, the $i$th data point of experiment $\alpha$ with point-to-point uncertainty $\sigma_{\alpha,i}$. 

		In the first term, we assign $s_\alpha$ a Gaussian distribution centered at $1$ with a width equal to the CME $\omega_\alpha$; for the GSI, MSU and RIKEN data shown in  Table~\ref{table:normsBayesian}, asymmetric Gaussians are used with $\omega_\alpha$ depending on the sign of $1-s_\alpha$. In our $\chi^2$ minimization, the $\chi^2$ definition is derived from equating the full exponents in Eq.~\ref{eq:cost} to $-\chi^2/2$. 
		In the absence of CMEs with $s_\alpha=1$, Eq.~\ref{eq:cost} leads to the usual uncorrelated least-squares minimization formula.

		In the Bayesian analysis, the terms in the first exponential are considered as part of the prior distribution  $P(\bm{\theta},\bm{s})$ that are conditioned on prior knowledge and assumptions such as the CMEs $\bm{\omega}$ in the scaling factors $\bm{s}$ or estimates of theoretical parameters $\bm\theta$ in the EFT. The posterior distribution for the theoretical parameters is then derived via Bayes's theorem, 
		\begin{align}\label{eq:bayes}
			P(\bm\theta, \bm{s}|D) = \frac{P_{\text{Bayes}}(D|\bm{\theta})P(\bm{\theta},\bm{s}) }{P(D)}\,,
		\end{align}
		where the likelihood function $P_{\text{Bayes}}(D|\bm{\theta})$ is defined without the first exponential in Eq.~(\ref{eq:cost}). However, since the GSI, MSU and RIKEN data have asymmetric CMEs, we draw priors from a uniform distribution spanning 50\% to 150\% of the central values for these  that are then used in the asymmetric normalization exponentials in the likelihood function. This was computationally simpler than generating asymmetric Gaussian priors that span the entire $(-\infty,\infty)$ range, which is neither necessary nor physical for a normalization constant.

	The constant $P(D)$ in Eq.~(\ref{eq:bayes}), known as the Bayesian evidence, guarantees the correct normalization of the posterior distribution. 
	It is useful in comparisons of different theories but does not affect the estimation of parameters.

	\subsection{Data Analysis and \texorpdfstring{$S_{17}$}{S17} Determination}\label{subsec:WG5:S17}
	
	We performed $\chi^2$ fits using several theoretical expressions over multiple energy ranges: $E \leq \SI{475}{\kilo\eV}$, $E \leq \SI{1250}{\kilo\eV}$, and over a ``non-resonant'' energy range  $E \leq \SI{490}{\kilo\eV}$ and  $\SI{805}{\kilo\eV}\leq E \leq \SI{1250}{\kilo\eV}$, in which the $M1$ contribution from the $1^+$ resonance can be ignored. Fits with and without Coulomb dissociation data were performed that resulted in overlapping $S_{17}(0)$ determinations within the estimated fitting uncertainties. The different low energy $E \leq \SI{475}{\kilo\eV}$ fits included one using the halo EFT  from 
	\citet{2018PhRvC..98c4616Z} and one using the halo EFT$_\star$ from \citet{2022PhRvC.106a4601H}, both including the excited $^7\text{Be}$ contribution. The simpler 
	halo EFT$_{\text{gs}}$ expressions, not including an excited component $^7\text{Be}^\star$, were also fitted in this low energy region with comparable results. For the fits in the other two energy ranges, halo EFT expressions with excited $^7\text{Be}^\star$ contributions from both \citet{2018PhRvC..98c4616Z} and \citet{2022PhRvC.106a4601H} were used. The \citet{2018PhRvC..98c4616Z} halo EFT expression, which does not include an $M1$ contribution, was supplemented with a Breit-Wigner resonance~\cite{1988ccna.book.....R} with its energy (around the $1^+$ resonance energy $E\approx\SI{630}{\kilo\eV}$), width, and proton-decay branching ratio determined by the fits. An R-matrix model was fitted to data in the three energy regions with the AZURE2 code~\cite{2010PhRvC..81d5805A}, again resulting in an $S_{17}(0)$ consistent with the halo EFT values. The results from the various fits are included in the supplemental material. 
	
	Bayesian fits were performed with halo EFT expressions from both \citet{2020JPhG...47e4002Z} and \citet{2018EPJA...54...89H} that are compatible with the various $\chi^2$ fits. 
	The Bayesian analysis had advantages over the $\chi^2$ fits in exploring parameter space and in determining realistic uncertainties \cite{PhysRevLett.122.232502}.

	The \pBe ~capture  proceeds through the $S=2$ and $S=1$ channels. The reaction is known to be peripheral, resulting in a $S_{17}(0)$ that has a sub-leading dependence on the strong interaction in the initial state \cite{2000PhRvC..62f5803B}. This behavior was shown to persist away from the threshold in halo EFT, in which for $E1$ capture the strong interaction in the incoming $s$-wave channel only contributes at NNLO. 

	Therefore, to high order, the capture is only sensitive to the sum of the squares of the ANCs in the $S=2$ and $S=1$ channels. 
	This results in an arbitrariness in the relative contributions from the two spin channels in the $\chi^2$ fits, without affecting the final $S_{17}$ determination. R-matrix, EFT$_{\text{gs}}$, and EFT$_\star$ fits confirm this behaviour. On the other hand, the ANCs have been extracted experimentally~\cite{2003PhRvC..67f2801T, 2006PhRvC..73b5808T} and also theoretically~\cite{2011PhRvC..83d1001N,2018PhRvC..98c4616Z} in \textit{ab initio} calculations. Although the determination of $S_{17}(0)$ is insensitive to the relative contributions of the two spin channels, knowledge of the relative contributions of the ANCs is important in the EFT framework. It establishes the hierarchy of different contributions in the perturbative expansion. This is crucial in developing a self-consistent expansion that is necessary to quantify the theoretical errors at any finite order of the perturbation. Prior knowledge of the ANCs and other parameters can be accounted for naturally in the Bayesian framework in which one specifies the prior probabilities $P(\bm\theta, \bm s)$. Similar constraints can be included in the $\chi^2$ fit by modifying the first exponential in Eq.~(\ref{eq:cost}) to center the theoretical parameters around known values.
	
	The Bayesian fits involved six theoretical parameters and 11 normalization constants $s_\alpha$ for the data sets. The priors for the normalization constants were as described earlier. The priors for the theoretical parameters are based on the underlying EFT assumptions~\cite{2022PhRvC.106a4601H} about the expansion in a small ratio $Q/\Lambda$ where $Q\sim\SI{40}{\mega\eV/c}$ represents the low momentum of infrared physics and $\Lambda\sim\SI{70}{\mega\eV}/c$ the cutoff of the theory. The $^5S_2$ and $^3S_1$ scattering lengths $a_0^{(2)}=-3.18^{+0.55}_{-0.50}$ fm and $a_0^{(1)}=17.34^{+1.11}_{-1.33}$ fm~\cite{2019PhRvC..99d5807P} are taken as inputs. For $E1$ capture at LO, the only fit parameter, the $^5P_2$ effective momentum $\rho$ is assumed to scale as $Q$ and accordingly we assume a uniform prior for $\rho$ between -100~MeV/$c$ and 1.5~MeV/$c$, where the upper limit is set by the physical constraint that the ANC squares have to be positive. At NLO, the $^3P_2$ ANC$^2$ $C_{1,1}^2$ is a fit parameter whose prior is taken from a normal distribution determined by experiment. At NNLO, we use the \textit{ab initio} calculation of \cite{2018PhRvC..98c4616Z} to draw the $^3P_2^\star$ ANC$^2$ from a normal distribution. A ratio of the $s$-wave scattering lengths in the coupled-channel $^3S_1$-$^3S_1^\star$ that is assumed to be $\mathcal O(1)$ in the EFT is drawn from a normal distribution with a mean of 0 and a standard deviation of 10 to cover a wide range. The $M1$ capture is dominated by the transition $^5P_1\rightarrow {}^5P_2$ and it requires three parameters: a $p$-wave scattering volume $a_1$, an effective momentum $r_1$ and a two-body current to regulate divergences. 
	\citet{2022PhRvC.106a4601H} determined $a_1=\SI{-108.13}{\femto\m^3}$ and $r_1=\SI{-111.23}{\mega\eV}$ from the the narrow $1^+$ resonance energy and width. Here, we kept $a_1$ fixed while drawing $r_1$ from a uniform distribution between -150 and -50~MeV. The Bayesian fit gave $r_1\sim \SI{-111.23}{\mega\eV}$. The two-body coupling was drawn from a uniform distribution and the fits gave a numerical value consistent with the EFT estimate.  
	We obtain the posterior distribution  $P(\bm\theta, \bm{s}|D)$ using a probabilistic integration method called Nested Sampling~\cite{Skilling06-BA} implemented in Python~\cite{2009MNRAS.398.1601F} that calculates both the posterior and the evidence.

\begin{figure}[htb]
		\begin{center}
			\includegraphics[width=0.48\textwidth,clip=true]{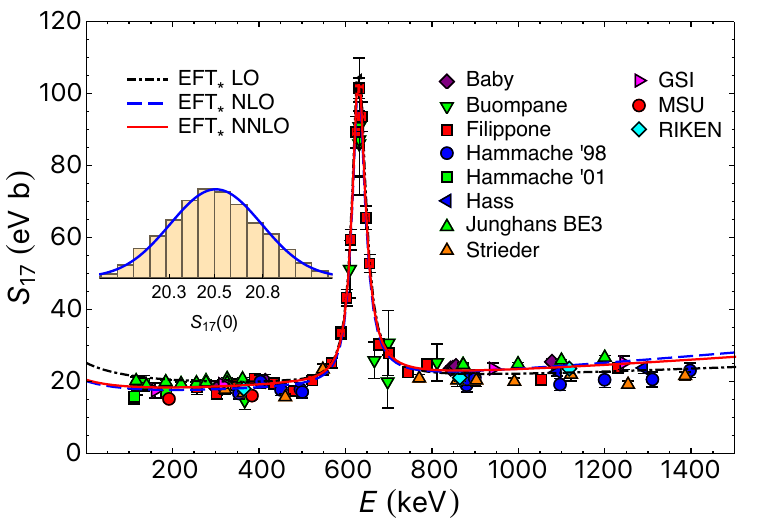}
		\end{center}
		\caption{\protect Astrophysical S-factor for \pBe. Data with $E\leq \SI{1250}{\kilo\eV}$ were included in the Bayesian fits shown here, which use the halo EFT of \citet{2022PhRvC.106a4601H} that includes an excited $^7$Be component. The dot-dashed (black) curve is the LO calculation, 
			the dashed (blue) curve is NLO, 
			and the solid (red) curve is NNLO. The inset shows the posterior probability distribution function for $S_{17}(0)$ in eV~b. The median as well as the 16th and 84th percentiles of the distribution are specified.
		}
		\label{fig:S17Bayesian}
	\end{figure}

	The experimental data and EFT$_\star$ fits at LO, NLO, and NNLO are shown in Fig.~\ref{fig:S17Bayesian}. $S_{17}$ and its derivatives at threshold are given in Table~\ref{table:S17Bayesian}. The quality of the fits, the consistency of the numerical values of the fit parameters with EFT assumptions, and the order-by-order improvement in the numerical value of the fitted capture cross section and S-factor give us confidence in the estimated theoretical uncertainty. We find the posterior probability distribution functions to be symmetric and well described by Gaussians, and as such we report the mean and standard deviation in Table~\ref{table:S17Bayesian}. Table~\ref{table:normsBayesian} gives the scaling factors for the experimental data sets determined from the NNLO fit along with their common mode errors.

	\begin{table}[htb]
		\centering
		\caption{$S_{17}$ and its first two energy derivatives at $E=50$~eV  determined from Bayesian fits of experimental data below 1250 keV using the halo EFT of \citet{2022PhRvC.106a4601H} that includes an excited $^7$Be component. The first error in each is from the fit. The second error is the estimated LO (30\%), NLO (10\%), and NNLO (3\%) EFT error, respectively, from higher order corrections.}
	\begin{ruledtabular}
		\begin{tabular}{lccc}
			Theory & $S_{17}$ (\si{\eV\barn})
			& $S'_{17}/S_{17}$ (\si{\mega\eV^{-1}})
			& $S''_{17}/S_{17}$ (\si{\mega\eV^{-2}})
			\\ \hline
			\begin{filecontents*}{WG5_tab01.csv}
Theory,S,dS,dSEFT,Sp,dSp,dSpEFT,Spp,dSpp,dSppEFT
LO,25.2,3,76,-2.44,4,73,35.8,6,108
NLO,20.2,2,20,-2.02,4,20,33.3,6,33
NNLO,20.5,3,6,-1.86,4,6,31.9,6,10
\end{filecontents*}
\csvreader[head to column names, late after line=\\]{WG5_tab01.csv}{}
			{\ \Theory 
				& \S(\dS)(\dSEFT)
				&\Sp(\dSp)(\dSpEFT)
				&\Spp(\dSpp)(\dSppEFT)
			}
		\end{tabular}
	\end{ruledtabular}
	\label{table:S17Bayesian}
    \end{table}

\begin{table}
	\centering
	\caption{Scaling factors for the data sets at NNLO, the number of data points  used in the $E\leq \SI{1250}{\kilo\eV}$ fits,  and the common mode errors.}
	\begin{ruledtabular}
		\begin{tabular}{lccc}
			Data set & Scaling &\ Data points & CME
			\\ \hline
			\begin{filecontents*}{WG5_tab02.csv}
Data,NNLO,dNNLO,nData,CME,CMEdown
Baby,0.97,0.01,8,2.22\%,5555
Buompane,1.08,0.03,9,4.0\%,5555
Filippone,1.03,0.02,25,11.9\%,5555
Hammache '98,1.11,0.03,12,5.7\%,5555
Hammache '01,1.06,0.04,3,11.5\%,5555
Hass,1.02,0.03,1,3.6\%,5555
Junghans BE3,0.93,0.01,18,2.67\%,5555
Strieder,1.13,0.02,9,8.3\%,5555
GSI,0.99,0.03,4,5.6\%,7.7\%
MSU,1.16,0.04,2,9.6\%,7.1\%
RIKEN,1.05,0.02,3,8.4\%,12.7\%
\end{filecontents*}
\csvreader[head to column names, late after line=\\]{WG5_tab02.csv}{}
			{\  \Data
				&\num{\NNLO\pm\dNNLO}
				&\nData
				&\ifthenelse{\equal{\CMEdown}{5555}}{\CME}
				{($-\CMEdown$,$+\CME$)}
			}
		\end{tabular}
	\end{ruledtabular}
	\label{table:normsBayesian}
\end{table}

\subsection{Theoretical Uncertainty}

The theoretical uncertainties in halo EFT are estimated from the perturbative expansion in $Q/\Lambda$ \cite{2022PhRvC.106a4601H}. $Q$ is associated with the physics at low momenta of interest: relative momentum $p\lesssim\SI{40}{\mega\eV}/c$, binding momentum $\gamma\sim\SI{15}{\mega\eV}/c$, inverse Bohr radius $\kappa_C\sim \SI{24}{\mega\eV}/\hbar c$, excitation momentum $\gamma_\Delta\sim\SI{27}{\mega\eV/c}$ and resonance momentum $p_R\sim\SI{32}{\mega\eV/c}$. The cutoff scale of $\Lambda\sim\SI{70}{\mega\eV}$ is estimated as described above. Except for a narrow momentum range around the $1^+$ resonance, capture is dominated by $E1$ transitions from $s$- and $d$-waves that preserve the channel spin. Thus the threshold theoretical estimates of $S_{17}(0)$ are based on $E1$ capture.  The ANC for the $^5P_2$ final bound channel is known to be about 5 times larger than the $^3P_2$ bound channel, so capture in the $S=2$ channel dominates. The LO contribution in the $Q/\Lambda$ expansion comes only from $s$-wave Coulomb interactions in the $S=2$ channel. The weakly bound $^8$B state enables peripheral capture without initial state strong interactions.

At NLO, $d$-wave Coulomb interactions in the $S=2$ channel and $s$-wave Coulomb interactions in the $S=1$ channel are added. Strong interactions in the initial $s$ and $d$ waves of the $S=1$ channel contribute at NNLO. Perturbatively, the $s$-wave scattering lengths $a_0^{(1)}$, $a_0^{(2)}$ contribute at this order. Though  $a_0^{(1)}$ is much larger than $a_0^{(2)}$, the $S=1$ channel is one order higher in the perturbation, making these contributions comparable. We also see mixing in the coupled $^3S_1$-$^3S_1^\star$ channels due to the excited core contribution, which is parameterized by scattering lengths. The mixing requires the $^3P_2^\star$ ANC in the capture cross section. The theory error is estimated as 30\% at LO, 10\% at NLO, and 3\% at NNLO from an estimated $Q/\Lambda\sim 1/3$. This is validated by the relative sizes of the cross section at different orders of the perturbation, and also by the sizes of the fitted parameters that are consistent with theoretical estimates.

The uncertainty from the higher order contributions is estimated as follows. Two-body currents make a relative contribution $\sim k_0 a_0 L_{\text{E1}}$ with photon energy $k_0=(p^2+\gamma^2)/(2\mu)\sim Q^3/\Lambda^2$. In the $S=2$ channel, $a_0^{(2)}\sim 1/\Lambda$ so this is a N$^3$LO contribution. In the $S=1$ channel, the larger $a_0^{(1)}\sim 1/Q$ makes this a $Q^2/\Lambda^2$ contribution; however, this spin channel is suppressed by one order in the perturbation. Thus two-body currents constitute a N$^3$LO 3\% uncertainty; $s$-wave effective range corrections would also constitute a  N$^3$LO uncertainty. Initial state $d$-wave interactions (suppressed by two powers of momentum) in the $S=2$ channel also constitute a N$^3$LO uncertainty. The uncertainty in the experimental determination of the $s$-wave scattering lengths is included in the 3\% N$^3$LO uncertainty.  

\subsection{Conclusions and Recommendation for \texorpdfstring{$S_{17}$}{S17}}\label{subsec:WG5:Conclusions}

R-matrix and EFT fits of direct and Coulomb dissociation data below 1250 keV, where the influence of resonances other than the $1^+$ can be neglected, were carried out. Many different theories and energy ranges were considered and consistent results were obtained. Our recommendation was obtained using the halo EFT of \citet{2022PhRvC.106a4601H} that includes an excited $^7$Be component. This theory was selected due to the generality and rigor of its treatment of short-range interactions in the initial and final states and of two-body currents, as well as its inclusion of $M1$ transitions and robust uncertainty quantification. The resulting recommendation is $S_{17}(0)=20.5(3)(6)$~eV~b, where the first error is due to fitting and the second due to neglected higher-order contributions; for the purposes of solar modeling, we intend that the two errors be added together in quadrature and recommend $S_{17}(0)=20.5(7)$~eV~b. 

The recommended $S_{17}(0)$ has a notably smaller uncertainty than the \citetalias{2011RvMP...83..195A} value for two primary reasons. Significant theoretical developments in halo EFT with a NNLO calculation have reduced the theoretical uncertainty by a factor of two. The use of floating data normalization factors allows discrepant data sets to be brought into agreement for low-energy  extrapolation without inflating the measurement uncertainties. In the Bayesian fits, the normalization factors are assigned priors, as is done for some of the theoretical parameters. Frequentist fits presented in the supplemental material include these in the definition of the $\chi^2$ cost function. Normalization factors different from unity increase the cost, but the overall cost is lowered by bringing precise measurements from different data sets into agreement.  This is effective because the number of scale factors is much less than the number of high precision data points. 
Due to the small uncertainties and large number of data points, the measurement of  \citet{2010PhRvC..81a2801J} exerts a strong influence on the fit. Planned measurements, such as those recently approved at $E=1$ and 1.2~MeV using a $^7$Be beam and the DRAGON recoil separator at TRIUMF, must have comparably small statistical and systematic uncertainties in order to compete effectively.



\section{The \texorpdfstring{$^{14}\text{N}{(p},{ \gamma}{)}^{15}\text{O}$ reaction ({\textit{S}}$_{{1\,14}}$)}{14N(p,γ)15O reaction (S1 14)} } 
\label{sec:S114}

The $^{14}$N$(p,\gamma)^{15}$O reaction proceeds through a number of bound-state transitions, but the total capture cross section is dominated by the capture to the 6.79~MeV state (3/2$^+$) in $^{15}$O. (In this chapter the reported level energies are taken from the most recent compilation \citep{1991NuPhA.523....1A}). The contribution of this transition to the total reaction cross section at solar energies is $71-78$\% \citep{2011RvMP...83..195A,2010PhRvC..81d5805A}. The second strongest transition with a contribution of $16-17$\% is the capture to the ground state of $^{15}$O, while the capture to the 6.18~MeV state (3/2$^-$) is responsible for about $7-8$\% of the total cross section. The remaining $5-7$\% is attributed to the capture to the other bound states (i.e. 5.18 MeV, 1/2$^+$; 5.24 MeV,  5/2$^+$; 6.86 MeV, 5/2$^+$ and 7.28 MeV, 7/2$^+$ ). 

In \citetalias{2011RvMP...83..195A} a lower value, with a greatly reduced uncertainty, for $S_{1 \,14}(0)$ of 1.66(12)~keV~b was recommended compared to that of \citetalias{1998RvMP...70.1265A}, 3.5$^{+0.4}_{-1.6}$~keV~b. This was the result of the following: the new implementation of the more rigorous R-matrix technique~\cite{2001NuPhA.690..755A} over the previously used Breit-Wigner analysis, higher precision low-energy capture data sets~\cite{2004PhLB..591...61F, 2005PhRvL..94h2503R, 2005EPJA...25..455I, Bem06, 2006PhLB..634..483L, 2008PhRvC..78b2802M} over the yields presented in \citet{1987NuPhA.467..240S}, transfer measurements of proton ANCs for the bound states~\cite{2002PhRvC..66e5804B, 2003PhRvC..67f5804M}, and constraints on the lifetime of the 6.79~MeV bound state~\cite{2001PhRvL..87o2501B, 2008PhRvC..77e5803S}. Since \citetalias{2011RvMP...83..195A}, several new experiments and R-matrix calculations have been completed, but our understanding of the ground state transition still remains incomplete.

\subsection{Current Status and Results} \label{sec:14N_pg_CSR}\label{subsec:WG3:Current_Status}

Since \citetalias{2011RvMP...83..195A}, several new measurements have been reported for the capture reaction~\cite{2016PhRvC..93e5806L, 2016PhRvC..94b5803D, 2018PhRvC..97a5801W, 2019PhRvC.100a5805G, 2020PhRvC.102b4308S, 2022PhRvC.105b2801G, 2022PhRvC.106f5803F}, the life-time of the 6.79~MeV state~\cite{2014PhRvC..90c5803G, 2020PhRvC.102b4308S, 2021PhRvC.103d5802F}, the proton bound state ANCs~\cite{2012PAN....75..291A}, as well as a measurement of the low energy $^{14}$N$(p,p)^{14}$N scattering cross section~\cite{2015PhRvC..91d5804D}. These and past works now include several R-matrix fits~\cite{2001NuPhA.690..755A, 2004PhLB..591...61F, 2005PhRvL..94h2503R, 2005EPJA...25..455I, 2016PhRvC..93e5806L, 2018PhRvC..97a5801W, 2022PhRvC.106f5803F}, resulting in a range of different approaches to the fitting, which gives insight into the systematic uncertainties of the model.

The recomended partial S(0) are discussed in sec.~\ref{sec:14N_pg_strongest_trans}.
The 6.79~MeV transition, being thre strongest, has seen a great deal of experimental attention. Its ease of measurement and relatively well understood reaction contributions result in a small uncertainty in the low energy extrapolation of the S-factor of $\approx$3\%. The ground state and 6.18~MeV transitions have uncertainties of $\approx$40\%, owing mainly to the complexity of the underlying reaction contributions and the much smaller cross section.

\subsection{\label{sec:259_res} Absolute strength of the 259 keV resonance} \label{subsec:WG3:278keV_resonance} 

The $E$~=~259~keV resonance in $^{14}$N$(p,\gamma)^{15}$O is often used as a reference for normalization in many experiments because it is a narrow resonance, has a precisely known strength, and has $\gamma$-ray transitions that span a wide energy range. The absolute strength of the resonance has been determined in three works~\cite{2016PhRvC..94b5803D, 2019PhRvC.100a5805G, 2020PhRvC.102b4308S} since \citetalias{2011RvMP...83..195A}. Based largely on the work of \citet{2016PhRvC..94b5803D}, a new resonance strength of $\omega\gamma_{259}~= 12.86~\pm~0.45$~meV is recommended, consistent with the previous value but with higher precision thanks to the new data sets. For the averaging we adopted the approach followed by \citetalias{2011RvMP...83..195A}. The common systematic uncertainty of the nitrogen stopping power of 2.9\% \cite{2010NIMPB.268.1818Z} was excluded from the weighted mean calculation, and summed in quadrature with the weighted mean uncertainty to obtain the final uncertainty. Additional to this, the stopping power and branching ratio corrected resonance strengths from~\cite{2016PhRvC..94b5803D} were used in the calculations instead of the originally quoted values. The resonance strength determination is summarised in Table \ref{tab:259_res}. For the branching ratios, those of \citet{2016PhRvC..94b5803D} are recommended.

\begin{table}
\centering
\caption{Summary of the $E$=259 keV resonance strength values along with their total uncertainties. The last row gives the recommended value with its total uncertainty. \label{tab:259_res}}
\begin{ruledtabular}
\begin{tabular}{lc}
 & $\omega\gamma_{259}$~(meV) \\ \hline
\citet{1982ZPhyA.305..319B}\footnote{As in ~\cite{2016PhRvC..94b5803D}} & 13.7~$\pm$~1.0 \\
\citet{2005PhRvL..94h2503R}\footnotemark[1] & 12.4~$\pm$~0.9 \\
\citet{2005EPJA...25..455I}   & 12.9~$\pm$~0.9 \\
\citet{Bem06}   & 12.8~$\pm$~0.6 \\
\citet{2016PhRvC..94b5803D}   & 12.6~$\pm$~0.6 \\
\citet{2019PhRvC.100a5805G}   & 13.4~$\pm$~0.8 \\
\citet{2020PhRvC.102b4308S}   & 12.8~$\pm$~0.9 \\ \hline
\noalign{\smallskip}
SF~III recommended value & 12.9~$\pm$~0.5 \\ 
\end{tabular}
\end{ruledtabular}
\end{table}

\subsection{\label{sec:14N_pg_indirect}Indirect Studies} \label{subsec:WG3:Indirect_Studies}

There are several different indirect techniques that have been used to help constrain the extrapolation of the low energy cross section.
 A new transfer study of proton ANCs has been reported by \citet{2012PAN....75..291A}.
 Since the 6.79~MeV level also acts a subthreshold resonance in the ground state transition, its lifetime is also needed. New lifetime measurements have been made by \citet{2014PhRvC..90c5803G}, \citet{2020PhRvC.102b4308S}, and \citet{2021PhRvC.103d5802F} in addition to the previously avaliable data by ~\cite{2001PhRvL..87o2501B} and \citet{2008PhRvC..77e5803S}. Becasue the lifetime is in a range that is very difficult to access experimentally ($t_{1/2} < 2$~fs) these works still report values either with very large uncertainties or upper limits only.
 Scattering cross sections can also be very helpful in constraining both bound and unbound level parameters for the R-matrix fit to the capture data. A consistent set of scattering measurements has been made by \citet{2015PhRvC..91d5804D} superseding the unutilisable data for $^{14}$N$(p,p)$ showing a large amount of discrepancy~\cite{2008NIMPB.266.1193G}.
 In the new work only yield ratios are reported in order to reduce the uncertainties associated with target degradation. This scattering data give added constraint to the broad, high energy, 3/2$^+$ resonance at $E \approx$~2.2~MeV.
 
\subsection{R-matrix considerations}
\label{subsec:WG3:R-matrix}

The phenomenological R-matrix has been the primary model used to fit and extrapolate cross section data, following the analysis of \citet{2001NuPhA.690..755A}. Many of the investigations have used the same multichannel R-matrix code, \texttt{AZURE2}, which was introduced by \citet{2010PhRvC..81d5805A}. The model has proven to be very popular because it is very flexible in its level structure description, yet at the same time retains enough physical constraints that it can be used for extrapolations into unobserved regions of the cross section. Of course this flexibility also allows for additional complications like over fitting, mistaking experimental resolution effects for inconsistencies between data sets or the need for additional reaction components~\cite{Wiescher2023}, to give just a few common examples.

\begin{figure*}
    \includegraphics[width=2.0\columnwidth]{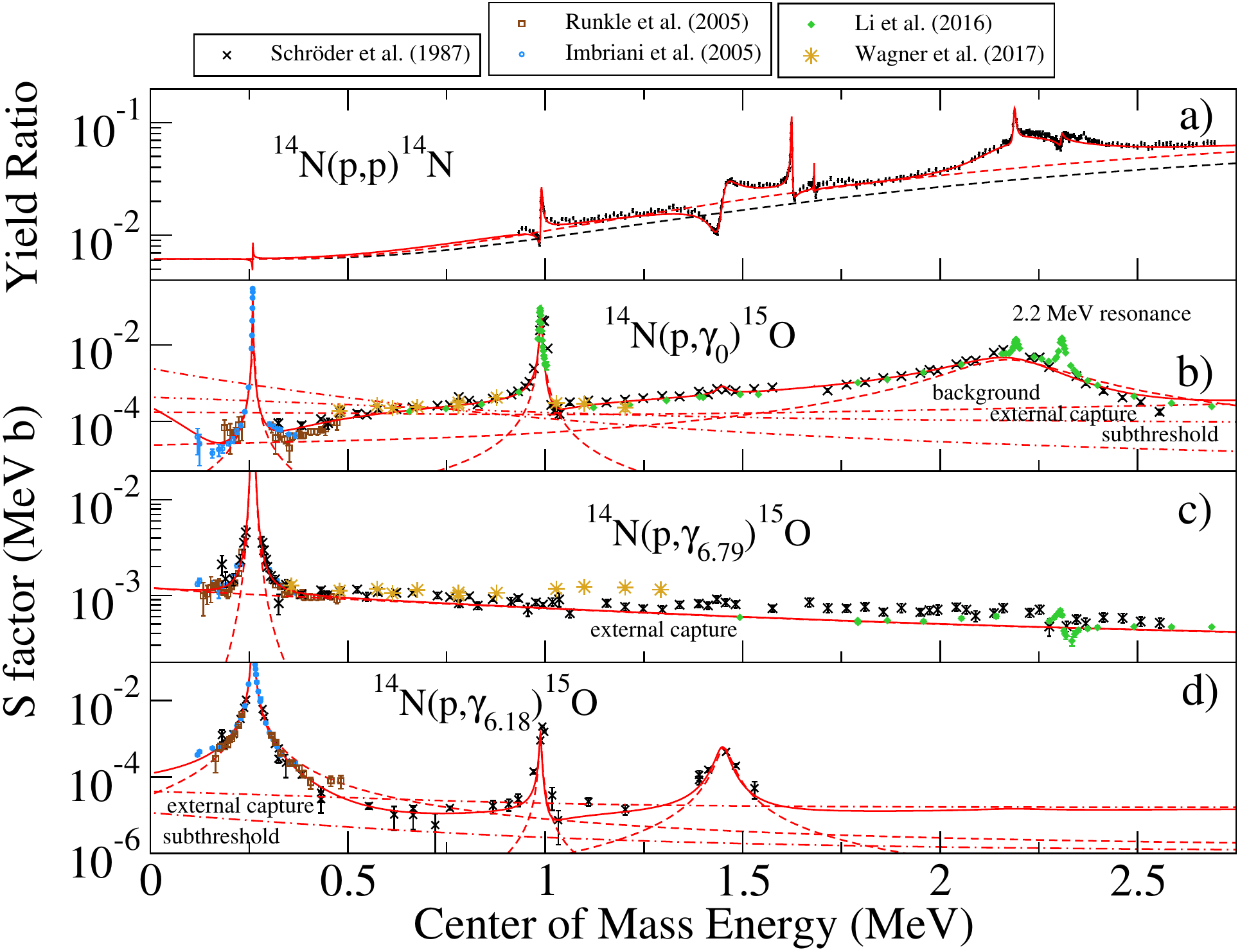}
    \caption{
    A representative yield ratio for the $^{14}$N--$p$ scattering cross section~\cite{2015PhRvC..91d5804D} (a), and low energy S-factors for the three strongest transitions in the $^{14}$N$(p,\gamma)^{15}$O reaction to the ground (b), 6.79~MeV (c), and 6.18~MeV (d) final states~\cite{1987NuPhA.467..240S,2005PhRvL..94h2503R,2005EPJA...25..455I,2016PhRvC..93e5806L,2018PhRvC..97a5801W}.
    An example R-matrix fit is shown where individual resonance contributions are indicated by dashed lines, external capture by dotted-dashed lines, and cross sections indicated by the solid lines.}
    \label{fig:14N_pg_Sfactors}
\end{figure*}

In the R-matrix framework, the cross section is described by including a number of reaction components, which includes resonances, subthreshold states, and background levels. For radiative capture reactions, an additional component is introduced, external capture~\cite{1978PhRvC..18.1962H, 1991AuJPh..44..369B, 2001NuPhA.690..755A}, which is used to model the non-resonant capture process. Fig.~\ref{fig:14N_pg_Sfactors} shows an example multichannel \texttt{AZURE2} R-matrix fit to scattering data and the three strongest transitions of the $^{14}$N$(p,\gamma)^{15}$O reaction: 6.79~MeV, ground state, and 6.17~MeV.

Using the alternative R-matrix parameterization of \citet{2002PhRvC..66d4611B}, the resonance components are characterized by observable energies and partial widths, while subthreshold resonances have their strength quantified by the state's asymptotic normalization coefficient (ANC). The strength of the external capture contributions are likewise given by ANCs.

The strongest transition, to the 6.79~MeV state, is dominated by external capture at solar fusion energies. As pointed out in \citetalias{2011RvMP...83..195A}, discrepancies among the data sets over the higher energy range make it unclear whether an additional contribution is needed from a background level. However, as \citet{2018PhRvC..97a5801W} showed, even when a background level is introduced, it has a very small effect on the low energy extrapolation owing to the dominance of the external capture component. 

The next strongest transition is to the ground state, where the R-matrix description of the low energy cross section is rather complicated. It is thought to have significant competing and interfering contributions from the 6.79 MeV subthreshold state, the narrow $E$ = 259~keV resonance, external capture, a background level, and the broad, higher energy, resonance at $E$ = 2.2~MeV. Despite all of these different contributions, which gives the R-matrix fit a considerable amount of flexibility for this transition, a satisfactory fit to the ground state transition data remains elusive~\cite{2016PhRvC..93e5806L, 2018PhRvC..97a5801W, 2022PhRvC.106f5803F}.

Finally, the 6.18 MeV transition is dominated by the narrow $E$ = 259 keV resonance and external capture contributions at low energy. A contribution from the 6.79 MeV subthreshold state may also be significant. 

\subsection{Partial S-factors}
\label{sec:14N_pg_strongest_trans}
\label{subsec:WG3:Partial_Sfactors}

In this compilation a different approach is adopted to determine the $S(0)$ value for the two strongest transitions than in \citetalias{2011RvMP...83..195A}. Because several are already available in recent publications, no new R-matrix calculation was performed. Instead the reported $S(0)$ values from the original publications \cite{2005PhRvL..94h2503R,2005EPJA...25..455I,2016PhRvC..93e5806L, 2018PhRvC..97a5801W, 2022PhRvC.106f5803F} are considered. In \citet{2005PhRvL..94h2503R}, \citet{2016PhRvC..93e5806L}, and \citet{2022PhRvC.106f5803F}, the data sets are normalized to the strength of the 259~keV resonance.  Therefore the reported S(0) values were renormalized by the ratios
of the resonance strengths used in each work and the present recommended one.  In constrast, 
\citet{2005EPJA...25..455I} and \citet{2018PhRvC..97a5801W} established an absolute normalization,
independent of the 259 keV resonance strength.

\subsubsection{\label{sec:14N_pg_GS}Ground State Transition}

The ground state transition cross section of the $^{14}$N$(p,\gamma)^{15}$O reaction is very challenging to measure accurately and to high precision because the cross section is very low and most experiments perform measurements with high efficiency $\gamma$-ray detectors in very close geometry. This increases the yield for these measurements but also greatly increases the effect of $\gamma$-ray summing into the ground state transition, with the largest contribution from the 6.79~MeV transition at low energies. In addition, the cross section on the high energy side of the strong, low energy, 259~keV resonance is rather inconsistent between different measurements, likely a result of nitrogen diffusion into the thick target backing material.

In \citetalias{2011RvMP...83..195A}, the R-matrix fit included the data of \citet{1987NuPhA.467..240S}, \citet{2005EPJA...25..455I}, \citet{2005PhRvL..94h2503R} and \citet{2008PhRvC..78b2802M}, below 2~MeV. A good description of the data was obtained over most of this energy region, with the exception of a consistent over estimation of the experimental data from $\approx$300 to 500~keV. The fit there favored the higher cross sections of \citet{2008PhRvC..78b2802M}, consistently overestimating the data of \cite{2005PhRvL..94h2503R} throughout this energy range. It should also be noted that the data of \citet{1987NuPhA.467..240S} were clerified to be yields that were uncorrected for experimental effects. Cross sections were deduced for these data in \citetalias{2011RvMP...83..195A} but where never pubished.

Since \citetalias{2011RvMP...83..195A}, there have been three new data sets reporting partial cross sections \cite{2016PhRvC..93e5806L, 2018PhRvC..97a5801W, 2022PhRvC.106f5803F}, which all used the \texttt{AZURE2} R-matrix code to derive $S(0)$. These values are considered here as an indication of the uncertainty in the low energy extrapolation. \citet{2016PhRvC..93e5806L} concentrated on differential cross section measurements between 0.5 and 3.5~MeV, while both \citet{2018PhRvC..97a5801W} and \citet{2022PhRvC.106f5803F} focused on bridging the gap between the measurements of \citet{2016PhRvC..93e5806L} and the lower energy LUNA~\cite{2005EPJA...25..455I} and Laboratory for Experimental Nuclear Astrophysics (LENA) measurements~\cite{2005PhRvL..94h2503R}.

The new data and accompanying R-matrix fits have repeatedly confirmed a discrepancy between the R-matrix fit and the experimental data for the ground state transition, especially when the high energy data of \citet{1987NuPhA.467..240S} and \citet{2016PhRvC..93e5806L} are included in the fit. On the R-matrix side, this could point to a missing reaction component (i.e. missing resonance or subthreshold state) or an incorrect spin-parity assignement. On the data side, the close geometry setups of most experiments result in large summing corrections. Inaccurate corrections for angular distributions could also play a role. Because of this unresolved inconsistency between data and model, a larger uncertainty has been attributed to this transition compared to \citetalias{2011RvMP...83..195A} (see Table~\ref{tab:14Npg_S0}). 

\subsubsection{\label{sec:14N_pg_679}6.79~MeV Transition}

In contrast to the ground state, repeated measurements of the 6.79~MeV transition continue to produce consistent results, especially at very low energies. The measurements of \citet{2018PhRvC..97a5801W} do show a more rapidly increasing cross section at high energies than those of previous works, but are consistent in the low energy region. In addition, the ANC determined from the normalization of the external capture cross section used to model the non-resonant capture in the R-matrix fit yields consistent results with those determined from transfer studies~\cite{2003PhRvC..67f5804M, 2002PhRvC..66e5804B}. Because of this continued consistency across measurements, the value adopted here is the weighted average of the reported values, where the uncertainty has been further reduced from that of \citetalias{2011RvMP...83..195A}.

\subsubsection{\label{sec:14N_pg_other} Other Transitions}
For the three weaker transitions reported in \citetalias{2011RvMP...83..195A}, there were no new experimental data. Here the values of $S(0)$ have been determined by scaling those from \citetalias{2011RvMP...83..195A} by the ratio of the 259~keV resonance strengths adopted here and from \citetalias{2011RvMP...83..195A}.

\begin{table}
\centering
\caption{Total S$_{1 \,14}$(0) as the sum of the partial transitions. \label{tab:14Npg_S0}}
\begin{ruledtabular}
\begin{tabular}{lccc}
Transition & $S_{1\, 14}(0)$~(keV~b) & $\Delta S_{114}(0)$ & Reference \\ \hline
tr~$\rightarrow$~0 & 0.30~$\pm$~0.11 & 37\% & Present \\
tr~$\rightarrow$~6.79 & 1.17~$\pm$~0.03 & 2.9\% & Present \\
tr~$\rightarrow$~6.17 & 0.13~$\pm$~0.05 & 38\% & \citetalias{2011RvMP...83..195A} \\
tr~$\rightarrow$~5.18 & 0.010~$\pm$~0.003 & 30\% & \citetalias{2011RvMP...83..195A} \\
tr(5.24)~$\rightarrow$~0 & 0.068~$\pm$~0.020 & 30\% & \citetalias{2011RvMP...83..195A} \\ \hline
R-matrix sum & 1.68~$\pm$~0.13\footnote{The uncertaity is the sum of the partial transition uncertainties.} & 7.6\% & \\
\footnotesize Additional syst. uncert.\footnote{From the normalisation to the $E$ = 259 keV resonance strength.}  &  & 3.5\% & \\ \hline
Total & 1.68~$\pm$~0.14 & 8.4\% & \\
\end{tabular}
\end{ruledtabular}
\end{table}

\subsection{\label{sec:14N_pg_total}Total S-factor and conclusions}
\label{subsec:WG3:Total_Sfactors}

Figure~\ref{fig:14N_pg_Sfactors} shows the low energy S-factors and R-matrix fits to the strongest capture transitions and the scattering yields at a representative angle and Table~\ref{tab:14Npg_S0} summarizes the uncertainties estimated for the zero energy S-factor, $S_{114}(0)$ for each of the significant capture transitions. 
Their sum gives the adopted total S(0) value. The absolute uncertainty of the R-matrix sum is the quadratic sum of the absolute uncertatinties of the partial transitions, while the recomended uncertainty conservatively includes an additional 3.5\% relative uncertainty from the resonance strength (see Table~\ref{tab:259_res}) to which most of the datasets were normalised.
As described in Section~\ref{sec:14N_pg_GS}, the uncertainty in the ground state transition has increased to 37\% over the 19\% estimate of \citetalias{2011RvMP...83..195A}, owing to repeated confirmation of a tension between different experimental data sets and the R-matrix description of the cross section. The uncertainty in the 6.79~MeV transition decreased from 4.0\% to 2.9\% due to the consistency of the various measurements for this transition.

In addition, the new total capture cross section of \citet{2022PhRvC.105b2801G}, performed using the activation technique over the energy range from 0.55 to 1.4~MeV, is larger than the sum of the partial R-matrix cross sections by $\approx$25\%, a significant margin compared to the systematic uncertainties of the different data sets. However, it should be emphasized that for several of the transitions the only data available are those of \citet{1987NuPhA.467..240S}, which are yields not cross sections and often give only upper limits. For example, the capture to the 7.56~MeV level in this energy range may be significant.

Given the above considerations, a somewhat larger uncertainty of 8.4\% has been estimated for $S_{1\, 14}(0)$, compared to the 7\% uncertainty recommended in \citetalias{2011RvMP...83..195A}. As one of the main sources of uncertainty remains the ground state transition, far geometry angular distribution measurements are recommended extending the differential measurements of \citet{2016PhRvC..93e5806L} to lower energies. These measurements will both reduce the effect of summing and provide more information on the underlying reaction mechanisms at these low energies. Further, as the data of \citet{1987NuPhA.467..240S} have significant uncorrected experimental effects, new measurements covering the same wide energy range should be undertaken. Such measurements are planned at both the LUNA and JUNA underground facilities. At very low energies, the LUNA and LENA measurements of the ground state transition by \citet{2005EPJA...25..455I} and \citet{2005PhRvL..94h2503R}, respectively, remain the only measurements. The discrepancy between the two measurements and the tension between the R-matrix fits make new measurements in this range also of high priority. Despite several additional attempts since \citetalias{2011RvMP...83..195A}, the lifetime of the 6.79~MeV transition still remains unmeasured. This remains a high priority, as having a well constrained value will greatly improve our understanding of the low energy behaviour of the ground state transition cross section. Finally, we recommend that the cross section to the 6.18~MeV and 7.56~MeV final states be measured at high energies in order to address the tension between the measured total cross section and the sum of the partial cross sections.



\section{Other CNO, \texorpdfstring{N\lowercase{e}-N\lowercase{a}}{Ne-Na} reactions}
\label{sec:OtherCNO}

In the following, we provide information on the proton-induced reactions involving the target nuclei $^{12,13}$C, $^{15}$N, $^{16,17,18}$O, $^{19}$F, $^{20,21,22}$Ne, and $^{23}$Na. Unless otherwise noted, we will focus on the low-energy region, i.e., below 150~keV in the center of mass. The reactions on lighter target nuclei up to and including $^{16}$O are of direct relevance for solar neutrinos; the other reactions play a role in hydrogen burning at somewhat higher temperatures, including asymptotic giant branch (AGB) stars. 

\begin{figure}
    \centering
    \includegraphics[width=\columnwidth]{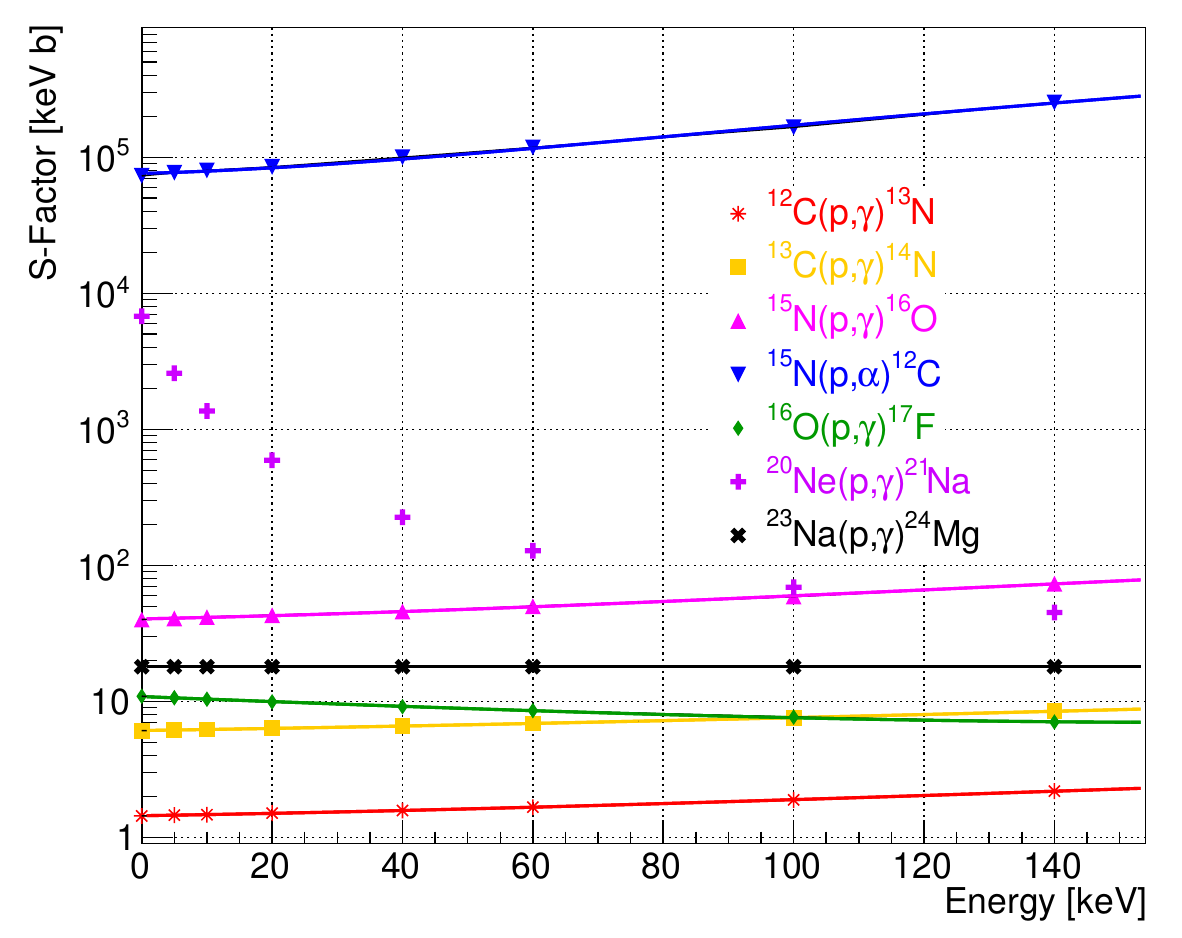}
    \caption{
    Astrophysical S-factors of CNO and NeNa cycle reactions. The circles represent the values listed in Table \ref{tab:wg6_sfac}. The solid lines represent polynomial (quadratic) fits of the S-factor values. The fit coefficients are listed in Table~\ref{tab:Sfactors_summary}. The S-factor of the $^{20}$Ne$(p,\gamma)^{21}$Na reaction cannot be accurately fit with a quadratic (or cubic) polynomial function.
    }
    \label{fig:sfactor}
\end{figure}

Table~\ref{tab:wg6} lists, where applicable, the S-factor at zero energy, the Gamow peak location, and references. Table~\ref{tab:wg6_sfac} provides recommended values of the S-factor and corresponding uncertainty on a grid of seven energy values. These results are displayed in Fig.~\ref{fig:sfactor}, together with their polynomial fits. The fit coefficients are listed in 
Table~\ref{tab:Sfactors_summary}.
Table~\ref{tab:wg6_narrow} gives values of recommended strengths of narrow resonances. A detailed discussion of each reaction is provided below.
%

\subsection{Reactions of the CNO-I cycle}
\label{sec:OtherCNO:CNO1}

\subsubsection{\texorpdfstring{$^{12}\text{C}(p,\gamma)^{13}\text{N}$}{12C(p,γ)13N}}
\label{sec:OtherCNO:CNO1:c12pg}

The low-energy reaction cross section is influenced by the tail of a broad resonance and the non-resonant capture to the ground state of $^{13}$N. The results of six new direct measurements have been published since \citetalias{2011RvMP...83..195A} \citep{2001PhRvC..64f5804N, 2023EPJA...59...59G, 2023PhRvC.107f2801S, 2023NuPhA103722705C, 2023PhRvL.131p2701S, 2023PhRvC.108c5805K}.

The first experiment by \citet{2001PhRvC..64f5804N} reports the rection cross section relative to that of the $^{13}$C($p,\gamma$)$^{14}$N reactions at $E_p^{lab} = 160$~keV to be $0.24\pm0.03$.

Two of the newest experiments were performed at ATOMKI 
\citep{2023EPJA...59...59G, 2023NuPhA103722705C}. The first employed the activation method in a wide laboratory bombarding energy range, $E_p^{lab} = 300-1900$~keV, to provide a dataset with an absolute normalization having independent uncertatinties, from that of the older data sets obtained using in-beam $\gamma$-ray spectroscopy.
The main focus of the second experiment was the precise determination of the energies and widths of the two low-energy resonances using in-beam $\gamma$-ray spectroscopy with detectors located at several angles. 

The fourth experiment, by \citet{2023PhRvC.107f2801S}, was performed at the Felsenkeller underground accelerator facility in Dresden, Germany, using in-beam $\gamma$-ray spectroscopy. The focus was on determining the parameters of the lowest-lying resonance employing a detector in close geometry with high solid-angle coverage. The data covered the energy range of $E_p^{lab} = 345-670$ keV, and allowed for an extrapolation to solar and stellar energies using R-matrix theory. The absolute normalization of this data set is about 25\% lower than the previous results. The low-energy slope of the S-factor also changed, resulting in an extrapolated value of $S(25~\mathrm{keV})$ $=$ $1.34\pm0.09$ keVb in the solar Gamow peak. 

The fifth experiment, by \citet{2023PhRvC.108c5805K}, was performed at Bochum and Notre Dame, and focused on the second resonance in the reaction. Differential cross sections at $0^\circ$ and $55^\circ$, thick target yields, and angular distributions at selected energies were measured in the energy range of $E_p^{lab} = 1000-2500$ keV. A comprehensive R-matrix fit of the available data was also performed, resulting in $S(25~\mathrm{keV})$ $=$ $1.48\pm0.09$ keVb.  

The sixth experiment, by the LUNA collaboration \citep{2023PhRvL.131p2701S}, obtained data in the range of $E_p$ $=$ $80-400$ keV, reaching the lowest bombarding energies to date. Both the activation technique and in-beam $\gamma$-ray spectroscopy were applied, with a consistent absolute normalization of the data, despite the different systematic uncertainties introduced by the two methods. Their S-factor in the solar Gamow peak, $S(25~\mathrm{keV}) = 1.53\pm0.06$~keVb, is in good agreement with \citet{2023PhRvC.108c5805K}, and is compatible within 2$\sigma$ with \citet{2023PhRvC.107f2801S}.

Apart from the direct measurements discussed above, the ground state ANC was determined in a recent particle transfer experiment \citep{2022EPJA...58...24A}. That work also reported a fit of the radiative capture data  available at the time. The R-matrix fits to the data of \citet{2023PhRvC.107f2801S,2023PhRvL.131p2701S,2023PhRvC.108c5805K} give ANC values consistent with the result of the transfer work.

Thanks to the high consistency of all recent datasets, the extrapolation of the cross section to solar energies has now improved uncertainty. Owing to the self-consistency between the used methods and the precise data coovering the lowest energies to date we adopt the S-factor recommended by \citet{2023PhRvL.131p2701S} (see Tables~\ref{tab:wg6} and \ref{tab:wg6_sfac}).

\subsubsection{\texorpdfstring{$^{13}\text{C}(p,\gamma)^{14}\text{N}$}{13C(p,γ)14N}}
\label{sec:OtherCNO:CNO1:c13pg}

This reaction proceeds to the ground state and five excited states in $^{14}$N. The low-energy cross section is influenced by the tail of a broad resonance at laboratory bombarding energy of $E_r^{lab} \approx 558$ keV. Previously, only a single work reported cross sections for all transitions \citep{1994NuPhA.567..354K}, which was the basis of the S-factor recommendations in both \citetalias{1998RvMP...70.1265A} and \citetalias{2011RvMP...83..195A}.

A comprehensive study by the LUNA collaboration \citep{2023PhRvL.131p2701S} provided data in the energy range of $E_p^{lab} = 80-400$ keV. One of their data sets was obtained with HPGe detectors to study partial cross sections. Their second data set was obtained with a $\gamma$-ray calorimeter to measure total reaction cross sections. The results from both data sets agree in the absolute cross section scale, despite major differences in the systematic uncertainties of the two methods. The new results are about 30\% lower than previous results of \citet{1994NuPhA.567..354K}. 
Despite this scale difference the calculated S-factor ratio for the $^{12}$C($p,\gamma$)$^{13}$N and $^{13}$C($p,\gamma$)$^{14}$N reactions at $E_p^{lab} = 160$~keV is $0.26\pm0.02$, in perfect agreement with the experimental reported one of $0.24\pm0.03$ by \citet{2001PhRvC..64f5804N}. 

The adopted S-factor given in Tables~\ref{tab:wg6} and \ref{tab:wg6_sfac} is based on the new result of \citet{2023PhRvL.131p2701S}, owing to its self-consistency between the different methods for measuring the cross sections, and the higher precision of the dataset compared to \citet{1994NuPhA.567..354K}.

\subsubsection{\texorpdfstring{$^{15}\text{N}(p,\alpha)^{12}\text{C}$}{15N(p,α)12C}}
\label{sec:OtherCNO:CNO1:n15pa}

Following \citetalias{2011RvMP...83..195A}, the experimental and phenomenological work on the $^{15}$N$(p,\alpha)^{12}$C reaction can be divided into three groups addressing different aspects: (i) new ($p,\alpha_0$)  data \citep{2012PhRvC..85c8801D}; (ii) new ($p,\alpha\gamma$) data \citep{2012PhRvC..85f5810I,2012PhRvC..86c9902I,2016NIMPB.381...58R,2019EPJA...55...41G,2021PhRvC.103f5801D}; and (iii) R-matrix analysis of existing data sets \citep{2013PhRvC..87a5802D,2020PhRvC.102b4628B}.

Other studies have been conducted for laboratory bombarding energies above 400~keV. In \citet{2016NIMPB.381...58R}, the strength of a resonance at $E_r^{c.m.} \approx$403~keV ($2^-$, E$_x$ $=$ $12530$~keV) is tabulated, and the value is in agreement with previous determinations. In \citet{2019EPJA...55...41G}, the thick-target yield for the $E_r^{c.m.} = 841$~keV resonance is shown, but no further details are reported. New cross section data for the ($p,\alpha_1 \gamma$) reaction down to about 900~keV center-of-mass energy are displayed in \citet{2021PhRvC.103f5801D}. They used the data for a comprehensive R-matrix fit, and reported parameters of resonances above about 1500~keV center-of-mass energy.

\citet{2012PhRvC..85f5810I,2012PhRvC..86c9902I} measured the $^{15}$N($p,\alpha \gamma$)$^{12}$C reaction down to about 140~keV laboratory bombarding energy, and showed that the S-factor of the ($p,\alpha_1 \gamma$) channel is orders of magnitude lower compared to the $(p,\alpha_0$) channel. 

Finally, \citet{2013PhRvC..87a5802D} contains a  detailed analysis of all available channels populating the $^{16}$O compound nucleus, expanding the results presented in \citet{2012PhRvC..85c8801D}. In this paper, a wide range of experimental data were considered leading to a single consistent R-matrix fit, accurately describing the broad level structure of $^{16}$O below $E_x = 13500$~keV. The resulting fit was used to extract an improved determination of the low-energy S-factor for the $^{15}$N($p,\alpha$)$^{12}$C reaction. In addition, \citet{2020PhRvC.102b4628B} have developed the partial-wave formalism for calculating the angular distributions of secondary $\gamma$-rays following particle emission, providing the framework needed for future studies that can directly utilize differential measurements of the $^{15}$N$(p,\alpha_1 \gamma)^{12}$C reaction.

The available evidence confirms that the ($p,\alpha_0$) channel dominates the total S-factor. The global R-matrix analysis in \citet{2013PhRvC..87a5802D} provides results that slightly exceed the experimental data of \citet{1982ZPhyA.305..325R} and also estimates a slightly larger value, $S(0) = (95\pm6) \times 10^3$~keVb, than previous works (see Table~VII of \citet{2013PhRvC..87a5802D}), calling for additional low-energy measurements. The extrapolated S-factor in \citet{2013PhRvC..87a5802D} agrees with the findings in \citet{2009PhRvC..80a2801L}, which is the most recent indirect study, determining the S-factor down to 20~keV center-of-mass energy. The global R-matrix analysis suggests a renormalization of the \citet{2009PhRvC..80a2801L} S-factor by 6\%, which is within the quoted uncertainty. Based on the information provided above, we adopt the S-factor of \citet{2009PhRvC..80a2801L}, $S(0)=(73\pm 5) \times 10^3$~keVb, which was also adopted in \citetalias{2011RvMP...83..195A}.

\subsection{Reactions of the NO, or CNO-II, cycle}
\label{sec:OtherCNO:CNO2}

\subsubsection{\texorpdfstring{$^{15}\text{N}(p,\gamma)^{16}\text{O}$}{15N(p,γ)16O}}
\label{sec:OtherCNO:CNO2:n15pg}

The low-energy $^{15}$N$(p,\gamma)^{16}$O cross section is dominated by its ground state transition~\citep{2010PhRvC..82e5804L, 2012PhRvC..85f5810I}. This transition's cross section is enhanced by the constructively interfering tails of two broad resonances at laboratory proton energies of $E^{lab}$~=~338 and 1028~keV, which correspond to 1$^-$ levels in the $^{16}$O compound system. In addition to ground-state proton and $\gamma$-ray de-excitation, these levels also decay strongly through the $\alpha_0$ and $\alpha_1$ channels. For the 338 keV 
resonance, $\Gamma_{\gamma} / \Gamma_{\alpha} = (6.7 \pm 0.3)\times 10^{-5}$
\cite{2013PhRvC..87a5802D}.  Because of interference effects, this ratio does not directly translate to the rate of passage from the CNO cycles to the NeNa cycle. Currently, the favored framework for the analysis and extrapolation of the available data is the phenomenological R-matrix. Here, the main components that are used to model the low-energy cross section of the $^{15}$N$(p,\gamma)^{16}$O reaction are the two broad 1$^-$ resonances, external capture, and a 1$^-$ background level. The strength of the external capture can be characterized by the ground state proton ANC of $^{16}$O, which has been experimentally determined via a proton transfer reaction~\citep{2008PhRvC..78a5804M}. Gamma-ray de-excitation to the ground state has also been observed through a 2$^+$ level in this energy region in the $^{12}$C+$\alpha$ reactions, but has been shown to be negligibly weak for $^{15}$N$(p,\gamma)^{16}$O~\citep{2013PhRvC..87a5802D}.

The cross section measurements used in \citetalias{2011RvMP...83..195A} were those of \citet{1960NucPh..15..289H}, \citet{1974NuPhA.235..450R}, and \citet{2009JPhG...36d5202B}. While the then newly measured data of \citet{2009JPhG...36d5202B} represented a significant improvement in accuracy over \citet{1974NuPhA.235..450R} and in precision over \citet{1960NucPh..15..289H}, they only cover a limited energy range. For \citetalias{2011RvMP...83..195A}, the two-level R-matrix fit of \citet{2008PhRvC..78a5804M} was adopted, which simultaneously fit $^{15}$N$(p,\alpha)^{12}$C data, giving $S(0)$~=~36~$\pm$~6~keVb for the $^{15}$N$(p,\gamma)^{16}$O reaction (see also \citet{2008PhRvC..78d4612B}). The fit shows that the extrapolation performed by \citet{1974NuPhA.235..450R}, yielding 64~$\pm$~6~keVb, overestimated S(0).

Following \citetalias{2011RvMP...83..195A}, a comprehensive measurement campaign was carried out at LUNA and the University of Notre Dame~\citep{2010PhRvC..82e5804L}, covering a laboratory energy range from $E^{lab}$~=~130 to 1800~keV. In that work, a three-level R-matrix fit was used, but the ground state proton ANC of \citet{2008PhRvC..78a5804M} was not considered, resulting in a somewhat larger extrapolated value of $S(0)$ $=$ $39.6(26)$~keVb than that adopted in \citetalias{2011RvMP...83..195A}. \citet{2011PhRvC..83d4604M} subsequently showed that a three-level R-matrix fit of equal quality could be obtained with the ANC, and reported an extrapolated value of $S(0)$ that ranged from 33 to 40~keVb.

\citet{2011A&A...533A..66C} expanded the energy range of \citet{2009JPhG...36d5202B} by measuring the cross section up to $E_{c.m.}$~=~370~keV. Their experiment scanned over the top of the lowest-energy broad resonance at $E_{c.m.}$~=~312~keV. The data were found to be consistent with the ground state data of \citet{2010PhRvC..82e5804L}.

\citet{2013PhRvC..87a5802D} performed a comprehensive R-matrix fit that included not only the $^{15}$N+$p$ data, but also a wide range of $^{12}$C+$\alpha$ data over a similar excitation energy range (see also Section~\ref{sec:OtherCNO:CNO1:n15pa}). The value from this analysis, $S(0)=$~40~$\pm$~3~keVb, was slightly above that of \citet{2010PhRvC..82e5804L} and agreed with the upper limit of \citet{2011PhRvC..83d4604M}. However, it is lower than the recent potential model extrapolation of $S(0)$~=~45~$^{+9}_{-7}$~keVb found in the NACRE II compilation~\citep{2013NuPhA.918...61X}. The difference between the two fits is likely due to the fact that NACRE II disregarded the higher-energy part of the \citet{2010PhRvC..82e5804L} data, which maps the interference pattern between the two 1$^{-}$ resonances more accurately than older data sets. In order to reproduce experimental data in that region, a 1$^{-}$ background level was added in the R-matrix fits of \citet{2010PhRvC..82e5804L} and later works.\\
Here we adopt the value of $S(0)=$~40~$\pm$~3~keVb from \citet{2013PhRvC..87a5802D} as it considers the most comprehensive set of available data.

While low-energy future measurements are recommended, experiments at high energy are highly desirable to better understand the reaction mechanisms responsible for the interference pattern that is observed between the two well known 1$^-$ resonances, which can not be completely reproduced unless some background contribution is also included.

\subsubsection{\texorpdfstring{$^{16}\text{O}(p,\gamma)^{17}\text{F}$}{16O(p,γ)17F}}
\label{sec:OtherCNO:CNO2:o16pg}

Below a laboratory energy of about $2700$~keV, i.e., the energy of the lowest-lying resonance \citep{1993NuPhA.564....1T}, the $^{16}$O($p,\gamma$)$^{17}$F reaction is a prime example of the non-resonant direct radiative capture process. This reaction has been measured many times using a variety of techniques, including activation, prompt $\gamma$-ray detection, and inverse-kinematics experiments. The $E1$ capture can proceed to the ground or first-excited state in $^{17}$F, with the latter being the dominant transition. In \citetalias{2011RvMP...83..195A}, a value of $S(0)$ $=$ $10.6 \pm 0.8$~keVb with a a $7.5$\% uncertainty was adopted. This result was obtained by normalizing the microscopic-model calculations of \citet{1998PhRvC..58..545B} to the data of \citet{Rolfs1973} and \citet{1997PhRvL..79.3837M}.

A recent statistically rigorous evaluation of the S-factor data of \citet{1958PhRv..111.1604H,1975CaJPh..53.1672C,1982ZPhyA.305..319B,1997PhRvL..79.3837M} is presented in \citet{2022PhRvC.106e5802I}. Their analysis, including a discussion of experimental uncertainties, used a combined fit of the transitions to the ground and first-excited states in $^{17}$F, and their sum, and was performed using a Bayesian model. The physical model was a single-particle model employing a Woods-Saxon potential for generating the radial bound-state wave function. The fit had three adjustable parameters: the radius parameter and diffuseness of the Woods-Saxon potential, and the ANCs for scaling the theoretical non-resonant capture S-factors. It was also found that a poor fit was obtained when using spectroscopic factors instead as scaling parameters. As the $^{16}$O($p,\gamma$)$^{17}$F reaction at low bombarding energies is peripheral, the analysis of the S-factor in terms of ANCs greatly reduced the sensitivity of the fit to the single-particle potential parameters. For the ANCs of the ground and first-excited state transitions, they found values of $C^2_\text{gs}$ $=$ $1.115$~fm$^{-1}$ ($\pm 4.0$\%) and $C^2_{fes}$ $=$ $7063$~fm$^{-1}$ ($\pm 4.0$\%), respectively. These results agree with those measured in the $^{16}$O($^3$He,d)$^{17}$F reaction by \citet{1999PhRvC..59.1149G}. It is not surprising that the uncertainties ($\approx 10$\%) in the latter reference are significantly larger than for the values quoted above, because the distorted wave Born approximation (DWBA) analysis of \citet{1999PhRvC..59.1149G} was subject to ambiguities in the choice of optical model potentials for the incoming and outgoing channels.

The total S-factor at zero energy reported by \citet{2022PhRvC.106e5802I} was $S_\text{tot}(0)$ $=$ $10.92$~keVb ($\pm 4.0$\%). The uncertainty is significantly smaller compared to the results published by \citet{2008PhRvC..77d5802I}, which were derived from an analysis that was not statistically rigorous. We adopt the results of \citet{2022PhRvC.106e5802I} in this work. Their S-factor is presented in Table~\ref{tab:wg6_sfac}.

We endorse the two recommendations made by \citet{2022PhRvC.106e5802I}. First, they advocated for a new low-energy measurement of the $^{16}$O($p,\gamma$)$^{17}$F reaction, specifically at center-of-mass energies below $200$~keV, which have so far only been reached in the experiment of \citet{1958PhRv..111.1604H}, albeit with a relatively large systematic uncertainty of $14$\%. Second, it would be interesting to compare their results to those from a future R-matrix analysis of the same data sets. 

\subsubsection{\texorpdfstring{$^{17}\text{O}(p,\gamma)^{18}\text{F}$}{17O(p,γ)18F}}\label{sec:OtherCNO:CNO2:o17pg}

In \citetalias{2011RvMP...83..195A}, only the non-resonant component of the $^{17}$O($p,\gamma$)$^{18}$F reaction cross section was discussed, which dominates the total cross section at solar temperatures. At the time, the results of \citet{2007PhRvC..75c5810C} were recommended. Since the publication of \citetalias{2011RvMP...83..195A}, several new S-factor measurements have been reported, both at low and high energies \citep{2010PhRvC..81d5801N, 2012PhRvC..86e5801K, 2012PhRvC..85c5803H, 2012PhRvL.109t2501S, 2014PhRvC..89a5803D, 2015PhRvC..91a5812B, 2017PhRvC..95c5805G}. Despite differences in the techniques exploited and energy ranges covered, the derived $S(0)$ values agree across different experiments, both for the non-resonant capture and the total S-factor, where the latter quantity includes the low-energy tails of two broad resonances ($E_r^{c.m.}$ $=$ 557 and 677~keV). Here we recommend the total S-factor from the most recent analysis of \citep{2017PhRvC..95c5805G}.  The $^{17}$O($p,\gamma$)$^{18}$F reaction cross section was measured in the laboratory bombarding energy range of 500-1800 keV with a typical total uncertainty of 10\% using the activation method. A comparison with other independent results, when possible, at $E$\textsubscript{p}$^{lab}$ $=$ $500$~keV resulted in good agreement with \citet{2012PhRvC..85c5803H} and \citet{2012PhRvC..86e5801K}, while a 20\% difference is observed with respect to \citet{2010PhRvC..81d5801N}. An R-matrix analysis was performed considering all the low energy data, resulting in a value of  $S(0)$\textsubscript{tot} = (4.7$\,\pm\,$1.0$)$~keVb\footnote{In the publication of \citet{2017PhRvC..95c5805G} the uncertainty is referred to as statistical only. However, it represents, in fact, the total uncertainty (Gy. Gy\"urky, priv. comm.)}.

At temperatures near $60$~MK, typical for hydrogen-shell burning in AGB stars, the $^{17}$O($p,\gamma$)$^{18}$F reaction rate is dominated by the $E_r^{c.m.}$ $=$ 65~keV resonance, which was not considered in \citetalias{2011RvMP...83..195A}. No direct measurements are reported for the resonance strength, $\omega\gamma$, which is presently estimated from experimental values of the partial and total widths. 
Partial widths of $\Gamma_{\alpha}$ $=$ $(130\,\pm\,5)$~eV and $\Gamma_{\gamma}$ $=$ $(0.44\,\pm\,0.02)$~eV were measured using the $^{14}$N+$\alpha$ reaction \citep{1980NuPhA.343...79M, 1977NuPhA.288..317B}. The most uncertain quantity is the proton partial width, $\Gamma_p$, which is estimated from the measured strength of the corresponding resonance in the ($p,\alpha$) channel. 
Our best estimate for the ($p,\alpha$) strength is $\omega\gamma_{p\alpha}$ $=$ $4.6_{-1.2}^{+5.4}\times10^{-9}$~eV (see Table~\ref{tab:wg6_narrow} and the following section for details), which leads to $\Gamma_p$ = $18.5^{+21.7}_{-4.7}\times10^{-9}$~eV.
Hence, our recommended value is $\omega\gamma_{p\gamma}\,=\,15.6^{+18.3}_{-4.0}\times10^{-12}$~eV (see Table~\ref{tab:wg6_narrow}).  This
value is consistent with the strength recommended by \citet{Palmerini2013THERA}, though we have inflated the error to account for the uncertain status of the proton 
width.\footnote{A direct measurement of the 65 keV resonance strength
\cite{2024PhRvL.133e2701G}, published after the SF III literature review was completed, supports the proton width from
\cite{2016PhRvL.117n2502B}.}
Moreover, the 65~keV (1$^-$) resonance is expected to interfere with the $-$2.0~keV (1$^-$) subthreshold resonance. We recommend the treatment of the interference contribution described in \citet{2015PhRvC..91a5812B}.
At typical temperatures of classical novae, the main  contributor to the $^{17}$O($p,\gamma$)$^{18}$F reaction rate is the $E_r^{c.m.}$ = 183.9 keV resonance, which was first observed by \citet{2004PhRvL..93h1102F}. It was subsequently remeasured by \citet{2007PhRvC..75c5810C, 2012PhRvL.109t2501S, 2014PhRvC..89a5803D, 2015PhRvC..91a5812B}.
Here we recommend the weighted mean adopted in \citet{2015PhRvC..91a5812B}, $\omega\gamma\,=\,$(1.77$\,\pm\,$0.09)$\times$10$^{-6}$~eV.\footnote{The article by \citet{2024PhRvL.133e2701G} reports an $\omega\gamma$ bare value of $(30 \pm 6 \text{(stat)} \pm 2 \text{(syst)}) \times 10^{-12}$ eV for the 64.5 keV resonance strength.}

\subsubsection{\texorpdfstring{$^{17}\text{O}(p,\alpha)^{14}\text{N}$}{17O(p,α)14N}}
\label{sec:OtherCNO:CNO2:o17pa}

The $^{17}$O($p,\alpha$)$^{14}$N reaction rate is determined by the contributions of narrow resonances. At solar energies, the dominant contributions originate from resonances at E$_r^{c.m.}$ $=$ $-$2 and 65 keV. At typical temperatures of AGB stars and classical novae, the main contributor to the rate is a resonance at E$_r^{c.m.}$ $=$ 183.9 keV. Concerning the subthreshold resonance, no new results have been published since \citetalias{2011RvMP...83..195A}. 

The E$_r^{c.m.}$ $=$ 65-keV resonance has been studied recently both in a direct experiment ($\omega\gamma = (10.0 \pm 1.4_{stat} \pm 0.7_{syst})\times10^{-9}$ eV; \citealt{2016PhRvL.117n2502B}) and with the Trojan Horse indirect Method ($\omega\gamma = (3.42 \pm 0.60)\times10^{-9}$ eV; \citealt{2015PhRvC..91f5803S}). The resonance strength reported by \citet{2016PhRvL.117n2502B} deviates by a factor of 2.5 from that of
\citet{2015PhRvC..91f5803S} and a factor of 2 from the pre-\citetalias{2011RvMP...83..195A} value of \citet{2005PhRvC..71e5801F}, which includes a revised analysis of the data from \citet{1995PhRvL..74.2642B}. Because the values disagree by more than their uncertainties, we recommend the value $\omega\gamma = 4.6^{+5.4}_{-1.2} \times 10^{-9}$ eV, with the expanded uncertainty encompassing all values in the literature.

The resonance at E$_r^{c.m.}$ $=$ 183.9 keV, on the other hand, is well known. The most recent determination of its strength is reported in \citet{2015EPJA...51...94B}, i.e., $\omega\gamma = (1.68 \pm 0.03_{stat} \pm 0.12_{syst})\times10^{-3}$~eV, which is in excellent agreement with previous results \citep{2007PhRvC..75c5810C, 2007PhRvC..75f5801M, 2007PhRvC..75e5808N}.

\subsection{Reactions of the CNO-III cycle}

\subsubsection{\texorpdfstring{$^{18}\text{O}(p,\gamma)^{19}\text{F}$}{18O(p,γ)19F}}
\label{sec:OtherCNO:CNO3:o18pg}

Important components for this reaction channel are a $E_r^{c.m.} = 20$ keV resonance, for which only indirect data exist \citep{1980NuPhA.349..165W,1986NuPhA.457..367C,2008PhRvL.101o2501L}, the resonances at $E_r^{c.m.} = 90$ keV and $E_r^{c.m.} = 143$ keV and the non-resonant capture process. Reported strengths of the $E_r^{c.m.} = 20$ kev resonance vary by one order of magnitude between $1.4 \times 10^{-22}$ \citep{1979NuPhA.313..346L} and $1.9 \times 10^{-21}$~eV \citep{1980NuPhA.349..165W}, depending on the assumed proton widths. We recommend a value of $\omega\gamma = 5.7 \times 10^{-22}$ eV, based on the results of \citet{1986NuPhA.457..367C}. A comprehensive measurement was done by \citet{1980NuPhA.349..165W}, covering the 80-2200 keV range at multiple angles and measuring branching ratios, the direct component (extrapolated from high-energy data), and strengths down to the $E_r^{c.m.} = 143$ keV resonance. The strength of the latter resonance, $\omega\gamma = (0.98 \pm 0.03)$ $\times 10^{-3}$~eV, has been remeasured multiple times with consistent results \citep{1980NuPhA.349..165W,1982ZPhyA.305..319B,1990PhRvC..42..753V,2016NIMPA.830..427D,2019PhLB..79734900B,2021PhRvC.104b5802P}. The strength of the $E_r^{c.m.} = 90$ keV resonance was recently controversial \citep{2012PhRvC..86f5804B, 2013PhRvC..88a5801F}, but a measurement by \citet{2019PhLB..79734900B} found it to be of insignificant strength ($\omega\gamma$ $=$ $(0.5 \pm 0.1)$ $\times 10^{-9}$~eV). The non-resonant component was determined in Wiescher et al. at an energy of $E^{c.m.}$ $=$ $1752$~keV and extrapolated to a value of $S(0)$ $=$ $15.7$~keVb. Buckner et al. normalized a non-resonant capture model calculation to the experimental value at $E^{c.m.}$ $=$ $1752$~keV, resulting in $S(0)$ $=$ $7.1$~keVb. No uncertainties were given for these two S-factors. \citet{2019PhLB..79734900B} extracted the non-resonant component from a cross section fit at center-of-mass energies between 85 and 150 keV, 
resulting in $S(0) = (23 \pm 3.8$) keVb. Here we adopt this value, but further investigation of the low-energy cross section is necessary to resolve these discrepancies.

\subsubsection{\texorpdfstring{$^{18}\text{O}(p,\alpha)^{15}\text{N}$}{18O(p,α)15N}}
\label{sec:OtherCNO:CNO3:o18pa}

The low-energy trend of the $^{18}$O($p,\alpha$)$^{15}$N cross section is determined by three resonances at 20, 90 and 143 keV (see Table \ref{tab:wg6_narrow}). The most recent determination of the strength of the 20 keV resonance comes from a Trojan horse method (THM) experiment \citep{2008PhRvL.101o2501L}. Their result was already recommended in \citetalias{2011RvMP...83..195A} and we maintain the same recommendation. The resonance at 90 keV was investigated in the past with direct \citep{1979NuPhA.313..346L} and indirect \citep{2008PhRvL.101o2501L} techniques, finding consistent results. Nevertheless, a new direct measurement reported a resonance strength one order of magnitude higher than the previously adopted value \citep{2019PhLB..790..237B}. Since no clear explanation can be found for such a discrepancy, here we adopt a weighted average of the three resonance strengths, but we inflate the uncertainty so that it embraces all three values. New experiments to solve this discrepancy are advised. The resonance at 143~keV is well known, and different experiments provided consistent results \citep{1979NuPhA.313..346L, 1995ZPhyA.351..453B, 2015EPJA...51...94B}. Here we adopt a weighted average of \citet{1995ZPhyA.351..453B} and \citet{2015EPJA...51...94B}, which have independent systematic uncertainties (while no information on individual contributions to the total uncertainty is given in \citet{1979NuPhA.313..346L}).
The non-resonant differential cross section has  recently been measured down to E$_{c.m.}$ = 55 keV by \citet{2019PhLB..790..237B}. At overlapping energies, the S-factor is consistent with the previous measurement by \citet{1979NuPhA.313..346L}. \citet{2019PhLB..790..237B} found a broad structure at $E \approx 110$ keV that they could only explain assuming the existence of a so-far unobserved resonance at E$_r^{c.m.} = 106$ keV. Further investigations are needed to shed light on this supposed resonance.

\subsubsection{\texorpdfstring{$^{19}\text{F}(p,\gamma)^{20}\text{Ne}$}{19F(p,γ)20Ne}}
\label{sec:OtherCNO:CNO3:f19pg}

The $^{19}$F$(p,\gamma)^{20}$Ne reaction has been challenging to access experimentally because it has a much smaller cross section than the competing $^{19}$F$(p,\alpha\gamma)^{16}$O reaction. The cross section of both reactions is dominated by a mixture of narrow and broad resonances corresponding to mostly unnatural parity states in $^{20}$Ne (mainly $J^\pi$~=~1$^+$ and 2$^-$). At very low energies, the cross section may be enhanced by a near threshold state at $E_{\text{c.m.}}\approx$11~keV and subthreshold and non-resonant capture contributions may also contribute significantly, but have not yet been well characterized~\citep{2021PhRvC.103e5815D}, as few particle transfer studies have been published~\citep{1975PhRvC..11...19B, Kio90}.

The first low-energy cross section measurement was that of \citet{2008PhRvC..77a5802C}, who measured only the $^{19}$F$(p,\gamma_1)^{20}$Ne transition. While this was likely a strong transition, there can certainly be other transitions with sizable contributions. \citet{2008PhRvC..77a5802C} covered the center-of-mass energy range from 200 to 760~keV, where three strong resonances were observed at $E_r^{c.m.}$~$=$~323.9, 564 and 634~keV. Only an upper limit was reported for an expected resonance at $E_r^{c.m.}=213$~keV. Also, no evidence of either a near-threshold state, subthreshold state, or non-resonant capture has been observed.

An investigation of the $\gamma$-ray branching ratios of the $E_r^{c.m.}=323.9$~keV resonance was undertaken by \citet{2021PhRvC.103e5805W}, using the DRAGON recoil separator at TRIUMF. Recoil detection provided an effective method of discriminating the large background from the $^{19}$F$(p,\alpha\gamma)^{16}$O reaction. It was found that the primary transitions to both the ground and the 4967~keV state dominate over the primary transition to the 1633~keV first-excited-state.

Finally, a new direct measurement of the low-energy cross section has been reported by \citet{2022Natur.610..656Z}, which overlaps with the previous low-energy data of \citet{2008PhRvC..77a5802C}. A new resonance was discovered at $E_r^{c.m.}$~=~225.2~keV in the $^{19}$F$(p,\gamma_1)^{20}$Ne cross section, attributable to a 3$^-$ state that corresponds to a weak resonance observed in the $^{19}$F$(p,\alpha\gamma)^{16}$O data of \citet{2000EPJA....7...79S} at the same energy. This low-energy resonance increases the $^{19}$F$(p,\gamma)^{20}$Ne reaction rate by about an order of magnitude at $T$ $\approx$ 0.1~GK. Like \citet{2021PhRvC.103e5805W}, \citet{2022Natur.610..656Z} also observed a sizeable ground state transition for the $E_r^{c.m.}=323.9$~keV resonance. As in the $^{19}$F$(p,\alpha\gamma)^{16}$O reaction (see Section~\ref{sec:OtherCNO:CNO3:f19pa}), an upper limit was also estimated for the proposed $E_r^{c.m.}$~=~11~keV resonance.

We have therefore adopted the resonance energies and strengths of \citet{2022Natur.610..656Z}, which are in good agreement with the previous measurements of \citet{2008PhRvC..77a5802C} and \citet{2021PhRvC.103e5805W} in the overlapping energy region (see Tables~\ref{tab:wg6} and \ref{tab:wg6_narrow}). 

\subsubsection{\texorpdfstring{$^{19}\text{F}(p,\alpha)^{16}\text{O}$}{19F(p,α)16O}}\
\label{sec:OtherCNO:CNO3:f19pa}

The $^{19}$F($p,\alpha$)$^{16}$O reaction was not included in \citetalias{2011RvMP...83..195A}. NACRE \citep{1999NuPhA.656....3A} pointed 
out that the total astrophysical S-factor was dominated by the non-resonant $(p,\alpha_0)$ channel, 
though underscoring the large uncertainties ($\approx 50\%$) affecting the reaction 
rate at low temperatures. After NACRE, the situation greatly changed following the publication
of a large number of papers addressing the measurement of the $\alpha_0$ channel 
\citep{2011ApJ...739L..54L,2015ApJ...805..128L,2017ApJ...845...19I,2013JPhG...40l5102L,2015PhLB..748..178L}, of the $\alpha\gamma$ 
channel \citep{2000EPJA....7...79S,2000NIMPA.450..353F,Ding2002,2019NIMPB.438...48Z,2021PhRvL.127o2702Z,2021NIMPB.499..118P}. In addition, 
several authors attempted to reach a consensus astrophysical factor carrying out refined 
analyses using, for instance, the phenomenological R-matrix \citep{2018ChPhC..42a5001H,2019AcPPB..50..393L,2019PhRvC.100d4307L,2021ApJ...913...51Z,2021PhRvC.103e5815D},
and carried out new theoretical calculations of the astrophysical S-factor \citep{2021IJMPE..3050102S}.

Regarding the measurement of the astrophysical factor for the $\alpha_0$ channel, populating the
$^{16}$O ground state, indirect (using the THM \citep{2014RPPh...77j6901T}) and direct methods were 
used, leading to consistent results. 
While the direct measurement \citep{2015PhLB..748..178L} could reach about 170~keV in the center-of-mass,
the THM measurements \citep{2011ApJ...739L..54L,2015ApJ...805..128L,2017ApJ...845...19I} could reach 5~keV, providing an extrapolated $S(0)=(17.4 \pm 2.7)\times 10^3$~keVb.
The most striking feature of both direct and indirect measurements is the evidence of 
resonances populating the $E_{c.m.}\lesssim 600$~keV energy region, at odds with the NACRE
extrapolation. 

Until the work of \citet{2021PhRvL.127o2702Z}, direct measurements of the $\alpha\gamma$ channel stopped at energies well above the Gamow energy. \citet{2021PhRvL.127o2702Z} reached energies
as low as $\approx 72$~keV in the center of mass, still above the Gamow window for solar fusion, using JUNA. Although the lowest energy
point is affected by a large uncertainty (87\%) and no correction for the electron screening effect is reported (for instance, \citet{1987ZPhyA.327..461A} estimated a 20\% enhancement at 70~keV center-of-mass energy for proton-induced reactions on oxygen), the R-matrix analysis the authors carried out indicates the presence 
of a $1^+$ resonance at $E_r^{c.m.} = 11$~keV, corresponding to a $^{20}$Ne excitation energy of 12.855~MeV.  This level
could significantly increase the S factor, yielding S(0)$\approx 200$~MeVb.
If this result is confirmed, the $\alpha\gamma$ channel would dominate the low-energy total
S-factor for $^{19}$F($p,\alpha$)$^{16}$O at center-of-mass energies below $\approx$ 50 keV.
Alternatively, if the  $E_r^{c.m.} = 11$~keV resonance is excluded from the fit, the deduced
value would be S(0) $\approx 1$~MeVb.

The presence of new data sets triggered the publication of review papers aiming to
provide the best fit of the available data. In particular, \citet{2019PhRvC.100d4307L} carried
out a comprehensive R-matrix analysis 
spanning an energy range from 
about $E^{c.m.}=200$~keV to $12000$~keV. The R-matrix fit predicts a S(0)$\approx 10$~MeVb for the $\alpha_0$ channel, with a 
conservative uncertainty of about 20\%, lower than the value from \citet{2017ApJ...845...19I}. 
An additional interesting result from this work is that the $\alpha_1$ channel, corresponding
to the population of the $0^+$ $^{16}$O first excited state, is negligible with respect to 
the $\alpha_0$ one below about $E^{c.m.}=50$~keV, yet if the $2^+$ state at 13095 keV is assumed to 
contribute, as supported by the $^{16}$O($\alpha,\alpha_1$)$^{16}$O$^*$ data, the $\alpha_1$ 
channel may significantly contribute to the total astrophysical factor above about $E^{c.m.}=50$~keV, 
calling for new data at low energies. 

The most comprehensive R-matrix analysis, however not including the results in \citet{2021PhRvL.127o2702Z}, 
is the one carried out in \citet{2021PhRvC.103e5815D}. For the $\alpha_1$  channel, the results are 
consistent with \citet{2019PhRvC.100d4307L}. For the $\alpha \gamma$ channel, the conclusions are 
that the $\alpha_2$ channel is the major one and, as shown by \citet{2021PhRvL.127o2702Z}, low- 
and sub-threshold resonances can enhance the astrophysical factor to dominate over the 
$\alpha_0$ channel at astrophysical energies. Finally, the result of the R-matrix analysis for the $\alpha_0$ channel supports a flat S-factor devoid of resonances, in contrast with 
\citet{2011ApJ...739L..54L,2015ApJ...805..128L,2017ApJ...845...19I,2013JPhG...40l5102L,2015PhLB..748..178L}, in agreement
with the analysis of \citet{1991PhRvC..44..952H}. However, this is based on the data from \citet{LW78}, which were excluded from the NACRE compilation due to a likely large underestimate of reported absolute cross sections, by about a factor of 2.

In summary, while the situation for the $\alpha_0$ and $\alpha_1$ channels is well constrained, the contribution of higher $^{16}$O excited states and, in particular, of the $\alpha_2$ channel, needs further studies to confirm the occurrence of the $E_r^{c.m.}=11$~keV resonance. This represents, at present, the largest source of uncertainty. Therefore, the recommended reaction rate lower limit is conservatively set by the \citet{2017ApJ...845...19I} one, while the upper limit is given by \citet{2021ApJ...913...51Z}, assuming the existence of the $E_r^{c.m.}= 11$~keV resonance.

\subsection{Reactions of the neon-sodium (NeNa) cycle}
\label{sec:OtherCNO:NeNa}

\subsubsection{\texorpdfstring{$^{20}\text{Ne}(p,\gamma)^{21}\text{Na}$}{20Ne(p,γ)21Na}}
\label{sec:OtherCNO:NeNa:ne20pg}

The low energy S-factor of the $^{20}$Ne$(p,\gamma)^{21}$Na reaction is dominated by the high energy tail of a subthreshold resonance only a few keV below the reaction threshold. In addition, the non-resonant capture into two additional states is relevant and provides the largest contribution in the energy range where data are available. All three transitions have been measured over a wide range of energies by two direct experiments \citep{1975NuPhA.241..460R, 2018PhRvC..97f5802L}, and the results can be used to arrive at S-factor recommendations. However, no consistent parametrizations have been published that could be used to compare the results. 

S-factor data from these two experiments were extracted either from their accompanying supplemental materials or from digitized forms of the published figures.  The results were normalized to the strength of the $E_r^{c.m.}$ $=$ $1113$~keV resonance \citep[$\omega\gamma$ $=$ $0.94$~eV;][]{2018PhRvC..97f5802L}. The resonant capture into the tail of the subthreshold resonance ($E_r^{c.m.} = -7.9$~keV; \citealt{2015NDS...127....1F}) can be described by the Breit-Wigner equation, as shown in \citet{1988ccna.book.....R,2015nps..book.....I}. To account, additionally, for the non-resonant capture to the ground state, a constant value of $S(E)$ $=$ $0.0463$~keVb, as recommended by the indirect experiment of \citet{2006PhRvC..73c5806M}, was included in the fit to the data of \citet{1975NuPhA.241..460R, 2018PhRvC..97f5802L}. For the two direct-capture transitions, the energy dependence of the S-factor was calculated with the program DIRCAP \citep{ILI22}, using a Woods-Saxon potential with a diffuseness of $a$ $=$ $0.65$~fm to achieve S-factor magnitudes comparable to the experimental data. These S-factors were fit individually to the experimental data and then summed to the total S-factor. The results are listed in Table~\ref{tab:wg6_sfac}. Uncertainties from the individual contributions were added in quadrature.

However, significant and unquantifiable uncertainties likely exist because the results of the two experiments differ systematically both in magnitude and energy dependence \citep{2018PhRvC..97f5802L}. This may be caused by experimental or data-analysis errors in either experiment, or in both. In addition, the experiments at low energies only measured in a small angular range and relied on sparse angular distribution measurements at higher energies to infer angular correlation effects. Also, the experiments did not reach sufficiently low energies to conclusively determine the potential parameters to be used in the DIRCAP calculation. Consequently, this will contribute to the S-factor uncertainty when the energy range of stellar burning, depending on the temperature, is located far below the experimental range.

A new measurement of the $^{20}$Ne$(p,\gamma)^{21}$Na direct-capture cross section down to $E^{c.m.}$ $=$ $250$~keV has recently been published by the LUNA collaboration \citep{2023PhRvC.108e2801M}. The potential-model results agree with the new experimental data.

\subsubsection{\texorpdfstring{$^{21}\text{Ne}(p,\gamma)^{22}\text{Na}$}{21Ne(p,γ)22Na}}\label{sec:21nepg}

At temperatures below 0.1 GK, the reaction rate is dominated by three narrow resonances at 16.6, 94.6 and 126 keV. The 16.6 and 94.6 keV  resonances have never been observed directly in the ($p,\gamma$) channel \citep{1982NuPhA.385...57G}. Upper limits on their strengths were derived from a $^{21}$Ne(d,$p)^{22}$Na experiment  \citep{1972PhRvC...6..885N, 2001ApJS..134..151I}. Results are reported in Table~\ref{tab:wg6_narrow}.
The non-resonant capture S-factor at zero energy was estimated by \citet{1975NuPhA.241..460R} using single-particle spectroscopic factors of the bound states in $^{22}$Na. According to the authors, the result, S(0) $\approx$ 20 keV b, is uncertain by a factor of three.

\subsubsection{\texorpdfstring{$^{22}\text{Ne}(p,\gamma)^{23}\text{Na}$}{22Ne(p,γ)23Na}}
\label{sec:OtherCNO:NeNa:ne22pg}
The $^{22}$Ne$(p,\gamma)^{23}$Na reaction used to be the most uncertain of the NeNa cycle, because of a number of poorly-known low-lying resonances. At temperatures below 30 MK, the reaction rate is dominated by a resonance at 36 keV. Its strength has been determined from a proton-transfer experiment \citep{2001PhRvC..65a5801H, 2010NuPhA.841..251I}. Two near-threshold resonances are expected at 5 and 28 keV, but the existing upper limits on their strengths are so low that they do not contribute significantly to the reaction rate \citep{2001PhRvC..65a5801H}. Two resonances, at 68 and 100 keV, were reported by \citet{1971PhRvC...4.2030P}, generating a great deal of discussion
in the literature. Subsequent searches included direct experiments \citep{2015PhRvL.115y2501C, 2016PhRvC..94e5804D, 2018PhRvL.121q2701F}, proton transfer reaction \citep{2001PhRvC..65a5801H} and inelastic scattering \citep{PhysRevC.108.045802}. No evidence for the resonances emerged from these searches. The latest direct upper limits on their strength suggest that, even if the two resonances exist, their contribution to the mean (recommended) thermonuclear reaction rate is negligible \citep{2018PhRvL.121q2701F}. 
A resonance doublet dominates the reaction rate at temperatures around 0.1 GK. The strength of the stronger resonance in the doublet was measured both with direct \citep{2018PhRvL.120w9901C, 2017PhRvC..95a5806K, 2020PhLB..80735539L, 2020PhRvC.102c5801W} and proton-transfer experiments (data from \citet{1971PhRvC...4.2030P, 2001PhRvC..65a5801H} re-analyzed in \citet{2020PhRvC.101b5802S}). The value adopted in Table~\ref{tab:wg6_narrow} is an average of all literature values mentioned above. The strength of the weaker resonance in the doublet was inferred from proton-transfer data \citep{2020PhRvC.101b5802S}.

At 0.2 GK, the reaction rate is mainly determined by a resonance at 181 keV. Its strength was measured in four independent experiments \citep{2018PhRvL.120w9901C, 2017PhRvC..95a5806K, 2018PhRvL.121q2701F, 2020PhRvC.102c5801W}. The strength recommended in Table~\ref{tab:wg6_narrow} is a weighted average of all four values. The most intense higher-lying resonances, relevant to nova and supernova explosions, have been recently addressed in \citet{2010PhRvC..81e5804L, 2015PhRvC..92d5807D, 2017PhRvC..95a5806K, 2020PhRvC.102c5801W}.
The non-resonant capture cross section has been measured by different groups \citep{2017PhRvC..95a5806K, 2018PhRvL.121q2701F, 2020PhRvC.102c5801W}. The low-energy trend of the ground state capture is described by a constant value of $S = (50 \pm 12)$ keVb, plus the contribution of a broad subthreshold state at $-$130 keV (E$_x$ $=$ 8664 keV) \citep{2018PhRvL.121q2701F}.  

\subsubsection{\texorpdfstring{$^{23}\text{Na}(p,\gamma)^{24}\text{Mg}$}{23Na(p,γ)24Mg}}
\label{sec:OtherCNO:NeNa:na23pg}

The $^{23}$Na$(p,\gamma)^{24}$Mg reaction has seen several recent studies~\citep{2011A&A...533A..66C, 2019PhLB..795..122B, 2021PhRvC.104c2801M, 2022PhRvC.106d5801B}. The cross section is dominated by three narrow resonances over the lowest experimentally measured energy range, located at $E_r^{c.m.}$~=~133, 240.6, and 295.9~keV. Below the lowest known resonance, non-resonant components are thought to dominate.

Several indirect measurements have been made to characterize energies and proton spectroscopic factors of bound and unbound states in $^{24}$Mg. The measurement of \citet{2004PhRvC..70d5802H} used the $^{23}$Na$(^3$He,d)$^{24}$Mg reaction to investigate unbound states (11690~$<E_x<$~12520~keV) near the proton separation energy ($S_p$~$=$~11693~keV). For lower-energy bound states that were not measured, the spectroscopic factors of \citet{1978PhRvC..18.2032G} where adopted. A direct capture model was used to calculate the non-resonant component of the cross section. This was followed by an investigation of the lowest energy resonance at $E_r^{c.m.}\approx$~133~keV by \citet{2013PhRvC..88f5806C}, where an upper limit for the resonance strength was determined. A subsequent study at the LUNA facility~\citep{2019PhLB..795..122B} was able to make a first measurement of the resonance strength, 1.46$^{+0.58}_{-0.53}\times 10^{-9}$~eV. A recent $^{23}$Na$(^3$He,d)$^{24}$Mg transfer study at the Triangle Universities Nuclear Laboratory (TUNL) revised the energy of the $\approx$133~keV resonance to $E_r^{c.m.}$~$=$~133(3)~keV \citep{2021PhRvC.104c2801M}. 

Direct cross section measurements of the $^{23}$Na$(p,\gamma)^{24}$Mg reaction have been made at the University of Notre Dame at energies between $E^{lab}$~=~500 and 1000~keV by \citet{2022PhRvC.106d5801B}. Two, broad, strong, resonances were observed in several $\gamma$-ray transitions at $E_r^{c.m.}$~$=$~841 and 981~keV whose low energy tails contribute in the low energy region. Note that the energy uncertainties presented here reflect the uncertainty in the beam calibration of the accelerator, which was neglected in \citet{2022PhRvC.106d5801B}. Further, measurements of the $E_x$~$=$~10740~keV primary transition in $^{24}$Mg, predicted to have the largest non-resonant capture component based on the spectroscopic factors of \citet{1978PhRvC..18.2032G}, observed a trend in the off-resonance cross section that was highly suggestive of non-resonant capture. Given the level of consistency observed between the non-resonant capture S-factors calculated from the spectroscopic factors of \citet{1978PhRvC..18.2032G} and the external capture S-factors from the R-matrix analysis, an uncertainty in the non-resonant S-factor of 50\% was recommended.
We have adopted the non-resonant component of \citet{2022PhRvC.106d5801B} in this work, which, like all previous calculations, relies heavily on the spectroscopic factors of \citet{1978PhRvC..18.2032G}.

\subsubsection{\texorpdfstring{$^{23}\text{Na}(p,\alpha)^{20}\text{Ne}$}{23Na(p,α)20Ne}}
\label{sec:OtherCNO:NeNa:na23pa}

The $^{23}$Na$(p,\alpha)^{20}$Ne reaction rate is dominated by narrow resonances, with a total of 52 resonances in the center-of-mass energy interval $6$ $-$ $2328$~keV.  

After NACRE, the state-of-the-art tabulation of the resonance strengths is the one in \citet{2010NuPhA.841..251I}. It is mostly based on the ($^3$He,d) proton transfer measurement by \citet{2004PhRvC..70d5802H}, supplying both the excitation energies and spectroscopic factors of the near-threshold states. In detail, below $E^{c.m.}=968$~keV, all the resonance strengths are taken from Table VI of \citet{2004PhRvC..70d5802H}, with the exception of the $E_r^{c.m.}=133$~keV resonance strength ~\citep{2021PhRvC.104c2801M}, for which the upper limit directly determined by \citet{2004ApJ...615L..37R,2005JPhG...31S1785I} is adopted, equal to $1.5 \times 10^{-8}$~eV. Above $E^{c.m.}=968$~keV, the \citet{2004PhRvC..70d5802H} strengths are rescaled to the one of the $E_r^{c.m.}=338$~keV resonance determined by \citet{2002PhRvC..65f4609R}. This result is considered as reference value since it provides the most recent directly-measured strength. 

Besides the above-mentioned works, two more should be considered. One is \citet{2005JPhG...31S1785I}, which is not used in \citet{2010NuPhA.841..251I} because the same upper limit for the $E_r^{c.m.}=133$~keV resonance as in \citet{2004ApJ...615L..37R} is reported. The other work is \citet{2013PhRvC..88f5806C}, which used the upper limit on the ($p,\gamma$) resonance strength and the upper limit on the $\alpha$ branching ratio (equal to 0.13 at the 95\% CL), to provide an upper limit on the $(p,\alpha)$ strength of the $E_r^{c.m.}=133$~keV resonance equal to $0.88 \times 10^{-9}$ eV. This implies a negligible contribution to the reaction rate and confirms the conclusions drawn in \citet{2010NuPhA.841..251I}. 

We note that below $10^9$K, the resonances dominating the total $^{23}$Na(p,$\gamma$)$^{24}$Mg S-factor differ from those determining the total $^{23}$Na(p,$\alpha$)$^{20}$Ne S-factor. Consequently, the S-factors of these two competing reactions are not correlated.

\begin{table*}[]
\begin{center}
\caption{Overview of CNO and NeNa reactions.}\label{tab:wg6}
\begin{ruledtabular}
\begin{tabular}{lcccll}
Reaction  & Category\footnotemark[1]      & $S(0)$ $\pm$ $\Delta S(0)$   & $E_0$\footnotemark[2] & Most recent reference  & Comments \\
                          &                               & (keVb)             & (keV)                 &                             &          \\
\hline
\vspace{0.1in}
$^{12}$C(\textit{p},$\gamma$)$^{13}$N    &   B           &  $1.44\pm0.06$                &  25   &  \citet{2023PhRvL.131p2701S}                     & \citetalias{2011RvMP...83..195A}: $S(0) =  1.34\pm0.21$ keVb \\
$^{13}$C(\textit{p},$\gamma$)$^{14}$N    &   B           &  $6.1\pm0.4$               &  25   &  \citet{2023PhRvL.131p2701S}                     &  \citetalias{2011RvMP...83..195A}: $S(0)=7.6\pm1.0$~keVb  \\
$^{15}$N(\textit{p},$\gamma$)$^{16}$O    &   B           &  $40\pm3$          &  28   &  \citet{2013PhRvC..87a5802D}     &  \citetalias{2011RvMP...83..195A}: $S(0)=36\pm6$~keVb      \\
$^{15}$N(\textit{p},$\alpha$)$^{12}$C    &   B           &  $73000\pm5000$    &  28   &  \citet{2009PhRvC..80a2801L}           &  \citetalias{2011RvMP...83..195A}: $S(0)$ $=$ $73000 \pm 5000$~keVb      \\
$^{16}$O(\textit{p},$\gamma$)$^{17}$F    &   B           &  $10.92\pm0.44$    &  30   &  \citet{2022PhRvC.106e5802I}             &  \citetalias{2011RvMP...83..195A}: $S(0)$ $=$ $10.6 \pm 0.8$~keVb  \\
$^{17}$O(\textit{p},$\gamma$)$^{18}$F    &   B           &  $4.7\,\pm\,1.0$   &  31   &  \citet{2017PhRvC..95c5805G}        &  \citetalias{2011RvMP...83..195A}: $S(0)$ $=$ $6.2\,\pm\,3.1$~keVb       \\
$^{17}$O(\textit{p},$\alpha$)$^{14}$N    &   A           &     $\cdots$            &  31   &  \citet{2016PhRvL.117n2502B}         &  near-threshold resonances \\
$^{18}$O(\textit{p},$\gamma$)$^{19}$F    &   B           &  $23.0\pm3.8$      &  31   &  \citet{2021PhRvC.104b5802P}         &  also in literature: S(0) = 7.1 and 15.7 keVb      \\
$^{18}$O(\textit{p},$\alpha$)$^{15}$N    &   B           &     $\cdots$            &  31   &  \citet{2019PhLB..790..237B}       &  low-energy resonances     \\
$^{19}$F(\textit{p},$\gamma$)$^{20}$Ne   &   A           &     $\cdots$            &  33   &  \citet{2022Natur.610..656Z}               & possible resonance near threshold      \\
$^{19}$F(\textit{p},$\alpha$)$^{16}$O    &   B           &     $\cdots$            &  33   & \citet{2021ApJ...913...51Z}               &  possible resonance near threshold       \\
$^{20}$Ne(\textit{p},$\gamma$)$^{21}$Na  &   B           &  $6776\pm550$      &  36   &  present work                    &          \\
$^{21}$Ne(\textit{p},$\gamma$)$^{22}$Na  &   A           &  $\approx$20       &  36   &  \citet{1975NuPhA.241..460R}             &  factor 3 uncertainty in $S(0)$      \\
$^{22}$Ne(\textit{p},$\gamma$)$^{23}$Na  &   A           &  $415\pm91$        &  36   &  \citet{2018PhRvL.121q2701F}                     &  sub-threshold resonance at $-130$~keV      \\
$^{23}$Na(\textit{p},$\gamma$)$^{24}$Mg  &   B           &  18~$\pm$~9        &  38   &  \citet{2022PhRvC.106d5801B}     &        \\
$^{23}$Na(\textit{p},$\alpha$)$^{20}$Ne  &   A           &     $\cdots$            &  38   &  \citet{2010NuPhA.841..251I}             &        \\
\end{tabular}
\end{ruledtabular}
\footnotetext[1] {\footnotesize Categories for the dominant rate contribution at the central temperature of the Sun: (``A'') Narrow resonances. (``B'') Broad resonant/non-resonant contributions. We define a ``narrow resonance'' by the criterion that its full contribution to the reaction rate can be calculated from the resonance energy and strength alone.}
\footnotetext[2] {\footnotesize Central energy location (center-of-mass system) of Gamow peak at the solar core.}
\end{center}
\end{table*}
%

\begin{table*}[]
\begin{center}
\caption{S-factors versus center-of-mass energy. Data are also plotted in Fig. \ref{fig:sfactor}.}\label{tab:wg6_sfac}
\begin{ruledtabular}
\begin{tabular}{lccccccc}
Reaction  & \multicolumn{7}{c}{$S(E)$~(keVb) (\% uncertainty)}  \\  
  \cline{2-8} \\
                                                            & 5~keV        & 10~keV       & 20~keV       & 40~keV       & 60~keV       & 100~keV    & 140~keV          \\
\hline
\vspace{0.1in}
$^{12}$C(\textit{p},$\gamma$)$^{13}$N\footnotemark[1]    & 1.46(4.1\%) & 1.47(4.1\%)   & 1.51(4.1\%)  & 1.58(4.1\%)  & 1.67(4.1\%)  & 1.89(4.0\%) & 2.19(4.1\%)            \\
$^{13}$C(\textit{p},$\gamma$)$^{14}$N\footnotemark[1]    & 6.16(6.0\%)  & 6.22(6.0\%)  & 6.34(6.1\%)  & 6.60(6.1\%)  & 6.89(6.1\%)  & 7.58(6.1\%)   & 8.47(6.1\%)            \\
$^{15}$N(\textit{p},$\gamma$)$^{16}$O                    & 40.8(7.5\%)  &  41.7(7.5\%) &  43.0(7.5\%) &  45.8(7.5\%) &  50.0(7.5\%) &  59.0 (7.5\%) &      73.4(7.5\%)           \\ 
$^{15}$N(\textit{p},$\alpha$)$^{12}$C                    & 7.7$\times10^4$(6.8\%)    &  8.0$\times10^4$(6.8\%)  &  8.5$\times10^4$(6.8\%)    & 1.00$\times10^5$(6.8\%)   & 1.18$\times10^5$(6.8\%)  &  1.66$\times10^5$(6.8\%)    &  2.53$\times10^5$(6.8\%)                \\   
$^{16}$O(\textit{p},$\gamma$)$^{17}$F\footnotemark[2]    & 10.64(4.0\%) & 10.38(4.0\%) & 9.92(4.0\%)  & 9.15(4.0\%)  &  8.56(4.0\%) & 7.67(4.0\%)   &  7.05(4.0\%)   \\   
$^{20}$Ne(\textit{p},$\gamma$)$^{21}$Na\footnotemark[3]  &  2586(8\%)   & 1365(8\%)    &  593 (7\%)   &   226(7\%)   &  128(7\%)    &   69(6\%)     &    45(6\%)               \\   
$^{23}$Na(\textit{p},$\gamma$)$^{24}$Mg                  &  18(50\%)    &   18(50\%)   &   18(50\%)   &   18(50\%)   &   18(50\%)   &   18(50\%)    &     18(50\%)              \\  
\end{tabular}
\end{ruledtabular}
\footnotetext[1] {\footnotesize From \citet{2023PhRvL.131p2701S}.}
\footnotetext[2] {\footnotesize From \citet{2022PhRvC.106e5802I}.}
\footnotetext[3] {\footnotesize From present work.}
\end{center}
\end{table*}
%

\begin{table*}[]
\begin{center}
\caption{ Recommended strengths and relative kinetic energies of narrow resonances. \label{tab:wg6_narrow} }
\begin{ruledtabular}
\begin{tabular}{lccl}
Reaction  & $E_r^{c.m.}$~(keV)      & $\omega\gamma$~(eV) \footnotemark[2]  &   Comments/references \\
\hline
\vspace{0.1in}
$^{17}$O(\textit{p},$\gamma$)$^{18}$F    &   65              &   $15.6^{+18.3}_{-4.0}\times10^{-12}$    &  see text for details                  \\
    &   183  &   ($1.77\,\pm\,0.09$)$\times10^{-6}$    &  see text for details                  \\
$^{17}$O(\textit{p},$\alpha$)$^{14}$N    &   65            &  $4.6_{-1.2}^{+5.4}\times10^{-9}$    &  see text for details   \\
                                            &   183.9           &   $(1.68 \pm 0.03_{stat} \pm 0.12_{syst})\times10^{-3}$  &  \citet{2007PhRvC..75c5810C,2015EPJA...51...94B}   \\
$^{18}$O(\textit{p},$\gamma$)$^{19}$F    &   20              &    $5.7 \times 10^{-22}$ &  \citet{1986NuPhA.457..367C}                 \\
                                            &   90              & $(0.53 \pm 0.07_{stat} \pm 0.07_{syst})\times10^{-9}$   &  \citet{2019PhLB..79734900B}                  \\
                                            &   143           & $(0.98 \pm 0.03)\times10^{-3}$  &  see text for details                  \\
$^{18}$O(\textit{p},$\alpha$)$^{15}$N    &   20              & $8.3^{+3.8}_{-2.6}\times10^{-19}$    &  \citet{2008PhRvL.101o2501L} \\
                                            &   90              &  $2.0^{+13.7}_{-0.4}\times10^{-7}$ & see text for details \\
                                            &   143           &  $(166 \pm 8)\times10^{-3}$  &  see text for details \\
$^{19}$F(\textit{p},$\gamma$)$^{20}$Ne   &   11.0
&   $\leq$8.5$\times$10$^{-29}$  &  \citet{1975PhRvC..11...19B, Kio90}                  \\         
                                            &   212.7
                                            &   $\leq$4.2$\times$10$^{-6}$    &  \citet{2022Natur.610..656Z, 2000EPJA....7...79S} \\
                                            & 225.2
                                            & (4.19~$\pm$~0.33)$\times$10$^{-5}$                             & \citet{2022Natur.610..656Z, 2000EPJA....7...79S} \\
                                            &   323.9
                                            &   (3.16~$\pm$~0.33)$\times$10$^{-3}$    &  \citet{2008PhRvC..77a5802C, 2021PhRvC.103e5805W, 2022Natur.610..656Z}   \\
$^{19}$F(\lowercase{p},$\alpha \gamma$)$^{16}$O    &   11.0
&   $(7.5\pm 3.0)\times$10$^{-29}$  &  \citet{2021ApJ...913...51Z}                  \\      
%
%
                                            &   212.7
                                            &   $0.0126 \pm 0.0013$    &  \citet{2000EPJA....7...79S,2021ApJ...913...51Z} \\
                                            & 225.2
                                            &   $ 0.0011 \pm 0.0004$                            & \citet{2000EPJA....7...79S,2021ApJ...913...51Z} \\
                                            &   323.9
                                            &   $ 23.5 \pm 0.6$    &  \citet{2021ApJ...913...51Z, 2022Natur.610..656Z}   \\
$^{21}$Ne(\textit{p},$\gamma$)$^{22}$Na  &   16.6            & $\leq$ 6.2$\times10^{-24}$ &  \citet{2001ApJS..134..151I} \\
                                            &   94.6            & $\leq$ 6.4$\times10^{-10}$ &  \citet{2001ApJS..134..151I} \\
                                            &   126             &   (3.75 $\pm$ 0.75)$\times10^{-5}$   & \citet{1982NuPhA.385...57G} \\
$^{22}$Ne(\textit{p},$\gamma$)$^{23}$Na  &   5               & $\leq$ 2.1 $\times10^{-51}$ &  \citet{2001PhRvC..65a5801H} \\
                                            &   28              &  $\leq$ 3.2 $\times10^{-25}$ & \citet{2001PhRvC..65a5801H}  \\
                                            &   35              & (3.1 $\pm$ 1.2) $\times10^{-15}$ & \citet{2010NuPhA.841..251I}  \\
                                            &   68              & $\leq$ 6$\times10^{-11}$ & \citet{2018PhRvL.121q2701F}  \\
                                            &   100             & $\leq$ 7$\times10^{-11}$ & \citet{2018PhRvL.121q2701F}  \\
                                            & 150               & (3.9$\pm$ 0.9) $\times10^{-9}$ & \citet{2020PhRvC.101b5802S}\\
                                            & 151               & (1.97$\pm$ 0.12) $\times10^{-7}$ & see text for details\\
                                            & 181              & (2.39$\pm$ 0.12) $\times10^{-6}$ & see text for details\\
$^{23}$Na(\textit{p},$\gamma$)$^{24}$Mg  &   133
&   1.46$^{+0.58}_{-0.53}\times$10$^{-9}$    &   \citet{2019PhLB..795..122B,2021PhRvC.104c2801M}                   \\
                                            &   240.6
                                            &  5.3(1)$\times$10$^{-4}$     &  \citet{1975AuJPh..28..141S, 1989NuPhA.493..124K}                  \\ 
                                            &   295.90
                                            &  1.05(19)$\times$10$^{-1}$     &  \citet{1975AuJPh..28..141S,1990NuPhA.510..209E}                  \\
%
$^{23}$Na(\textit{p},$\alpha$)$^{20}$Ne  &   133
&   $\leq$ 1.5$\times10^{-8}$               &   \citet{2004ApJ...615L..37R,2005JPhG...31S1785I, 2021PhRvC.104c2801M}                   \\
                                            &   240.6
                                            &  $\leq$ 0.1    &  \citet{2001PhRvC..65a5801H}                  \\ 
                                            &   295.90
                                            &  1.03(26)$\times$10$^{-2}$     &  \citet{2010NuPhA.841..251I}                  \\  
\end{tabular}
\end{ruledtabular}
\footnotetext[1] {\footnotesize We define a ``narrow resonance'' by the criterion that its full contribution to the total rate can be calculated from the resonance energy and strength alone.}
\footnotetext[2] {\footnotesize The resonance strength, in the center-of-mass system, for the reactions listed here is defined as $\omega\gamma$ $\equiv$ $(2J+1)(2j_t+1)^{-1}(2j_p+1)^{-1}\Gamma_p \Gamma_x /\Gamma$, with $J$, $j_t$, and $j_p$ the spins of the resonance, target, and projectile, respectively, and $\Gamma_p$, $\Gamma_x$, and $\Gamma$ the proton partial width, $\gamma$-ray or $\alpha$-particle partial width, and total resonance width, respectively.}
\end{center}
\end{table*}
%



\section{Electron screening of nuclear reactions} 
\label{sec:Screening}

Nuclear fusion reactions measured in the laboratory and those occurring in the solar core are both affected by the
electronic environments in which they take place. The effects are, however, different requiring special care in deriving the appropriate solar reaction rates. As long as these rates cannot be measured under solar plasma conditions in the laboratory, the usual two-step strategy is to remove the laboratory screening effects from the data to obtain the bare nuclear cross section $\sigma^\mathrm{b}$, 
which is then modified to incorporate the solar plasma screening modifications. Thus,    one must address the differential effects of screening for the solar cross section
$\sigma^\mathrm{solar}$  and the laboratory results, $\sigma^\mathrm{lab}$. 

\subsection{Screening in laboratory experiments} 

Screening diminishes the Coulomb barriers that retard interactions between bare nuclei, thus enhancing cross sections, with the effects becoming more pronounced with decreasing center-of-mass energy.  
Nuclear astrophysics has advanced rapidly over the last two decades with the deployment of underground accelerators and other low-background techniques \cite{2010ARNPS..60...53B} that have enabled low-energy, low-counting-rate experiments. In particular, laboratory measurements of the pp-chain reactions $^2$H($p,\gamma$)$^3$He and $^3$He($^3$He,2$p)^4$He have been made at energies corresponding to the solar Gamow peak.  
A treatment of the distorting effects of screening then becomes quite important in extracting the bare S-factor $S^b(E)$. 

The bare cross section is given by
\begin{equation}
\sigma^b(E) = \frac{1}{E} S(E) P(E); \quad P(E) \equiv \mathrm{exp}\left(-2 \pi \eta(E) \right)
\label{eq:sc1}
\end{equation}
with $\eta(E) = Z_1Z_2\alpha \sqrt{\mu /2E}$, where $\mu$ is the reduced mass and $E$ the center-of-mass
energy. The effects of screening can be incorporated into the Gamow penetration factor $P(E)$ through an energy shift,
$E \rightarrow E+U_e$, where the screening potential $U_e$ has the net effect of lowering the Coulomb barrier.
At the lowest accessible energies in laboratory measurements, the separation of target and projectile during tunnelling is much smaller than the atomic radius: thus the electrons, originally bound to the target (assuming a projectile beam of bare nuclei),
are attracted by the joint charge of target and projectile during the fusion. The screening potential can then be replaced
by a constant screening energy $U_e$. This picture was confirmed in simulations of the electron dynamics in low-energy fusion reactions
\cite{1993PhRvC..48..837S}.
After expanding the argument of the exponential in Eq. \ref{eq:sc1} to first order in $U_e$ one obtains the following relation between the
laboratory and bare S-factors \cite{1987ZPhyA.327..461A}
\begin{equation}
  S^\mathrm{lab}(E)=S^\mathrm{b}(E) \, \mathrm{exp} \left({\frac{\pi \eta(E) U_e }{E}} \right) 
  \label{eq:sc2}
\end{equation}
This result neglects a small normalization correction associated with the effects $U_e$ on the matching of
the external wave function to that in the Coulomb region.

The $1/E$ dependence of the exponent is responsible for the growing importance of the screening correction 
at low energies. Although the screening potential $U_e$ in principle can be calculated from the electron charge
distribution, in most applications it is treated as a free constant parameter, determined along with $S^\mathrm{b}(0)$ and its
derivatives from a fit to data.  Figure 4 of \citetalias{2011RvMP...83..195A} shows the results for $S_{33}(E)$, where the fit 
yielded $U_e = 305 \pm 90$ eV. The exponential
dependence of the screening correction leads to a rapid increase in  $S_{33}^\mathrm{lab}(E)$ relative
to $S_{33}^b$, reaching $\sim$ 40\%, at the lowest data point measured.  This example also underscores the
potential for screening to complicate the extraction of $S^\mathrm{b}(E)$ for energies relevant to the solar core, even in cases where measurements
are restricted to higher energy than the Gamow window. 

In principle one can test the adequacy of Eq. \ref{eq:sc2} by doing laboratory experiments at low energies --
even if the reactions studied are not directly relevant to nuclear astrophysics. Such studies began more
than 30 years ago \cite{1987ZPhyA.327..461A}, using gaseous targets to limit energy loss effects and
other systematics.  Subsequent studies have varied greatly in their configurations, e.g., targets ranging from atomic and molecular
gases to metals,\footnote{In this section, the word ``metals'' is used with the traditional meaning from chemistry, not the astrophysical one.} thin and thick targets, and use of direct and inverse kinematics \cite{2022JPhG...49a0501A}. 
The experiments can be very challenging at the energies needed, subject to backgrounds from both intrinsic
activities in detectors and external sources such as cosmic rays.  %

The values of $U_e$ extracted from experiment using gaseous targets have often exceeded the adiabatic limit, defined in atomic physics as the difference between the electron binding energies of the separate atoms in the entrance channel and that of the composite atom \citep{1995NIMPB..99..297R,2001PrPNP..46...23R,1988PhLB..202..179E,1992ZPhyA.342..471E,1993ZPhyA.345..231A,1994ZPhyA.350..171P,1995ZPhyA.351..107G,2001NuPhA.690..790A}. This in turn has generated unease about the reliability
of the values for $S^\mathrm{b}(E)$ extracted from laboratory measurements for use in astrophysics \cite{1997NuPhA.627..324B}.

Several suggestions have been made to resolve this apparent discrepancy.
For example, an error in the estimated stopping power could result in a smaller reaction effective energy, so that the experimentally deduced screening would
be larger than the true value \cite{1996PhLB..369..211L,1996PhRvC..53...18B}.
Indeed such deviations have been found between theoretical calculations of stopping powers and the tabulation which is
traditionally being used in the analysis of low-energy fusion data
\cite{2000PhRvC..62d5802B,2004PhLB..585...35B,1991PhRvL..66.1831G}. An accurate experimental determination of stopping powers
for gaseous hydrogen and helium targets would be very desirable.
Corrections associated with the non-uniform distribution of the electron cloud around the nucleus \cite{1988ApJ...331..565C} and the effects of electron-electron interactions \cite{1993PhRvC..48..837S} have
also been suggested as potential generators of unidentified systematic error. For molecular targets
the screening energy shows a significant dependence on the scattering angle, being smallest if the projectile
passes the spectator nucleus before fusion \cite{1996NuPhA.605..387S}. Furthermore, the experimental deduction of screening energies requires assumptions about the bare nuclear cross sections, which is another source of uncertainty. This was demonstrated by \citet{2021ARNPS..71..345T,2014ApJ...785...96T}, who, after constraining fits by the addition
of higher energy data from a Trojan Horse measurement, obtained a lower value of $U_e$.
Clusterization and polarization effects have also been proposed to effect the deduced screening energies in specific low-energy reactions
\cite{2016PhLB..755..275S,2017PhRvC..95c5801S}

For resonances, both the energy and width can be affected by screening
(\citealp{1954AuJPh...7..373S}; \citealp{2007NuPhA.781...81Z}). Particular care
has to be taken for narrow, low-energy resonances
where the assumption of a constant screening energy is not valid and the radial dependence of the screening potential
has to be considered which leads to a significant reduction
of screening on the resonance width
(\citealp{2023PhRvC.107d4610I}).

Particularly large screening energies were observed in deuteron- and proton-induced fusion reactions
if the projectile was implanted into metals 
(\citealp{2001EL.....54..449C,2002EPJA...13..377R,2002PhLB..547..193R}; 
\citealp{2002JPSJ...71.2881K}; \citealp{2006EPJA...27S..83C}; 
\citealp{2015PhRvC..92f5801C,2020EPJWC.22701012L}).
While the d+d fusion data obtained with a gaseous target \cite{1995ZPhyA.351..107G}
were compatible with the adiabatic limit ($U_e=20$ eV, \citealp{1990PhLA..146..128B,1991PhLA..153..456B}),
screening energies obtained for various metal hosts varied among themselves and could reach values of several 100 eV. As the adiabatic limit
derived for the fusion on individual atoms in a gas, does not apply to reactions with targets implanted in a host material,
an appropriate model has been developed \cite{2004EL.....68..363C,2008PhRvC..78a5803H}
which agrees nicely with recent screening energies
obtained under high-vacuum conditions 
\cite{2016EL....11322001C,2022PhRvC.106a1601C}.
\citet{2008PhRvC..78a5803H}
also identified oxygen and carbon contamination as the source for some particularly large deduced screening energies. We note that the studies of d+d low-energy fusion
were motivated by the quest to enhance nuclear fusion rates by environmental effects
\cite{2022PhRvC.106a1601C}.

The potential discrepancy between observed and theoretical screening energies would become astrophysically irrelevant if it were possible
to measure reaction rates directly under solar plasma conditions. This may prove possible using high-intensity lasers.
First measurements of light-ion reactions in plasma have been performed, though under plasma conditions for which screening effects are expected to be negligible 
\cite{2013PhRvL.111h2502B,2016PhRvL.117c5002Z,2020PhRvC.101d2802Z,Casey23-FP}.
A direct measurement of bare nuclear cross section, thus avoiding screening effects, is proposed for storage ring experiments, in which a stored beam of ions can collide with a transverse beam \cite{2023EPJA...59...81G}.

\subsection{Screening in the solar core}

The energy-generating reactions that occur in the solar core involve nuclei that are almost completely ionized.
Consequently the atomic environment differs substantially from that of the terrestrial experiments in which these same reactions are measured.
Once $S^\mathrm{b}(E)$ is determined from these laboratory measurements, the screening effects of the plasma must be folded in to yield $S^\mathrm{solar}(e)$.
Adopting the notation of \citetalias{2011RvMP...83..195A}, the correction factor is defined as
\begin{equation}
f_0(E) = \frac{\sigma^\mathrm{solar}(E) }{ \sigma^\mathrm{b}(E)}  
\end{equation}
In the weak screening approximation \cite{1954AuJPh...7..373S}, the ion-ion Coulomb potential is screened
on a length scale given by the Debye radius $R_D$
\begin{equation}
   V(r) = \frac{\alpha Z_1 Z_2  }{ r} \mathrm{exp} \left( -\frac{r  }{ R_D} \right)
\end{equation}
yielding
\begin{equation}
    f_0 \sim \mathrm{exp} \left( \frac{ \alpha Z_1 Z_2  }{ R_D k T} \right)
\end{equation}
We refer the reader to \citetalias{2011RvMP...83..195A} for the functional dependence of $R_D$ on temperature,
density, and the mass fractions $X_i$, and for discussions of corrections for electron degeneracy and
incomplete ionization.

\citetalias{2011RvMP...83..195A} discusses the conditions under which the weak screening approximation is valid:
under solar core conditions this requires $Z_1Z_2 < 10$, a condition satisfied by pp-chain and CNO bi-cycle reactions.
Yet corrections are expected at some level, as the inter-ion potential is not completely negligible, in all collisions,
compared to the relative kinetic energy of the ions \cite{1954AuJPh...7..373S,1969ApJ...155..183S,
1979ApJ...234.1079I,1988ApJ...331..565C,1997ApJ...481..883O,2003PhRvC..67a4603F,2002A&A...383..291B,1997RvMP...69..411B}. 

Dynamic corrections --nonadiabatic effects that arise
when the velocities of reacting nuclei momentarily exceed typical plasma velocities-- have 
been a point of some controversy that both \citetalias{1998RvMP...70.1265A} and \citetalias{2011RvMP...83..195A} addressed.  However, the absence of such dynamic corrections even for
large Gamow energies has been shown to be a consequence of the nearly exact thermodynamic equilibrium of the
solar plasma
\cite{1997RvMP...69..411B, 1998ApJ...496..503G, 1998ApJ...504..996G}.
Specifically, \citet{2002A&A...383..291B} found that corrections to the Salpeter formula under solar conditions would be at the $\sim$ few per cent level, and pointed to specific errors in several papers that had come to contrary conclusions.

There have been a few recent papers advocating changes in solar rates due to dynamical corrections, including 
\citet{2019A&A...623A.126V}, \citet{2018SoPh..293..111W},
\citet{2013ApJ...764..118S}, \citet{2011Ap&SS.336..111M}, 
\citet{2010Ap&SS.328..153M}, and \citet{2009ApJ...701.1204M}. 
The approaches employed are based on modeling the plasma, and in general the authors do not relate their conclusions to
earlier work, particularly \citet{2002A&A...383..291B}, making them difficult to evaluate.
The essential point of \citet{2002A&A...383..291B} is the conceptual simplifications that result from recognizing
that screening in the solar plasma can be formulated as a problem in equilibrium statistical mechanics.
This approach removes all need to classified rates as fast or slow relevant to some
plasma timescale, as there is no time in equilibrium statistical mechanics.  Consequently we continue
to regard \citet{2002A&A...383..291B} as the most realistic estimate of corrections to the Salpeter formula.

\section{Radiative opacities} 
\label{sec:Opacities}

\subsection{Introduction}
\label{subsec:WG9:Intro}

Opacity is a measure of the photon absorption of matter and is an essential quantity for understanding radiative heat transfer in the Sun. 
Radiative heat transfer occurs through the absorption and emission processes taking place within the material that the radiation traverses.  
In local thermodynamic equilibrium (LTE)\footnote{Local thermodynamic equilibrium can be defined as the situation where the thermodynamic properties of a microscopic volume of matter are the same as their thermodynamic equilibrium values corresponding to the local electron temperature and density. 
The assumption of LTE is commonly used in stellar modeling, and the discussion of opacities in this section will be limited only to LTE conditions. 
In some circumstances (such as solar coronal modeling), non-LTE conditions should be given careful consideration, but this will not be discussed in this paper.}, the emission and absorption are straightforwardly related through the Planck function. 
In this scenario, determining the absorption coefficients or opacity, denoted as $\kappa_{\nu}$, provides a complete description of how radiation is transported through the material. 

For example, in the equilibrium diffusion limit, the radiation heat flux $F_R$ is directly related to the Rosseland-mean opacity (RMO) $\kappa_R$ through the following equation \cite{huebner_book} 
\begin{equation}
F_R = - \frac{16\pi}{3}\frac{\sigma T^3 }{\kappa_R \,  \rho} \nabla T \ ,
\label{eq:heat_flux}
\end{equation}
where $T$ represents temperature, $\nabla T$ denotes its gradient, $\rho$ is the mass density, and $\sigma$ is the Stefan-Boltzmann constant. 
The quantity $\kappa_R$ is the solar Rosseland mean opacity (RMO). 
Thus, any error in this quantity introduces error in the radiative heat transport and the simulated solar evolution. 
Quantifying the error in $\kappa_R$ is very challenging because $\kappa_R$ is a weighted mean of a complex solar mixture opacity: 
\begin{equation}
\frac{1}{\kappa_R} = \int^{\infty}_0 \frac{1}{\kappa_{\nu}} w_{\nu} d\nu \ ,
\label{eq:rosseland_mean}
\end{equation}
where $w_{\nu}$ is a weighting function related to the temperature derivative of the Planck function $B_{\nu}$:
\begin{equation}
B_{\nu} = \frac{2h\nu^3}{c^2} \left( e^{h\nu/kT} \right)^{-1} \ ,
\label{eq:planck_function}
\end{equation}
where $\nu$ denotes the photon frequency, $h$ the Planck constant, $k$ the Boltzmann constant, and $c$ the speed of light.
Moreover, the frequency-resolved solar opacity $\kappa_{\nu}$ is a sum of elemental opacities $\kappa_{\nu, i}$ weighted by their mass fractions ($b_i$): 
\begin{equation}
\kappa_{\nu} = \sum_i{b_i \kappa_{\nu, i}} \label{eq:solar_opacity}
\end{equation}

Equation \ref{eq:rosseland_mean} reveals that the accuracy of $\kappa_{\nu}$ is especially important over the spectral range where $w_{\nu}$ is high and $\kappa_{\nu}$ is low (due to the inverse weighting). Equation \ref{eq:solar_opacity} shows that accuracy in $\kappa_{\nu}$ requires good knowledge of both element opacities $\kappa_{\nu, i}$ and mass fractions $b_i$.

\begin{figure}
\includegraphics[width=1.0\columnwidth]{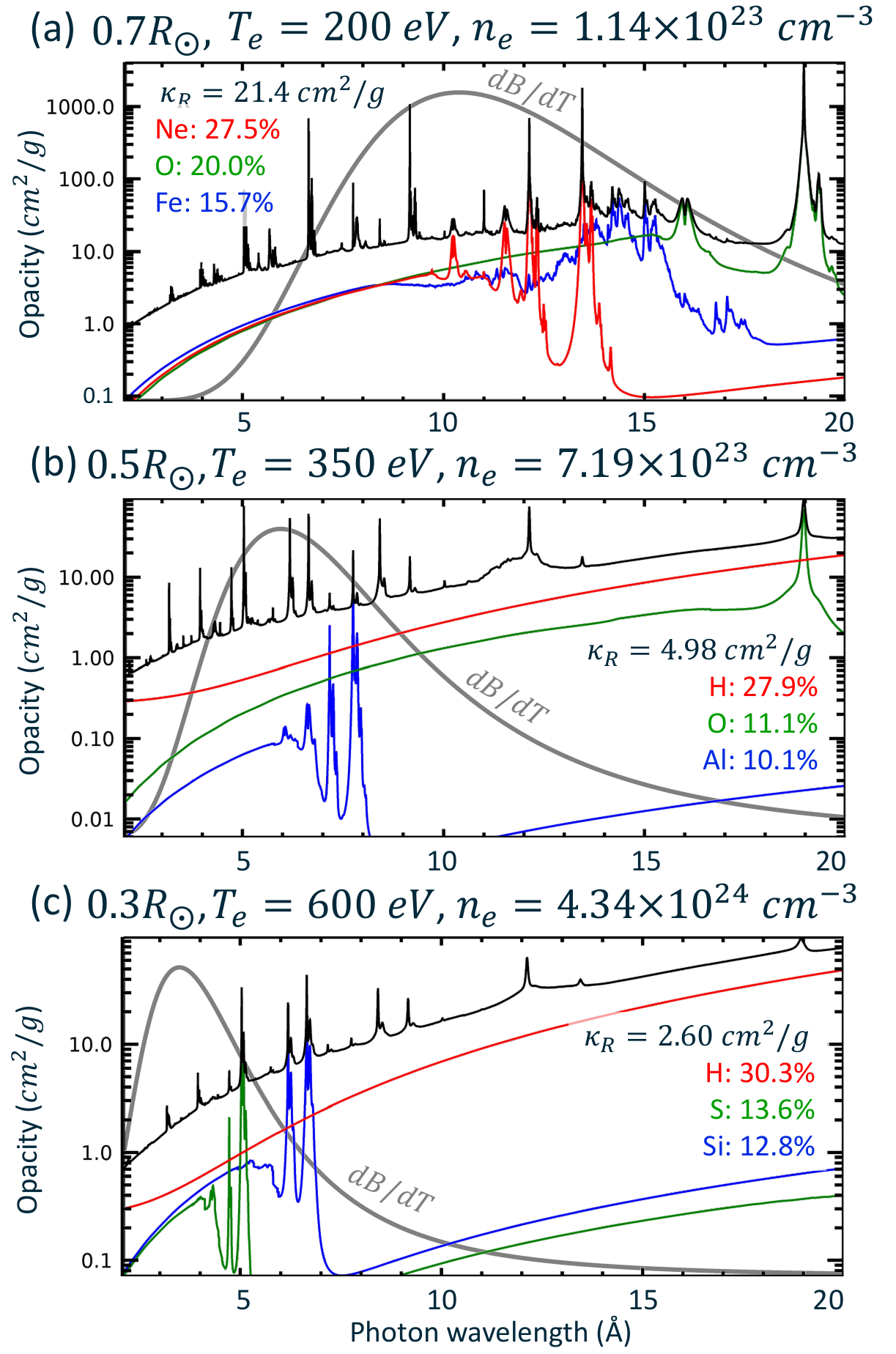}
\caption{Monochromatic opacity of the Sun at three different solar radii as indicated. The temperature and electron density at each solar radius are displayed, as well as some of the major elemental opacity contributions at these conditions. Note that not all contributions are plotted for clarity. The percentages of three elements displayed indicate the leading elemental contributors to the Rosseland mean opacity (RMO) at these conditions. These are roughly quantified by calculating how much the RMO is reduced if each element is completely removed. This figure was made using the GS98 solar abundances and OPLIB opacities, and is mainly intended to show the spectral complexity of the solar opacity.}
\label{fig:WG9-solar-opacity}
\end{figure}

Figure~\ref{fig:WG9-solar-opacity} shows the total monochromatic opacity of the Sun at three different solar radii, highlighting the variations in $w_{\nu}$ (gray), $\kappa_{\nu}$ (black), and some of the leading individual element contributions $\kappa_{\nu,i}$ (red, green, blue) to the RMO determination (Eq. \ref{eq:rosseland_mean}), as a function of radius. 
This figure was made assuming the GS98 solar abundances and using Opacity Library (OPLIB) opacities \cite{Colgan_2016}, and is included primarily for illustrative purposes. 
As shown in the figure, the solar opacity varies considerably in both magnitude (note the log axes) and shape across different solar radii. Moreover, each element contributes distinctively to the overall monochromatic opacity, with either a dominant continuous opacity spectrum, or strong lines that arise from bound-bound transitions. 
If there is an error in the calculated solar opacity $\kappa_{\nu}$, it would stem from inaccuracies in the calculated element opacities and/or abundances. Additionally, the error is expected to be a complex function of photon energy and would vary at each radius due to changes in temperature and density. As a result, it is not possible to provide a single factor for uncertainty or correction. It is crucial to assess the accuracy of the intricate spectral features depicted in Fig.~\ref{fig:WG9-solar-opacity} in order to better understand the nature of the errors. 

These complex spectral features arise from microscopic physics at the atomic level – that is, the absorption of radiation through quantum mechanical processes. If one knows the atomic energy levels and transition probabilities of all the relevant atomic states (of each atom or atomic ion of each material in the Sun), then the opacity of that material may be computed.\footnote{We note that the opacity of molecules and dust, or grains, can be very important in some circumstances \cite{huebner_book}; however they play a minor role in stellar modeling and so will not be discussed here.} 
Since atomic structure and transition probabilities can be predicted (under various approximations), starting from the Schr\"odinger (or Dirac) equation, this seems like an achievable goal. However, complications quickly arise from a multitude of factors, such as: how to deal with the infinity of atomic states that may be populated in each atom or atomic ion; how to account for the perturbations caused by the plasma environment on the atomic states; and how to handle the large quantities of data that arise with such calculations, particularly for multi-electron atoms and ions. In solar modeling, one requires the opacity of all relevant elements (around 30) at all (plasma) conditions that are found at all radii throughout the solar evolution.
Thus, the opacity must be tabulated as a function of species, plasma temperature, plasma density, and photon energy. 

Nevertheless, steady progress has been made in determining the opacity of solar mixtures over the last five decades. 
In the 1980s, insufficient accuracy in stellar-envelope opacity was hypothesized through multiple astrophysical puzzles \cite{Simon_1982}, and opacity models were subsequently improved by incorporating more detailed atomic physics and a more complete set of states \cite{Iglesias_1996, Seaton_1987,Berrington_1987, 1995ASPC...78...51M}. The improved accuracy was confirmed through a few benchmark experiments \cite{PhysRevLett.67.3784, PhysRevLett.69.3735, PhysRevE.53.R1332, Foster_1988}
and improved agreement with some astrophysical observations 
\cite{1991ApJ...371L..73I,10.1111/j.1365-2966.2009.15229.x,10.1111/j.1365-2966.2009.16141.x}. However, the calculated solar opacities are not experimentally validated at most radii, and in particular more deeply into the Sun where the conditions (temperature and density) are more extreme. 

In the last decade, experimental methods have steadily improved, in concert with the availability of powerful x-ray sources, such as the Sandia Z facility. Monochromatic opacities were successfully measured at multiple conditions relevant to solar convection-base conditions \cite{PhysRevLett.99.265002,Bailey_Nature_2015}.
Systematic experimental studies performed on chromium, iron, and nickel helped narrow down hypotheses for the discrepancies \cite{PhysRevLett.122.235001}. 
All of these experimental results  provided essential clues for testing various approximations for solar interior opacity calculations; however, the reported model-data disagreements also raised significant controversy. This led to various theoretical investigations and the development of independent experimental methods \cite{PERRY2020100728,NIF_opacity}. 
Ongoing theoretical and experimental investigations are primarily focused on understanding the origins of reported
disagreements, as a first step toward reaching consensus.  Our aim here is to provide a status report for the field, summarizing past work on solar opacity calculations
and experiments, while identifying some of the open questions that are the focus of current work.

\subsection{Opacity models} 

\label{subsec:WG9:Models}

Over the years, a wide range of approximations has been used for the various components of opacity calculations. To help navigate these approximations, it may be useful to briefly summarize the key components of a frequency-dependent opacity.
Opacities are often broken down into various components. The first of these is ``bound-bound'' absorption (a photoexcitation transition between two levels of an atomic ion), which manifests as strong line features in the opacity spectrum (such as the black `spikes' in Fig.~\ref{fig:WG9-solar-opacity}). The ``bound-free'' absorption (photoionization of the ground or excited level of an atomic ion) is normally a continuous (wavelength-dependent) contribution (such as the smooth (red) hydrogen opacity contribution in the middle and lower panels of Fig.~\ref{fig:WG9-solar-opacity}). The ``free-free'' absorption (also known as inverse Bremsstrahlung) is where a free electron absorbs a photon, and finally ``scattering'' (the scattering of a photon by a free electron) contributions can be important at high photon energies. Other contributions may be important in molecules or other material. 
Equation~\ref{eq:rosseland_mean} shows that due to the harmonic mean inherent in the RMO, the most important photon energy range is when the spectral opacity is low and when the weighting function $w_{\nu}$ is high. In the solar case, at a given solar radius, it is important to know which elements dominate the opacity at such a region, which will of course depend on the elemental abundance and its contribution at that distance, as illustrated in Fig.~\ref{fig:WG9-solar-opacity}. 

The determination of opacities usually starts from one of two fundamental approaches. The first approach, which was used in the first  opacity computations \cite{huebner_book}, starts from a ``mean ion'' model, where a (fictitious) ``average'' atom or ion of a given element is conceived, with fractional occupation numbers, which is consistent with the given temperature and density of the species under consideration. The fractional occupation numbers, usually determined through a self-consistent iterative approach, inform the real (physical) ion stages of importance at such conditions. The average atom can then be unfolded through various procedures to the physical ion stages of relevance, and the opacity of such ions can be determined. In this model the plasma effects (in particular the perturbation of the average atom by the fields produced by the surrounding plasma ions and electrons) can be included in a natural way, and this approach is often most useful at high densities, where plasma perturbations on the atoms or ions are most important. The infinity of atomic states are naturally truncated by plasma effects that remove bound states from consideration. The old Inferno model \cite{PhysRevB.20.4981} and the more recent STAR tables \cite{starsolaropacityapj,atoms6030035} are based on this method. We note that very recent work \cite{Gill_2023} has considerably improved some of the approximations used in related approaches.

The second approach is often referred to as detailed configuration (or term) accounting, where one determines all possible ion stages of the species and then determines all possible ground and (multiply) excited states of each ion, each with a corresponding population that is condition dependent. This method is often more spectroscopically accurate than the mean ion model, because it considers the structure of each ion separately, but can require enormous computational resources for complex species. The structure of each ion is determined from solution of the Schr\"odinger (or Dirac) equation, which can then also produce transition probabilities. Within this procedure, many computational difficulties must be overcome to produce acceptably accurate data for all ion stages of all relevant elements. A few examples are given for illustration. Within the solution of the Schr\"odinger equation it is desirable to include ``configuration-interaction'' (CI), a two-body term in the Hamiltonian that accounts for electron-electron interactions \cite{cowan1981theory}, which results in a quite accurate description of atomic energy levels. However, the inclusion of this term for atomic ions with large numbers of configurations can quickly result in a computationally intractable problem. Therefore, approximations in which CI is included in only a limited manner are often employed. Furthermore, careful consideration has to be given to the number of states that should be included in the calculations. A compromise is always necessary between including as many states as possible versus the computational resources available for the calculation. Convergence checks are necessary to ensure that sufficient states are included. However, the rate of convergence of the opacity with respect to the number of states included will depend on the plasma conditions of interest, so this in itself is not necessarily a straightforward matter. The computer codes that are used to generate the opacity must also be able to efficiently process large amounts of atomic data.

Similarly, when considering photoexcitation and photoionization of atomic ions, it is desirable to include coupling of the bound electrons to the continuum electron(s). Often termed ``close-coupling'', this may be accomplished through techniques such as the R-matrix approach \cite{burkeRmatrix,Seaton_1987} or other related methods. However, such approaches are also computationally intensive, and can be severely limited in the number of terms that can be included in the resulting close-coupling expansions. It is also not clear how to include plasma perturbations within such approaches. As a result, perturbative approaches, such as distorted-wave methods \cite{cowan1981theory, 2005MNRAS.360..458B, SAMPSON2009111}, are often employed, which are much more computationally tractable, and of sufficient accuracy for mid- and highly-ionized ions.

In concert with an atomic model, the thermodynamics of the system must be considered. That is, an equation-of-state (EOS) description of the material is required so that the thermodynamic properties and atomic state populations can be determined for a given material temperature and density. In the mean ion approach the thermodynamic quantities arise naturally through the description of the atom/ion in a plasma. In a Detailed Configuration Accounting approach, a common approach to EOS calculations starts from a chemical picture based on a minimization of the chemical free energy \cite{rogers1992radiative, hummer1988equation, dappen1987statistical, hakel2004chemeos}. Such a model has to treat a wide range of physical conditions, ranging from ideal gas through high density conditions where pressure ionization may rapidly strip ions of all their electrons. The number of atomic states retained in the partition function is a key quantity in determining if the EOS is complete at the conditions of interest. This assessment of completeness will of course strongly depend on the plasma conditions.

The final important aspect of opacity models that we wish to briefly discuss concerns line broadening. Photoabsorption lines in a plasma become broadened due to a number of factors, such as natural and Doppler broadening. Collisional broadening (often known as Stark broadening) is often a dominant broadening mechanism. This is broadening due to the plasma microfields caused by the motion of electrons and ions in a plasma. As before, line broadening is an important part of any opacity model, but any treatment of this effect has to be tractable for all ions and all conditions encountered when building an opacity table. While quite accurate line broadening models exist for one- and two-electron systems \cite{lee1988plasma,stambulchik:2019a,Gomez_2022}, the treatment of broadening for multi-electron systems often requires a  number of approximations \cite{dimitrijevic1986simple,dimitrijevic1987simple}.

\subsection{Production of tables} \label{subsec:WG9:Tables}

Some of the earliest determinations of opacity for stellar physics started by the pioneering work of Seaton and collaborators in the UK \cite{Seaton_1987,Berrington_1987}. This development led to the ``Opacity Project” (OP), which in time produced a set of opacity tables for stellar modeling that are still in common use today 
\cite{2005MNRAS.360..458B}
At around the same time, OPAL (OPacity Astrophysical Library) opacity tables were published by Lawrence Livermore National Laboratory (LLNL) \cite{Iglesias_1996}. Both of these sets of tables were widely adopted by the stellar modeling community. Some early theoretical opacity work was performed at Los Alamos National Laboratory (LANL) \cite{huebner_book}, starting from pioneering work by Mayer and later by Arthur Cox in the 1960s. In the 1990s, the LANL work resulted in “OPLIB” (OPacity LIBrary) tables produced from the LEDCOP code \cite{1995ASPC...78...51M}; in the mid 2010s the modern ATOMIC code was used to produce a new generation of opacity tables that improved the computation of opacities in a number of ways \cite{Colgan_2016}. All of these tables are available electronically in various formats. Other important opacity tables have been produced in France (the OPAS and SCO-RCG opacity efforts) \cite{Blancard_2012,Mondet_2015,pain2021super}, Israel (using the STAR code) \cite{starsolaropacityapj,atoms6030035}, and elsewhere. Differences amonmg these tables and their underlying opacity models have been the focus of dedicated workshops and conferences, for example \citet{WAO_2017}.

Many of these tables (primarily OP and OPAL) were quickly adopted by the solar modeling community and were useful in solving the Cepheid variability puzzle \cite{Simon_1982}, where updated opacities helped resolve a previous disagreement between observation and modeling of Cepheid variables. Detailed comparisons between these two sets of tables have been made on numerous occasions, both on comparisons of individual opacities \cite{2005MNRAS.360..458B} and on how solar models that use these tables compare \cite{2006ApJ...649..529D}. 
Overall, the agreement between the various opacity tables was deemed to be ``satisfactory'' (within 5 \%) for the accepted solar models of the late 1990s \cite{1995ASPC...78...51M}.  This paradigm was shaken when new determinations of the elemental solar abundances were reported \cite{2005ASPC..336...25A}, which implied that solar models disagreed with helioseismology observations. It was suggested \cite{Serenelli_2009} that changes in the opacity of a few species (mainly Fe, Ne, O) could reconcile the solar models with the observations. Since then, intensive work has been performed in exploring some of the approximations made in the older opacity tables. The OP effort has focused on new and larger R-matrix calculations for important ions of Fe \cite{PhysRevLett.116.235003,10.1093/mnras/stab2016}, primarily Fe$^{17+}$. At Los Alamos, the modern ATOMIC code was used to create a new generation of opacity tables for H through Zn \cite{Colgan_2016}. Although significant improvements in many aspects of the calculations were incorporated, the resulting opacity tables did not significantly improve the comparison of solar modeling with helioseismology. The study of \citet{starlinewidthpaper} showed that one possible way to produce larger opacities under solar conditions would be if line broadening effects were significantly increased -- by several orders of magnitude.  Opacity increases of $\approx$ 10\% near the solar convective zone and $\approx$ 2\% in the solar core have been attributed to ionic correlations \cite{starionioncorrpaper}.  There is need for further investigation of such topics.

\subsection{Experimental testing of calculated iron opacity} \label{subsec:WG9:Experimental}
Errors in calculated solar opacity can arise from inaccuracies in both the abundance and calculated element opacities in a complex way. Therefore, it is important to investigate the accuracy of abundance and elemental opacities separately. The accuracy of elemental opacities can be best evaluated by examining the frequency-resolved (monochromatic) opacity. Since each spectral feature of the element opacity often relies on different physics and approximations, disagreements between model predictions and experimental data on a frequency-resolved basis can help identify which approximations are invalid and guide refinements in opacity theory. Here, we provide an overview of the fundamentals 
as well as previous investigations into measurements of solar-interior opacity.
A more general and tutorial discussion of the experimental methods and a historical overview can be found elsewhere \cite{bailey_experimental_2009}.

The material opacity $\kappa_{\nu}$ is related to its transmission $T_{\nu}$ by the following equation: 
\begin{equation}
T_{\nu} = \frac{I_{\nu}-\epsilon_{\nu}}{B_{\nu}-\epsilon_{\nu}} = e^{-\kappa_{\nu} \rho L}
\end{equation}
where $B_{\nu}$ is the backlight spectrum, $I_{\nu}$ is the backlight transmitted through the sample, $\epsilon_{\nu}$ is a sum of the sample plasma emission and other backgrounds, and $\rho L$  is the areal density of the opacity sample. 
Thus, sample opacity $\kappa_{\nu}$ can be experimentally determined by heating the sample and accurately measuring the following quantities: the backlight with and without the heated sample ($B_{\nu}$, $I_{\nu}$), the plasma self-emission and background ($\epsilon_{\nu}$), and the sample areal density ($\rho L$). 

Several challenging criteria must be met if opacity measurements are to be both interpretable and useful [Perry 1996, Bailey 2009]. First, the elements, conditions, and spectral ranges of interest must be identified. Second, a macroscopic opacity sample must be uniformly heated to the desired conditions to achieve local thermodynamic equilibrium. Third, the heated sample must be backlit with a bright and spectrally smooth radiation, $B_{\nu}$, to accurately determine frequency-resolved absorption. It is critical that the backlight is significantly brighter than the sample plasma emission or other background signals, $\epsilon_{\nu}$. Fourth, the absorption spectrum must be recorded with spectrometers that provide sufficient signals and spectral resolving power for the opacity study. Fifth, the condition of the heated sample ($T_e$, $n_e$, $\rho L$) must be diagnosed independently of the opacity in question. Simulated conditions are not appropriate for this purpose since their accuracy depends heavily on the accuracy of the opacity in question. Finally, the opacity spectrum and its uncertainty must be accurately determined from the measured $I_{\nu}$, $B_{\nu}$, $\epsilon_{\nu}$, and $\rho_{\nu}$ and their uncertainties.

After the revision of solar abundance in 2005 \cite{2005ASPC..336...25A}, solar models and helioseismology disagreed, and the accuracy of solar-interior opacity was called into question. This disagreement was primarily due to a significant reduction in solar opacity resulting from the reduced metallicity, and it was the greatest at the base of the solar convection zone (hereafter convection-zone base or CZB). Since then, solar abundance has been continuously revised, with the latest abundances determined by two groups being Z/X=0.0225 \cite{2021A&A...653A.141A} and Z/X=0.0187 \cite{2022A&A...661A.140M}, which still disagree with each other, leaving significant uncertainty in the solar abundance. For determining the accuracy of calculated solar opacity, the accuracy of calculated element opacities must also be experimentally scrutinized. Since Fe and O are the two dominant sources of opacity at the CZB, their opacities must be experimentally tested at CZB conditions (i.e., $T_{e}=182\  \text{eV}$, $n_{e}=9\times 10^{22}\ \text{e cm}^{-3}$). Unfortunately, previous experimental approaches were not suitable for this purpose because their backlighters were not bright enough to mitigate the bright self-emission produced at these conditions.

\begin{figure}
    \includegraphics[width=\columnwidth]{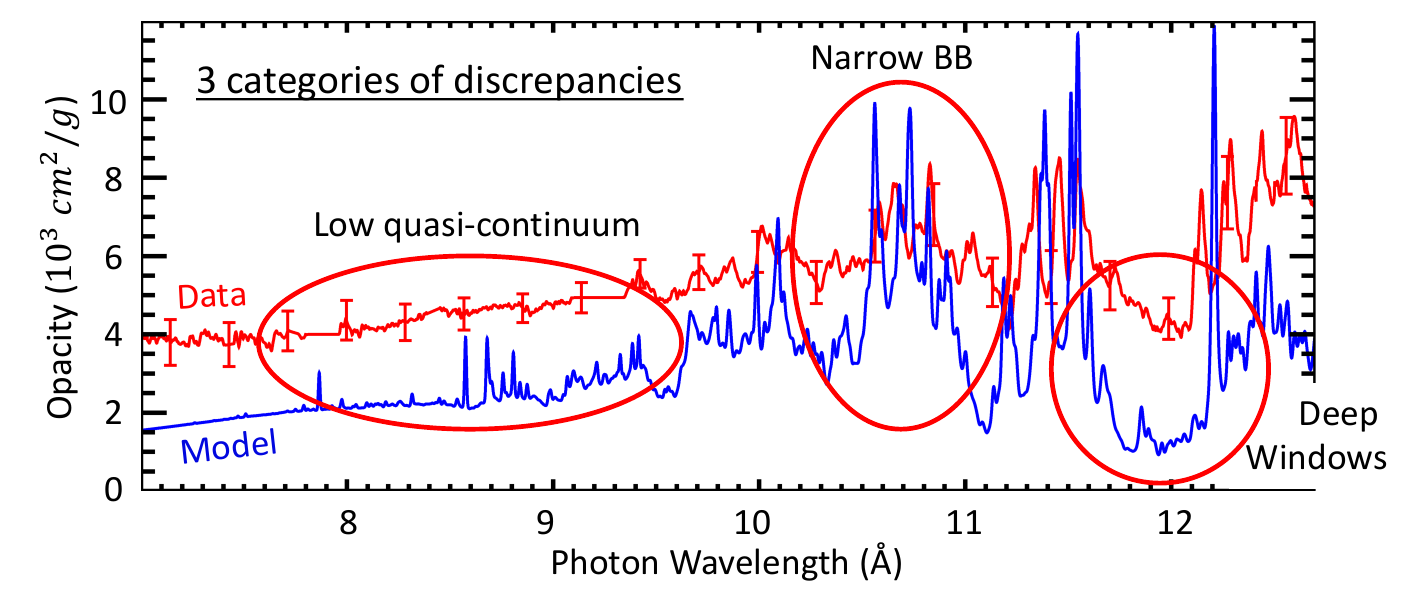}
    \caption{
    Comparison of measured opacity (red online) and calculated opacity (blue online, \citealp{Badnell_2003}),
    at a temperature of 182 eV and an electron density of $3.1\times 10^{22} \ \text{e cm}^{-3}$. 
    Three types of measured-vs-modeled opacity disagreements are shown:
    (i) a lower quasi-continuum, (ii) narrower bound-bound line features, and (iii) deeper opacity windows. These trends were observed in all opacity models compared in
    \citet{Bailey_Nature_2015}
    }
    \label{fig:WG9-fe-opacity}
\end{figure}

Over the past two decades, experimental methods have been refined using one of the brightest x-ray sources, Z-pinch. In 2007, iron opacities were successfully measured at 150~eV and $7\times10^{21}\ \text{e cm}^{-3}$, providing support for the accuracy of calculated Fe opacities under these specific conditions. In 2015, iron opacities were measured at larger temperatures and densities. This study revealed a significant discrepancy between the calculated and measured opacities as the temperature and density approached the CZB conditions, as shown in Fig.~\ref{fig:WG9-fe-opacity}. This is a significant concern since the experimental density was still approximately three times lower than the actual density at the CZB. Opacity models employ approximations to account for density effects such as line broadening and pressure ionization. If these approximations are incorrect, the disagreement between calculated and measured opacities could be even more pronounced at the densities found at the CZB. 

The disagreement observed in the iron opacity was complex, as depicted in Fig.~\ref{fig:WG9-fe-opacity}. The modeled opacities were lower in the quasi-continuum region at short wavelengths ($<10$ \AA), narrower in bound-bound lines, and deeper in the opacity valleys known as windows. These discrepancies are likely caused by various factors, but it is difficult to disentangle different sources of discrepancies based solely on the disagreement in iron opacity alone. 

In 2019, a comprehensive study was conducted to measure opacities for Cr, Fe, and Ni at solar interior temperatures \cite{PhysRevLett.122.235001}. This systematic study helped narrow down hypotheses on the sources of disagreement by observing how the discrepancies change as a function of atomic number. The measurements revealed that measured opacities for all three elements had narrower line features than the model predictions, suggesting potentially inaccurate density effects in opacity models. Disagreement in the opacity windows was observed for Cr and Fe, while Ni did not show significant disagreement. At the experimental conditions used in the study, Cr and Fe were in an open L-shell configuration, whereas Ni was closer to a closed L-shell configuration. It is known that the calculations of both population and equation of state (EOS) are more challenging in open L-shell configurations. Therefore, the observed trend in window disagreement may suggest a connection to the difficulties associated with EOS calculations in such configurations. 

One intriguing finding from this systematic study was the disagreement observed in the quasi-continuum region. Unlike the discrepancies observed in bound-bound line-width and window opacities, no clear trend was observed in the quasi-continuum disagreement. The model and data agree well for Cr and Ni, but not for Fe, at high temperatures. Two possible explanations have been proposed for this unexpected behavior. The first hypothesis suggests the presence of missing physics in the opacity models that becomes significant under the conditions encountered by high-temperature Fe-opacity experiments. The second hypothesis proposes that the opacity experiments or analyses are flawed, but only for Fe at high temperatures. Until opacity experiments and theories are reconciled, neither possibility can be definitively ruled out.

Despite numerous investigations, the discrepancies between experimental and theoretical iron opacities remain unresolved \cite{Bailey_Nature_2015,10.1063/1.4889776,10.1063/1.4872324, NAGAYAMA201617,PhysRevE.93.023202,PhysRevE.95.063206,Colgan_2016, PhysRevLett.116.235003, PhysRevLett.122.235001,PhysRevLett.117.249501, KILCREASE201536,Mancini_2016}. On the experimental side, measurements of sample spatial gradients have been conducted and found to be negligible \cite{10.1063/1.4872324}. Numerical tests have also shown that temporal gradients, self-emission, and the presence of tamping material do not have significant impact on the results 
\cite{10.1063/1.4889776,PhysRevE.93.023202,PhysRevE.95.063206}.  Furthermore, the uncertainties in temperature and density diagnostics resulting from the choice of spectral model have been determined, and it has been found that these uncertainties are too small to account for the reported discrepancies \cite{NAGAYAMA201617}. Additionally, background measurements have also been performed and found to be negligible \cite{10.1063/5.0057225}.

The persistent differences between experiment \cite{Bailey_Nature_2015, PhysRevLett.122.235001} and opacity models has led to a sustained effort to reconcile these discrepancies. The experimental efforts have been previously discussed. On the theoretical side, tests of some of the approximations used in the opacity tables have been conducted (\citealp{Colgan_2016}, \citealp{PhysRevLett.116.235003}, \citealp{PhysRevLett.117.249501}, \citealp{10.1093/mnras/stab2016}), particularly in assessing the convergence of opacity models with respect to the number of configurations included in the calculations. However, no significant change in the final opacity has been reported from such studies. 

Several groups have postulated new physical effects that could produce more opacity contributions than in previous works. Of particular note is the possibility of two-photon opacity contributions, where an atom or ion could simultaneously absorb two photons (possibly of different frequency) \cite{MORE201744,MORE2020100717}. All current opacity tables omit such contributions, as they are believed to be a minor contribution to the overall opacity. Preliminary studies by More and co-workers \cite{MORE201744,MORE2020100717} have suggested this could be a factor in the comparison of theory with the Z measurements, but other studies \cite{IGLESIAS20154, KRUSE201938, KRUSE2021100976, PAIN201823} find that such a contribution should be small. 

Another intriguing study has reported that transient spatial localization may result in increased opacities \cite{liu2018transient,Zeng2022}. This is a density-dependent effect where the plasma effects localize the continuum wavefunctions near the absorbing ion, potentially resulting in increased broadening and bound-free cross sections. While the work of \cite{Zeng2022} suggests that this effect is significant, other studies cast doubt on this claim \cite{IGLESIAS2023101043}.

Opacity theorists and experimentalists continue to collaborate closely to address discrepancies and uncertainties in the calculated solar RMO. They carefully examine each other's work, seeking to resolve the discrepancies and accurately quantify the uncertainty associated with calculated solar opacity.

\subsection{Future opacity work} \label{subsec:WG9:Future}

The continuing disagreement between the Z Fe experiments \cite{Bailey_Nature_2015, PhysRevLett.122.235001} and theoretical models is by far the biggest concern of the opacity community. New theoretical efforts, while intriguing, have not yet resolved these differences.  New experimental results, particularly the opacity-on-NIF (National Ignition Facility) campaign \cite{PERRY2020100728,NIF_opacity}, are eagerly anticipated to shed some light on the current impasse.
New approaches are also being developed for opacity measurements using the APOLLON laser at LULI (France) and on the GEKKO XII laser (Japan) \cite{PhysRevLett.95.235004}. There has been a suggestion for measuring heat conductivity under thermodynamic conditions resembling those found deeper in the radiative zone, using hydro-carbon foams doped to imitate the solar composition \cite{starionioncorrpaper}.
Such progress will have implications for solar modeling, as the combination of opacity and elemental abundances is key to resolving the significant discrepancy between helioseismology and solar models. As previously noted, the abundance of solar elements is also the subject of renewed scrutiny \cite{2021A&A...653A.141A,2022A&A...661A.140M}.



\section{Experimental facilities for solar fusion studies} 
\label{sec:Facilities}

Improved experimental facilities have led to significant progress in nuclear astrophysics research in recent years, and future instrumental developments will likely continue to push boundaries in the next decade.
Key goals driving technological developments include, among others, tests of the weak interactions and of solar properties that make use of high precision solar neutrino measurements, helioseismological mappings of the core metallicity, and detailed solar modeling. 
There is also a host of related open questions in different areas, such as Big Bang nucleosynthesis, red giant evolution, the evolution of supernova progenitors, and a variety of transient explosive phenomena in astrophysics 
where a quantitative understanding of the nuclear physics is essential.
Due to the small cross sections of low-energy charged-particle reactions, experiments must be designed to achieve signal rates significantly lower than the background rates from cosmic rays, natural radioactivity in the laboratory, and induced activity from beam interactions with target impurities. In the following, we outline current experiences and future facilities—either in development or planned—that will pave the way for possible major breakthroughs in the coming years.

\subsection{Above-ground facilities} 
Above-ground facilities always play a crucial role in investigating nuclear reactions of astrophysical interest. Various methods are applied to minimize background. A prevalent approach involves employing passive shielding around the detection area, typically utilizing a layered configuration of lead, copper, and polyethylene. This setup helps to decrease unwanted signals and neutron background in detectors with relatively small capacities. However, there are additional techniques that can be utilized to further diminish background interference, enabling measurements at energy levels closer to those pertinent to astrophysics. 

\subsubsection{LENA}
The Laboratory for Experimental Nuclear Astrophysics (LENA) is part of the Triangle Universities Nuclear Laboratory, located on the campus of Duke University, in North Carolina, US. LENA features two accelerators: a 230-keV ECR accelerator that produces the world's most intense low-energy proton beams \cite{10.1063/1.5024938} and a 2-MV Singletron from High Voltage Engineering Europa B.V. \cite{10.1063/5.0150982}. These machines deliver ion beams with intensities ranging up to 20 mA DC proton current at lower energies and 2 mA at several MeV—along with advanced beam pulsing capabilities. Each accelerator has dedicated transport systems and control mechanisms for simultaneous operation. A key component of LENA is the $\gamma$-ray coincidence spectrometer, which allows measurements with sensitivities comparable to underground facilities \cite{PhysRevC.91.015812}. The detector setup includes a 130$\%$ p-type HPGe detector, HPGe clover detectors, and a 16-segment NaI(Tl) annulus, along with the APEX detector featuring 24 position-sensitive NaI(Tl) bars for enhanced measurements. Recently, LENA added a cosmic-ray veto system with nine plastic scintillators to reduce cosmic-ray muon background, conveniently positioned due to its wheeled assembly. 

\subsubsection{NSL}
The Nuclear Science Laboratory (NSL) at the University of Notre
Dame, hosts a number of low energy accelerators and the TriSOL
radioactive beam facility to maintain a rigorous experimental
program in nuclear astrophysics, nuclear structure physics,and
fundamental symmetries as well as in a broad range of nuclear
physics and nuclear chemistry applications. The 5MV single ended
pelletron provides high intensity proton, helium, and A$\le$40
heavy ion beams for nuclear astrophysics related experiments in
forward kinematics using solid and gas target technologies and in
inverse kinematic using the St. George recoil separator. The 10 MV
FN pelletron tandem is the main machine for nuclear structure
physics experiments, but also serves the accelerator mass spectrometry program on
long-lived radioactivities in terrestrial and meteoritic samples.
The FN is also used for indirect measurements of reactions of astrophysical interest and serves as a driver for the production of light radioactive isotopes beams (A$\le$40) for nuclear astrophysics and
fundamental symmetry studies.

\subsection{Underground Facilities}
\label{subsec:WG8:Underground}

Underground facilities are essential to push the boundary of direct measurements towards the lowest energies of astrophysical interest (10s-100s keV).
In particular, capture-reaction measurements leading to neutron- or $\gamma$-ray emissions are severely hampered in surface laboratories because of the overwhelming background associated with cosmic rays. 
This background can be suppressed by many orders of magnitude by exploiting the natural shielding provided by the rock overburden in underground sites. 
The improvements possible with this strategy have been demonstrated by pioneering work at the Laboratory for Underground Nuclear Astrophysics (LUNA) at INFN-LNGS, first with a 50-kV accelerator (LUNA I) and then with a 400-kV one (LUNA II) still in operation today \cite{2022ARNPS..72..177A}.
Since the last Solar Fusion review, a new accelerator has been installed at INFN-LNGS, and other underground laboratories have become operational in China (JUNA),  the US [Compact Accelerator System for Performing Astrophysical Research (CASPAR)], and Germany (Felsenkeller). 

\subsubsection{Bellotti Ion Beam Facility} \label{subsec:WG8:LUNA}
\begin{figure}
    \centering
    \includegraphics[width=0.9\columnwidth]{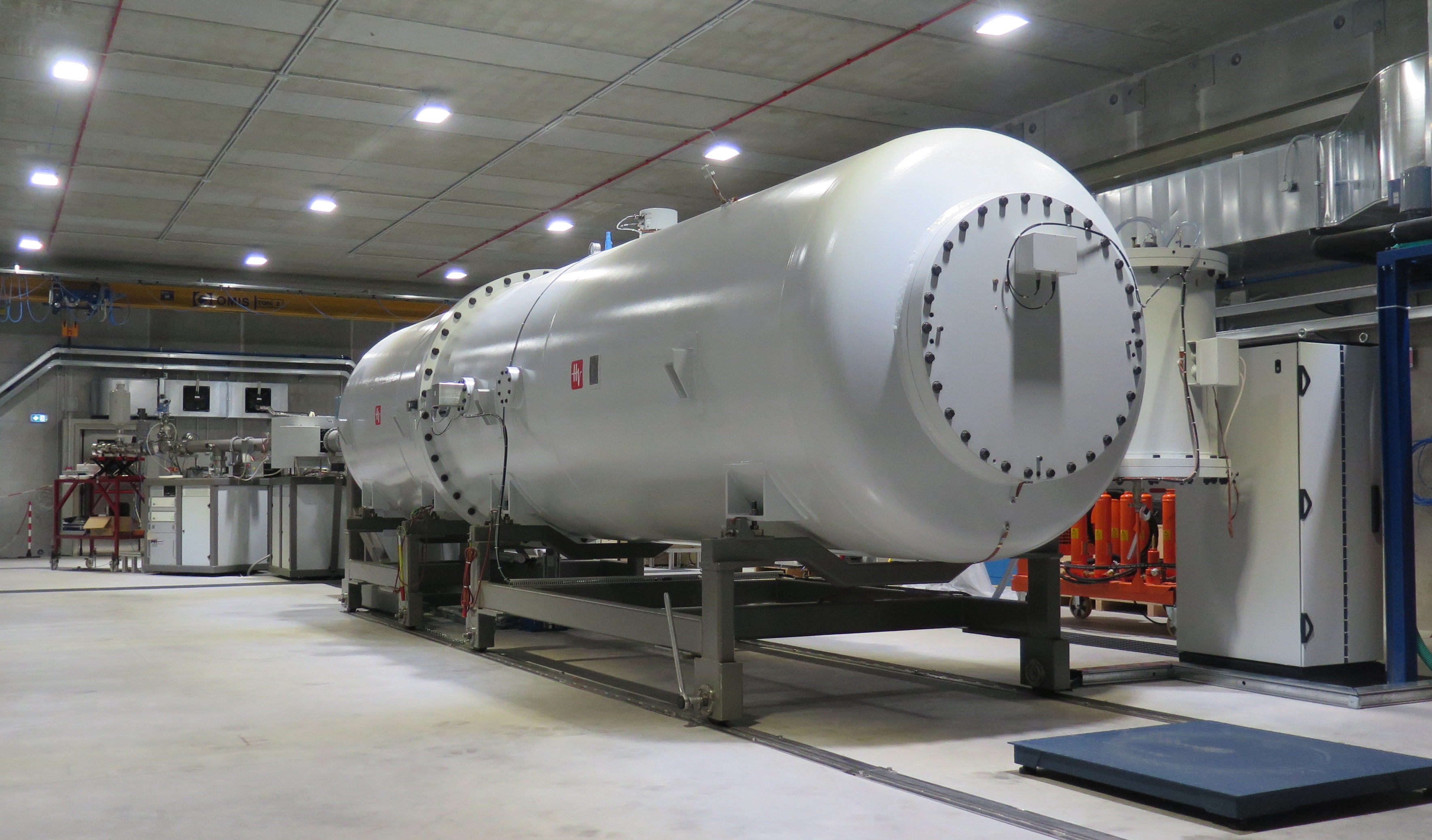}
    \caption{
    Photo of the 3.5~MV Singletron{\textsuperscript{TM}} accelerator recently installed at LNGS (credit: Matthias Junker).}
    \label{fig:LUNA-MV}
\end{figure}

The INFN-LNGS has recently expanded its accelerator capabilities
with the installation of a new 3.5~MV Singletron{\textsuperscript{TM}} machine designed and set up by High Voltage Engineering Europe (HVEE)~\cite{2019NIMPB.450..390S}, see Fig.~\ref{fig:LUNA-MV}. 
The 3.5~MV machine is equipped with two independent beam lines which can be operated with  solid and gas target systems.  
Acceptance tests at HVEE demonstrated that the machine can deliver intense proton, helium and carbon beams (1, 0.5 and 0.15~mA, respectively) with well defined energy resolution (0.01\% of a teravolt) and stability (0.001\%h$^{-1}$ of a teravolt) (Di Leva, 2020 - private comm.).
The 1.4km of rock overburden at LNGS, equivalent to $\approx$ 3000 m.w.e. in a flat site, affords significant background reductions for $\gamma$-ray, neutron-, and charged-particles detection (see \cite{2022ARNPS..72..177A} for a recent review).
A first experimental proposal presented by the LUNA-Collaboration focuses on measurements of the reactions  $^{14}$N(p,$\gamma$)$^{15}$O, $^{12}$C+$^{12}$C, $^{13}$C($\alpha$,n)$^{16}$O and $^{22}$Ne($\alpha$,n)$^{25}$Mg, the latter being in the context of the ERC Starting Grant SHADES. 
The 3.5MV accelerator is now part of the Bellotti Ion Beam Facility \cite{junker2023deep} and open to external users.\footnote{
\url{https://www.lngs.infn.it/en/pagine/bellotti-facility-en}
}

\subsubsection{JUNA}\label{subsec:WG8:JUNA}

JUNA is part of the China Jinping Underground Laboratory, established on the site of hydropower plants in the Jinping mountain, Sichuan, China \cite{2022FBS....63...43L}.
The laboratory, located near the middle of a traffic tunnel, is shielded by 2400~m of mainly marble overburden (6720 m.w.e.).
In December 2020, the JUNA collaboration installed a 400 kV accelerator capable of delivering H$^+$ and He$^+$ beams with intensities of up to 10 particle mA, and a He$^{2+}$ beam with an intensity of up to 1 particle mA. Some important results have already been published. For example, the lowest energy study to date of the $^{19}\text{F}(p,\gamma)^{20}\text{Na}$ reaction advocates a breakout from the CNO cycle as a possible route for calcium production in population III stars \cite{2022Natur.610..656Z}.

\subsubsection{CASPAR}\label{subsec:WG8:CASPAR}

The Compact Accelerator System for Performing Astrophysical Research (CASPAR) laboratory is the only US-based deep underground accelerator  and is operated by a collaboration of the University of Notre Dame and the South Dakota School of Mines and Technology \cite{2016APS..DNP.JJ003R}. 
The accelerator system has been fully operational since 2018 and is located 
1480 m
below surface at the Sanford Underground Research Facility (SURF)\footnote{See \url{http://www.sanfordlab.org}} in Lead, South Dakota, formerly the Homestake gold mine. 
The rock overburden results in a 4300 m.w.e shielding effect, significantly decreasing cosmic-ray induced background with a muon flux level of $4 \times 10^{-9}$ cm$^{-2}$ s$^{-1}$. 
The residual neutron flux consists of primarily low-energy ($<$10 MeV) neutrons generated by ($\alpha,n$) reactions induced by the decay of naturally occurring uranium and thorium 
radionuclides in the surrounding rock, and is generally of the order of 10$^{-6}$ neutrons cm$^{-2}$ s$^{-1}$ \cite{2006PhRvD..73e3004M,2009NIMPA.606..651M}.   
The CASPAR accelerator is a 1~MV Van de Graff style JN accelerator with a 150 to 1100~kV operational range, well suited for overlap with higher energy measurements. 
The accelerator provides proton and $\alpha$ beams with up to $\sim 250~\mu$A on target.
The scientific program continues to explore stellar neutron sources and expands the present studies into the magnesium range probing $\alpha$-capture reactions on  $^{24}$Mg, $^{25}$Mg, and $^{26}$Mg isotopes. 
A new program has been initiated to explore the endpoint of nova nucleosynthesis, studying proton capture reactions in the Ar to Fe range. 
CASPAR is well suited for these measurements, but will be complemented by a new low energy machine, presently under development at Notre Dame. 

\subsubsection{Felsenkeller} \label{sec:Felsenkeller} 
\label{subsec:WG8:Felsenkeller}

The Felsenkeller laboratory is hosted within a network of tunnels, excavated into hornblende monzonite rock,  previously used as a cool storage place for the homonymous nearby brewery.
Jointly funded and built by TU Dresden and HZDR, the laboratory operates a 5~MV Pelletron-type accelerator built by NEC. 
With the addition of a RF ion source in the high-voltage terminal, the accelerator can be operated in tandem mode or as a single-ended machine and is capable of delivering proton, $\alpha$-particle, and carbon beams with 
currents up to several tens  of $\mu$A.
Unlike the other sites mentioned in this section, Felsenkeller is a shallow underground laboratory, with a rock overburden of only 140 m.w.e.{\footnote{See \citet{2025EPJA...61...19B} for a more complete description of this laboratory.}
Such a depth is sufficient to shield all components of cosmic-ray induced radiation, except muons, for which further active shielding is generally required \cite{2019APh...112...24L}. Detailed studies of muon, neutron, and $\gamma$ backgrounds have been carried out \cite{2019APh...112...24L,2019EPJA...55..174S,2020PhRvD.101l3027G,2023APh...14802816T}, and the $^{12}$C($p,\gamma$)$^{13}$N reaction has been studied \cite[]{2023PhRvC.107f2801S}.
The further scientific program of Felsenkeller foresees the study of several reactions of astrophysical interest, including $^3$He($\alpha$,$\gamma$)$^7$Be and $^{12}$C($\alpha$,$\gamma$)$^{16}$O.

\subsection{Indirect methods} 

Indirect methods are also pivotal in nuclear astrophysics due to the challenges associated with directly measuring reactions at extremely low energies. They can have systematic uncertainties that are different from those of direct measurements, and they provide supplementary information that can constrain R-matrix and other models used in the extrapolation of data from direct measurements. For the present review, it is worth further exploring here the ANC (asymptotic normalization coefficient) and THM (Trojan Horse) methods. For a detailed description of the methods, refer to \cite{Tribble2014,Tumino2021}. 

The ANC method determines the zero-energy cross-section for radiative-capture reactions by exploiting their peripheral nature. This means the reaction depends mostly on the long-distance behavior of the wave function. The ANC is extracted from transfer reactions using the DWBA, with uncertainties primarily arising from the optical model description, still much smaller than those affecting spectroscopic factors. While the method is focused to zero energy, it complements direct measurements and can be applied to loosely bound nuclei. The method has been applied to the $^{3}$He($\alpha$,$\gamma$)$^{7}$Be (section VI.C). 

The THM is a powerful technique for indirectly measuring astrophysical S(E) factors of reactions involving charged particles. By studying a related reaction with a spectator particle, the THM allows for the extraction of the desired cross section without the need for extrapolation. The THM relies on the assumption of quasifree kinematics, where the spectator particle has a minimal effect on the reaction of interest. The method requires careful selection of beam energy and momentum transfer to ensure that the Coulomb barrier is overcome and the spectator particle remains relatively undisturbed. Since Solar Fusion II, significant advancements have been made, focused on improving the model's accuracy and assessing its systematic uncertainties \cite{Tribble2014,Tumino2021}. 
For reactions dominated by broad resonances, the modified R-matrix approach \cite{Lacognata2015,Trippella2017} allows for the incorporation of half-off-energy-shell and energy resolution effects within a well-established framework. This method enables multi-channel descriptions of reactions, such as in the $^{12}$C+$^{12}$C fusion studies \cite{Tumino2018}, and includes a DWBA-based normalization procedure that does not rely on direct data \cite{Lacognata2009}. The modified R-matrix framework has been applied to the $^{15}$N(p,$\alpha$)$^{12}$C (section XI.A.3), $^{19}$F(p,$\alpha$)$^{16}$O (section XI.C.4), and $^{23}$Na(p,$\alpha$)$^{20}$Ne (section XI.D.5) reactions.
For narrow resonance reactions, a simplified approach has been introduced \cite{Lacognata2022} to deduce resonance strengths. This method reduces systematic errors from normalization and theory to the percent level through multi-resonance normalization and covariance in error propagation. The narrow resonance approach has been used in analyzing the $^{17}$O(p,$\gamma$)$^{18}$F (section XI.B.3), $^{17}$O(p,$\alpha$)$^{14}$N (section XI.B.4), and $^{18}$O(p,$\alpha$)$^{15}$N (section XI.C.2) reactions.

\subsection{Plasma facilities}
\label{subsec:WG8:Plasma}

\subsubsection{NIF/OMEGA}
\label{subsec:WG8:NIF/OMEGA}
Several efforts have recently been devoted to studying nuclear reactions in a plasma environment reminiscent of stellar conditions \cite{2017PhPl...24d1407G} 
using the two large laser facilities: OMEGA at the Laboratory for Laser Energetics, University of Rochester \cite{1997OptCo.133..495B}, and the National Ignition Facility (NIF) at Lawrence Livermore National Laboratory \cite{Wonterghem16}. 
Both facilities use high-power lasers to implode spherical capsules containing reactants of interest to high temperatures (of order 1-20 keV) and densities (up to 10$^3$ g cm$^{-3}$ in the extreme case of implosions with a fuel ice layer \cite{2022PhRvL.129g5001A}). 
The OMEGA laser has the capability of delivering 30 kJ of laser energy divided between 60 laser beams to the target; the NIF can deliver up to 2.0 MJ using 192 laser beams, and can thus be used to implode larger amounts of materials to more extreme conditions compared to OMEGA. 
Nuclear experiments at these two facilities are enabled by an extensive suite of nuclear diagnostics, originally developed to do inertial confinement fusion experiments \cite{2018JPhG...45c3003C}. 
Initial results have been obtained on the \dpg reaction \cite{2020PhRvC.101d2802Z,2022FrP....10.4339M}, the $^3$He($^3$H, $\gamma$)$^6$Li reaction \cite{2016PhRvL.117c5002Z}, and the $^3$H($^3$H,2n)$\alpha$ reaction \cite{2017NatPh..13.1227C}.
Proton spectra have also been measured for the $^3$He($^3$He,2p)$\alpha$ reaction \cite{2017PhRvL.119v2701Z}. 
In addition to low-Z reaction nuclear S-factor studies, the NIF and OMEGA facilities are also promising for studies of plasma effects on nuclear reactions, including screening \cite{Aliotta22-FP,Casey23-FP}; for studies of charged-particle-induced reactions \cite{Wiescher22-FP}; and, thanks to the high neutron fluxes achievable (up to $5\times 10^{27}$ neutrons cm$^{-2}$ s$^{-1}$ \cite{2022PhRvL.129g5001A}), for studying reactions on excited states \cite{Thompson22-FP}. 
However, there are challenges remaining to be addressed in fully developing this new platform for nuclear experiments, including the impact on the results of rapid gradients in space and time \cite{Crilly22-FP}.

\subsubsection{PANDORA}
\label{subsec:WG8:PANDORA}
PANDORA (Fig.~\ref{fig:pandora}) is a device conceived for multidisciplinary studies, including many of  astrophysical interest \cite{2022Univ....8...80M}. 
Two of its main objectives are: (a) to perform the first measurements of $\beta$ decays in plasmas of astrophysical relevance, in order to verify the results obtained in storage rings with $^{187}$Re (a lifetime reduction by 9 orders of magnitude), and (b) to measure the opacities of plasmas of astrophysical interest (kilonova ejecta). 
PANDORA will mainly consist of three subsystems: 
\begin{itemize}
\item{an innovative superconducting magnetic plasma trap, able to produce and confine plasmas with electron-ion density up to 10$^{13}$ cm$^{-3}$ and electron temperature of $T_e \sim 0.1-30$~keV;}
\item{an advanced plasma multi-diagnostic system, consisting of a set of non-invasive diagnostic tools capable of operating simultaneously, for non-intrusive monitoring of the thermodynamic plasma properties and parameters;} 
\item{an array of 14 HPGe (High-purity Germanium) detectors for $\gamma$-ray spectroscopy, surrounding the plasma trap}.
\end{itemize}

\begin{figure}
    \centering
    \includegraphics[width=0.95\columnwidth]{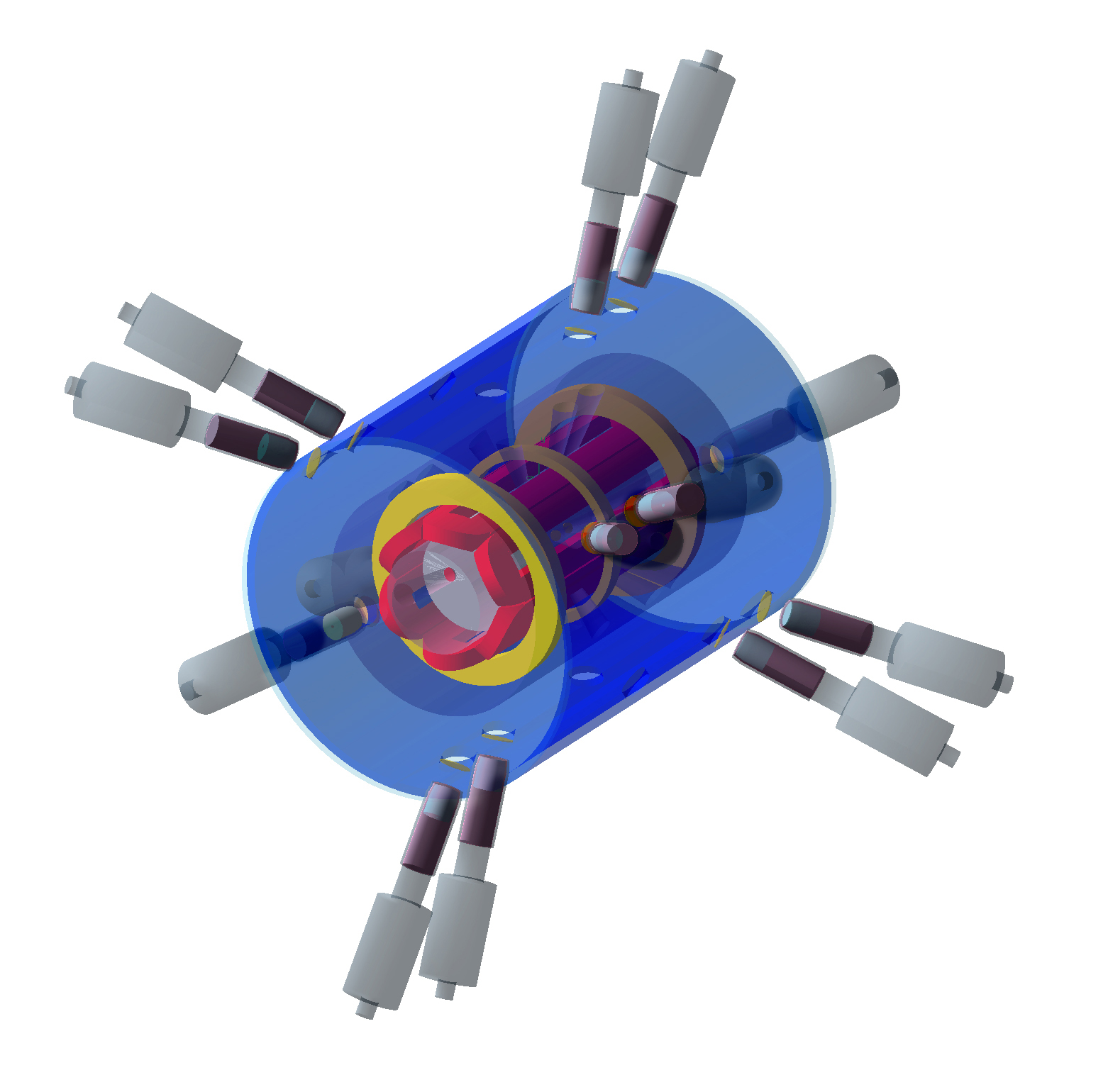}
    \caption{Schematic drawing of the PANDORA trap with holes to house the HPGe detectors. (credit: Domenico Santonocito).}
    \label{fig:pandora}
\end{figure}

The next development steps will involve:
preliminary numerical simulation studies to
assess the possibility of measuring opacities in astrophysical plasmas  (relevant for kilonova ejecta);
study of expected abundances and constraints in AGB stars for some (key) nuclides of interest; 
experimental investigation of magnetic confinement and turbulence in plasma, using an existing trap at ATOMKI-Debrecen; 
definition of the analysis algorithm for X-ray imaging and spatially resolved spectroscopy.

Initial physics cases to be investigated include $^{176}$Lu (a potential cosmo-chronometer), $^{134}$Cs (to reproduce adequately the observed abundance ratio of the two s-only isotopes, $^{134}$Ba and $^{136}$Ba) and $^{94}$Nb (to solve the puzzle about the exact contribution of s-processing to $^{94}$Mo).

\subsubsection{Sandia Z}
\label{subsec:WG8:SandiaZ}
The Sandia National Laboratories Z machine is the world's largest pulsed power accelerator \cite{sinars_pop2020}. 
It charges 22 MJ of electrical energy into a capacitor and then discharges it all at once (over approximately 100 ns) into targets ranging from millimeters to centimeters in size. 
The peak power reached is 80 terawatts, 15 times the steady-state electrical power generated by all of the world's power plants. 
By concentrating this massive power into small targets, the Z machine can convert them into a state of high energy density (HED), which refers to an extreme state of matter with a pressure exceeding $10^6$ times atmospheric pressure. 

The Z machine has been used for a wide range of HED science, including inertial confinement fusion, dynamic material properties, and laboratory astrophysics.
In particular, the Sandia Z machine can convert this electrical power into X-ray power using a scheme called the Z-pinch dynamic hohlraum \cite{rochau_zapp_2014}. 
This is the most energetic X-ray source on earth and has been used by academic collaborations such as the Wooton Center for Astrophysical Plasma Properties and the Center for Laboratory Astrophysics.
For the stellar opacity project (see Sect. \ref{subsec:WG9:Experimental}), they heat iron or oxygen to the conditions relevant to the base of the convection zone and measure their frequency-resolved opacities to test hypotheses for the solar opacity-abundance problem 
\cite{2008PhR...457..217B, bailey_experimental_2009, Bailey_Nature_2015, PhysRevLett.122.235001}. 
For the accretion disk projects, they heat silicon, iron, and neon photoionized plasmas to conditions similar to those at black hole accretion disks to experimentally test the validity of calculated atomic data, kinetics, and spectral formation \cite{loisel_benchmark_2017}. 
This is important for accurately interpreting high-resolution data from upcoming X-ray space telescopes such as XRISM. 
For the white dwarf photosphere project, they reproduce its photosphere conditions and validate spectroscopic methods used for understanding the age of the universe as well as constraining its evolutionary paths and supernova remnants \cite{falcon_laboratory_2015, montgomery_experimental_2015, schaeuble_apj2019, schaeuble_experimental_2021}. 
These experiments reach temperatures ranging from $10^4$ to $2 \times 10^6$ K and densities ranging from $10^{16}$ to $10^{23}$ electrons cm$^{-3}$ and are performed simultaneously by placing their samples at different locations from the Z-pinch dynamic hohlraum.

\subsection{Storage rings for nuclear astrophysics studies}
\label{subsec:WG8:Storage_rings}

\subsubsection{CRYRING} \label{subsec:WG8:CRYING}

A main limitation of experimental studies involving unstable nuclei arises from the difficulty of producing radioactive ion beams of adequate intensity and purity at the energies of interest for astrophysical applications.
Storage rings offer a key advantage over traditional ISOL and in-flight approaches as they allow for storing and re-circulating radioactive ions over and over again, thus allowing for multiple interactions of un-reacted beam particles with an in-ring target \cite{2020PrPNP.11503811S}.
Beam re-circulation significantly improves the quality of the beam as it results in orders-of-magnitude increased intensity (typical boosting factors of $\sim 10^5$), limited only by the duty cycle of the measurement \cite{2023NIMPA104868007B}, and in improved purity of the stored beam, as only beam particles with the right mass-to-charge ratio will survive in the beam orbit.

One of the main technical challenges of storage rings comes from the requirement of ultra high vacuum conditions ( $<10^{-10}$~mbar) necessary to guarantee that ions can be recirculated with minimal losses.
This, in turn, translates into a requirement for sufficiently thin targets (i.e. typically $\leq 10^{11}-10^{14}$ atoms cm$^{-2}$) to minimise beam ion losses through scattering or electron capture (see \cite{2023NIMPA104868007B} for further details).

A dedicated low-energy storage ring, CRYRING \cite{2016EPJST}, inherited from Stockholm University, was installed at GSI in 2016, as a Swedish in-kind contribution to the FAIR facility. The facility allows nuclear astrophysics
studies to be done with radioactive nuclei.
After its initial recommissioning \cite{herfurth2018commissioning}, CRYRING now serves as a low-energy extension for experimental storage ring beams, as well as a standalone machine with a local ion source \cite{2017HyInt.238...13G}. 
After in-flight production in the Fragment Recoil Separator at relativistic energies, rare ions can now be cooled, post-decelerated and stored in the full range down to about 100 keV/$u$. A study is underway to reach lower energy by means of a transverse low-energy beamline from the the FISIC project \cite{Schury_2020,Glorius2023}.
CRYRING is now equipped with an in-ring micro-droplet gas target, whose low density ($\leq 10^{14}$~atoms cm$^{-2}$) ensures minimal beam energy loss and straggling through the target, a unique advantage for charged particle spectroscopy at storage rings compared to standard techniques.  

\begin{figure}
    \centering
    \includegraphics[width=0.95\columnwidth]{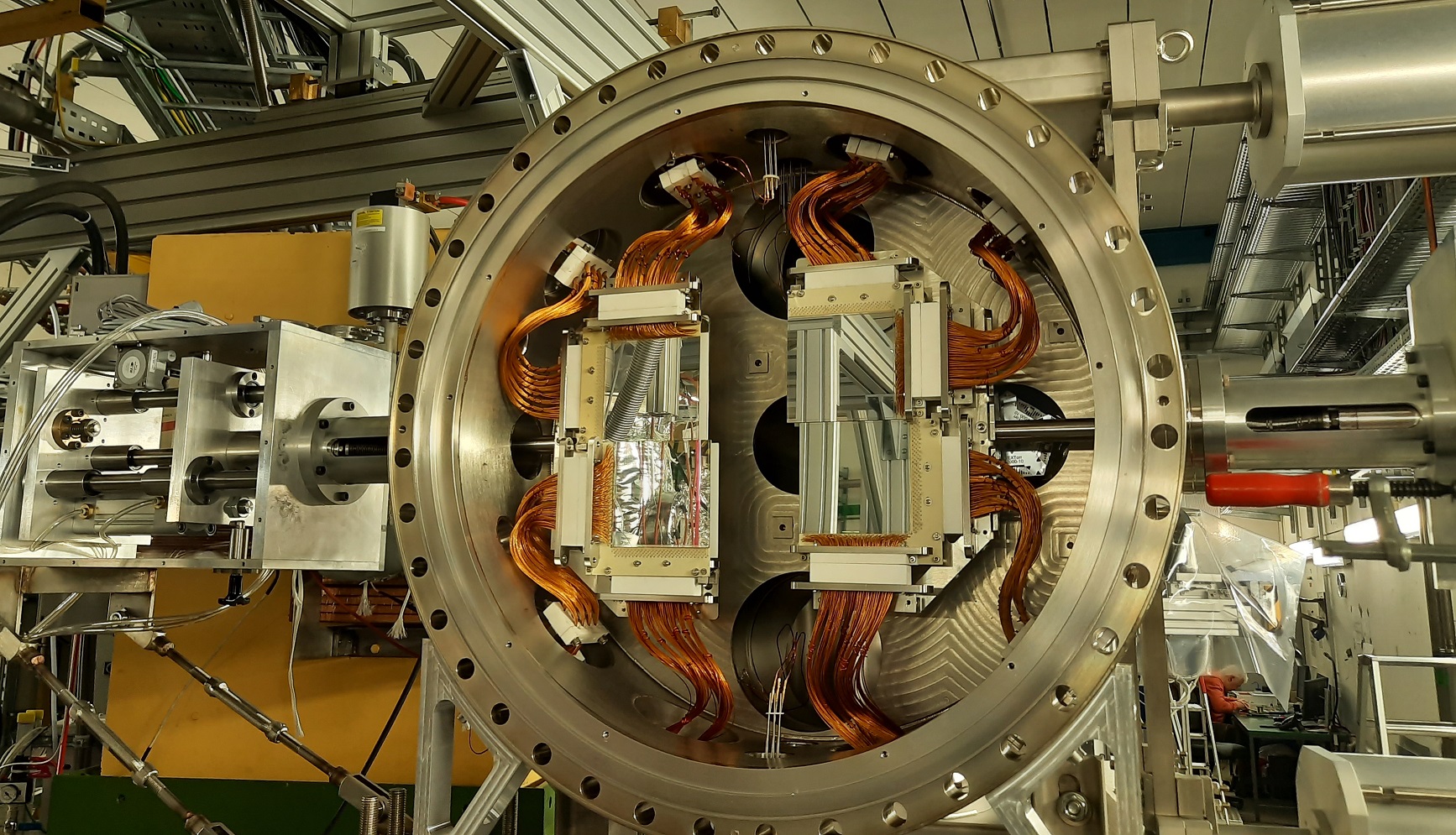}
    \caption{
    Photo of the CARME chamber recently installed at CRYRING. Four silicon strip detectors are visible at the centre of the chamber (credits: Carlo Bruno).}
    \label{fig:carme}
\end{figure}

In addition, a new detection chamber (Fig. \ref{fig:carme}), CARME (CRYRING Array for Reaction MEasurements), was recently installed at CRYRING and fully commissioned in early 2022.
Specifically designed to fulfil requirements of stored beam experiments and operating at a pressure of $10^{-12}$~mbar, CARME can host up to eight double-sided-silicon-strip detectors with excellent angular and energy resolution.

Finally, the upcoming installation of the FISIC transverse beam line \cite{2020JPhCS1412p2011S} will allow to intercept the beam stored in CRYRING, thus opening up unprecedented opportunities for crossed-beam experiments. 
An exciting application may be the study of nuclear reactions between ion beams, i.e. unaffected by the electron screening, directly at the energies of astrophysical interest.
Initial proposals for the study of nuclear astrophysics reactions have already been approved.


\section{Closing remarks}
\label{sec:Summary}

\subsection{Summary}
\label{sec:Summary:Summary}

This review summarizes the significant progress that has been made over the last decade in understanding the nuclear reactions that govern energy production in solar and stellar hydrogen burning.  It also describes some of the plasma and atomic physics that effects the solar environment in which these reactions take place, as well as the diagnostics tools -- solar neutrinos and helioseismolgy -- by which we can probe that environment. 

For $S_{11}$ and $S_{12}$, cross sections have been derived from first principles, using both potential theory and effective field theory. In the case of $S_{12}$, new experimental data agree well with the theoretical predictions. $S_\mathrm{hep}$, which governs a minor branch
of solar fusion, must also be taken from theory.

The S-factors for other pp-chain reactions are based on laboratory measurements. In some cases, including notably $S_{17}$, most of the studies reviewed here are relatively far from the solar Gamow peak energies, underlining the need for theoretical and experimental tools to bridge the gap. 

The nuclear physics of the pp chain remains a significant uncertainty in SSM predictions of the individual neutrino fluxes - pp, pep, hep, $^7$Be,
$^8$B, and CNO.
The nuclear errors are comparable to, and in the case of the hep neutrinos greatly exceed, the ``environmental'' errors that are generated by
other uncertainties in the SSM \cite{annurev:/content/journals/10.1146/annurev-nucl-011921-061243}.  

For several reactions including $S_{33}$ and $S_{1 \, 14}$, the SF~III recommended values are less precise than those of \citetalias{2011RvMP...83..195A}. This fact should not be taken as a sign of regression, but instead as indicative of a more cautious treatment of uncertainties. In some of the error
estimates made in this review, we have been mindful of past lessons on the potential impact of unidentified systematics.

The status of our understanding of radiative opacities has been reviewed here, the first time the subject has been included in the Solar Fusion series. The
debate generated by the solar composition problem has brought renewed attention to the complicated interplay between opacities and composition in the SSM. Given new experimental opportunities, especially luminous X-ray sources coupled with plasma targets, there is the expectation of rapid progress in this field over the next decade.

There has been only very limited progress on electron screening since \citetalias{2011RvMP...83..195A}. This field has important practical implications not only for the Sun but also for laboratory experiments, and may merit additional attention in the coming decade. 
Progress may depend on achieving a better understanding of low-energy stopping powers.

\subsection{Recommended values for S-factors and their derivatives}
\label{sec:Summary:RecommendedValues}

The SF~III recommendations for nuclear reaction S-factors and their derivatives are presented in Table \ref{tab:Sfactors_summary} of the Introduction. In the sections that follow the Introduction, we presented the details leading to these results and identified opportunities for future work, which are summarized directly below.

\subsection{General recommendations for future work}
\label{sec:Summary:Recommendations}

For all of the solar fusion reactions, obtaining high-precision experimental data in Gamow window remains an elusive goal, one that may not be  reached in the foreseeable future. However, such direct measures are just one of many avenues for improving our understanding of the nuclear physics of the Sun.  
The recommendations below are in most cases summaries of more detailed discussions presented in this review; see
the relevant sections for additional details.

\begin{enumerate}
\item For all of the experimentally accessible pp-chain reactions and for $S_{1 \, 14}$ which determines the CNO cycle rate, there is a need for at least one experimental data set spanning the entire energy range, encompassing the more limited data sets that currently exist.  This will provide an important crosscheck on the normalizations of those data sets. Experimentalists can now tackle this challenge due to the new generation of underground accelerators that have the dynamic range needed for such studies. 
\item We recommended continued effort on reaction theory relevant to pp-chain reactions. The needs go beyond rates to include
both astrophysical (e.g., precise neutrino spectra) and laboratory (e.g., $\gamma$-ray angular distributions) observables. 
\item We urge the community to set an ambitious goal for the precision of future S-factor measurements: reduce the nuclear physics uncertainties 
to a factor of two below current SSM ``environmental'' uncertainties, so that they are no longer a significant contributor to SSM neutrino flux uncertainties. With the exception of the hep neutrinos,
this goal could be reached by a future reduction of a factor $\sim$ two, typically, in the individual S-factor uncertainties \cite{annurev:/content/journals/10.1146/annurev-nucl-011921-061243}.
As solar neutrino data are an important input into global neutrino oscillation analyses, nuclear uncertainties will continue to feed into that program until this goal is achieved.  Improvements in $S_{1 \, 14}$ and other CNO cycle S-factors are needed, as the current 8.4\% uncertainty would be a limiting factor in
extracting the solar core's metallicity from a future large-volume CN neutrino experiment.  The progress reported here on $S_{17}$, with the uncertainty reduced significantly from \citetalias{2011RvMP...83..195A} to 3.4\%, should be continued, given the current and future $^8$B neutrino programs of Super-K
and Hyper-K: these neutrinos are our best solar core thermometer.  Finally, ongoing Super-K efforts to detect hep neutrinos provide strong motivation for improving estimates of $S_\mathrm{hep}$ and the associated neutrino spectrum. (Uncertainties in the high-energy tail of the $^8$B spectrum must also be reduced, as discussed in Section \ref{sec:Observations}.)
\item New experimental methods are becoming available that could help us better understand electron screening in terrestrial and solar reactions.  Plasma conditions resembling those of the Sun can now be produced in the laboratory.  Reactions can be studied in rings, using ions stripped of their atomic shells.  The level of theory activity in this field has declined over the past decade, so we hope experimental progress will help renew interest.
\item In order to break the degeneracy between solar abundances and solar opacities, efforts should be made to improve our understanding of radiative opacities for the Sun's principal metals.  A high priority is the resolution of the current discrepancy between measured and modeled iron opacities. Solar physicists need access to state-of-the-art, open-source opacity codes.
\item We recommend that resources be available to extend the use of current and next-generation dark matter detectors to solar neutrino detection.  This includes designing new detectors so that they are highly capable for both applications. Some of the high-priority
goals of solar neutrino physics -- such as a 1\% measurement of the pp neutrinos to check the luminosity constraint -- might be achieved with a dual-purpose detector.
\item The Solar Fusion program to periodically review the nuclear physics of the Sun and other hydrogen-burning stars should be continued, with the separation
between studies being no more than ten years.  The format should continue to be open, giving all researchers working in the field an
opportunity to contribute, and broadly international.
\item Future Solar Fusion studies should continue to strengthen the connections between this community
and others interested in main sequence stars.  SF~III has placed increased emphasis on hydrogen burning at higher
temperatures, plasma
physics, opacities, and asteroseimology.  However, with observations now being made of solar-like stars in their formation stages, we anticipate additional connections
developing, well beyond the limits of this study.
\end{enumerate}

\subsection{Outlook}
\label{sec:Summary:Outlook}

Six decades after the initial observations of solar neutrinos by Ray Davis \cite{2003RvMP...75..985D}, significant advances have been made in our understanding of the SSM and indeed of hydrogen-burning stars in general. The driver of this progress has been experiment: the observations made of all of the principal solar neutrino sources; our improved understanding of the flavor physics of those neutrinos; the advances made in helioseismology, a second quantitative probe of the solar interior; and our improved knowledge of the input microphysics of the SSM, including nuclear cross sections and opacities.    Despite this progress, improvements are still needed.  
In several cases the dominant uncertainty in our solar neutrino
flux predictions stems from the limited precision
of the nuclear S-factors. Similarly, uncertainties remain in our understanding of radiative transport in the Sun, including the continuing debate about the Sun's primordial composition. As changes in composition often can be mimicked by changes in radiative opacities, our imperfect knowledge of the latter has slowed resolution of the solar composition problem.

The prospects for substantial progress in the next decade are bright. Whereas at the time of \citetalias{2011RvMP...83..195A}, there was only one underground ion accelerator worldwide, now there are five, opening up possibilities for independently cross-checking important nuclear data. The low background that can be achieved underground has led to the first measurements of reactions in the Gamow window. Other new experimental capabilities have come on line that are advancing our understanding of solar atomic physics, including low-energy storage rings and new plasma facilities, the latter made possible in part by advances in high-power lasers. The field is approaching the point where opacity measurements can be made under conditions closely approximating those of the solar interior. 

Theory has progressed as well, including advances in neutrino flavor physics that have made neutrinos once again a precise probe of solar physics.  Techniques with a firmer footing in first principles, such as effective field theory and lattice QCD, are being employed in the extraction of pp chain S-factors. Conventional approaches like the R-matrix can now be applied to complex nuclear systems where the inclusion of multiple reaction channels is important.

On the observational side, several very large neutrino detectors are under construction, including Hyper-Kamiokande, JUNO, and DUNE. Certain direct-detection dark matter experiments have reached a sensitivity where nuclear recoils due to the coherent solar neutrino scattering have become a background, leading to discussions of how such detectors might be optimized for solar neutrino detection \cite{PhysRevD.97.095029}. This underscores the remarkable radiopurity of these underground detectors. Finally, the helioseismic studies that proved so important in establishing the credibility of the SSM are now being extended to other stars -- a major reason SF~III was undertaken at this time.  Asteroseismic studies have the potential to tell us what is special and what is common about the Sun as a star.

Given current excitement about the emerging field of
multi-messenger astrophysics, it is perhaps fitting to remember
that the Sun was the prototype for this field.  An array
of physical measurements of the Sun -- its mass, radius, luminosity, abundances -- were combined with neutrino
flux and helioseismic measurements to constrain and then test the SSM.  The result was the discovery of neutrino mass
and flavor mixing.

This success was a multi-disciplinary effort that has not
yet run its course.  We still have uncertainties in our
characterization of the Sun's nuclear, atomic, and
weak interaction physics that, if improved, would make the
Sun an even more powerful laboratory.  As our best
known and nearest star, the Sun is the test of our 
understanding of the structure and evolution of other
main-sequence stars.  If the past is any indicator
of the future,
the efforts we make over the next
decade to make the Sun an even more precisely calibrated reference will yield a rich return in our understanding of physics.

\section*{List of Symbols and Abbreviations}
\begin{description}[itemsep=0pt]\setlength\itemsep{-0.5ex}
     \item[ANC] asymptotic normalization coefficient.
     \item[BBN] Big Bang nucleosynthesis.
     \item[CME] common-mode error.
     \item[EFT] effective field theory, and variants: chiral ($\chi$EFT), pion-less ($\slashed{\pi}$EFT), hybrid (EFT*), etc.
     \item[LO, NLO, NNLO] leading order, next to (next to) leading order, etc.
     \item[SSM] Standard Solar Model.
 \end{description}

\section*{Acknowledgements}
We gratefully acknowledge discussions with many colleagues.  The help they
provided contributed significantly to this review.

We thank the following institutions for their support of the SF~III effort:

\begin{itemize}[leftmargin=.4cm]
    \item
    Physics Frontier Center N3AS (Network for Neutrinos, Nuclear Astrophysics, and Symmetries), UC Berkeley, 
    host of the SF~III workshop.  N3AS is supported by the U.S. National Science Foundation under cooperative agreement 2020275.
    
    \item
    Institute for Nuclear Theory, University of Washington. The INT is supported by the US Department of Energy
    under grant DE-FG02-00ER41132.
    
    \item
    European Union, ``Chemical Elements as Tracers of the Evolution of the Cosmos -- Infrastructures for Nuclear Astrophysics'' (ChETEC-INFRA), HORIZON2020 project number 101008324.
    
    \item
    The Heising Simons Foundation, grant 2017-228.
    
    \end{itemize}
    
\noindent
The authors of SF~III acknowledge the following support:
\begin{itemize}[leftmargin=.4cm]
    \item
    U.S. National Science Foundation, through grants 
    \begin{itemize}[leftmargin=.4cm]
        \item PHY-1430152 (Joint Institute for Astrophysics) 
        and OISE-1927130 (IReNA), Michigan State University 
        \item PHY-1913620 and PHY-2209184, Mississippi State University
        \item PHY-2011890 and PHY-2310059, Notre Dame University 
        \item PHY-2111426, University of Tennessee
        \item PHY-2108339, University of Wisconsin
    \end{itemize}
  
    
    
    \item
    U.S. Department of Energy, through contracts
    \begin{itemize}[leftmargin=.4cm]
    \item DE-AC02-07CH11359, Fermi National Accelerator Laboratory. Fermilab is operated by Fermi Research Alliance LLC for the Office of Science.
    \item DE-AC02-05CH11231, Lawrence Berkeley National Laboratory. LBNL is operated by the University of California for the Office of Science.
    \item DE-AC52-07NA27344, Lawrence Livermore National Laboratory.  LLNL is operated by the Lawrence Livermore National Security LLC for NNSA.
    \item 89233218NCA000001, Los Alamos National Laboratory.  LANL is operated by Triad National Security LLC for NNSA.
    \item DE-NA0003525, Sandia National Laboratories.  Sandia National Laboratories are managed and operated by NTESS for the NNSA.
    \item DE-AC05-00OR22725, Oak Ridge National Laboratory. ORNL is operated by UT-Battelle, LLC for the Office of Science.W
    \end{itemize}
    

    \item U.S. Department of Energy, through  awards
    \begin{itemize}[leftmargin=.4cm]
    \item DE-FG02-00ER41138 and DE-SC0004658, University of California, Berkeley
    \item DE-FG02-93ER40789, Colorado School of Mines
    \item DE-FG02-97ER41033 (TUNL), Duke University 
    \item DE-SC0013617 (FRIB Theory Alliance), Michigan State University
    \item DE-FG02-97ER41041, University of North Carolina 
    \item DE-FG02-88ER40387, DE-NA0003883, and DE-NA0004065, Ohio University
    \item DE-FG02-08ER41533, Texas A\&M University, Commerce    
    \item DE-FG02-97ER41020 (CENPA), University of Washington
    \item DE-SC0019465, University of Wisconsin
    \item DE-SC0023692, Iowa State University
    \item DE-SC0019257, San Diego State University
    \end{itemize}


    
    


    \item 
    Helmholtz Research Academy Hesse for FAIR, Germany

    \item 
    Hungarian Scientific Research Fund grant FK134845, National Research, Development, and Innovation Office
    
    \item
    Israel Science Foundation grant 889/23

    \item  Italian Ministero dell'Universit\`a e della Ricerca (MUR) grants 2022E2J4RK (program PRIN 2022) and 2017W4HA7S (program PRIN 2017)

    \item
    The Natural Sciences and Engineering Research Council of Canada (NSERC)
  
    \item
    Spanish Unidad de Excelencia Mar\'{i}a de Maeztu, grant CEX2020-001058-M, Generalitat de Catalunya, grant 2021-SGR-1526, and MICINN, grant PID2023-149918NB-I00

    \item
    U.K. Science and Technology Facilities Council, grant ST/V001051/1

\end{itemize}
    
This paper describes objective technical results and analysis. Any subjective views or opinions that might be expressed in the paper do not necessarily represent the views of the U.S. Department of Energy, United States Government, or any other agency listed above.

For the purpose of open access, the authors have applied a Creative Commons Attribution (CC BY) licence to any Author Accepted Manuscript version arising from this submission.\\

\begin{appendix}

 \section{Gallium Neutrino Source Cross Sections}

Here we describe in more detail the constraint that the electron capture rate for $^{71}$Ge 
 places on the $^{51}$Cr and $^{37}$Ar neutrino source cross sections for $^{71}$Ga,  expanding on the discussion of Section \ref{sec:galliumanomaly}.  We discuss both the early treatment of Bahcall and the more recent analysis of \citet{PhysRevC.108.035502,ELLIOTT2024104082},
 performed in support of the BEST experiment
 
 Following Bahcall~\cite{1978RvMP...50..881B}, the electron (EC) rate can be written in terms of a dimensionless phase space factor $f_{\rm EC}$:
\begin{eqnarray}
f_{\rm EC}&=&2\pi^2\left(\frac{\hbar}{mc}\right)^3\left(\frac{q}{mc^2}\right)^2|\psi(R)|^2,
\end{eqnarray}
where $q$ is the energy available to the neutrino and we have omitted atomic shell effects for brevity. The last factor is the electron density at the nucleus within the radius $R$.  Multiplying this by the experimentally determined half-life $t_{1/2}$ yields the traditional $ft$ value needed to calculate the neutrino cross section. \citet{1978RvMP...50..881B}  defines a characteristic scale $\sigma_0$ for neutrino cross sections on $^{71}$Ga in the form
\begin{eqnarray}
    \sigma_0&=&1.2429 \times 10^{-47}\left(\sum_i q_i^2g_i^2\right)^{-1} \\
    &=& 8.611 \times 10^{-46}  {\rm \  cm}^2 \label{eq:bahcallconstant}
\end{eqnarray}
where $g_i=\psi(R)$.  \citet{1997PhRvC..56.3391B}  reports the number given above and states that ``it is about 0.5\% less than the value $8.8012 \times 10^{-46}$ I have used since 1984.''  Since these numbers differ by 2.2\%, there is clearly an error in one of the 3 numbers.  

 Recently \citet{PhysRevC.108.035502,ELLIOTT2024104082} have recast the derivations with fundamental constants replacing numerical values and correction terms explicitly stated. They conclude that the ground-state cross section is 2.5\% lower, and the excited-state contributions somewhat larger, than values used in the anomaly analysis. The adjusted net anomaly is minimally reduced.  
 
 They find the following expression for the neutrino cross section on $^{71}$Ga leading to the ground state (gs) of $^{71}$Ge:
\begin{widetext}
\begin{eqnarray}
 \sigma_{\mathrm{gs}}  &=& \frac{G_F^2 \cos^2{\theta_C}~ g_A^2}{  \pi}  ~ p_e E_e~{\cal F}(Z_f,E_e)  ~ {\mathrm{B}}^{(\nu,e)}_\mathrm{GT}(\mathrm{gs}) ~{[1+g_{v,b(\nu,e)}]}~[ 1 + \epsilon_q], 
 \label{eq:sigmaor}
 \end{eqnarray}
 where $E_e=q-Q_{\rm EC}+m_e-\Delta_0$ and $\Delta_0\simeq90$ eV is a small correction introduced by Bahcall for the energy lost to electronic rearrangement in the charge-changing reaction.  The Gamow-Teller matrix element ${\mathrm{B}}^{(\nu,e)}_\mathrm{GT}(\mathrm{gs})$ is determined from the measured half-life $\tau_{1/2}$ of the EC decay of $^{71}$Ge:
 \begin{equation}
 \omega = \frac{\mathrm{ln}[2]}{\tau_{1/2} } = \frac{G_F^2 \cos^2{\theta_C}~ g_A^2}{2 \pi} ~|\phi_{1s}|_\mathrm{avg}^2 ~ q_{1s}^2~ { \textstyle \left[2(1+\epsilon_o^{1s}) (1+ \frac{P_L+P_M}{ P_K} )\right]}~ [2 ~\mathrm{B^{(\nu,e)}_{GT}}(\mathrm{gs})]~[1+g_{v,b(\mathrm{EC}})] ~[ 1 + \epsilon_q].  
 \label{eq:sigmao}
 \end{equation}
 \end{widetext}
In this expression, $\epsilon_o^{1s}\simeq0.013$ is a correction for overlap and exchange to account for the differences in the wave functions of the initial and final atomic states, $g_{v,b}$ is a radiative correction, and $\epsilon_q\simeq5\times10^{-4}$ is a correction for the weak magnetism contribution to the transition probability.  The factor 2 with $\mathrm{B^{(\nu,e)}_{GT}}(\mathrm{gs})$ is a spin statistical factor.  Hence,
\begin{widetext}
\begin{eqnarray}
\sigma_{\mathrm{gs}} ~ &=&  \omega ~ \frac{~ p_e E_e~{\cal F}(Z_f,E_e)}{|\phi_{1s}|_\mathrm{avg}^2~q_{1s}^{2}}~[1+g_{v,b(\mathrm{EC}})]^{-1}~{\textstyle\left[2(1+\epsilon_o^{1s}) (1+ \frac{P_L+P_M}{ P_K} )\right]^{-1}} \\
 &\equiv& \sigma_0  ~ \frac{p_e E_e~{\cal F}(Z_f,E_e)}{m_e^{2} c^{3}}\frac{1}{2\pi\alpha Z}.
 \label{eq:sigma2}
 \end{eqnarray}
The last line extracts the Bahcall cross-section scale $\sigma_0$ and inserts the dimensionful parameters needed to scale the energy and momentum.  The atomic factors of \citet{PhysRevC.108.035502} are equivalent to those of Bahcall:
\begin{eqnarray}
     \sum_i |\phi_i |_\mathrm{avg}^2 q_{i}^2&=& 0.01443~ \frac{(m_ec^2)^5}{4\pi(\hbar c)^3} \label{eq:bahcalldensity}
\end{eqnarray}
 Unfortunately, none of the three most advanced calculations referenced in \cite{1997PhRvC..56.3391B} of the densities $g_i$  has been published so the numerical factor 0.01443 is taken from the ratio in Eq.~\ref{eq:bahcallconstant}.  (The computer code GRASP \cite{DYALL1989425} may still be available.) Rather than use the theoretical electron density of each subshell at the nucleus, Elliott \etal\  use the experimentally measured subshell ratios, which has the advantage that only a single overlap and exchange correction and a single energy are needed,  for the $1s$ state. As Bahcall has stated, overlap and exchange have little effect on the {\em total} EC rate, so Elliott \etal\ write Eq.~\ref{eq:bahcalldensity} as follows, with the inclusion of a small Q-value update,
\begin{eqnarray}
     \sum_i |\phi_i |_\mathrm{avg}^2 q_{i}^2(1+\epsilon_0^i)&=&|\phi_{1s}|_\mathrm{avg}^2~q_{1s}^{2}{\textstyle\left[(1+\epsilon_o^{1s}) (1+ \frac{P_L+P_M}{ P_K} )\right]} \label{eq:haxtondensity} 
\end{eqnarray}
The resulting expression is,
\begin{eqnarray}
    \sigma_0~&=&\frac{8\pi^2\ln{(2)}\alpha Z}{|\phi_{1s}|_\mathrm{avg}^2~q_{1s}^{2}~t_{1/2}}~[1+g_{v,b(\mathrm{EC}})]^{-1}~{\textstyle\left[2(1+\epsilon_o^{1s}) (1+ \frac{P_L+P_M}{ P_K} )\right]^{-1}}. \\
    &=&~\left(\frac{\hbar c}{m_ec^2}\right)^3\frac{4\pi^2\ln{(2)}\alpha Z}{0.01440~c~t_{1/2}}\left(\frac{0.2221\rm\ MeV}{q_{1s}}\right)^2~[1+g_{v,b(\mathrm{EC}})]^{-1} \\
    &=&~8.63 \times 10^{-46}~ [1+g_{v,b(\mathrm{EC}})]^{-1} {\rm\ cm}^2,
\end{eqnarray}
 \end{widetext}
 which is only 0.3\% larger than the value given in \citet{1997PhRvC..56.3391B}. 
 
 The above cross-section expressions give a ground-state cross section 2.6\% lower than in the cross section table, Table II in \citet{1997PhRvC..56.3391B}.  Whether this extends to the solar neutrino rates has not yet been evaluated, but in any scenario some reconsideration of the low-energy solar neutrino data may be required once a resolution to the Ga anomaly has emerged.

\section{Bayesian Methods}
\label{AppendixWG4}

The treatment of uncertainties when dealing with multiple data sets was addressed in the appendix of \citetalias{2011RvMP...83..195A}, including the Inflation Factor Method used by the Particle Data Group that was adopted by most of the \citetalias{2011RvMP...83..195A} 
working groups. Readers are directed there.  We now extend that discussion to include the Bayesian methods used by several of the SF~III work groups.  For erxample, in the analyisis of the \dpg reaction presented in Section~\ref{subsec:WG4:Reviews}, both Bayesian analysis and Bayes model averaging were utilized.

Consider an optimization that utilizes Gaussian distributed priors in a constrained Bayesian analysis.  
With such priors, maximizing the likelihood amounts to minimizing an augmented $\chi^2$
\begin{equation}\label{eq:chisq_aug}
\chi^2_{\rm aug} \equiv \sum_{D, i} 
    \frac{[y_{D_{i}} - f_D\, S(E;\lambda)]^2}
    {\sigma_{D_{i}}^2 + (\sigma_{D_{i}}^{\rm ext})^2 }
    +\sum_p \left( \frac{\lambda_p - \tilde{\mu}_p}{\tilde{\sigma}_p} \right)^2 .
\end{equation}
The first term in the above equation is the standard $\chi^2$ and the second term derives from the Bayesian constraint on the parameters ($\lambda_p$) with prior mean ($\tilde{\mu}_p$) and width ($\tilde{\sigma}_p$), which are chosen with some prior knowledge or can be optimized as described below.
Of course, care must be taken when choosing values for the priors ($\tilde{\mu}_p$ and $\tilde{\sigma}_p$).  In the absence of prior information, it is common to set $\tilde{\mu}_p=0$.  Similarly, one should not make $\tilde{\sigma}_p$ too small, unless one has prior knowledge to do so, else this will bias the determination of $\lambda_p$.

In the first term in Eq.~(\ref{eq:chisq_aug}), the double sum runs over the datasets ($D$) and individual results from each dataset ($D_i$), with the mean value and stochastic uncertainty of each data point given by $y_{D_{i}}$ and $\sigma_{D_{i}}$, respectively.

The theoretical model describing the S-factor data, $S(E; \lambda)$, is a function of the energy ($E$) and a set of parameters ($\lambda$) that must be determined. The quoted systematic uncertainties are parameterized by the normalization factors $f_D$, with a prior of unit normalization and a width characterized by the quoted systematic uncertainty.  
It is straightforward within a Bayesian framework
to utilize distribution functions for $f_D$ that are not Gaussian, such as a log-normal or other distributions. 
The parameters $f_D$ represent a normalization of the model function for a given dataset, which, viewed from a non-Bayesian perspective, can also be interpreted as normalization factors that must be applied to the data to match the ``true'' underlying distribution.

Finally, $\sigma_{D_{i}}^{\rm ext}$ are unknown \textit{extrinsic uncertainties}~\cite{2019PhRvC..99a4619D}.
If one assumes that a smooth function of energy can accurately describe the data, then the data in a given experimental set may scatter about this presumed ``true'' value by more than is reflected by the quoted statistical uncertainties. It was suggested in \citet{2019PhRvC..99a4619D} that this extra scatter might be explained by some additional source of statistical uncertainty unbeknownst to the experimenter, 
or by some unknown systematic uncertainty that is different for each data point in the same data set, as opposed to a correlated systematic that affects all data points similarly.
In either case, this extrinsic uncertainty can be accommodated in a Bayesian analysis framework by adding the additional uncertainty as a normal-distributed source of noise with a width that is constrained by the data.  
In \citet{2019PhRvC..99a4619D} and \citet{2021ApJ...923...49M}, $\sigma_{D_{i}}^{\rm ext.}$ was added as an absolute uncertainty, independent of the energy. For some data,
\citet{2022PhRvC.105a4625O} suggested adopting instead a relative uncertainty, such that the scale of the extrinsic fluctuations is proportional to the mean value of $S(E)$. This strategy of adding extra extrinsic uncertainty to the data sets can be viewed as an alternative to that of inflating the quoted statistical uncertainties by $\sqrt{\chi^2_\nu}$, which is often used for seemingly incompatible data sets~\cite{2022PTEP.2022h3C01W}, where $\chi^2_\nu$ is the $\chi^2$ per degree of freedom.
An advantage of the extrinsic uncertainty method is that it uses the observed scatter within a given data set as a measure of the possible size of unreported uncertainties, as opposed to uniformly increasing the uncertainty in all data sets by the same relative amount.

One can further perform a Bayes model averaging, which we implement and describe here.
For a given model, after optimizing the posterior parameter distributions, the Bayes factor (BF) is proportional to the probability of the model given the data \cite{2017bmad.book.....H}.
Therefore, for a fixed data set, the BF can be used as a relative probability of each model, thus enabling a weighted model-averaging procedure.
If we assume a uniform likelihood for each model, the expectation value and variance of a quantity $Y$ is given by
\begin{align}
{\rm E}[Y] &= \sum_k {\rm E}[Y|M_k] P[M_k|D]
\\
\label{eq:BMA_Var}
{\rm Var}[Y] &=  
    \sum_k {\rm Var}[Y|M_k] P[M_k|D]
\nonumber\\&\phantom{=}
    +\sum_k {\rm E}^2[Y|M_k] P[M_k|D] - {\rm E}^2[Y]\, ,
\end{align}
where ${\rm E}[Y|M_k]$ denotes the expectation of $Y$ given the model $M_k$ and $P[M_k|D]$ denotes the probability of the model $M_k$ given the data ($D$), which is given by
\begin{equation}\label{eq:P(M|D)}
P[M_k|D] = \frac{{\rm BF}_{M_k}}{\sum_l {\rm BF}_{M_l}}\, ,
\end{equation}
with ${\rm BF}_{M_l}$ the Bayes factor of model $l$.
Similarly, for a given model, in the absence of prior information on the size of an unknown parameter, the optimal width of its prior can be estimated by finding the value of $\tilde{\sigma}_p$ that maximizes the BF, which typically provides a reasonable approximation to marginalizing over the prior width.
\newline

\end{appendix}

\setcounter{secnumdepth}{0}
\addcontentsline{toc}{section}{References}
\bibliography{SF3}


\ifarXiv
    \clearpage
    \phantomsection
    \setcounter{secnumdepth}{0}
    \addcontentsline{toc}{section}{Supplementary material}
    \foreach \x in {1,...,\numbersupplementpages}
    {
        \clearpage
        \includepdf[pages=\x,pagecommand=\thispagestyle{empty}]{\supplementfilename}
    }
\fi


\end{document}